\documentclass[12pt]{article}
\usepackage[margin = 1in]{geometry}
\usepackage{amsmath,amssymb}
\usepackage{graphicx,psfrag,epsf}
\usepackage{enumerate}
\usepackage{natbib}
\usepackage{url} % not crucial - just used below for the URL 
\usepackage{algorithm2e}

\usepackage{xcolor}
\usepackage{subfig}

\usepackage{etoolbox}
\newtoggle{preprint}

\togglefalse{preprint} % preprint

\newcommand{\blind}{1}

\newcommand{\thetitle}{Exact Gradient Evaluation for Adaptive Quadrature Approximate Marginal Likelihood in Mixed Models for Grouped Data}

\newcommand{\figurepath}{figures}

\newcommand{\lcrx}[4][{-1}]{ 
  \IfEq{#1}{-1}{\left #2 {{{{#3}}}} \right #4}{
    \IfEq{#1}{0}{#2 {{{{#3}}}} #4}{
  \IfEq{#1}{1}{\bigl #2 {{{{#3}}}} \bigr #4}{
  \IfEq{#1}{2}{\Bigl #2 {{{{#3}}}} \Bigr #4}{
  \IfEq{#1}{3}{\biggl #2 {{{{#3}}}} \biggr #4}{
  \IfEq{#1}{4}{\Biggl #2 {{{{#3}}}} \Biggr #4}{
    \GenericWarning{"4th argument to lcrx must be -1, 0, 1, 2, 3, or 4"}
    }}}}}}} % specify size with {-1,...4} as optional argument

\newcommand{\Reals}{\mathbb{R}}
\newcommand{\Nats}{\mathbb{N}}
\newcommand{\Tr}{^{\scriptscriptstyle\text{T}}} % transpose
\newcommand{\dee}{\mathrm{d}} % for integrals \int f(x) \dee x
\DeclareMathOperator*{\argmax}{\arg\max} % argmax
\newcommand{\abs}[1]{\vert {#1} \vert }
\newcommand{\inv}{^{-1}}
\newcommand{\one}{\mathbf{1}}

\newcommand{\zero}{\mathbf{0}}

\newcommand{\EE}{\mathbb{E}}

\newcommand{\bias}{\text{Bias}}
\newcommand{\relbias}{\text{Rel. Bias}}

\newcommand{\setdelim}{\ \vert \ }
\newcommand{\AQ}{{\texttt{\tiny \upshape AQ}}}

\newcommand{\mb}[1]{\boldsymbol{#1}}

\newcommand{\hessidx}{H}
\newcommand{\hess}{\mb{\hessidx}}
\newcommand{\hessDidx}{M}
\newcommand{\hessD}{\mb{\hessDidx}}
\newcommand{\hessDall}{\hessD(\params,\re;\obsall)}
\newcommand{\jointhess}{\widehat{\hess}(\params)}
\newcommand{\jointhessi}[1]{\widehat{\hess}_{#1}(\params,\condmodei{#1}(\params))}

\newcommand{\cholscalar}{L}
\newcommand{\chol}{\mb{\cholscalar}}
\newcommand{\D}{\mb{D}}
\newcommand{\cholD}{\mb{F}}
\newcommand{\cholinvvec}{\mb{v}}
\newcommand{\jointcholbase}{\widehat{\chol}}
\newcommand{\jointchol}{\widehat{\chol}(\params)}
\newcommand{\jointcholi}[1]{\widehat{\chol}_{#1}(\params,\condmodei{#1}(\params))}

\newcommand{\varparam}{\mb{\delta}}
\newcommand{\corparam}{\mb{\phi}}
\newcommand{\varparamapprox}{\widetilde{\varparam}}
\newcommand{\corparamapprox}{\widetilde{\corparam}}
\newcommand{\corparamdim}{r}
\newcommand{\Su}{\mb{S}(\re)}
\newcommand{\varcomp}{\sigma}
\newcommand{\varcompapprox}{\widetilde{\varcomp}}

\newcommand{\Jac}{\mb{J}}

\newcommand{\param}{\theta}

\newcommand{\responsespace}{\Yy}

\newcommand{\respace}{\Uu}
\newcommand{\meanspace}{\Mm}

\newcommand{\response}{Y} 
 
\newcommand{\paramdim}{p}

\newcommand{\obs}{y}
\newcommand{\obsall}{\mb{\obs}}
\newcommand{\covi}{x}
\newcommand{\cov}{\mb{\covi}}
\newcommand{\recovi}{v}
\newcommand{\recov}{\mb{\recovi}}
\newcommand{\reidx}{u}
\newcommand{\re}{\mb{\reidx}}

\newcommand{\regparamidx}{\beta}
\newcommand{\regparam}{\mb{\beta}}
\newcommand{\resd}{\mb{\sigma}}
\newcommand{\obssdi}{\phi}
\newcommand{\obssd}{\mb{\obssdi}}
\newcommand{\params}{\mb{\param}}

\newcommand{\parammle}{\widehat{\param}}
\newcommand{\paramsmle}{\widehat{\params}}
\newcommand{\AQmle}[1]{\paramsmle^{\AQ}_{#1}}
\newcommand{\AQmleidx}[1]{\parammle^{\AQ}_{#1}}

\newcommand{\linkfunc}{h}

\newcommand{\responsedist}{F}
\newcommand{\redist}{G}
\newcommand{\responsedens}{f}
\newcommand{\redens}{g}
\newcommand{\margdens}{\pi}
\newcommand{\marglikall}{\margdens(\params;\obsall)}

\newcommand{\approxmargdens}{\widetilde{\pi}}
\newcommand{\approxmargdensAQ}{\approxmargdens^{\AQ}}

\newcommand{\approxmarglikAQall}[1]{\approxmargdensAQ_{#1}(\params;\obsall)}

\newcommand{\approxloglik}{\widetilde{\ell}}
\newcommand{\approxloglikik}[2]{\approxloglik_{#1}^{#2}}
\newcommand{\approxloglikAQ}{\approxloglik^{\AQ}}
\newcommand{\approxloglikAQall}[1]{\approxloglikAQ_{#1}(\params)}
\newcommand{\approxloglikAQMLE}[1]{\approxloglikAQ_{#1}(\AQmle{#1})}

\newcommand{\approxloglikAQfull}[2]{\approxloglikik{#1}{#2}\left(\params,\condmodei{i}(\params),\jointcholi{i}\right)}

\newcommand{\gfun}{g}
\newcommand{\gfunbasei}{\gfun^i_{\quadnum}\left(\params,\re,\chol;\quadpointvec\right)}
\newcommand{\gfunbaseiprime}{\gfun^i_{\quadnum}\left(\params,\re,\chol;\quadpointvec^{\prime}\right)}

\newcommand{\jointdens}{\pi}
\newcommand{\jointlik}[2]{\jointdens(#1,#2;\obsall)}
\newcommand{\jointlikall}{\jointlik{\params}{\re}}
\newcommand{\gradjointlogdens}{\mb{g}}
\newcommand{\gradjointloglik}[2]{\gradjointlogdens(#1,#2;\obsall)}
\newcommand{\gradjointloglikall}{\gradjointloglik{\params}{\re}}
\newcommand{\hessjointlogdens}{\mb{H}}
\newcommand{\hessjointloglik}[2]{\hessjointlogdens(#1,#2;\obsall)}
\newcommand{\hessjointloglikall}{\hessjointloglik{\params}{\re}}

\newcommand{\condmode}{\widehat{\re}}
\newcommand{\condmodeidx}{\widehat{\reidx}}
\newcommand{\condmodei}[1]{\condmode_{#1}}

\newcommand{\approxhess}{\widetilde{\boldsymbol{H}}}
\newcommand{\approxhesscol}[1]{\widetilde{\boldsymbol{d}}_{#1}}

\newcommand{\Varmat}{\mb{\Sigma}}

\newcommand{\Varmatapprox}{\widetilde{\Varmat}}

\newcommand{\normalchol}{\boldsymbol{A}}
\newcommand{\normalD}{\boldsymbol{D}}
\newcommand{\normalcholapprox}{\widetilde{\normalchol}}
\newcommand{\normalDapprox}{\widetilde{\normalD}}

\newcommand{\indsim}{\overset{\text{ind}}{\sim}}
\newcommand{\iidsim}{\overset{\text{iid}}{\sim}}

\newcommand{\quadnum}{k}

\newcommand{\quadweight}[1]{w_{#1}}
\newcommand{\quadpoint}{z}
\newcommand{\quadpointvec}{\mb{\quadpoint}}

\newcommand{\quadpointset}{\mathcal{Q}}
\newcommand{\quadpointsetonedim}{\quadpointset(1,\quadpointnum)}
\newcommand{\quadpointsetmultidim}{\quadpointset(\redim,\quadpointnum)}

\newcommand{\quadpointnum}{k}
\newcommand{\weight}{\omega}
\newcommand{\weightk}{\weight_{\quadpointnum}}

\newcommand{\numgroups}{m}
\newcommand{\numpergroup}{n}
\newcommand{\numtotal}{N}
\newcommand{\redim}{d}
\newcommand{\regparamdim}{q}

\newcommand{\obssddim}{w}
\newcommand{\resddim}{s}

\newcommand{\mean}{\mu}
\newcommand{\pred}{\eta}
\newcommand{\predvec}{\boldsymbol{\pred}}

\newcommand{\regfun}{\Psi}

\newcommand{\gradvec}{\mb{g}}

\newcommand{\Mm}{\mathcal{M}}

\newcommand{\Uu}{\mathcal{U}}

\newcommand{\Yy}{\mathcal{Y}}

\newcommand{\fd}{\epsilon}

\begin{document}

\def\spacingset#1{\renewcommand{\baselinestretch}%
{#1}\small\normalsize} \spacingset{1}

%%%%%%%%%%%%%%%%%%%%%%%%%%%%%%%%%%%%%%%%%%%%%%%%%%%%%%%%%%%%%%%%%%%%%%%%%%%%%%

\if1\blind
{
  \title{\bf \thetitle}
  \author{Alex Stringer\thanks{
    The author gratefully acknowledges funding from the Natural Sciences and Engineering Research Council of Canada, and helpful comments from Drs. Art Owen and Audrey B{é}liveau.}\hspace{.2cm}\\
    Department of Statistics and Actuarial Science, University of Waterloo\\}
  \maketitle
  \date{}
} \fi

\if0\blind
{
  \bigskip
  \bigskip
  \bigskip
  \begin{center}
    {\LARGE\bf \thetitle}
\end{center}
  \medskip
} \fi

\bigskip
\begin{abstract}
  A method is introduced for approximate marginal likelihood inference via adaptive Gaussian quadrature in mixed models with a single grouping factor.
  The core technical contribution is an algorithm for computing the exact gradient of the approximate log-marginal likelihood.
  This leads to efficient maximum likelihood via quasi-Newton optimization that is demonstrated to be faster
  than existing approaches based on finite-differenced gradients or derivative-free optimization.
  The method is specialized to Bernoulli mixed models with multivariate, correlated Gaussian random effects; here
  computations are performed using an inverse log-Cholesky parameterization of the Gaussian density that involves no
  matrix decomposition during model fitting, while
  Wald confidence intervals are provided for variance parameters on the original scale.
  Simulations give evidence of these intervals attaining nominal coverage if enough quadrature points are used,
  for data comprised of a large number of very small groups exhibiting large between-group heterogeneity. 
  The Laplace approximation is well-known
  to give especially poor coverage and high bias for data comprised of a large
  number of small groups.
  Adaptive quadrature mitigates this, and the methods in this paper improve the computational feasibility of this more accurate method.
  All results may be reproduced using code available at \url{https://github.com/awstringer1/aghmm-paper-code}.
\end{abstract}

\noindent%
\vfill

\tableofcontents

\newpage
\spacingset{1.45} % DON'T change the spacing!

\section{Introduction}\label{sec:intro}

\subsection{Background}\label{subsec:background}

Grouped data, including longitudinal or repeated measures data, are common in modern practice. A popular class of models for grouped data are mixed models in which observations are assumed independent conditional on latent group-specific characteristics. 
This conditional independence yields a joint likelihood, but because this joint likelihood includes latent variables, it cannot be used directly for inference. 
Likelihood inferences in mixed models are instead based on the marginal likelihood of the observed data, defined as an integral of the joint likelihood over the latent variables. 
In general, this integral is intractable and must be approximated, and in practice inferences about parameters are made by maximizing an approximate marginal likelihood. 
This paper discusses a method for accurate and efficient approximate marginal likelihood inference in mixed models for grouped data with a single grouping factor.

Intractable integrals occur routinely in statistical inference. The unique aspect of the models considered in this paper is the factoring of the marginal likelihood into a product of $\numgroups$ 
low-dimensional integrals, where $\numgroups$ is the number of groups present in the data. 
This structure is computationally convenient, as it enables accurate approximation of the marginal likelihood using low-dimensional (adaptive) quadrature. 
However, it also introduces a critical \emph{requirement} for high accuracy that has not been 
fully appreciated.
The marginal likelihood is a product of $\numgroups$ integrals, and hence any approximation with fixed
relative accuracy will incur error tending to $\infty$ as $\numgroups\to\infty$.
However, $\numgroups\to\infty$ is \emph{necessary} for 
consistent estimation of parameters \citep{nie_convergence_2007,jiang_usable_2022}. 
This yields the result that as more data are obtained,
estimation quality will \emph{degrade} 
unless each of the $\numgroups$ integrals are approximated with increasingly high accuracy.
An integral approximation attaining arbitrarily high accuracy is required for inference about parameters in mixed models for grouped data. 

Adaptive Gauss-Hermite quadrature (AGHQ; simply AQ in what follows) has been used in statistical problems for decades \citep{naylor_applications_1982,tierney_accurate_1986,liu_note_1994,jin_note_2020},
including for fitting mixed models \citep{pinheiro_approximations_1995,pinheiro_efficient_2006,bates_fitting_2015,rizopoulos_glmmadaptive_2020}.
The technique has recently been shown to provide arbitrarily accurate low-dimensional integral approximations in statistical problems \citep{bilodeau_stochastic_2021},
number of groups, $\numgroups$. 
and this motivates AQ as a method for fitting mixed models.
However, computing the AQ approximation to the marginal likelihood even for a single parameter value is non-trivial, 
and hence maximizing it using
numerical optimization is challenging.
Current state-of-the-art approaches to fitting mixed models either use AQ with finite-differenced gradients 
(\texttt{R} package \texttt{GLMMadaptive}, \citealt{rizopoulos_glmmadaptive_2020})
or derivative-free optimization (\texttt{R} package \texttt{lme4}, \citealt{bates_fitting_2015}; one-dimensional random effects only).
The latter approach suffers from requiring a potentially excessive number of function evaluations \citep[section 9.2]{nocedal_optimization} while
the former suffers from the same problem as well as potential inaccuracies in the gradient evaluation \citep[section 9.1]{nocedal_optimization},
leading to computational challenges compared to methods depending on exact gradients.
The core technical contribution of the present paper is an algorithm for
computing the \emph{exact} gradient of the approximate log-marginal likelihood for any number of quadrature points and random effects dimension $\geq1$.
Quasi-Newton optimization with this novel exact gradient is shown to outperform these existing approaches in terms of speed (usually by around $2-4\times$) while matching or improving stability.

When run with a single quadrature point AQ is the Laplace approximation.
The Laplace approximation is used extensively in marginal likelihood approximation
for latent variable models and forms the basis of many classical approaches to fitting mixed models; see \citet{breslow_approximate_1993} and \citet{wolfinger_1993}.
Exact gradient evaluation algorithms exist for the Laplace approximate marginal likelihood in general \citep{kristensen_tmb_2016} and for mixed models \citep{brooks_glmmtmb_2017},
and both Newton \citep{wood_smoothing_2016} and quasi-Newton \citep{stringer_fast_2021} optimization based on these
exact gradients have been shown to be fast and stable.
Unfortunately, the Laplace approximation is often not sufficiently accurate for inference in mixed models; 
see \citet{joe_accuracy_2008,kim_smoking_2013} and Section \ref{subsec:simulations} of the present paper for empirical evidence.
In cases where the Laplace approximation is not sufficiently accurate, these same references provide evidence that AQ can be
sufficiently accurate,
and the evidence from the Laplace case suggests that
use of quasi-Newton optimization with an exact gradient could mitigate the computational challenges associated with this more accurate method. 
This observation motivates the development of the exact gradient evaluation algorithm for the AQ-approximate marginal likelihood 
presented in this paper.

In this paper we derive an algorithm for computing the exact gradient of the AQ-approximate marginal likelihood for mixed models
with a single grouping factor.
This leads to efficient quasi-Newton optimization
that substantially outperforms existing approaches based on finite-differenced gradients and derivative-free optimization;
specifically, in sections \ref{sec:bernoulli} and \ref{sec:data} it is shown to yield results 
as or more satisfactory than these existing approaches with a typical factor of $2-4\times$ reduction in computation time for binary mixed models with correlated random intercepts and slopes.
The procedure applies to the broad class of mixed models defined in Section \ref{sec:prelims} that
includes generalized linear and non-linear mixed effects models with response distributions
not limited to the exponential family and random effects distributions not limited to the Gaussian.
In Section \ref{sec:bernoulli} we specialize the method to Bernoulli mixed models with multivariate correlated Gaussian random effects,
a specific type of mixed model of substantial practical importance for which obtaining accurate inferences is known to be challenging.
We use an inverse log-Cholesky parameterization of the multivariate Gaussian, such that evaluating the density requires no
matrix decomposition or inversion, yielding efficient computation and closed-form derivative expressions. 
We then back-transform to provide Wald confidence intervals for variance parameters on the scale on which they are defined in the model, yielding interval estimates for quantities reported in practice.
Simulations demonstrate that these intervals attain nominal coverage when enough quadrature points are used,
even for large numbers of small groups exhibiting large heterogeneity. 
In contrast, as is well-known, the Laplace approximation performs poorly, exceptionally so in cases with very small groups that are common in practice.
Our simulation results corroborate this observation and highlight it by presenting it in tandem with results from the AQ approximation, which
is acceptably accurate.
In Section \ref{sec:data} we apply the new procedure to two previously reported data examples, 
showing that the approach based on exact gradients out-performs existing approaches in computation time
while returning indistiguishable inferences.
This includes a comparison to the computationally demanding (but accurate) profile likelihood
and bootstrap confidence intervals for variance parameters favoured by \texttt{lme4}; our Wald intervals for the variance
parameters are very similar to these but $1$ to $3$ orders of magnitude faster to compute for a data example exhibiting
very large between-subject variability.

\section{Mixed Models and Approximate Marginal Likelihood}\label{sec:prelims}

\subsection{Mixed Models for Grouped Data}\label{subsec:prelims-models}

We consider the following class of two-level mixed models:
\begin{equation}\begin{aligned}\label{eqn:mixedmodeldef}
  \response_{ij} \setdelim \re_i &\indsim \responsedist(\mean_{ij},\obssd), \ \re_i \iidsim \redist(\resd), \\
  \pred_{ij} = \linkfunc(\mean_{ij}) &=  \regfun(\cov_{il},\recov_{ij},\re_i;\regparam).
\end{aligned}\end{equation}

Here $i=1,\ldots,\numgroups$ indexes groups and $j=1,\ldots,\numpergroup_i$ indexes observations within groups which have size $\numpergroup_i$; when all groups are the same size, this size is denoted by $\numpergroup_i\equiv\numpergroup$.
The $\response_{ij}\in\responsespace\subseteq\Reals$ 
and $\re_i\in\respace\subseteq\Reals^{\redim}$ are the observed and unobserved stochastic components---data and random effects---of the model, 
having distributions $\responsedist(\mean_{ij},\obssd)$ and $\redist(\resd)$ respectively.
The corresponding densities, $\responsedens(\mean_{ij},\obssd)$ and $\redens(\resd)$, are assumed to 
be at least three times continuously differentiable with respect to $\mean_{ij},\obssd$, and $\resd$.
The covariates, $\cov_{ij},\recov_{ij}$, are assumed fixed and known, and the regression function, $\regfun$, is assumed fixed and known up to unknown parameters $\regparam$, and to be at least three times continuously differentiable.
The mean parameter, $\mean_{ij}$, takes values in a space $\meanspace\subseteq\Reals$ and
the link function, $\linkfunc:\meanspace\to\Reals$, is assumed monotone and at least three times continuously differentiable.
The unknown parameters, $\params = (\regparam,\obssd,\resd)$, are referred to as regression coefficients, dispersion parameters, and variance parameters respectively.

The full model (Eq \ref{eqn:mixedmodeldef}) is general, and restrictions yield more familiar models.
With $\redist$ a zero-mean Gaussian, $\responsedist$ from the exponential family, and linear regression function,
$\regfun(\cov_{il},\recov_{ij},\re_i;\regparam) = 
\cov_{ij}\Tr
\regparam + \recov_{ij}\Tr\re_i$,
the model (Eq. \ref{eqn:mixedmodeldef})
is a generalized linear mixed model (GLMM; \citealt{breslow_approximate_1993}). 
If in addition $\responsedist$ is also Gaussian, model (Eq. \ref{eqn:mixedmodeldef}) is a linear mixed model \citep{bates_fitting_2015}.
With $\responsedist$, and
$\redist$ both Gaussian
and $\regfun$ a fixed, known, nonlinear function, 
the model (Eq. \ref{eqn:mixedmodeldef}) is a Gaussian non-linear mixed model (NLMM; \citealt{pinheiro_approximations_1995,wolfinger_1993,vonesh_note_1996}). 
The methods in Sections \ref{sec:prelims} and \ref{sec:methods} of this paper apply to any such mixed model, while
the experiments of Section \ref{sec:bernoulli} and data analyses of Section \ref{sec:data} focus on Bernoulli
generalized linear mixed models with $\responsedist$ a Bernoulli distribution, $\redist$ a (multivariate) zero-mean Gaussian,
and linear $\regfun$.

Inferences about $\params$ are ideally based on a 
marginal likelihood, $\marglikall=\int\jointlikall\dee\re$,
where $\jointlikall = \prod_{i=1}^{\numgroups}\margdens_i(\params,\re_i;\obsall_i)$,
$\margdens_i(\params,\re_i;\obsall_i) = \prod_{j=1}^{\numpergroup_{i}}\responsedens(\obs_{ij};\mean_{ij},\obssd)\redens(\re_{i};\resd),$
$\obsall = (\obsall_1\Tr,\ldots,\obsall_\numgroups\Tr)\Tr\in\responsespace^{\numtotal}$ with $\obsall_i = (\obs_{i1},\ldots,\obs_{i\numpergroup_i})$, $\obs_{ij}$ the observed value of $\response_{ij}$, $\numtotal=\numpergroup_1+\cdots+\numpergroup_\numgroups$,
and $\re = (\re_1\Tr,\ldots,\re_\numgroups\Tr)\Tr\in\respace^{\redim\numgroups}$.
The marginal independence of the $\numgroups$ components of $\re$ gives special structure to the integral defining the marginal likelihood:
\begin{equation}\label{eqn:marglik}
  \marglikall = \int\jointlikall\dee\re = \prod_{i=1}^{\numgroups} \int\margdens_i(\params,\re_i;\obsall_i)\dee\re_{i}.
\end{equation}
We focus on cases in which inferences cannot be based on the marginal likelihood (Eq. \ref{eqn:marglik}) because the $\numgroups$ integrals defining it are intractable. 
This occurs for most models, the exception being those with conjugate response/random effects pairs that are chosen specifically to make these integrals
tractable; see \citet{lee_hierarchical_1996}. 
The Gaussian/Gaussian case is by far the most common such case and has been researched extensively; see \citet{bates_fitting_2015}.
We restrict focus in this paper to models in which the integral defining (Eq. \ref{eqn:marglik}) is intractable and must be approximated.

\subsection{Approximate Marginal Likelihood Inference}\label{subsec:approxmarglikinference}

Without conjugacy,
approximations to these integrals (Eq. \ref{eqn:marglik}) are required, and this leads to inferences being based instead on an approximate marginal likelihood, as follows.
Under the assumption that $\redim
=\text{dim}(\re_i)$ is small, the marginal likelihood involves only low-dimensional integrals. 
Quadrature techniques usually incur computational cost
that is exponential in dimension, 
so evaluating $\numgroups$ separate $\redim$-dimensional integrals is far more efficient than evaluating a single $(\redim\numgroups)$-dimensional integral. This factoring of the marginal likelihood therefore enables the use of accurate quadrature techniques for approximation, and this is well-recognized in the literature. However, this structure also \emph{requires} the use of a highly accurate integral approximation: as $\numgroups\to\infty$, although the data contain more information about $\params$, the number of integrals being approximated grows, and the accuracy of the approximation to the entire marginal likelihood
will decrease, leading to inaccurate inferences.
A more accurate numerical method therefore may be required for data sets which have more information about $\params$,
and this illustrates the need to choose this approximation carefully.
This makes AQ especially attractive, as it can be made arbitrarily accurate as more points are added \citep{bilodeau_stochastic_2021}.
Adding more points comes at an increased computational burden, and the exact gradient evaluation algorithm presented in this paper reduces this burden substantially.
Let $\condmode(\params) = 
\argmax_{\re}
\jointlikall = (\condmodei{1}(\params)\Tr,\ldots,\condmodei{\numgroups}(\params)\Tr)\Tr$, 
$$
\jointhess = -\partial^{2}_{\re}\log\jointlik{\params}{\condmode(\params)} = \text{diag}\left( \jointhessi{1},\ldots,\jointhessi{\numgroups}\right),
$$
and 
$$
\jointhessi{i} = \jointcholi{i}\jointcholi{i}\Tr,
$$ 
where $\jointcholi{i}$ is the lower Cholesky triangle. 
Let $\quadpointnum\in\Nats$, and let $\quadpointsetonedim\subset\Reals$ be the set of nodes from a $\quadpointnum$-point Gauss-Hermite quadrature rule in one dimension, 
$\quadpointsetmultidim = \quadpointsetonedim^{\redim}$ be the product extension of this rule to $\redim$ dimensions, 
and $\weightk:\quadpointsetmultidim\to\Reals$ be the corresponding quadrature weights. 
The adaptive quadrature approximation to the marginal likelihood (Eq. \ref{eqn:marglik}) is:
\begin{equation}\label{eqn:aqapproximatemarginallikelihood}
  \approxmarglikAQall{\quadnum} = 
  \prod_{i=1}^{\numgroups}
  \left[\abs{\jointcholi{i}}^{-1}
  \sum_{\quadpointvec\in\quadpointset(\redim,\quadnum)}
  \weightk(\quadpointvec)\jointdens_i
  \left\{\params,\jointcholi{i}\inv\quadpointvec + \condmodei{i}(\params);\obsall_i\right\}\right].
\end{equation}
The approximation (Eq. \ref{eqn:aqapproximatemarginallikelihood})
has been used in generalized linear \citep{pinheiro_efficient_2006} and non-linear \citep{pinheiro_approximations_1995} mixed models, and is available in modern software (\citealt{bates_fitting_2015} for $\redim=1$ only; \citealt{rizopoulos_glmmadaptive_2020}).

The case $\quadnum=1$ is called a Laplace approximation, and is considerably simpler. Here $\quadpointset(\redim,1) = \{\zero\}$ and $\weight_1(\zero) = (2\pi)^{\redim/2}$, and hence
\begin{equation}\label{eqn:laplaceapproximatemarginallikelihood}
  \approxmarglikAQall{1} = 
  (2\pi)^{(\redim\numgroups)/2}\abs{\jointchol}^{-1}\jointdens
  \left\{\params,\condmode(\params);\obsall\right\}
\end{equation}
is recognized as the usual Laplace-approximate marginal likelihood (e.g. \citealt{wood_fast_2011}).
In contrast to Eq. \ref{eqn:aqapproximatemarginallikelihood}, Eq. \ref{eqn:laplaceapproximatemarginallikelihood} does not involve a sum over the $\quadnum^\redim$-dimensional set $\quadpointset(\redim,\quadnum)$. 
The computations required to use the Laplace approximation therefore scale well with $\redim$, being at most the $O(d^3)$ operations required
to compute the Cholesky factor, and even this can be reduced substantially for models with a sparse Hessian; see \citet{rue_fast_2001,kristensen_tmb_2016}.
For this reason, Eq. \ref{eqn:laplaceapproximatemarginallikelihood} is used ubiquitously in latent variable modeling, including for fitting mixed models \citep{breslow_approximate_1993,wolfinger_1993}.
However, in some mixed models, the Laplace approximation may fail dramatically; see
\citet{joe_accuracy_2008,kim_smoking_2013,breslow_approximate_1993}, and 
 Section \ref{subsec:simulations} of the present paper.
Note that when the response and random effects are both Gaussian
the Laplace approximation (Eq. \ref{eqn:laplaceapproximatemarginallikelihood}) is exact, providing a connection
between the methods discussed here and methods for linear mixed models \citep{bates_fitting_2015}.
Again, we focus only on cases where the marginal likelihood is not tractable
and must be approximated.

\subsection{Computation of the maximum approximate marginal likelihood estimator}\label{subsec:computationprelims}

Inferences about $\params$ are based on the maximum approximate marginal likelihood estimator,  
$$
\AQmle{\quadnum}=\argmax\approxmarglikAQall{\quadnum}.
$$
Despite its frequent applied use, few 
details are available in the literature regarding the computation of $\approxmarglikAQall{\quadnum}$ and $\AQmle{\quadnum}$. 
The approximation, $\approxmarglikAQall{\quadnum}$, is a smooth, log-concave, many-times-continuously-differentiable function of $\params$.
Gradient-based quasi-Newton optimization is a well-established framework for such problems \citep{nocedal_optimization}, and is widely implemented in readily-available, open-source software, including the popular \texttt{optim} function in the \texttt{R} 
language \citep{r_core_team_r_2021}.
It seems clear that a gradient-based approach to computing $\AQmle{\quadnum}$ should be preferable to other approaches.
However, computing the
gradient exactly---as we do in Section \ref{sec:methods}---is challenging, and a method for doing so was not available prior to the present work.
A result of this lack of gradient information about $\log\approxmarglikAQall{\quadnum}$ is that 
a variety of alternative approaches to computation of $\approxmarglikAQall{\quadnum}$---or otherwise to make
inferences about $\params$---have been considered in the literature.

In the Laplace case ($\quadnum=1$), for generalized linear mixed models with Gaussian random effects, \citet{breslow_approximate_1993} 
derive a modified set of estimating equations for $\regparam$ for fixed $\resd$, solve them iteratively via Fisher scoring, and 
maximize a profile likelihood to estimate $\resd$; \citet{pinheiro_efficient_2006} provide a recurrence relation and related least-squares interpretation. 
For $\quadnum\geq1$, \citet{pinheiro_efficient_2006} 
and \citet{pinheiro_approximations_1995} give formulas for the approximation for multi-level generalized linear and non-linear mixed models, respectively.
However, neither latter framework includes details on how the necessary optimization should be performed, nor provide expressions or algorithms for gradient computation. 
\citet{mcculloch_maximum_1997} gives a 
Monte Carlo approximation to (Eq. \ref{eqn:marglik}) and corresponding versions of the EM and Newton-Raphson algorithms, 
and \citet{booth_maximizing_1999} expand on and motivate the use of the Monte Carlo EM algorithm where the intractable integral defining the E-step is approximated using random sampling.
\citet{rizopoulos_glmmadaptive_2020} implements an EM algorithm in software, suggesting its use for finding initial values to further pass to a quasi-Newton optimization with finite-differenced gradients. 

A common motivation that unites these previous approaches is not a preference for alternatives to gradient-based optimization, but 
rather that a lack of gradient information about $\log\approxmarglikAQall{\quadnum}$ simply precludes the use of this favourable approach. 
For the Laplace approximation ($\quadnum=1$) only, \citet{kristensen_tmb_2016} provide an algorithmic gradient of $\log\approxmarglikAQall{1}$, 
and \citet[Section 5]{stringer_fast_2021} demonstrate its efficiency over contemporary approaches in fitting a (Bayesian) mixed model to a large set of data.
We develop an algorithm for exact computation of the gradient of $\log\approxmarglikAQall{\quadnum}$ in Section \ref{sec:methods}, leading to efficient quasi-Newton optimization for finding $\AQmle{\quadnum}$ for any $\quadnum\geq1$.

\section{Computations with the approximate marginal likelihood}\label{sec:methods}

We give an algorithm for exact computation of the gradient, $\nabla_{\params}\approxloglikik{\quadnum}{}(\params)$, of the approximate log-marginal likelihood, $\approxloglikik{\quadnum}{}(\params) = \log\approxmarglikAQall{\quadnum}$, in Section \ref{subsec:gradientcomputation}. 
The technical difficulties are to
differentiate ``through'' (a) the ``inner'' optimization required to find $\condmode(\params)$, and 
(b) the matrix decompositions, linear system solves,
and log-determinant calculations all required to compute $\approxmarglikAQall{\quadnum}$ for each fixed $\params$.
The main technical tools used are
(a) implicit differentiation to account for $\condmode(\params)$, and
(b) algorithmic differentiation of the Cholesky decomposition, $\jointcholi{i}$, and the forward substitution required to compute $\jointcholi{i}\inv\quadpointvec$. Applying these tools effectively requires a careful organization of the computations.
We discuss the use of the new gradient computations for point estimation 
in Section \ref{subsec:pointestimation}, and confidence intervals in Section \ref{subsec:confidenceintervals}.
The important special case of scalar random effects (the ``random intercepts'' model) is treated in Section \ref{subsec:scalar}.

\subsection{Exact gradient computation}\label{subsec:gradientcomputation}

We seek the gradient, $\nabla_{\params}\approxloglikik{\quadnum}{}(\params)$, of the approximate log-marginal likelihood, $\approxloglikik{\quadnum}{}(\params) = \log\approxmarglikAQall{\quadnum}$. To begin, let 
$$
\approxloglikik{\quadnum}{}(\params) = \sum_{i=1}^{n}\approxloglikik{\quadnum}{i}(\params),
$$
 where
\begin{equation*}
  \begin{aligned}
    \approxloglikik{\quadnum}{i}(\params) &= \approxloglikAQfull{\quadnum}{i}, \\
    \approxloglikik{\quadnum}{i}(\params,\re,\chol) &= \log\left\{\sum_{\quadpointvec\in\quadpointset(\quadnum,\redim)}\gfunbasei \right\}, \\
    \gfunbasei &= \abs{\chol}\inv\quadweight{\quadnum}(\quadpointvec)\margdens_i(\params,\chol\inv\quadpointvec+\re). \\
  \end{aligned}
\end{equation*}
To keep notation concise, we suppress dependency of these functions on $\obsall$,
for example writing $\margdens_i(\params,\re) \equiv \margdens_i(\params,\re;\obsall)$.
We use the following convention for treating partial derivatives with respect to arguments of multi-variable functions. 
If $f:\Reals^{m+p}\to\Reals$ takes arguments $\mb{x}\in\Reals^m,\mb{y}\in\Reals^p$,
then $f(\mb{x};\mb{y})$ is the function $g:\Reals^m\to\Reals$ defined by $g(\mb{x}) = f(\mb{x},\mb{y})$. This notation will help communicate the
order in which terms are differentiated. For brevity we use $\partial_{\mb{x}}g(\mb{x})$ to mean $\partial g/\partial\mb{x}$.
The remainder of this Section is dedicated to the details of the gradient computation, but note that the full algorithm is given in Algorithms 1 and 2 in Section \ref{supp:algorithms} of the supplement.

The computations are organized as follows.
We have:
\begin{align*}
  \nabla_{\params}\approxloglikik{\quadnum}{i}(\params) &= \left.\partial_{\params}\approxloglikik{\quadnum}{i}\left(\params,\re,\jointcholbase_i(\params,\re)\right)\right\vert_{\re=\condmodei{i}(\params)} +  \partial_{\params}\condmodei{i}(\params)\cdot\left[\left.\partial_{\re}\approxloglikik{\quadnum}{i}\left(\params,\re,\jointcholbase_i(\params,\re)\right)\right\vert_{\re=\condmodei{i}(\params)}\right],
\end{align*}
and
\begin{align*}
  \partial_{(\params,\re)}\approxloglikik{\quadnum}{i}\left(\params,\re,\jointcholbase_i(\params,\re)\right) &= \left.\partial_{(\params,\re)}\approxloglikik{\quadnum}{i}\left(\params,\re;\chol\right)\right\vert_{\chol=\jointcholbase_i(\params,\re)} + \left.\partial_{(\params,\re)}\approxloglikik{\quadnum}{i}\left(\chol(\params,\re);\params,\re\right)\right\vert_{\chol=\jointcholbase_i(\params,\re)}.
\end{align*}
Computation of $\partial_{\params}\condmodei{i}(\params)$ and $\partial_{(\params,\re)}\approxloglikik{\quadnum}{i}\left(\chol(\params,\re);\params,\re\right)$ are the subjects of Sections \ref{subsec:modediff} and \ref{subsec:choldiff}. The remaining terms are as follows: 
\begin{align*}
  \partial_{(\params,\re)}\approxloglikik{\quadnum}{i}\left(\params,\re;\chol\right) &= \frac{\sum_{\quadpointvec\in\quadpointset(\quadnum,\redim)}\gfunbasei\partial_{(\params,\re)}\log\gfunbasei}{\sum_{\quadpointvec^{\prime}\in\quadpointset(\quadnum,\redim)}\gfunbaseiprime}, \\
  \partial_{(\params,\re)}\log\gfunbasei &= \partial_{(\params,\re)}\log\margdens_i(\params,\re).
\end{align*}
The $\partial_{(\params,\re)}\log\margdens_i(\params,\re)$ term is the gradient of the \emph{joint} log-likelihood, which is assumed tractable, albeit always model-specific. 
Calculations for Bernoulli mixed models and multivariate Gaussian random effects are given in Sections \ref{ref:bernoullilikelihood} and \ref{sec:normal}, and these are generalized to any exponential family distribution in Section \ref{supp:exponentialfamily} of the supplement.

\subsubsection{Implicit differentiation of $\widehat{u}_i(\theta)$}\label{subsec:modediff}

Implicit differentiation to obtain $\partial_{\params}\condmodei{i}(\params)$ is standard; see \citet{kristensen_tmb_2016} and \citet{stringer_fast_2021}. By definition we have $\partial_{\re}\log\margdens_i(\params,\condmodei{i}(\params))=0$, and differentiating
this equation gives
$$
\jointhessi{i}\partial_{\params}\condmodei{i}(\params) = -\partial^{2}_{\params,\re}\log\margdens_i(\params,\condmodei{i}(\params)). 
$$
At the point in Algorithm 2 (see the supplement, Section \ref{supp:algorithms}) when $\partial_{\params}\condmodei{i}(\params)$ is needed, the mode, $\condmodei{i}(\params)$, and the Cholesky, $\jointcholi{i}$, of the Hessian, $\jointhessi{i}$,
have already been computed, which enables efficient computation of $\partial_{\params}\condmodei{i}(\params)$ via a single application of each of forward and backward substitution.
In the implementation $\partial_{\params}\condmodei{i}(\params)$ is never formed explicitly, but rather the matrix-vector
product in which it appears is computed directly using this method.

\subsubsection{Algorithmic differentiation of $L(\theta,u)$}\label{subsec:choldiff}

\citet{smith_differentiation_1995} describes a reverse-mode algorithmic differentiation of any function, $f(\chol)$, of a Cholskey decomposition, $\chol\equiv\chol(\params,\re)$, of a positive-definite matrix, $\hess(\params,\re)=\chol(\params,\re)\chol(\params,\re)\Tr$, with respect to the underlying parameters, $(\params,\re)$. 
We apply this algorithm to $f(\chol)\equiv \approxloglikik{\quadnum}{i}\left(\chol(\params,\re);\params,\re\right)$, treating the parts of $\approxloglikAQ_i$ that depend on $(\params,\re)$ through $\chol(\params,\re)$
as variable, and all other instances of $(\params,\re)$ as fixed.
The required inputs to the algorithm are: (a) the computed Cholesky decomposition, $\chol$; 
(b) an array of derivatives, $\hessD = \partial_{(\params,\re)}\hess(\params,\re)$, of the original matrix, $\hess(\params,\re)$; 
and (c) the lower-triangular matrix of derivatives, $\cholD = \partial f(\chol)/\partial\chol$, of the function, $f(\chol)$, with respect to the elements of $\chol$.
In our setting, $\chol$ is already computed, and $\hessD$ is model-specific; again, calculations for Bernoulli mixed models and multivariate Gaussian random effects are given in Sections \ref{ref:bernoullilikelihood} and \ref{sec:normal}, and generalized to the exponential family in Section \ref{supp:exponentialfamily} of the supplement.
For $j\geq l=1,\ldots,\redim$, we have:
\begin{align*}
  \cholD_{jl} = 
  \partial_{L_{jl}}\approxloglikik{\quadnum}{i}\left(\chol;\params,\re\right) &= 
  \frac{\sum_{\quadpointvec\in\quadpointset(\quadnum,\redim)}\gfunbasei\partial_{L_{jl}}
  \log\gfunbasei}{\sum_{\quadpointvec^{\prime}\in\quadpointset(\quadnum,\redim)}\gfunbaseiprime}, \\
  \partial_{L_{jl}}
  \log\gfunbasei &= -\frac{1}{L_{jj}}
  \one(j=l)+
  \left(\frac{\partial\chol\inv\quadpointvec}{\partial L_{jl}}\right)\Tr\left.\partial_{\re}
  \log\margdens_i(\params,\re^{\prime})\right\vert_{\re^{\prime}=
  \chol\inv\quadpointvec+\re}.
\end{align*}
The $\redim$-dimensional vector $\cholinvvec\equiv\chol\inv\quadpointvec$ is obtained for each $\quadpointvec\in\quadpointset(\quadnum,\redim)$ by solving the equation $\chol\cholinvvec=\quadpointvec$ via forward substitution \citep[Algorithm 4.1-1]{golub_1983}. Accordingly, we obtain its derivative by differentiation of this algorithm with respect to $L_{jk}$; see Algorithm 3 in Section \ref{supp:algorithms} of the supplement. 

\subsection{Point Estimation}\label{subsec:pointestimation}

With the gradient, $\nabla_{\params}\approxloglikik{\quadnum}{}(\params)$, of the approximate log-marginal likelihood, $\approxloglikik{\quadnum}{}(\params) = \log\approxmarglikAQall{}$, available,
we compute the adaptive quadrature approximate maximum likelihood estimator,
$
\AQmle{\quadnum} = \text{argmax}_{\params} \ \approxloglikik{\quadnum}{}(\params),
$
using limited-memory BFGS (L-BFGS) quasi-Newton optimization, of the type
described by \citet[Section 6.1]{nocedal_optimization}.
Although L-BFGS is a standard algorithm, its implementation is nontrivial, specifically the
step-length selection algorithm run at each quasi-Newton iteration. 
One option is to pass separate functions which compute $\approxloglikik{\quadnum}{}(\params)$ and $\nabla_{\params}\approxloglikik{\quadnum}{}(\params)$ 
into a pre-existing implementation of this algorithm, such as in the \texttt{optim} function in the \texttt{R} language \citep{r_core_team_r_2021}; this strategy is taken by the \texttt{GLMMadaptive} package of \citet{rizopoulos_glmmadaptive_2020}, and the \texttt{lme4} package of \citet{bates_fitting_2015} implements its derivative-free approach in a similar manner.
However, we note two opportunities for efficiency that this leaves unrealized:
\begin{enumerate}
  \item[(a)] Computation of $\approxloglikik{\quadnum}{}(\params)$ occurs as a byproduct of computation of $\nabla_{\params}\approxloglikik{\quadnum}{}(\params)$,
    so the log-likelihood is available at no additional cost once the gradient has been computed. 
    This is not the case when derivatives are computed by finite-differences (\texttt{GLMMadaptive}), and not relevant when using derivative-free optimization (\texttt{lme4}).
  \item[(b)] Each evaluation of $\approxloglikik{\quadnum}{}(\params)$ requires $\numgroups$ inner optimizations to find $\condmodei{1}(\params),\ldots,\condmodei{\numgroups}(\params)$.
    Within an iterative outer optimization, the values of these modes from the previous iteration
    should be used as starting values for the subsequent iteration, reducing the number of inner optimization steps. 
\end{enumerate}
To implement these efficiencies without completely re-implementing L-BFGS and the step-size selection algorithms, 
we use the flexible implementation of L-BFGS provided at \url{https://github.com/yixuan/LBFGSpp/} (accessed 05/2023), which implements
the step-length selection strategy of \citet[Section 3.5]{nocedal_optimization}.

Starting values are required to run the L-BFGS optimization; these are somewhat model-specific, and we offer some general guidance here that we used in the experiments and data analyses of Sections \ref{sec:bernoulli} and \ref{sec:data},
where we fit Bernoulli generalized linear mixed models with correlated multivariate Gaussian random effects.
% In our experiments with Bernoulli mixed models in Sections \ref{sec:bernoulli} and \ref{sec:data}, we: 
We recommend to 
(a) fit an ordinary generalized linear model and use the estimated $\widehat{\regparam}$ as a starting value for $\regparam$; 
(b) set $\condmodei{i}(\params) = \zero$ (the mean of $\redist$) for each $i$; and (c) set $\resd=1$ for variance parameters and $0$ for covariance parameters (i.e. if the random effects are mean-zero Gaussian with covariance matrix $\Varmat(\resd)$, we set $\Varmat(\resd) = I_\redim$). 
We did not observe any sensitivity to this choice of starting values in our experiments, although caution should be taken if attempting to generalize these recommendations to other models.

The availability of $\nabla_{\params}\approxloglikik{\quadnum}{}(\params)$ also leads immediately to an approximate Hessian, computed as a simple finite-differenced Jacobian of the gradient.
For $\fd>0$, define the $\paramdim\times\paramdim$
matrix $\approxhess = [\approxhesscol{1}:\cdots:\approxhesscol{\paramdim}]$,
where $:\cdots:$ denotes column-wise concatenation, $\paramdim = \text{dim}(\params)$, and 
$$
\approxhesscol{j} = \frac{\nabla_{\params}\approxloglikik{\quadnum}{}(\param_1,\ldots,\param_j+\fd,\ldots,\param_\paramdim) - \nabla_{\params}\approxloglikik{\quadnum}{}(\params)}{\fd}, j = 1,\ldots,\paramdim. 
$$
At termination of the L-BFGS algorithm, we run Newton's method from the terminal point, using $\approxhess$
in place of the Hessian of $\approxloglikik{\quadnum}{}(\params)$.
We found that the terminal point from L-BFGS is usually satisfactory, in which case a single iteration of Newton's method
is run, at a marginal computational cost of $\paramdim$ additional gradient evaluations to
evaluate $\approxhess$ (which is also required for confidence intervals; see Section \ref{subsec:confidenceintervals}), plus the $O(\paramdim^3)$ cost of a system solve involving it (recall that $\paramdim=\text{dim}(\params)$ is generally small). In some cases,
however, the terminal point from L-BFGS is not satisfactory, and we observe that in almost all such cases, 
the Newton iterations starting from it
produce a satisfactory estimate. 
This was the strategy used to produce all the simulation and data analysis results
in Sections \ref{sec:bernoulli} and \ref{sec:data}.

\subsection{Confidence intervals}\label{subsec:confidenceintervals}

With $\approxhess$
available, an approximate $(1-\alpha)100\%$ Wald confidence interval for the $j^{th}$ component of $\params$, $\param_j$, for $j=1,\ldots,\paramdim$ is given by
$$
(\AQmle{\quadnum})_j \pm z_{1-\alpha/2}\left(\approxhess^{-1}\right)_{jj}^{1/2},
$$
where $z_{\alpha}$ satisfies $P(Z<z_{\alpha}) = \alpha$ for $Z\sim\text{N}(0,1)$ and $0<\alpha<1$.

Computing $\approxhess$ involves $\paramdim$ evaluations of $\nabla_{\params}\approxloglikik{\quadnum}{}(\params)$,
using the novel gradient calculations from Section \ref{subsec:gradientcomputation}.
The evaluation speed is orders of magnitude faster than that of computing $\AQmle{\quadnum}$, and hence the cost of computing the Wald intervals is negligible compared to that of the point estimates.
For moderate $\paramdim$, it is efficient to simply directly invert $\approxhess$ and take
its diagonal elements; for larger $p$, the methods of e.g. \citet{rue_approximate_2007} for
determining the diagonal of a matrix inverse from its Cholesky decomposition could be applied,
although we did not have need for this.
The value of $\fd=10^{-8}$ used in our experiments (Section \ref{subsec:simulations}) was sufficient to achieve the nominal empirical coverages reported there.
We remark that the use of a single finite-difference operation
for computing the Hessian as the Jacobian of the gradient is now
possible due to the gradient being available exactly, reducing the computational
cost of this step.

In cases where the final objects of inferential interest are not the elements of $\params$ directly, but rather some transformation of them, $\approxhess\inv$ is used as an input to a Delta-method confidence interval.
Such situations are model-specific; see Section \ref{sec:normal} for the Bernoulli/Gaussian mixed model case.

\subsection{Scalar random effects}\label{subsec:scalar}

When $\redim=1$, implementation of the approximate marginal likelihood and gradient computations
simplifies considerably. 
When $\re\equiv\reidx$ is a scalar
the computations can be performed using simple floating point types---rather than vector and matrix types---for all $\re$-dependent quantities, improving implementation efficiency. 
The $\redim=1$ case includes the random-intercepts model, which is of considerable interest in its own right, and is the only case for which AQ is implemented in the popular \texttt{lme4} software \citep{bates_fitting_2015}; efficient implementation of the present procedure is therefore required for comparison against this established method, 
as well as being of independent interest.

The steps of the novel gradient evaluation procedure remain the same. 
Because $\chol$ is now a scalar, $\chol\equiv\cholscalar = (\hessidx)^{1/2}$ where $\hessidx = -\partial^2_{\reidx^2}\log\jointlik{\params}{\condmodeidx(\params)}$.
The algorithm of \citet{smith_differentiation_1995} is no
longer needed, and the corresponding term simplifies to
$$
\partial_{(\params,\reidx)}\approxloglikAQ_i\left(\chol;\params,\reidx\right) = (F_{11})(\hessD) / (2L),
$$
where $\hessD = \partial_{(\params,\reidx)}\hessidx$ is now a $(\paramdim+1)$-dimensional vector.
The $\chol$-derivative of Section \ref{subsec:choldiff} simplifies to:
$$
\partial_{\cholscalar}\log\gfunbasei = -\frac{1}{\cholscalar} - \frac{\quadpoint}{\cholscalar^2}\left.\partial_{\reidx}
  \log\margdens_i(\params,\reidx^{\prime})\right\vert_{\reidx^{\prime}=
  \quadpoint/\cholscalar+\reidx}.
$$

\section{Bernoulli mixed models with multivariate Gaussian random effects}\label{sec:bernoulli}

An important special case of Eq. \ref{eqn:mixedmodeldef} is the Bernoulli mixed model:
\begin{align}\label{eqn:bernoullimodel}
  \response_{ij} \setdelim \re_i \indsim \text{Bern}(p_{ij}), \ \re_i \iidsim \text{N}\{\zero,\Varmat(\resd)\}, \ 
  \log\frac{p_{ij}}{1-p_{ij}} = \cov_{ij}\Tr
\regparam + \recov_{ij}\Tr\re_i.
\end{align}
The model (Eq. \ref{eqn:bernoullimodel}) is commonly used for longitudinal/repeated measures binary outcomes \citep{kim_smoking_2013,breslow_approximate_1993,hedeker_note_2018} and is prominent in ecology \citep{bolker_generalized_2008} and psychology \citep{bono_report_2021}.
Fitting the model (Eq. \ref{eqn:bernoullimodel}) is challenging:
beyond the challenges with fitting any mixed model described in Sections \ref{sec:prelims}
and \ref{sec:methods},
a parameterization of $\Varmat(\resd)$ that leads to stable and efficient optimization, but produces estimates and confidence intervals for those estimates on an interpretable scale is required.
In this Section, we give the details of applying the methods of Sections \ref{sec:prelims} and \ref{sec:methods} to this important model.
This both serves to give the details for one important case, and also to
illustrate the steps required to apply the methods in this paper to specific models.
We note that the multivariate Gaussian calculations given in Section \ref{sec:normal}
apply to any mixed model with correlated multivariate Gaussian random effects,
and are hence of independent interest.

\subsection{Likelihood calculations}\label{ref:bernoullilikelihood}

The approximate log-marginal likelihood is a sum over groups,
$\approxloglikik{\quadnum}{}(\params)=\sum_{i=1}^{\numgroups}\approxloglikik{\quadnum}{i}(\params)$, and same for its gradient.
Accordingly, we give calculations for a single group.
The likelihood derivative calculations up to order 2 are standard, however we
repeat them here to illustrate the form of the third-order derivatives
required, and for consistency of notation.

The log-likelihood calculations corresponding to Eq. \ref{eqn:bernoullimodel} are
\begin{equation*}
\begin{aligned}
\ell_i(\params,\re) = \log\margdens_i(\params,\re_i;\obsall_i) &= \sum_{j=1}^{\numpergroup_i}\obs_{ij}\pred_{ij} - \log\left(1+e^{\pred_{ij}}\right), \\
\frac{\partial\ell_i(\params,\re)}{\partial(\regparam,\re)} &= \left(\frac{\partial\ell_i}{\partial\predvec_i}\right)\Tr\left[\mb{X}_i:\mb{V}_i\right], \\
 \frac{\partial^2\ell_i(\params,\re)}{\partial(\regparam,\re)\partial(\regparam,\re)\Tr} &= \left[\mb{X}_i:\mb{V}_i\right]\Tr\left(\frac{\partial^2\ell_i}{\partial\predvec_i\partial\predvec_i\Tr}\right)\left[\mb{X}_i:\mb{V}_i\right], \\
\end{aligned}
\begin{aligned}
 \\
\frac{\partial\ell_i}{\partial\pred_{ij}} &= \obs_{ij} - \frac{e^{\pred_{ij}}}{1+e^{\pred_{ij}}} \\
\frac{\partial^2\ell_i}{\partial\pred_{ij}^2} &= -\frac{e^{\pred_{ij}}}{1+e^{\pred_{ij}}}\left(1-\frac{e^{\pred_{ij}}}{1+e^{\pred_{ij}}}\right),
\end{aligned}
\end{equation*}
where $\predvec_i = (\pred_{i1},\ldots,\pred_{i\numpergroup_i})\Tr\in\Reals^{\numpergroup_i}$, $\mb{X}_i = [\cov_{i1}:\cdots:\cov_{i\numpergroup_i}]\in\Reals^{\numpergroup_i\times\regparamdim}$, and $\mb{V}_i = [\recov_{i1}:\cdots:\recov_{i\numpergroup_i}]\in\Reals^{\numpergroup_i\times\redim}$.
Define $\hess^{\obs}_{i} = -\partial^2\ell_i(\params,\re)/\partial\re\partial\re\Tr$; we seek the third-order derivative matrices $\partial\hess^{\obs}_{i}/\partial\param_{l}$. These are zero for components of $\params$ corresponding to $\resd$, and for $\re$ and $\regparam$ are given by
\begin{align*}
\frac{\partial\hess^{\obs}_{i}}{\partial\re_l} &= \sum_{j=1}^{\numpergroup_i}\frac{\partial^3\ell_i}{\partial\pred_{ij}^3}\recov_{ij}\recov_{ij}\Tr\recovi_{ijl}, \
\frac{\partial\hess^{\obs}_{i}}{\partial\regparam_l} = \sum_{j=1}^{\numpergroup_i}\frac{\partial^3\ell_i}{\partial\pred_{ij}^3}\recov_{ij}\recov_{ij}\Tr\covi_{ijl}, \\
\frac{\partial^3\ell_i}{\partial\pred_{ij}^3} &= \frac{e^{\pred_{ij}}}{1+e^{\pred_{ij}}}\left(1-\frac{e^{\pred_{ij}}}{1+e^{\pred_{ij}}}\right)\left(1-2\frac{e^{\pred_{ij}}}{1+e^{\pred_{ij}}}\right).
\end{align*}
These are a special case of the calculations for general exponential family distributions, which follow similarly; see Section \ref{supp:exponentialfamily} of the supplement.

\subsection{Multivariate Gaussian random effects}\label{sec:normal}

We require a parameterization of the Normal density that leads to efficient and stable optimization.
In particular, $\Varmat$ is required to be symmetric and positive-definite,
and it is preferable to impose this via re-parameterization rather than
employing constrained optimization techniques; see \citet{pinheiro_unconstrained_1995}.
However, the final desired output of the procedure is point and interval estimates for variance components on a scale that is interpretable in the context of the original problem.
While point estimates for the unique elements of $\Varmat$ are straightforward to obtain via invariance of maximum likelihood to general transformations, obtaining interval estimates via the Delta method requires innovation.

We use the following ``inverse log-Cholesky'' parameterization of $\Varmat$:
\begin{equation*}
\Varmat\inv(\resd) = \normalchol\normalD\normalchol\Tr, \ \normalD = \text{diag}\left(e^{\delta_1},\ldots,e^{\delta_\redim}\right), 
\texttt{lower.tri}(\normalchol) = \corparam,
\ \resd = (\varparam,\corparam),
\end{equation*}
where $\normalD$ is diagonal and $\normalchol$ is unit lower-triangular so that
$\text{dim}(\corparam) = \corparamdim = \redim(\redim-1)/2$ and hence $\dim(\resd) = \resddim = \redim(\redim+1)/2$. 
The covariance parameters, $\corparam$, represent the lower triangle of $\normalchol$ in
column-major order excluding the diagonal, so $\phi_1 = \normalchol_{21},\phi_2 = \normalchol_{31}$, and so on.

Because $\normalD>0$, $\Varmat\inv$ is symmetric and positive-definite, and hence $\Varmat$ is as well. The free parameters, $\resd$,
are unconstrained, and---again, because $\normalD>0$---uniquely determine $\Varmat\inv$ \citep{golub_1983}. 
In this parameterization the random effects density is
\begin{equation}\label{eqn:normaldensity}
\log\redens(\re;\resd) = \text{const} + \frac{1}{2}\sum_{j=1}^{\redim}\delta_j - \frac{1}{2}\re\Tr\normalchol\normalD\normalchol\Tr\re.
\end{equation}
Note that our log-Cholesky parameterization differs from those of \citet{rizopoulos_glmmadaptive_2020}, \citet{bates_fitting_2015}, and \citet{pinheiro_unconstrained_1995} in that we transform $\Varmat\inv$ instead of $\Varmat$.
The reason to do this is because evaluation of the density and its derivatives in this parameterization requires only basic matrix operations, avoiding inverses, determinants, system solves, and any such computation that would involve performing a matrix decomposition at every evaluation.
This structure leads to efficient computations involving the density,
and hence improves efficiency of the optimization.

We seek the three derivatives required to implement Algorithm 2 (see the supplement, Section \ref{supp:algorithms}), and it will be convenient to
express Eq. \ref{eqn:normaldensity} directly
in terms of $\re$, $\varparam$, and $\corparam$.
We first note that the product, $\normalchol\Tr\re$, can be expressed explicitly in terms of $\corparam$, $\re$, and a unique $\redim\times\corparamdim$ \emph{selection matrix}, $\Su$, as $\normalchol\Tr\re = \re + \Su\corparam$; see Section 1.3 of the supplement for full details. This yields the following derivative expressions:
\begin{equation*}
\begin{aligned}
  \frac{\partial\log\redens}{\partial\re} &= -\re\Tr\normalchol\normalD\normalchol\Tr, \\ 
  \frac{\partial\log\redens}{\partial\delta_j} &= \frac{1}{2}\left[1 - \exp(\delta_j)\left\{\re+\Su\corparam\right\}^2_j\right], \\
  \frac{\partial\log\redens}{\partial\corparam} &= -\left\{\re+\Su\corparam\right\}\Tr\normalD\Su.
\end{aligned}
\qquad
\begin{aligned}
  \frac{\partial^2\log\redens}{\partial\re\partial\re\Tr} &= -\normalchol\normalD\normalchol\Tr. \\
  \frac{\partial^2\log\redens}{\partial\delta_j^2} &= -\frac{1}{2}\exp(\delta_j)\left\{\re+\Su\corparam\right\}^2_j, \\
  \frac{\partial^2\log\redens}{\partial\corparam\partial\corparam\Tr} &= -\Su\Tr\normalD\Su.
\end{aligned}
\end{equation*}
The cross terms are as follows:
\begin{equation*}
\begin{aligned}
  \frac{\partial}{\partial\delta_j}\frac{\partial\log\redens}{\partial\re} &= -\exp(\delta_j)(\normalchol\Tr\re)_j\normalchol_j, \\
  \frac{\partial^2\log\redens}{\partial\corparam\partial\reidx_j} &= -\left[\re\Tr\normalD\left\{\frac{d\Su}{d\reidx_j}\right\} + (\normalD\Su)_j + \corparam\Tr\Su\Tr\normalD\left\{\frac{d\Su}{d\reidx_j}\right\} + \corparam\Tr\left\{\frac{d\Su\Tr}{d\reidx_j}\right\}\normalD\Su \right], \\
  \frac{\partial}{\partial\delta_j}\frac{\partial\log\redens}{\partial\corparam} &= -\exp(\delta_j)\left\{\re+\Su\corparam\right\}\Tr_j\Su_j. 
\end{aligned}
\end{equation*}
Given these novel expressions involving $\Su$ for the density and its derivatives, the third-order derivatives required to implement the procedure in the present paper are now
straightforward to obtain.
Let
$$
\hess^{\reidx} = -\frac{\partial\log\redens}{\partial\re\partial\re\Tr} = \normalchol\normalD\normalchol\Tr.
$$
We seek derivatives of the elements of $\hess^{\reidx}$ with respect to $\re,\varparam,\corparam$.
Clearly $(\partial/\partial\re)\hess^{\reidx} = 0$ as Eq. \ref{eqn:normaldensity} is a quadratic function of $\re$.
For the variance parameters,
\begin{align*}
    \frac{\partial}{\partial\delta_l}\hess^{\reidx} = \exp(\delta_l)\normalchol_l\normalchol_l\Tr, \qquad
    \frac{\partial}{\partial\phi_l}\hess^{\reidx} &= \normalchol\normalD\frac{\partial\normalchol\Tr}{\partial\phi_l} + \frac{\partial\normalchol}{\partial\phi_l}\normalD\normalchol\Tr.
\end{align*}
The remaining computational details pertaining to these expressions are given in Section \ref{supp:algorithms} of the supplement.

While this log-Cholesky parameterization of $\Varmat\inv$ is useful for model fitting, 
point and interval estimates for elements of $\Varmat$ are required on the
original scale for interpretation.
Point estimates are obtained in a straightforward manner by invoking
the invariance of maximum likelihood to transformations, and simply computing
$
\Varmatapprox = \normalcholapprox\normalDapprox\normalcholapprox\Tr,
$
where $\normalcholapprox$ and $\normalDapprox$ are $\normalchol$ and $\normalD$ evaluated at 
the approximate MLEs, $\corparamapprox$ and $\varparamapprox$, obtained by indexing the appropriate
components of $\AQmle{\quadnum}$.
Interval estimates require innovation, and here we provide delta method Wald intervals for $\Varmat_{ij}$ based on Wald intervals for $\corparamapprox$ and $\varparamapprox$.
Define $\sigma_{ij} = \Varmat_{ij}$ and $\varcompapprox_{ij} = \Varmatapprox_{ij}$. We use the approximation
$$
\text{Var}(\varcompapprox_{ij})\approx \Jac_{ij}\Tr\text{Var}(\varparamapprox,\corparamapprox)\Jac_{ij},
$$
where $\text{Var}(\varparamapprox,\corparamapprox)$ is the appropriate block of $\approxhess\inv$ (the inverse of the approximate Hessian; see Section \ref{subsec:confidenceintervals}), and 
$$
\Jac_{ij}\Tr = \frac{\partial\varcompapprox_{ij}}{\partial(\varparamapprox,\corparamapprox)}
$$
is the Jacobian of the map $(\varparamapprox,\corparamapprox)\to\varcompapprox_{ij}$.
The dimension of $(\varparamapprox,\corparamapprox)$ is $\resddim = \redim(\redim+1)/2$,
the dimension of $\Jac_{ij}$ is $\resddim\times1$, 
and the dimension of $\text{Var}(\varparamapprox,\corparamapprox)$ is $\resddim\times\resddim$.
We require an expression for $\Jac\Tr_{ij} = (\partial\varcompapprox_{ij} / \partial(\varparamapprox,\corparamapprox)_1,\ldots,(\partial\varcompapprox_{ij} / \partial(\varparamapprox,\corparamapprox)_\resddim)$.
This is given as follows:
\begin{equation*}
\frac{\partial\Varmat\inv_{ij}}{\partial\delta_l} = \exp(\delta_l)\normalchol_{il}\normalchol_{jl}, 
\frac{\partial\Varmat\inv_{ij}}{\partial\phi_l} = \sum_{t=1}^{\redim}\exp(\delta_t)\left\{\frac{d\normalchol_{it}}{d\phi_l}\normalchol_{jt} + \normalchol_{it}\frac{d\normalchol_{jt}}{d\phi_l}\right\}, 
\frac{\partial\Varmat}{\partial(\varparamapprox,\corparamapprox)_l} = -\Varmat\frac{\partial\Varmat\inv}{\partial(\varparamapprox,\corparamapprox)_l}\Varmat.
\end{equation*}
Because $\redim$ is assumed small enough for quadrature to be computationally efficient in the first place, $\redim$ and $\resddim$ are small enough that matrix computations, including explicit inversion, can be performed naively at minimal computational cost.
For example, a random slopes model (see Eq. \ref{eqn:simmodel1}) has $\redim=2$ and $\resddim=3$, leading to
negligible computational burden relative to that required to obtain $\AQmle{\quadnum}$.

The only quantities for which further detail is required are the selection matrix, $\Su$, its derivative-vector/matrix products, $\normalD\{d\Su/du_j\}$ and $\{d\Su\Tr/du_j\}\normalD$, and the derivative matrices, $d\normalchol_{it}/d\phi_l$.
We offer Algorithms 4--6 in Section \ref{supp:algorithms} of the supplement which describe computation of these quantities.
Our algorthmic approach to these computations avoids complicated higher-order tensor algebra, despite the procedure depending on third-order derivatives of multivariable functions. 
The result is an efficient implementation of the whole procedure that involves only matrix and vector algebra.

Finally, for the variances, $\varcompapprox_{ii}$, we compute Wald intervals on the log-scale as $\log\varcompapprox_{ii}\pm z_{1-\alpha/2}\text{se}(\log\varcompapprox_{ii})$,
where $\text{se}(\log\varcompapprox_{ii}) = \text{Var}(\log\varcompapprox_{ii})^{1/2}$ and
$$
\text{Var}(\log\varcompapprox_{ii}) \approx \text{Var}(\varcompapprox_{ii}) / \varcompapprox_{ii},
$$
and then obtain an interval for $\varcomp_{ii}$ by exponentiating this interval. This ensures that the lower bounds
always remain positive, and generally appears to give more accurate intervals.

\subsection{Empirical Evaluation}\label{subsec:simulations}

Two sets of simulations offer empirical evidence that: (a) inferences about $\params$ using $\AQmle{\quadnum}$ are accurate for some $\quadnum$, and that there are compelling cases where $\quadnum>1$ is
required, motivating the need for novel methods to make such inferences efficiently; and (b) the novel methods of Section \ref{sec:methods} and the setup of the present Section yield computations that are fast and stable relevant to existing, established methods. On the former point, $\quadnum>1$ is seen to be most impactful for cases involving any or all of large $\numgroups$, small $\numpergroup$, or large $\resd$; the Laplace approximation ($\quadnum=1$) gives exceptionally poor inferences in these cases, and this is mitigated by choosing large enough $\quadnum$. 
On the latter point, we find evidence that the proposed method yields results at least as favourable as existing methods in shorter amounts of time. We strongly emphasize that the proposed methods do not \emph{compete} with existing methods that already give favourable results, but rather that they \emph{complement} existing methods by replacing cumbersome derivative-free optimization or finite difference-based gradient calculations with exact gradient calculations, typically yielding a factor of $2-4$ reduction in computing time.

\subsubsection{Absolute performance}\label{subsubsec:absolute}

We use a similar simulation setup to \citet{breslow_approximate_1993}, considering the following model:
\begin{equation}\label{eqn:simmodel1}
  \response_{ij} \setdelim \re_i \indsim \text{Bern}(p_{ij}), \ \re_i \iidsim \text{N}\{\zero,\Varmat(\resd)\}, \ 
  \log\frac{p_{ij}}{1-p_{ij}} = \beta_0 + \beta_1\covi_i + \beta_2 t_j + \beta_3\covi_i t_j + \reidx_{i1} + \reidx_{i2}t_j.
\end{equation}
This is model (\ref{eqn:bernoullimodel}) with $\cov_{ij} \equiv (\covi_{i},t_j)\Tr$ and $\recov_{ij} = (1,t_j)\Tr$.
The group-specific covariate, $\covi_i$, takes value $1$ for half the groups and value $0$ for the other half.
The measurement-specific covariate, $t_j$, takes values on an equally-spaced grid of length $\numpergroup_i$ from $-3$ to $3$.
We consider all combinations of $\numgroups = \{100,200,500,1000\}$, $\numpergroup = \{3,5,7,9\}$, and $\quadnum = \{1,3,\ldots,23,25\}$.
\citet{breslow_approximate_1993} considered only $\numgroups = 100$, $\numpergroup=7$, and $\quadnum=1$, concluding that
the PQL approach based on the Laplace approximation appeared negatively biased for binary data; they do not assess coverage of confidence intervals. 
Our simulations recover the claim about bias and provide additional evidence that confidence intervals based on the Laplace approximation achieve very poor coverage,
but that $\quadnum\gg1$ mitigates these shortcomings.

We deliberately focus on small $\numpergroup$ because this situation appears to be common in practice but yields the least accurate
integral approximations, and so is of importance in the present context.
We would expect that for larger $\numpergroup$, the Laplace approximation ($\quadnum=1$) would be adequate,
and hence we do not focus on this case. 

For the variance matrix, we choose
$$
\Varmat(\resd) = \begin{pmatrix} \sigma^2_{1} & \sigma_{12} \\ \sigma_{12} & \sigma^2_{2} \end{pmatrix} = \begin{pmatrix} 2 & 1 \\ 1 & 1 \end{pmatrix},
$$
which gives $\text{Corr}(u_{i1},u_{i2}) \approx 0.71$. These variance components are on the logit scale, and 
hence this can be considered a very difficult situation of random effects having high variance and high correlation.
\citet{breslow_approximate_1993} consider $\text{Corr}(u_{i1},u_{i2}) = 0$ only, and smaller $\sigma^2_1 = 0.5,\sigma^2_2 = 0.25$.
Finally, we use $\regparam = (-2.5,-.15,.1,.2)\Tr$, which yields imbalanced responses due to the low value for $\beta_0$. This further increases the difficulty of making inferences. 
Overall, this is a challenging simulation setup.

We report complete and detailed performance results for all seven parameters in Section \ref{supp:simulations} of the supplement.
Here we report results for $\beta_0$, $\sigma_1$, and $\sigma_{12}$, which 
were parameters for which we observed the largest differences in inferences for different
values of $\quadnum$.
We conclude that for all parameters and combinations of $\numgroups$ and $\numpergroup$, it appears
possible to choose a large enough $\quadnum$ such that inferences based on $\AQmle{\quadnum}$
have low bias and nominal interval coverage.
This is especially relevant to practice, as---unlike $\numgroups$ and $\numpergroup$---the
practitioner may always increase $\quadnum$, and the novel methods in the present paper
make doing so much faster and easier.
The simulation study in Section \ref{subsubsec:relative} reports computational comparisons to existing methods.

\begin{figure}[t]
\centering
\subfloat[Bias, $\widehat{\beta}_1$]{
  \centering
  \includegraphics[width=2in]{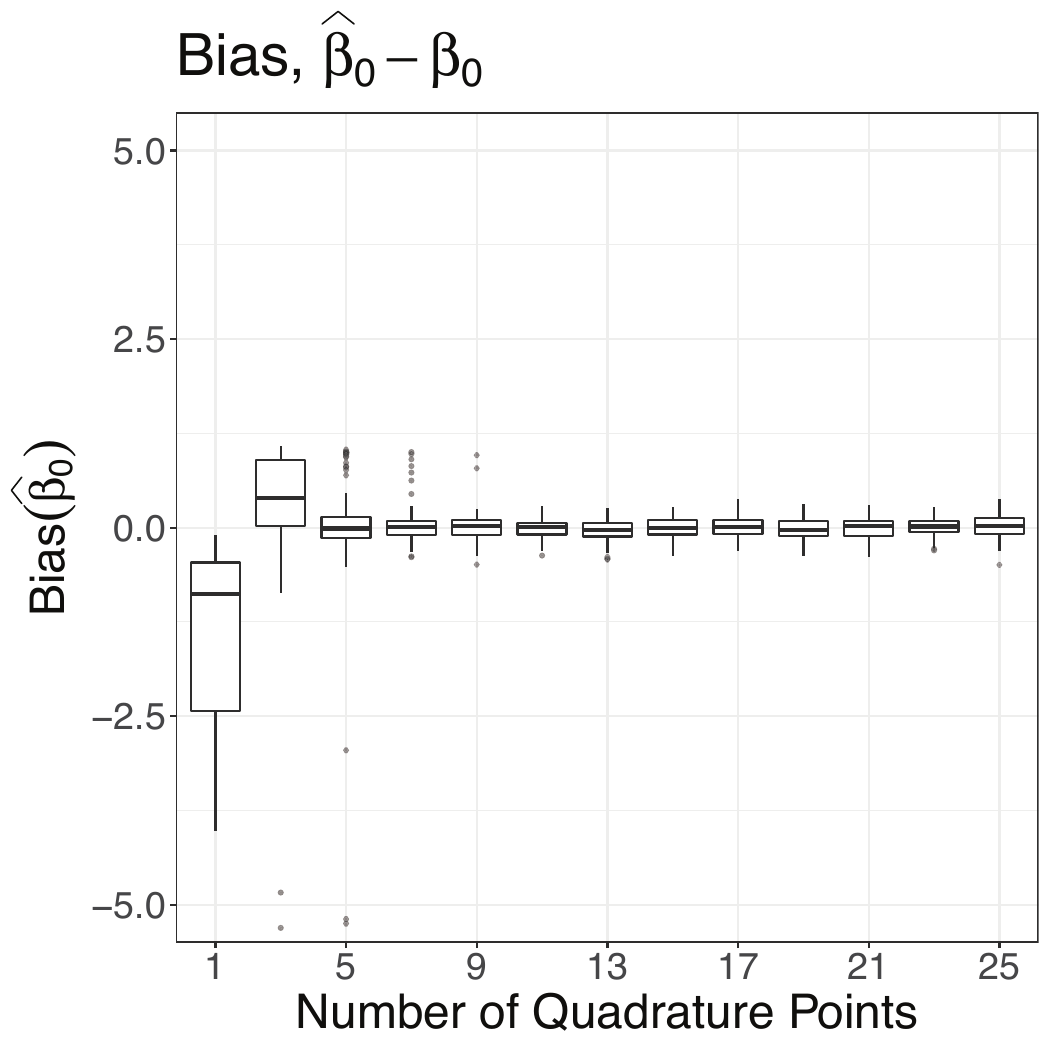}
}
\subfloat[Bias, $\widehat{\sigma}^2_1$]{
  \centering
  \includegraphics[width=2in]{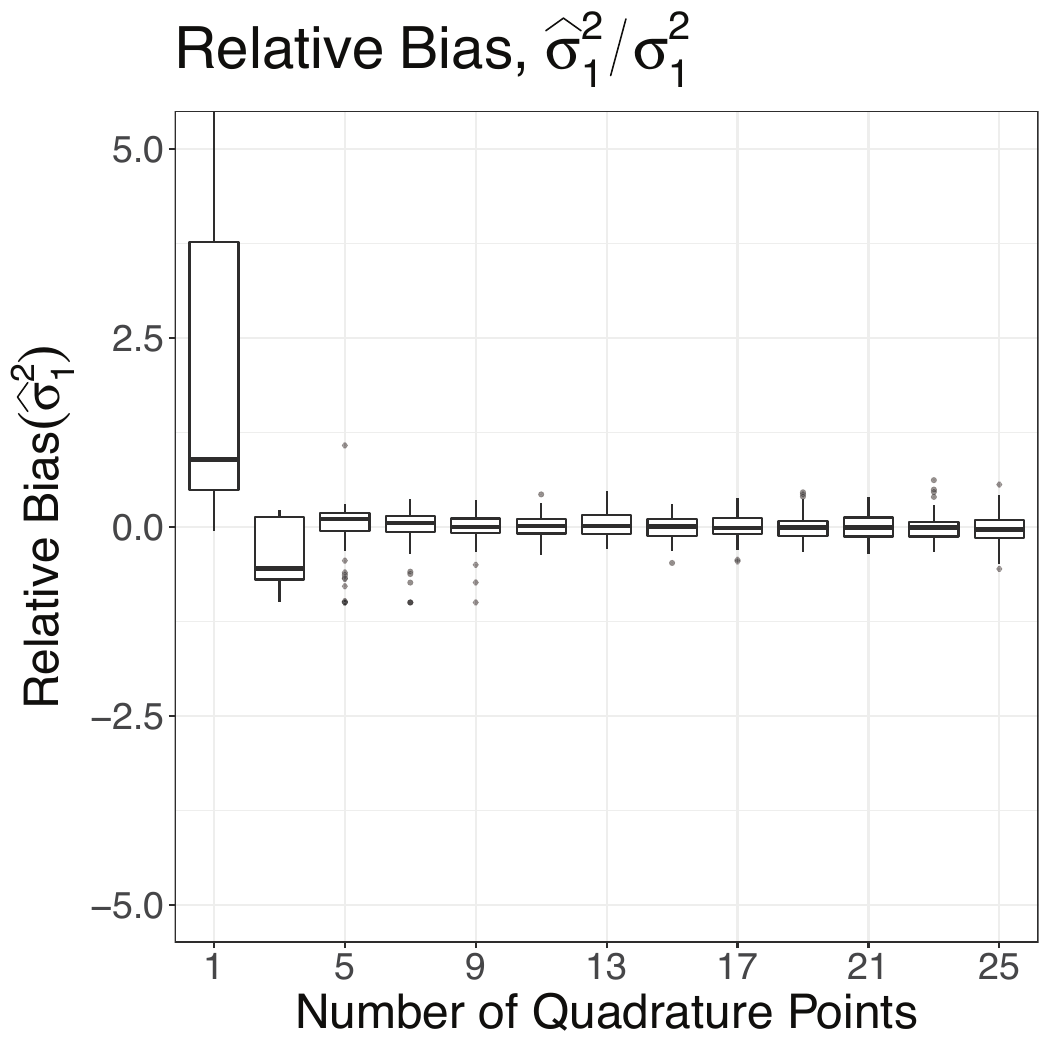}
}
\subfloat[Bias, $\widehat{\sigma}_{12}$]{
  \centering
  \includegraphics[width=2in]{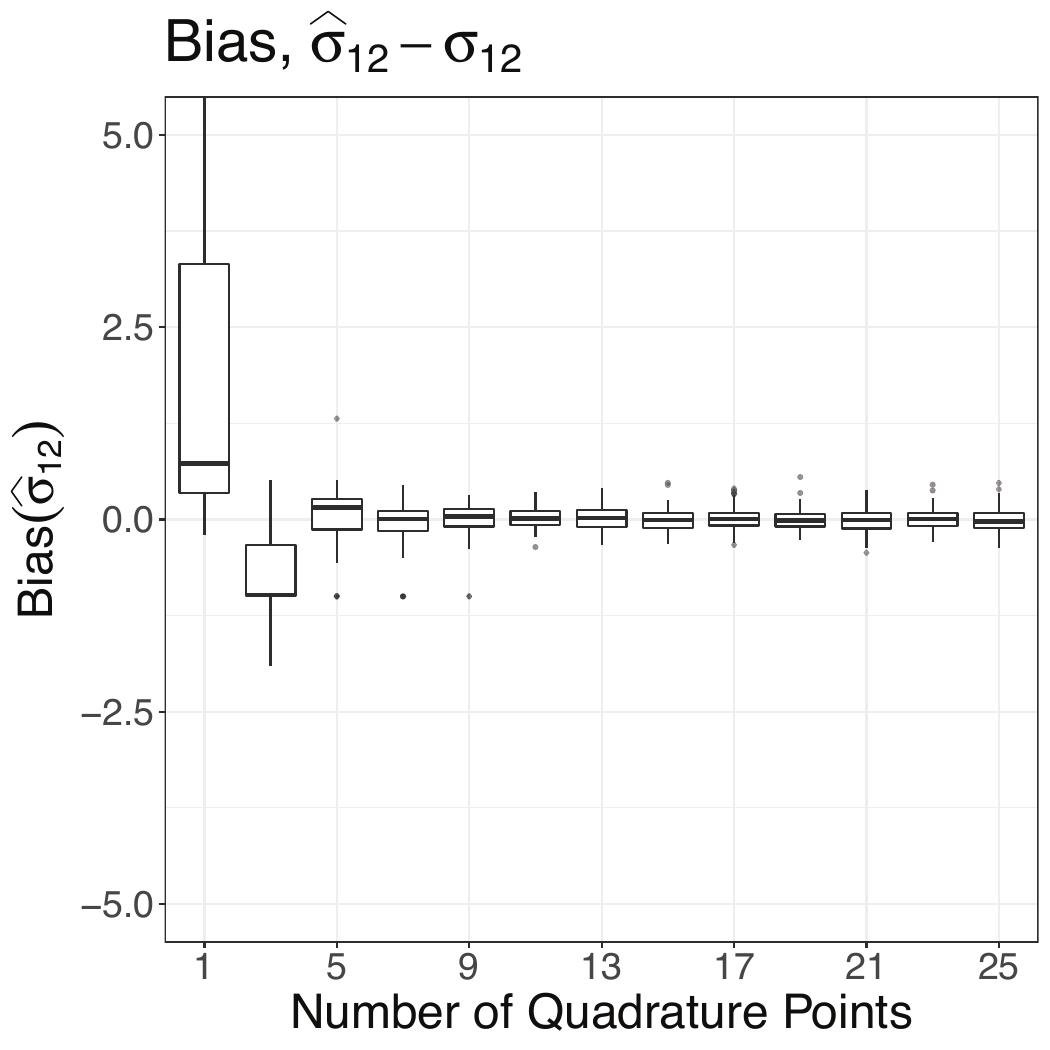}
}
\\
\subfloat[Empirical Coverage Proportion, $\widehat{\beta}_1$]{
  \centering
  \includegraphics[width=2in]{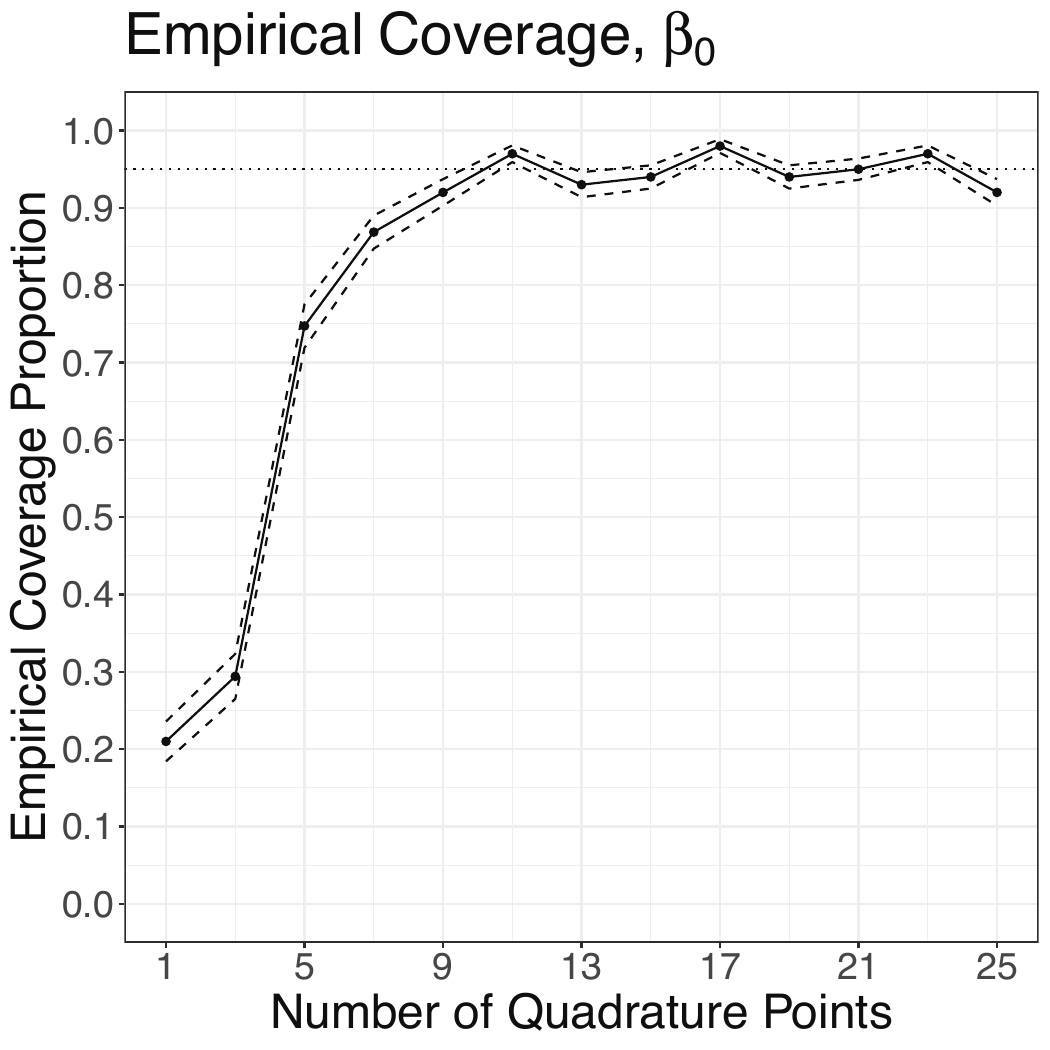}
}
\subfloat[Empirical Coverage Proportion, $\widehat{\sigma}^2_1$]{
  \centering
  \includegraphics[width=2in]{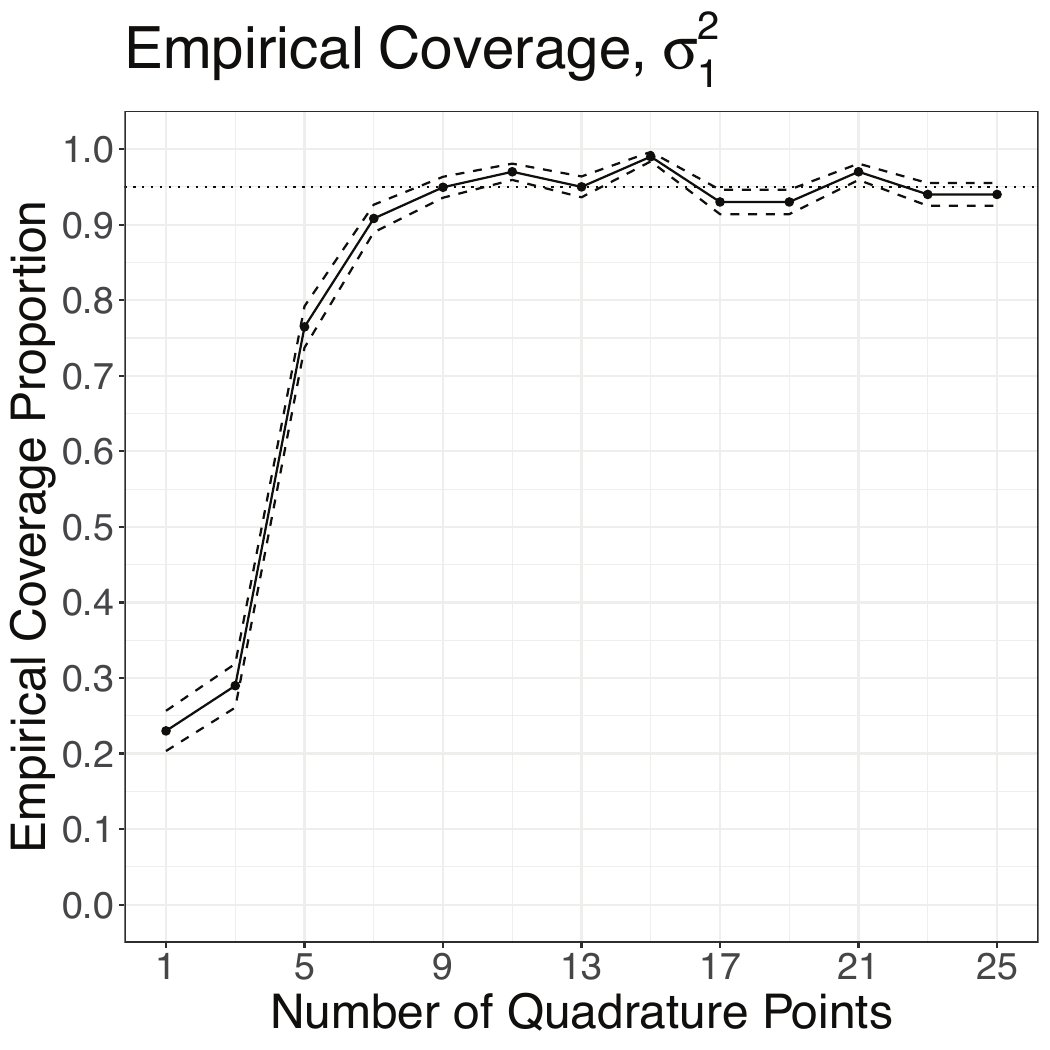}
}
\subfloat[Empirical Coverage Proportion, $\widehat{\sigma}_{12}$]{
  \centering
  \includegraphics[width=2in]{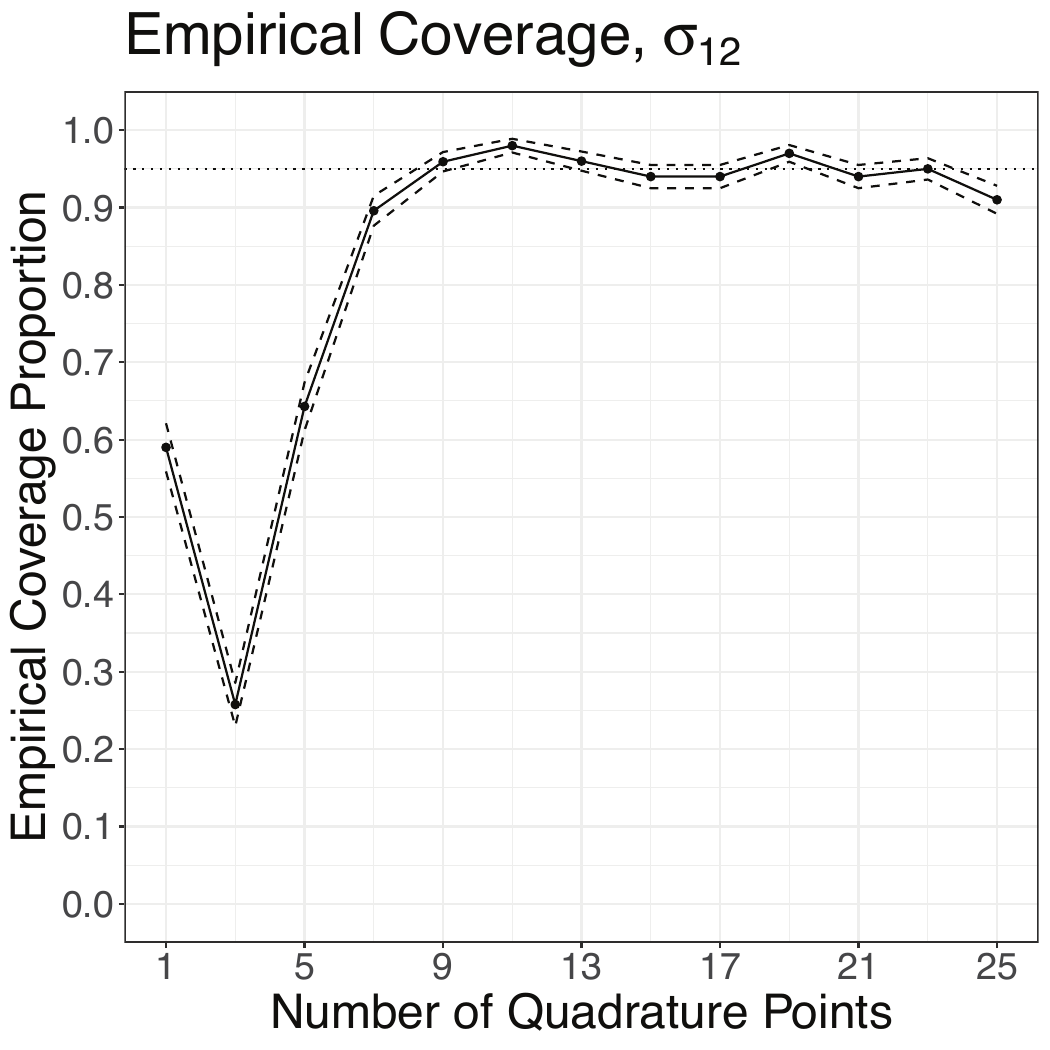}
}
\\
\subfloat[Interval Lengths, $\widehat{\beta}_1$]{
  \centering
  \includegraphics[width=2in]{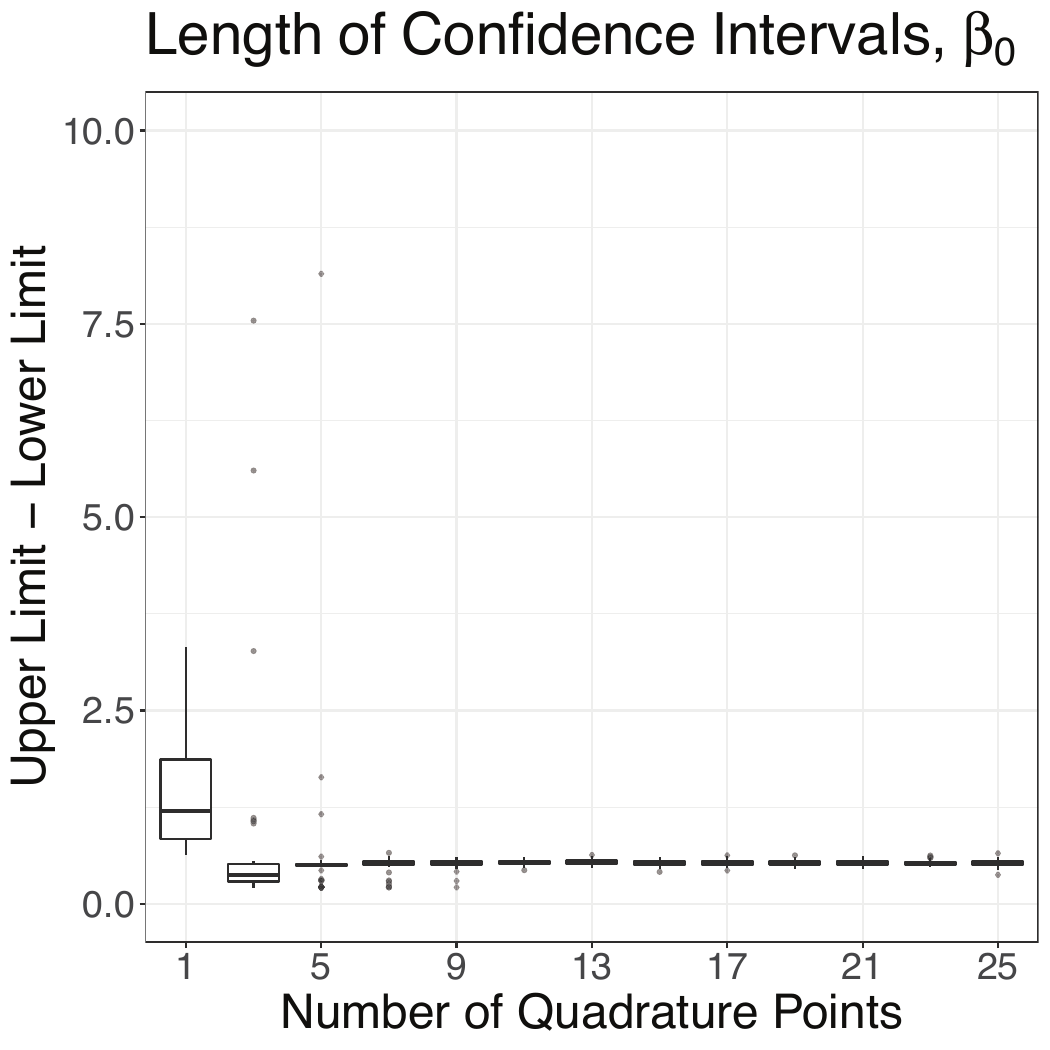}
}
\subfloat[Interval Lengths, $\widehat{\sigma}^2_1$]{
  \centering
  \includegraphics[width=2in]{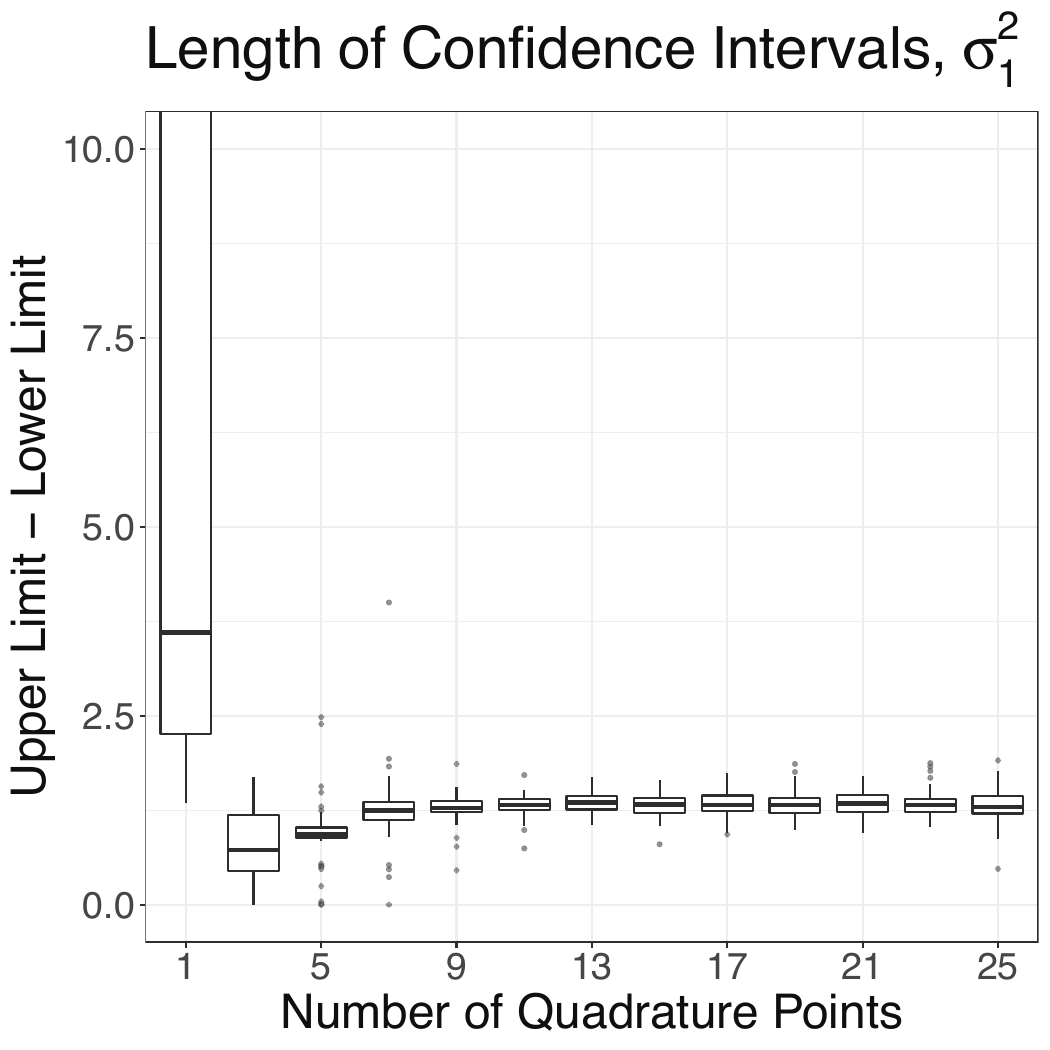}
}
\subfloat[Interval Lengths, $\widehat{\sigma}_{12}$]{
  \centering
  \includegraphics[width=2in]{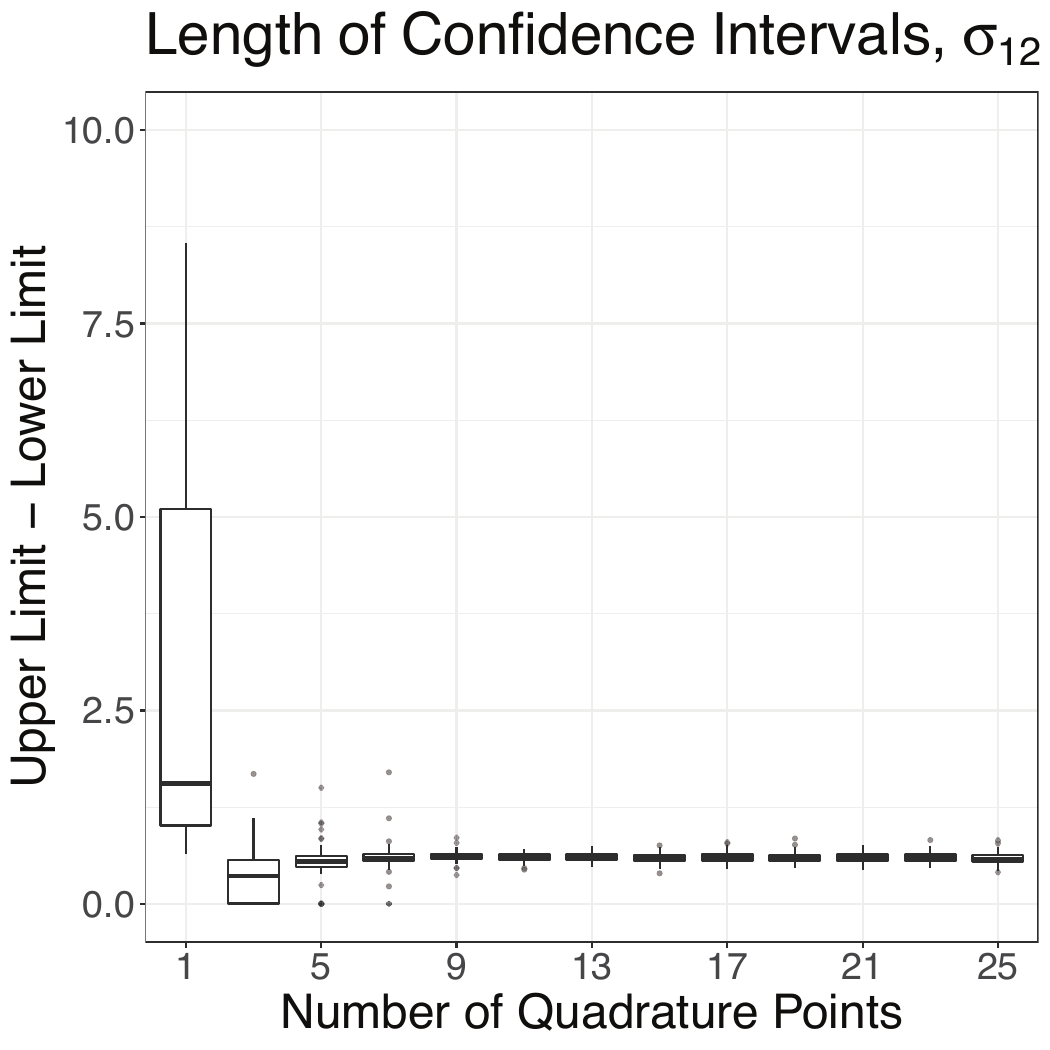}
}
\caption{Simulation results for $1000$ sets of data generated from model (\ref{eqn:simmodel1})
using $\numgroups=1000$ and $\numpergroup=5$, and with large variance components ($\sigma^2_1 = 2,\sigma_{12} = 1$) and imbalanced
binary response ($\beta_0 = -2.5$, logit scale). In all cases, taking $\quadnum$ large enough yields low bias and intervals
attaining nominal coverage. Lower $\quadnum<11$ or so yields higher bias, low coverage,
and wide intervals, with the Laplace approximation $(\quadnum=1)$ yielding exceptionally poor inferences.}
\label{fig:mainsimulationresults}
\end{figure}

Results are presented in Figure \ref{fig:mainsimulationresults}. We report the absolute bias, $\bias(\param;\quadnum) = \EE\AQmleidx{\quadnum} - \param$, for $\beta_0$ and $\sigma_{12}$, and the relative bias, $\relbias(\param;\quadnum) = \EE\AQmleidx{\quadnum} / \param$, for $\sigma_1^2$. The bias is exceptionally high for the Laplace approximation ($\quadnum=1$), which is consistent with comments made by
\citet{breslow_approximate_1993}. The distribution of the bias appears to converge to one centred at $0$ as $\quadnum$ is increased. We report the empirical coverage proportions and lengths (which are of course simply scalar multiples of the standard errors) of the Wald intervals (Sections \ref{subsec:confidenceintervals} and \ref{sec:normal}) for the same parameters. Similar to the bias, the coverages appear to converge to nominal as $\quadnum$ is increased, with poor performance for lower $\quadnum$. For $\sigma_{12}$, the improved performance of $\quadnum=1$ relative to $\quadnum=3$ is attributed to the $\quadnum=1$ intervals being far wider than those for $\quadnum=3$, which
appear too narrow relative to higher $\quadnum$. 
We emphasize that the conclusion here is that
we appear to be able to always choose a large enough $\quadnum$ to provide valid inferences.
The novel methods in the present paper make such computations more convenient for the practitioner.

\subsubsection{Relative performance}\label{subsubsec:relative}

We perform a second simulation study to attempt to isolate the impact of using the new, exact gradient computations on
computational time and estimation quality. To this end, we simulate datasets from the model (\ref{eqn:simmodel1}) and fit
this model using both the new approach and the \texttt{GLMMadaptive} \texttt{R} package \citep{rizopoulos_glmmadaptive_2020,r_core_team_r_2021}.
Because the \texttt{lme4} software does not fit models with multivariate random effects by AQ with $\quadnum>1$, we also consider the following
random intercepts model:
\begin{equation}\label{eqn:simmodel2}
  \response_{ij} \setdelim \reidx_i \indsim \text{Bern}(p_{ij}), \ \reidx_i \iidsim \text{N}(0,\sigma^2), \ 
  \log\frac{p_{ij}}{1-p_{ij}} = \beta_0 + \beta_1\covi_i + \reidx_{i}.
\end{equation}
Model (\ref{eqn:simmodel2}) is model (\ref{eqn:simmodel1}) with $\beta_2 = \beta_3 = \sigma_2 = 0$, and we set $\regparam = (-2.5,-.15)$ and $\sigma^2 = 2$.
We use model (\ref{eqn:simmodel2}) to compare the new approach to \texttt{lme4}, where we use the alternative
efficient scalar implementation of the new approach briefly described in Section \ref{subsec:scalar}.
We include only the results for the combination of $\numgroups=1000$ and $\numpergroup=5$ shown also in Figure \ref{fig:mainsimulationresults}; again, the complete analysis presented in Section \ref{supp:dataanalysis} of the supplementary materials.

Figure \ref{fig:compsimulationresults} shows boxplots of (a) the relative computation times, (b) the difference in minimized approximate base-$10$ average negative log-likelihood values, $-\approxloglikAQMLE{\quadnum}/(\numtotal\log10)$, and (c) the difference in base-$10$ logarithms of the $2$-norm of the exact gradient of the minimized average approximate negative log-likelihood values, $\log_{10}\|\nabla\approxloglikAQall{\quadnum}(\AQmle{\quadnum})\|_2/(\numtotal\log10)$, for the new approach against each of \texttt{GLMMadaptive} and \texttt{lme4}. Positive values indicate that each existing approach returned an MLE, $\AQmle{\quadnum}$, having a higher minimized negative log-likelihood
value, or higher gradient norm at the minimum than the new approach; in all metrics, a higher value therefore means that the new method performed favourably compared to an existing method.
We find that the new approach tends to give results that are at least as favourable as existing methods with a typical factor of $2-4$ reduction in computation time than each existing method, with slight variations for different values of $\quadnum$.
The expanded simulations in Section \ref{supp:simulations} of the supplementary materials show that this tends to hold across different values of $\numgroups$ and $\numpergroup$.

Both \texttt{GLMMadaptive} and \texttt{lme4} return
comparable satisfactory results to the new method.
The new method returns these results in a shorter computation time.
This suggests that the use of exact gradients
has the potential to speed up optimization for mixed models, compared to finite-differenced gradients (\texttt{GLMMadaptive}) or derivative-free optimization techniques (\texttt{lme4}).
The absolute computation times of the chosen examples are mostly a few seconds on the hardware used; however, the relative computation times are expected to hold for larger sets of data for which the absolute computation
times are also larger.

We remark as well that our implementation of this procedure is young and potentially naive, 
while the other two are quite mature and efficient; it is implausible that these gains in speed are due to a superior
implementation of the methods, since it is implausible that the present implementation is in any way superior to these robust and mature
software packages.
The most plausible explanation for the difference in empirical performance is that the use of exact gradients and quasi-Newton L-BFGS minimization is 
appreciably computationally favourable to the use of derivative-free optimization methods or finite-differenced gradient-based quasi-Newton optimization in approximate likelihood inference for mixed models.

\begin{figure}[t]
\centering
\subfloat[Comp. Time, \texttt{GLMMadaptive}]{
  \centering
  \includegraphics[width=2in]{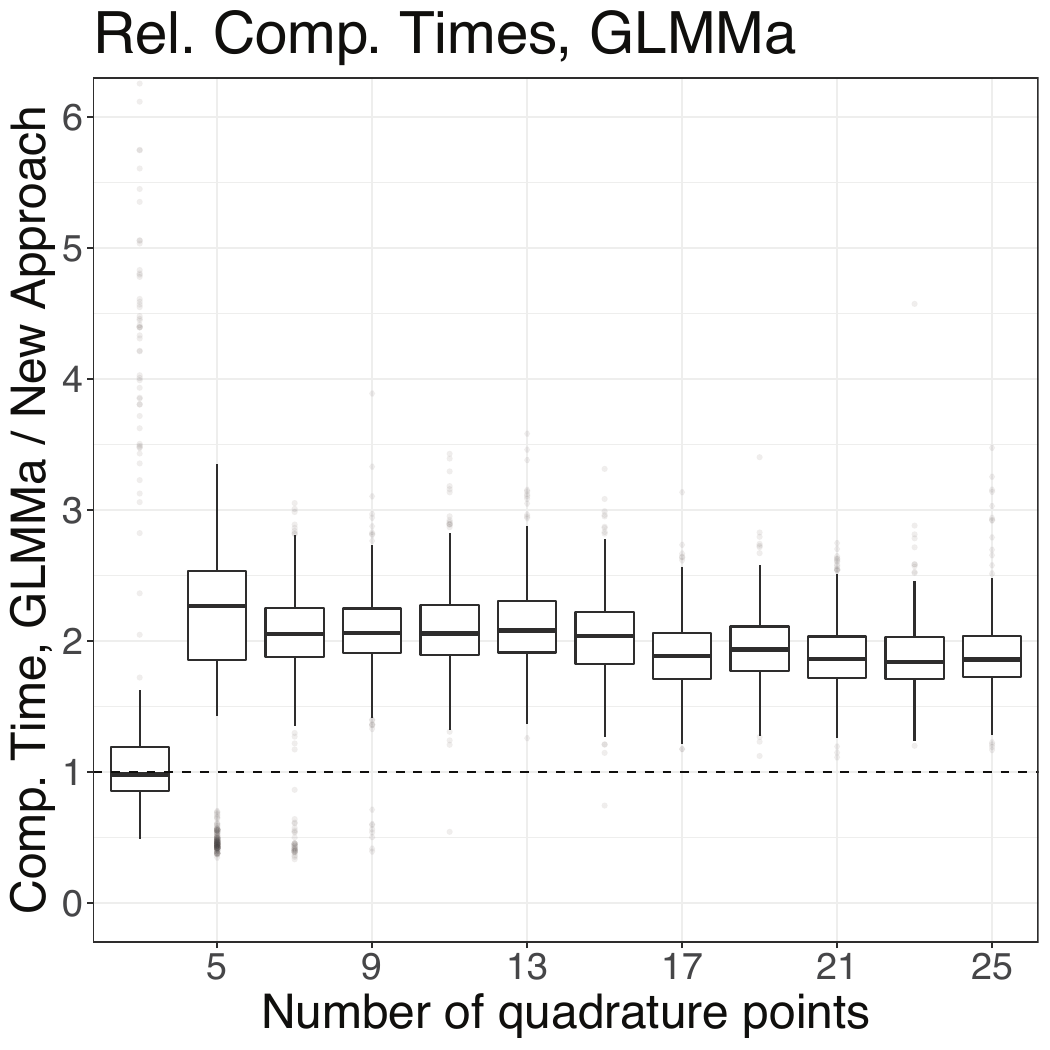}
}
\subfloat[$-\approxloglikAQall{\quadnum}(\AQmle{\quadnum})/(\numtotal\log10)$, \texttt{GLMMadaptive}]{
  \centering
  \includegraphics[width=2in]{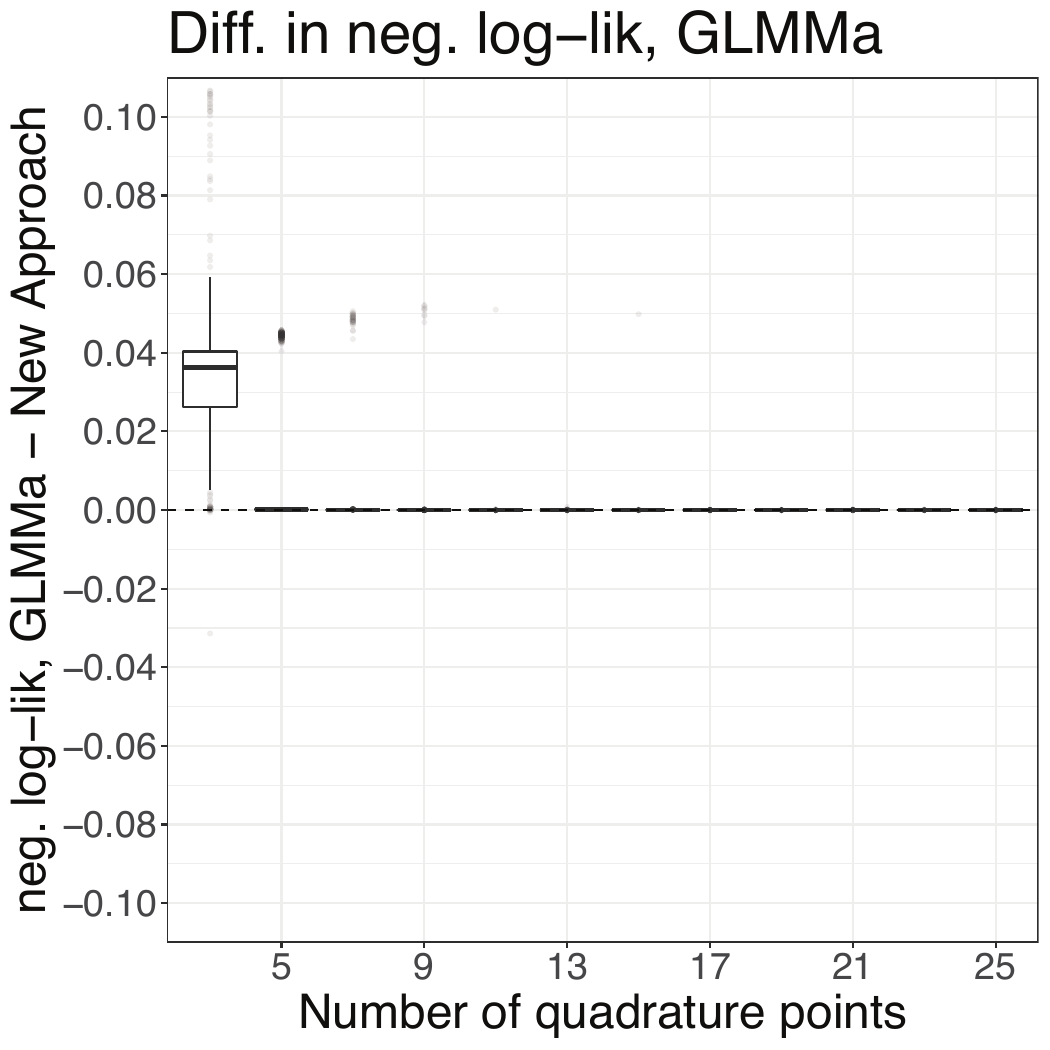}
}
\subfloat[$\log_{10}\|\nabla\approxloglikAQall{\quadnum}(\AQmle{\quadnum})\|_2/(\numtotal\log10)$, \texttt{GLMMadaptive}]{
  \centering
  \includegraphics[width=2in]{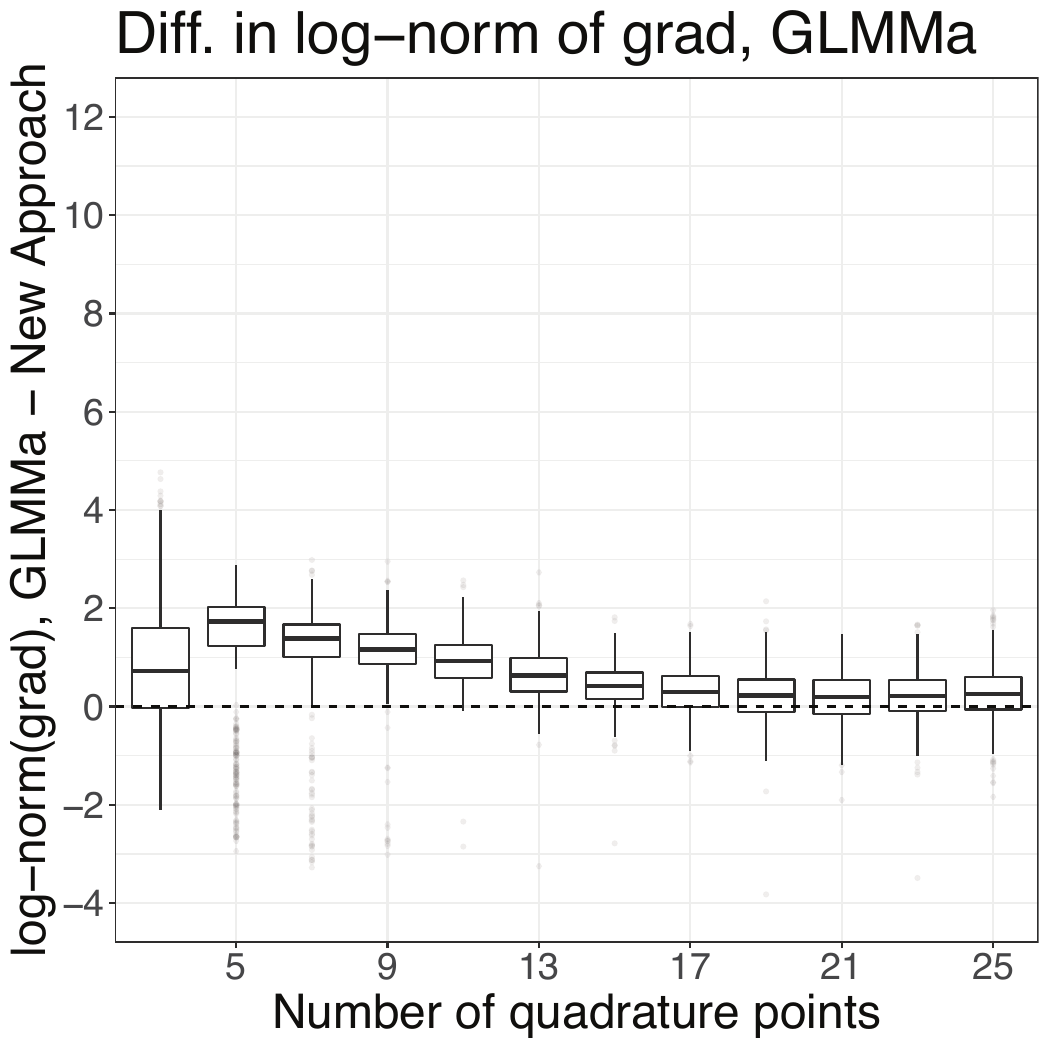}
}
\\
\subfloat[Comp. Time, \texttt{lme4}]{
  \centering
  \includegraphics[width=2in]{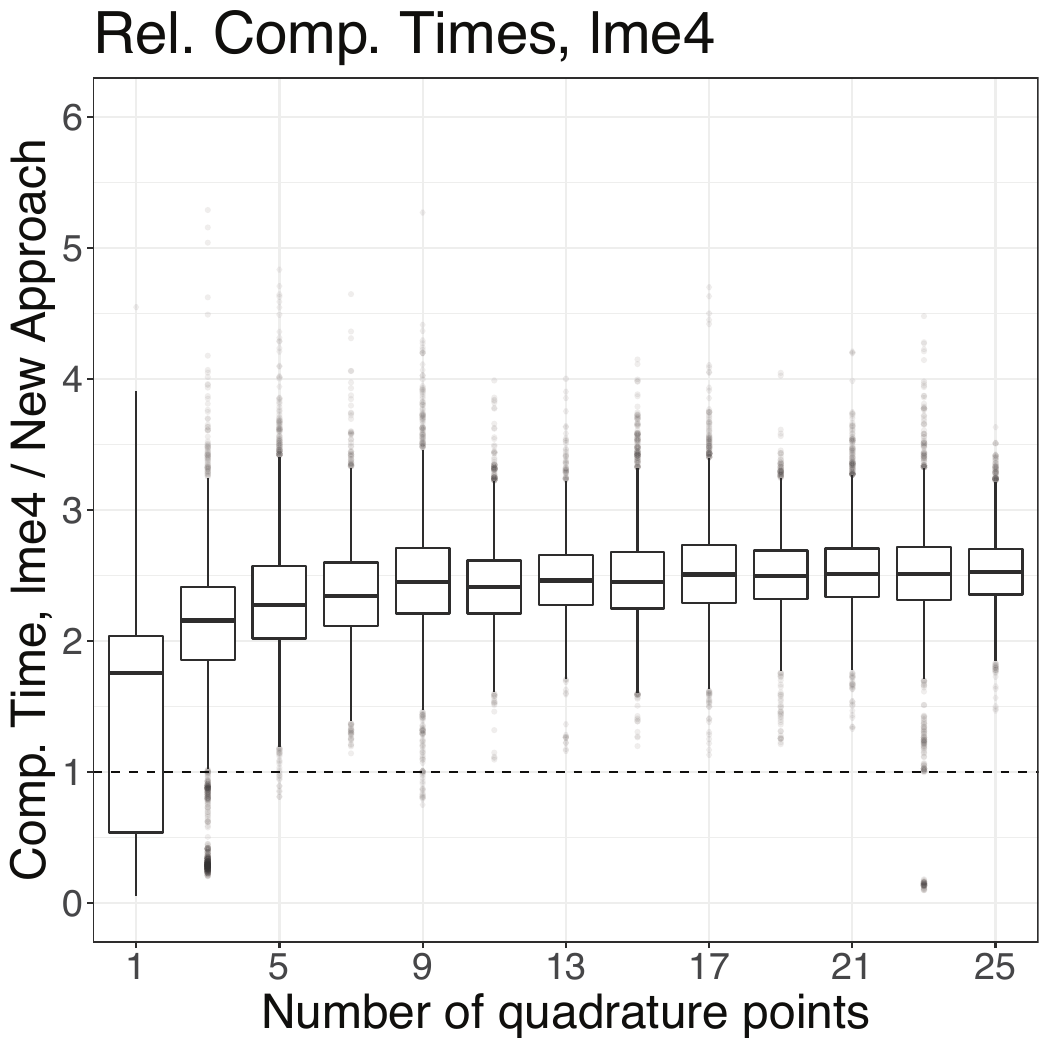}
}
\subfloat[$-\approxloglikAQall{\quadnum}(\AQmle{\quadnum})/(\numtotal\log10)$ difference, \texttt{lme4}]{
  \centering
  \includegraphics[width=2in]{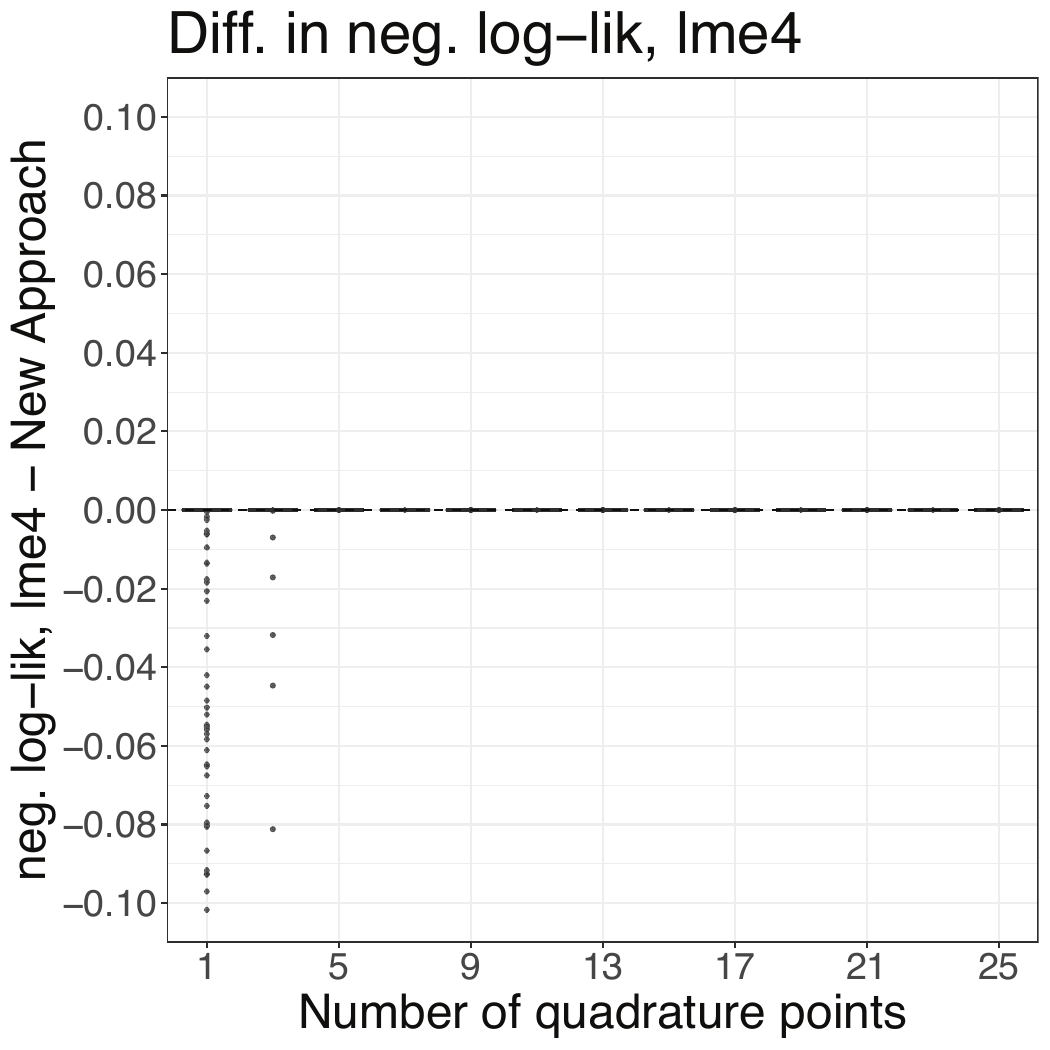}
}
\subfloat[$\log_{10}\|\nabla\approxloglikAQall{\quadnum}(\AQmle{\quadnum})\|_2/(\numtotal\log10)$ difference, \texttt{lme4}]{
  \centering
  \includegraphics[width=2in]{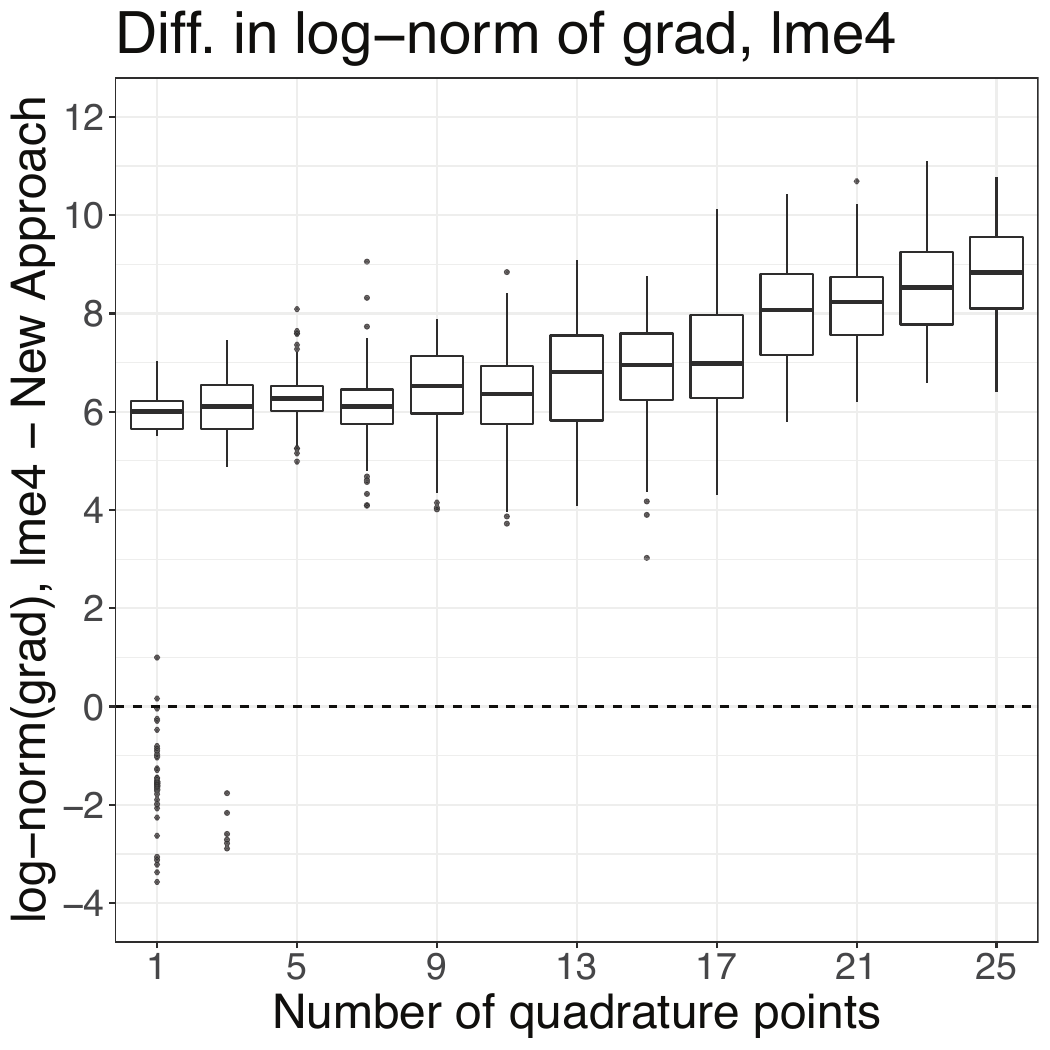}
}
\\
\caption{Simulation results for $500$ sets of data generated from model (\ref{eqn:simmodel1}) (\texttt{GLMMadaptive}, a-c) and $1000$ sets of data generated from model (\ref{eqn:simmodel2}) (\texttt{lme4}, d-f)
using $\numgroups=1000$ and $\numpergroup=5$, and with large variance components ($\sigma^2_1 = 2,\sigma_{12} = 1$) and imbalanced
binary response ($\beta_0 = -2.5$, logit scale).
Plots (b,e) and (c,f) show the differences in base-$10$ average minimized negative log-likelihood, $-\approxloglikAQall{\quadnum}(\AQmle{\quadnum})/(\numtotal\log10)$, and base-$10$ log of the $2$-norm of the gradient of the average log-likelihood at the minimum, 
$\log_{10}\|\nabla\approxloglikAQall{\quadnum}(\AQmle{\quadnum})\|_2/(\numtotal\log10)$, for each existing method minus the new method.
Positive values indicate that the new approach performed favourably. Because both \texttt{GLMMadaptive} and \texttt{lme4} give satisfactory
performance in absolute terms (see the supplement, Section \ref{supp:simulations}), the fact that the new method performs mostly as well or better can be interpreted as the new method not sacrificing performance
in order to achieve its $2-4$ times speed-up in computation time (a,d).
}
\label{fig:compsimulationresults}
\end{figure}

\section{Data Analysis}\label{sec:data}

We present two re-analyses of mixed models previously reported in the literature.
The first provides an example of multivariate random effects, where small $\numpergroup$
with moderate $\numgroups$ leads to very different inferences for different $\quadnum$.
The second provides the same for scalar random effects, in the presence of very large
between-subject heterogeneity.

\subsection{Smoking cessation}\label{subsec:smoking}

\citet{hedeker_note_2018} provide data from a study by \citet{gruder_1993} on the associations between
social support and smoking cessation. The data consist of binary indicators, $\response_{ij}$, with $\response_{ij}=1$
indicating smoking cessation, from
$i = 1,\ldots,m=489$ subjects at $j = 1,\ldots,n_i\in\{1,2,3,4\}$ follow-up times after a televised intervention.
Subjects were randomly assigned to receive social support ($\covi_i=1$) or not ($\covi_i=0$) and followed up at
$t_j = 0,6,12$, or $24$ months post-intervention (note the $t_j$ values in the data are normalized).
The following model is considered:
\begin{equation}\label{eqn:smokingmodel}
    \response_{ij} \setdelim \re_i \indsim \text{Bern}(p_{ij}), \ \re_i \iidsim \text{N}\{\zero,\Varmat(\resd)\}, \ 
    \log\frac{p_{ij}}{1-p_{ij}} = \beta_0 + \beta_1\covi_i + \beta_2 t_j + \beta_3\covi_i t_j + \reidx_{i1} + \reidx_{i2}t_j.
\end{equation}
The random intercept, $\reidx_{i1}$, is included to account for potentially different baseline propensity for cessation
among subjects and the random slope, $\reidx_{i2}$, is included because the correlations between cessation at each
time point appear nonzero; see \citet{hedeker_note_2018} for details.

We fit this model using our new approach and \texttt{GLMMadaptive}; recall that \texttt{lme4} does not fit the random slopes
model with $\quadnum>1$. 
Inspection of their SAS code reveals that \citet{hedeker_note_2018} used $\quadnum=11$; this choice of estimator is not discussed by them.
We fit the model with $\quadnum$ between $1$ and $25$, and observe that
results appear to stop changing for $\quadnum\geq17$ or so.

\begin{figure}[p]
\centering
\subfloat[Estimates, $\widehat{\sigma}^2_1$]{
  \centering
  \includegraphics[width=3in]{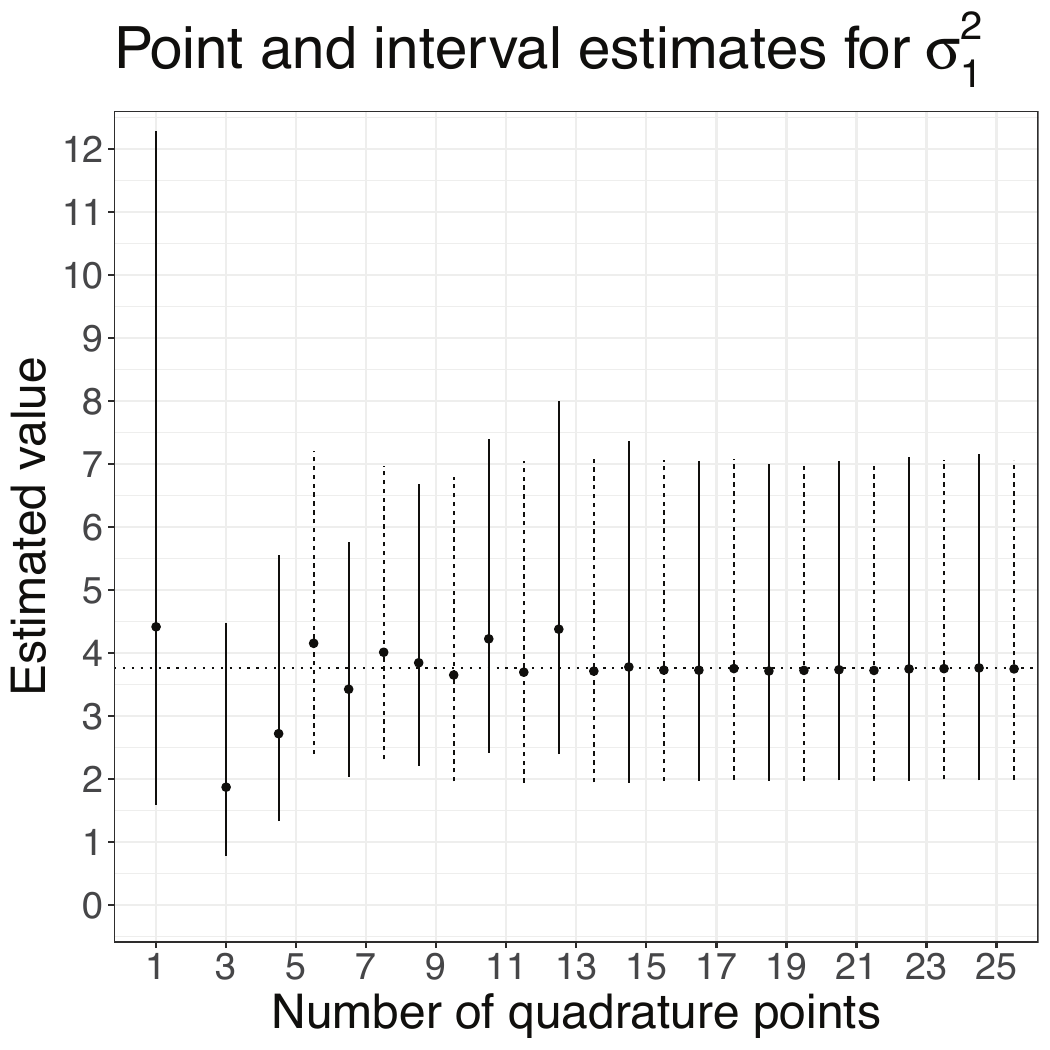}
}
\subfloat[Estimates, $\widehat{\sigma}^2_2$]{
  \centering
  \includegraphics[width=3in]{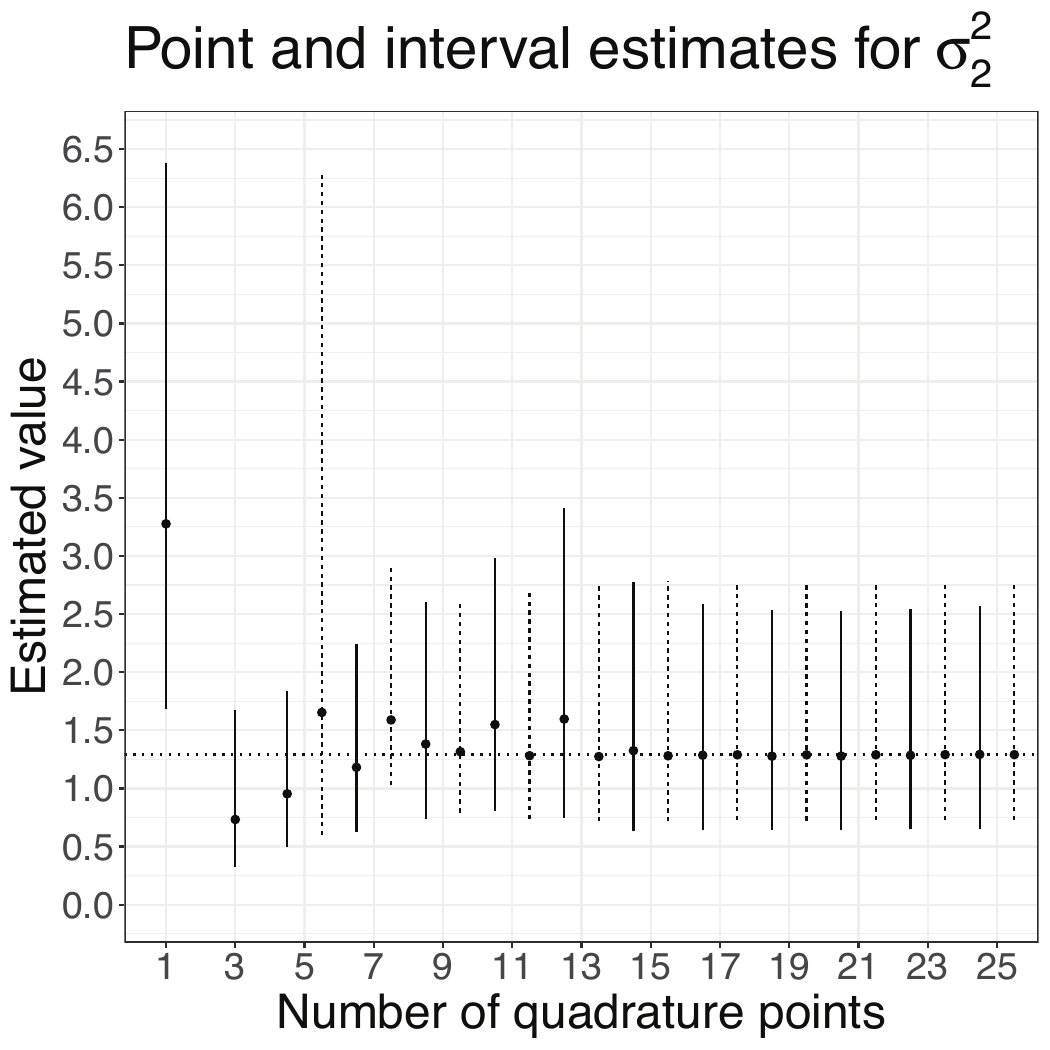}
}
\\
\subfloat[Estimates, $\widehat{\sigma}_{12}$]{
  \centering
  \includegraphics[width=3in]{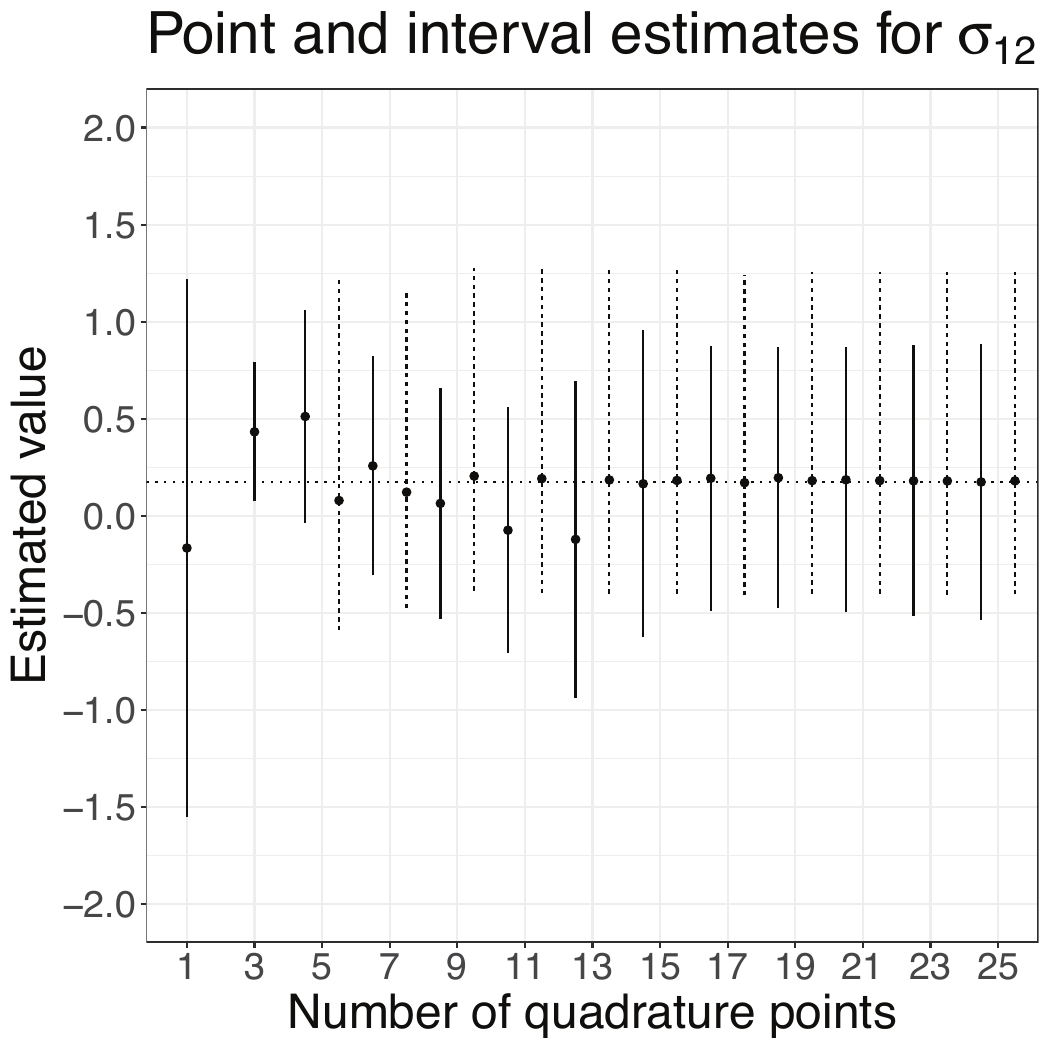}
}
\subfloat[Rel. Comp. Times, \texttt{GLMMa}/new.]{
  \centering
  \includegraphics[width=3in]{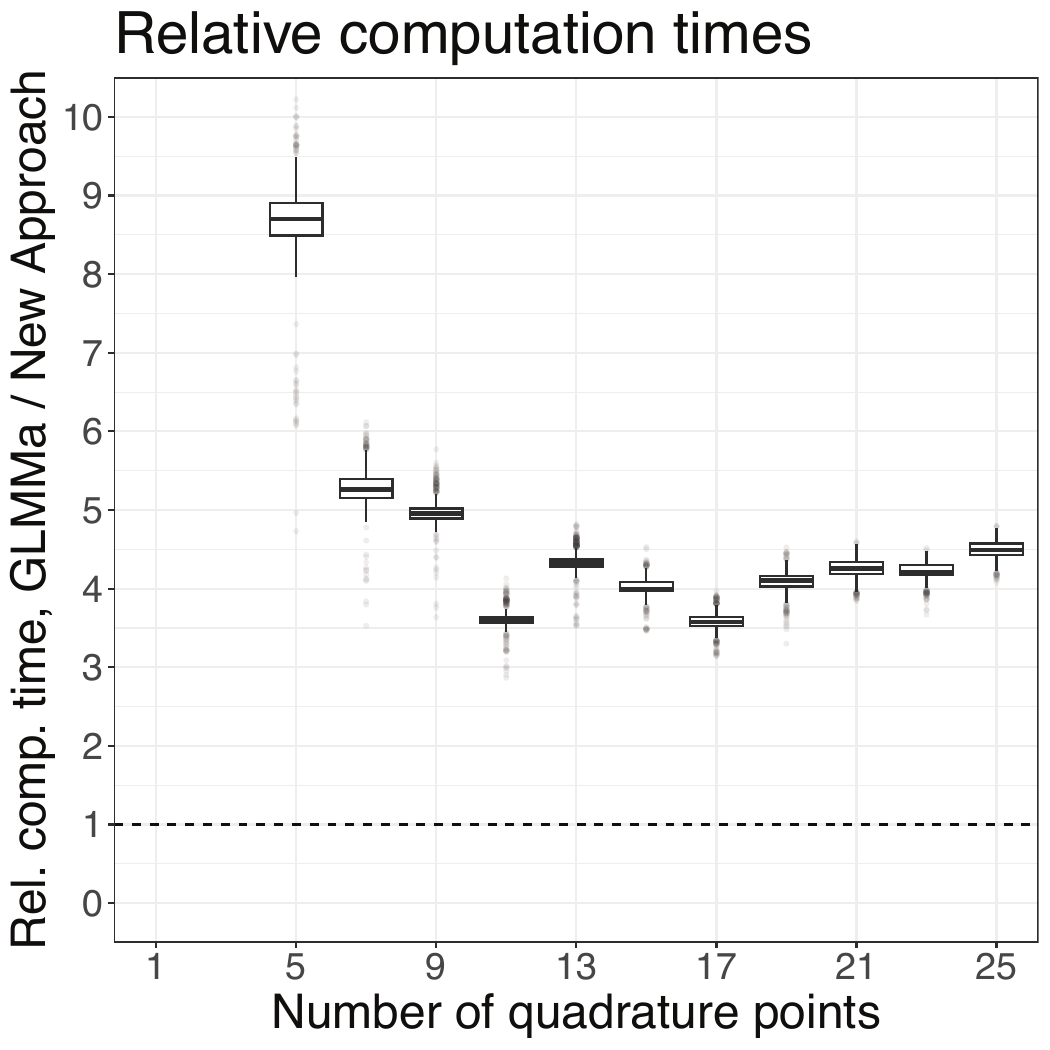}
}
\caption{
Results for the smoking cessation data of Section \ref{subsec:smoking}. (a) -- (c): Parameter estimates for variance components from the new approach (---) and \texttt{GLMMadaptive} (- - -).
In all cases, the estimates stop changing after some $\quadnum$ for both methods. The new approach returns a non-singular
Hessian for all $\quadnum$, while \texttt{GLMMadaptive} does so for $\quadnum>3$.
For $\sigma^2_2$ and $\sigma_{12}$, the intervals for large $\quadnum$ are narrower for the new method than for \texttt{GLMMadaptive},
much so in the case of $\sigma_{12}$.
(d): relative computation times for \texttt{GLMMadaptive} relative to the new approach, based on $500$ repetitions of the fit.
The new approach is between $2.5$ and $10$ times faster than \texttt{GLMMadaptive} for this problem, with some variability across
values of $\quadnum$.
}
\label{fig:smoking-results}
\end{figure}

Figure \ref{fig:smoking-results} (a) -- (c) shows the estimated variance components from both methods; regression coefficients were essentially
the same from both procedures and are shown in Section \ref{supp:dataanalysis} of the supplement.
Of particular note is the very large estimate for $\widehat{\sigma}_1^2$.
In all cases, when $\quadnum$ is large enough, inferences stop changing. 
For $\sigma_{12}$ and to a lesser extent
$\sigma^2_2$, the large-$\quadnum$ intervals are smaller from the new method than \texttt{GLMMadaptive}; combined
with the nominal coverage of these intervals for this parameter in the simulations, this can be regarded as an advantage
of the new method.

Figure \ref{fig:smoking-results} (d) shows the relative computation times of \texttt{GLMMadaptive} compared to the new method
over $500$ repetitions of the fit. Although it varies with $\quadnum$, the new method is usually around $4-5$ times, and in uncommon cases up to 10 times faster
than \texttt{GLMMadaptive}.
It was never less than $2.5$ times faster.

\subsection{Toenail fungus treatment}\label{subsec:toenail}

\citet{lesaffre_effect_2001} report measurements of a binary indicator, $\response_{ij}$, with $\response_{ij}=1$ indicating the absence of toenail
infection, from $i=1,\ldots,\numgroups=294$ subjects who were given an oral treatment for toenail infection $(\covi_{i}=1)$ or not ($\covi_i=0$)
and followed up at $j=1,\ldots,\numpergroup_j\in\{1,\ldots,7\}$ times $t_j = -3,\ldots,3$.
They fit the random intercepts model:
\begin{equation}\label{eqn:toenailmodel}
    \response_{ij} \setdelim \reidx_i \indsim \text{Bern}(p_{ij}), \ \reidx_i \iidsim \text{N}(0,\sigma^2), \ 
    \log\frac{p_{ij}}{1-p_{ij}} = \beta_0 + \beta_1\covi_i + \beta_2 t_j + \beta_3\covi_i t_j + \reidx_{i1}.
\end{equation}
The purpose of the analysis by \citet{lesaffre_effect_2001} was to investigate the sensitivity of inferences to
the choice of $\quadnum$; although focus was on non-adaptive Gauss-Hermite quadrature, they also use AQ, concluding
that inferences appear to stop changing around $\quadnum=10$ or so.
We fit this model to these data using our new method, \texttt{lme4}, and \texttt{GLMMadaptive} to investigate performance
and choice of $\quadnum$.

\begin{figure}[p]
\centering
\subfloat[Estimates, $\widehat{\beta}_0$]{
  \centering
  \includegraphics[width=3in]{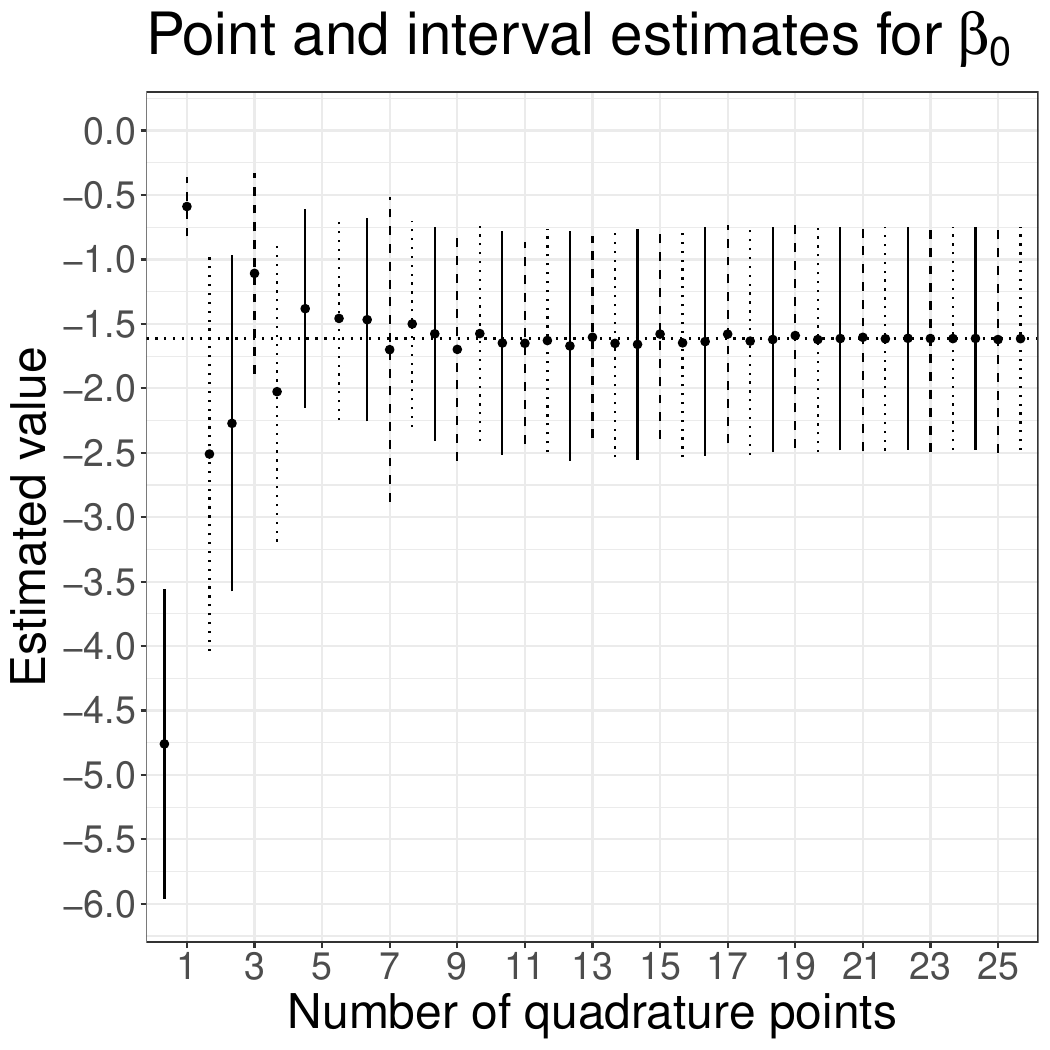}
}
\subfloat[Estimates, $\widehat{\beta}_1$]{
  \centering
  \includegraphics[width=3in]{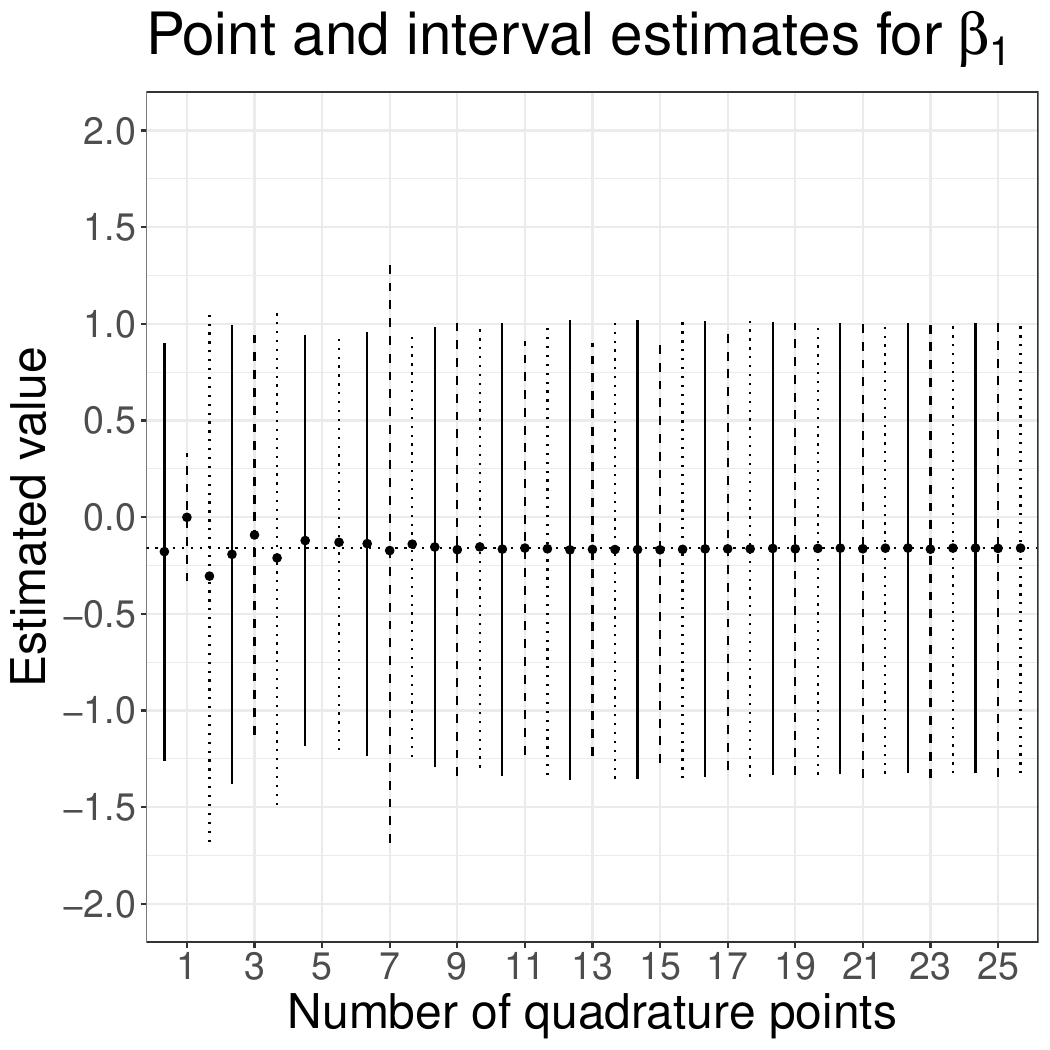}
}
\\
\subfloat[Estimates, $\widehat{\sigma}$]{
  \centering
  \includegraphics[width=3in]{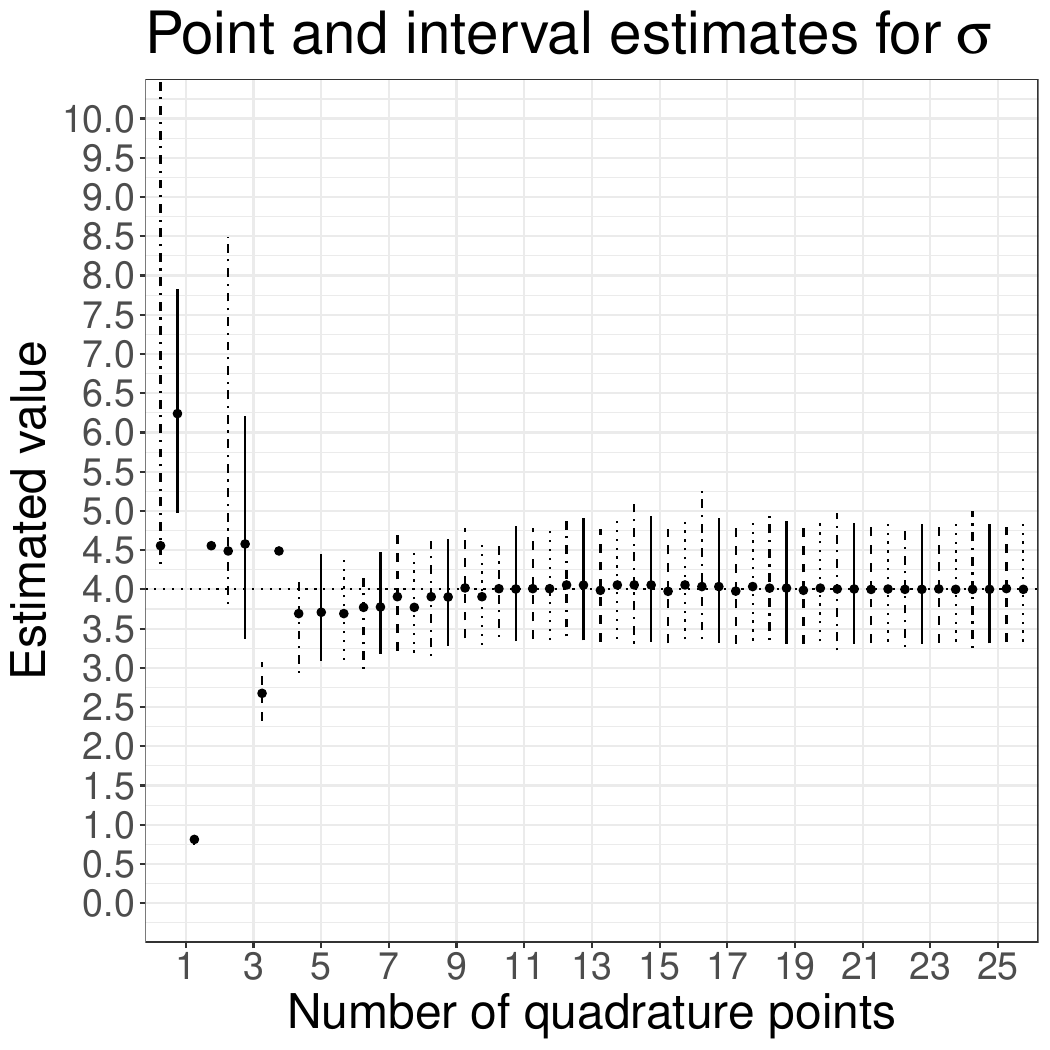}
}
\subfloat[Rel. Comp. Times, existing/new]{
  \centering
  \includegraphics[width=3in]{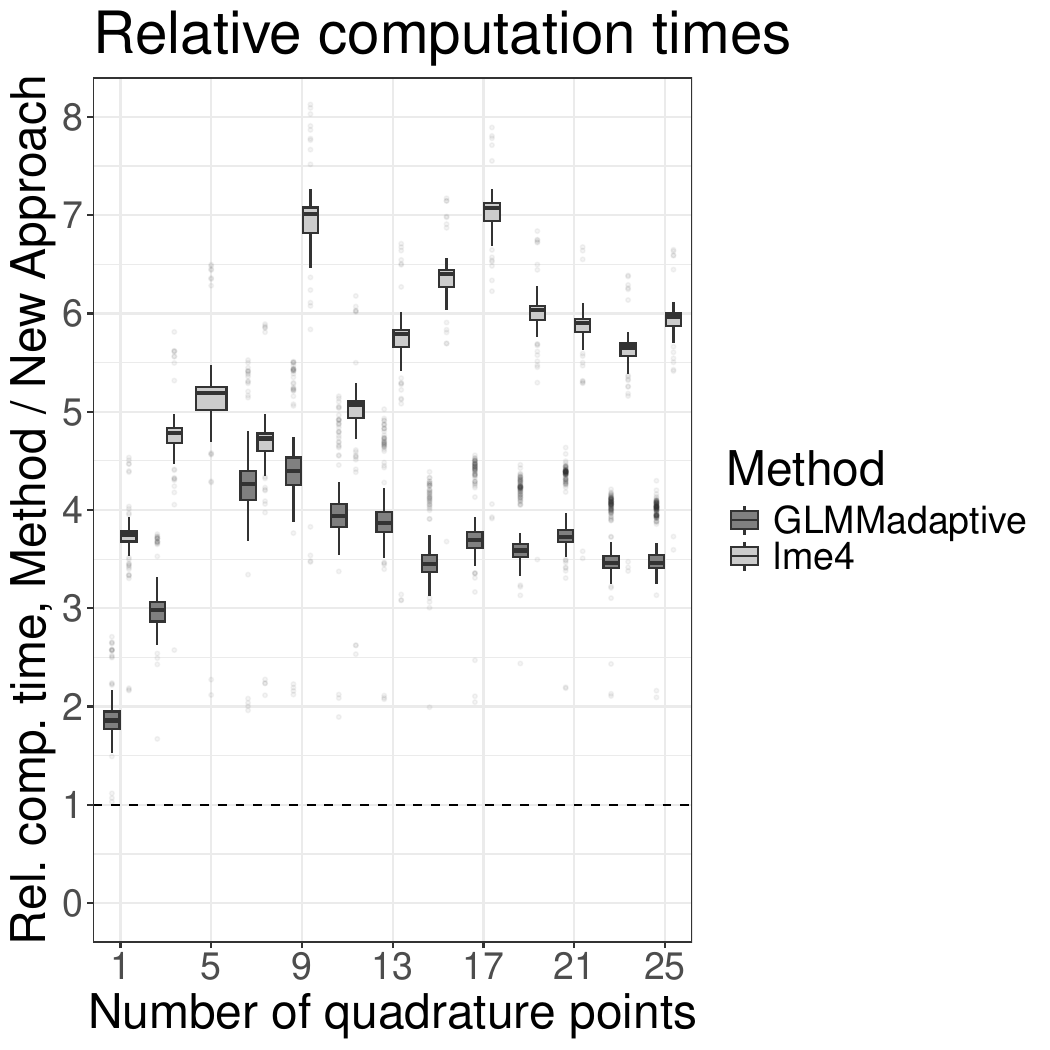}
}
\caption{
Results for the toenail infection treatment data of Section \ref{subsec:toenail}. (a) -- (b): Parameter estimates for regression coefficients from the new approach (---), \texttt{GLMMadaptive} (- - -),
and \texttt{lme4} ($\cdots$), with Wald confidence intervals.
(c): Parameter estimates for the between-subject standard deviation, $\sigma$, from the new approach (---), \texttt{GLMMadaptive} (- - -), \texttt{lme4} with profile likelihood ($\cdots$), and \texttt{lme4} with bootstrap intervals based on $200$ fits ($\cdot-\cdot$).
For large enough $\quadnum$, there is very little difference between the Wald intervals 
and the profile and bootstrap intervals for $\sigma$.
(d): relative computation times for \texttt{GLMMadaptive} and \texttt{lme4} (point estimation only) relative to the new approach, based on $500$ repetitions of the fit.
The new approach is between $2.5$ and $10$ times faster than \texttt{GLMMadaptive}, and up to $7$ times faster than \texttt{lme4} for this problem, with some variability across
values of $\quadnum$.
}
\label{fig:toenail-results}
\end{figure}

Figure \ref{fig:toenail-results} shows the results of fitting the model (\ref{eqn:toenailmodel}) to the toenail infection treatment data
of \citet{lesaffre_effect_2001}.
For $\regparam$, the point estimates and Wald confidence intervals for all methods level off for large enough $\quadnum$
and are in agreement with each other. The point where inferences stop changing is around $\quadnum=17$ or so, higher than the $\quadnum=10$ suggested by \citet{lesaffre_effect_2001}.
For $\sigma$, the point estimates for all three methods are also in agreement for large $\quadnum$.
Note the exceptionally large estimate of $\widehat{\sigma}\approx4$, indicating massive between-subject heterogeneity.

The \texttt{lme4} software includes options to compute
accurate bootstrap and profile likelihood intervals for
the variance components.
The bootstrap in particular should be accurate at any sample size and for any values of $\varcomp$.
While the Wald intervals provided by our approach and \texttt{GLMMAdaptive} are found to be accurate in the examples shown here, they may be inaccurate for smaller $\numtotal$.
However, the difference in computation time is large:
the profile approach takes around $60$ times longer than the point estimates
and Wald intervals computed using the new approach.
A reviewer points out that the availability of the exact gradient could potentially be used to speed up the computation of profile likelihood intervals.
The bootstrap with $200$ samples took between $1300-1600$ times longer than the new approach.
Figure \ref{fig:toenail-results} (c) shows all the intervals; for large enough $\quadnum$ the Wald, profile, and bootstrap agree closely for these data.

Figure \ref{fig:toenail-results} (d) shows the relative computation times of \texttt{lme4} and \texttt{GLMMadaptive}
compared to the new approach, across $500$ repetitions of the fit. 
The new approach is mostly between $2$ and $8$ times faster than these established approaches,
and is never slower. 
Note that these time comparisons for \texttt{lme4} do
not include confidence intervals for $\sigma$.

\section{Discussion}\label{sec:discussion}

The restriction to mixed models with a single grouping factor
is necessary for the developments in this paper.
The core aspect of the model that was considered is the factoring of the marginal likelihood
into a product of $\numgroups$ low dimensional integrals;
any model in which this does not occur will require further innovation
in order to apply accurate low-dimensional quadrature techniques.
\citet{pinheiro_efficient_2006} apply adaptive quadrature
in multi-level mixed models, and it would be interesting
to derive the gradient algorithm presented in this paper to this case.
It remains
unclear whether it is possible to derive a procedure for
fitting mixed models with more general random effects structure
using AQ.

We conjecture here that it is not possible to develop a quadrature-based procedure that works for \emph{all} mixed model
random effects structures
by showing an example where it does not appear possible, as follows.
Consider a model with crossed random effects as defined by \citet[Eq. 1]{ghosh_2022}:
\begin{equation}\label{eqn:crossedmodel}
\begin{aligned}
  \response_{ij} \setdelim u_i,v_j &\indsim \responsedist(\mean_{ij},\obssd), \ u_i \iidsim N(0,\sigma_u^2), v_j \iidsim N(0,\sigma_v^2), \\
  \pred_{ij} = \linkfunc(\mean_{ij}) &= \beta_0 + \cov_{ij}\Tr\regparam + u_i + v_j.
\end{aligned}
\end{equation}
The marginal likelihood corresponding to Eq. \ref{eqn:crossedmodel} is \citep[Eq. 5]{ghosh_2022}
% \begin{equation}\label{eqn:crossedlikelihood}
$
\marglikall=\int\margdens(\params,\mb{u},\mb{v})\dee\mb{u}\dee\mb{v},
% \end{equation}
$
where $\mb{u} = (u_1,\ldots,u_{R})\Tr$ and
$\mb{u} = (v_1,\ldots,v_{C})\Tr$ with $R,C$ very large and total sample size $N\ll RC$.
% Their total sample size is $N\leq RC$, and hence their setup guarantees that
% dimension of the integral grows at least as $O(N^{1/2})$.
Crucially, 
$\marglikall$
% Eq. \ref{eqn:crossedlikelihood} 
does not factor over the random effects, and low-dimensional quadrature cannot be applied as it can for the model considered in this paper.

Every mixed model analysis in which inferences are obtained by maximizing an approximate marginal likelihood
where the approximation is adaptive quadrature must specify a number of quadrature points, $\quadnum$,
for which estimates will be reported.
There is currently no clear principle upon which to base 
this important choice. 
However, the simulations of the previous
Section demonstrate that it does have a substantial effect on the quality of inferences.
Choosing a $\quadnum$ that is too high was never observed to lead to substantially worse inferences than
one that is too low, and simply increases computation time.
Choosing a $\quadnum$ that is too low, however, can clearly lead to dramatically inappropriate inferences.
We therefore recommend to choose a $\quadnum$ high enough that inferences appear to stop changing as $\quadnum$ is further increased.
We are certainly not the first to suggest this as a strategy for choosing $\quadnum$; in fact, we were unable to find a published simulation study or data analysis
which used these methods and reported choosing $\quadnum$ in any other way.
Development of a more principled strategy for choosing $\quadnum$ may be regarded as an open problem.

Automatic differentiation (AD) is an established tool for obtaining exact gradients, and \texttt{C++} implementations are accessible in 
\texttt{R} via the Stan Math library \citep{carpenter2015stan,carpenter_stan_2017} and the Template Model Builder (TMB; \citealt{kristensen_tmb_2016}).
Naive automatic differentiation of $\log\approxmarglikAQall{\quadnum}$ is 
challenging due to the inner optimization required to compute $\condmode(\params)$,
the third derivatives involved in differentiating the Hessian determinant, 
and the matrix algorithms required to compute the Hessian determinant and solve the triangular system required to scale the quadrature nodes.
The \texttt{glmmTMB} software \citep{brooks_glmmtmb_2017}, through the \texttt{TMB}
software \citep{kristensen_tmb_2016}, provides efficient automatic differentiation of $\log\approxmarglikAQall{1}$;
we include a brief comparison in Section \ref{supp:simulations} of the supplementary materials.
The proposed gradient evaluation algorithm in Section \ref{sec:methods} uses algorithmic differentiation
to differentiate through the Cholesky and forward solve algorithms,
as well as implicit differentiation for $\condmode(\params)$.
These contributions could plausibly lead to an efficienct implementation of
automatic differentiation for  $\log\approxmarglikAQall{\quadnum}$.

\bibliographystyle{apalike}
\bibliography{references}

\newpage
\appendix

\newpage

\section{Algorithms}\label{supp:algorithms}

This section gives algorithms for:
\begin{enumerate}
    \item Evaluating the approximate log-marginal likelihood, $\approxloglikik{\quadnum}{}(\params)$, and its (exact) gradient, $\nabla_{\params}\approxloglikik{\quadnum}{}(\params)$.
    \item Evaluating the derivative of the solution, $\boldsymbol{v}$, to the lower-triangular system, $\chol\boldsymbol{v} = \quadpointvec$, with respect to the elements of (the lower triangle of) $\chol$.
    \item Evaluating vector products and their derivatives involving the matrix quantities $\Su$ and $\normalchol$ required to implement models with the parameterization of the multivariate Gaussian given in Section 4.2 of the main manuscript.
\end{enumerate}

\subsection{Approximate marginal log-likelihood and exact gradient}

\begin{algorithm}[H]
    \label{alg:gradient-input}
    \caption{
    Quantities required to compute the approximate log-marginal likelihood, $\approxloglikik{\quadnum}{}(\params)$, and its exact gradient, $\nabla_{\params}\approxloglikik{\quadnum}{}(\params)$; see Algorithm \ref{alg:gradient-compute} for computations.
    }
    \KwData{
    \\
        $\obsall_i\in\Reals^{\numpergroup_i},i=1,\ldots,\numgroups$; \\
        $\numtotal = \numpergroup_1 + \cdots + \numpergroup_\numgroups$; \\
        $\mb{X}\in\Reals^{\numtotal\times\regparamdim}$; 
        $\mb{Z}\in\Reals^{\numtotal\times\redim}$; \\
        $\regparam\in\Reals^{\regparamdim}$; \\
        $\obssd\in\Reals^{\obssddim}$; \\
        $\resd\in\Reals^{\resddim}$; \\
        $\jointlikall$ joint likelihood, $\obsall = (\obsall_1,\ldots,\obsall_\numgroups)\in\Reals^{\numtotal}$; \\
        $\gradjointloglikall = \partial\log\jointlikall / \partial(\params,\re)$; \\
        $\hessjointloglikall = \partial^2\log\jointlikall / \partial(\params,\re)\partial(\params,\re)\Tr$; \\
        $\hessDall = (\hessD_1,\ldots,\hessD_\paramdim)\in\Reals^{\redim\times\redim\times\paramdim}$, $\hessD_l = \partial\hessjointloglikall / \partial(\params,\re)_l, l=1,\ldots,\paramdim$; \\
        $\quadnum\in\Nats$, $\quadpointset(\quadnum,1) = \{\quadpoint_1,\ldots,\quadpoint_\quadnum\}\subset\Reals$, $\weight_\quadnum:\quadpointset(\quadnum,1)\to\Reals^{+}$.
    }
    \textbf{Set}: \\
        $\params = (\regparam,\obssd,\resd)\in\Reals^{\paramdim}$; \\
        $\re = \zero\in\Reals^{\redim}$; \\
        $\approxloglikik{\quadnum}{}(\params)=0$; \\
        $\nabla_{\params}\approxloglikik{\quadnum}{}(\params) = \zero\in\Reals^{\paramdim}$ \\
        $\mb{\pred}_i = \mb{\pred}^\prime_i = \mb{\pred}^{\prime\prime}_i = \zero\in\Reals^{\numpergroup_i}, i\in\{1,\ldots,\numgroups\}$; \\
        $\quadpointset(\quadnum,\redim) = \quadpointset(\quadnum,1)^\redim$; \\
        $\weight_\quadnum(\quadpointvec) = \weight_\quadnum(\quadpoint_1)\times\cdots\times\weight_\quadnum(\quadpoint_\redim)$; \\
        $\mb{v} = \zero\in\Reals^{\quadnum^\redim}$, 
        $\mb{g}_{1} = \zero\in\Reals^{\paramdim+\redim}$,
        $\mb{g}_{2} = \zero\in\Reals^{\paramdim+\redim}$,
        $\mb{g}_{3} = \zero\in\Reals^{\paramdim+\redim}$,
        $\mb{g}_{4} = \zero\in\Reals^{\paramdim}$; \\
        $\cholD\in\Reals^{\redim\times\redim}$ lower-triangular; \\
        $\mb{w}^L\in\Reals^{\quadnum^\redim\times\redim(\redim+1)/2}$; \\
        $\mb{w}^u\in\Reals^{\quadnum^\redim\times\paramdim}$. \\
        \KwResult{Input to Algorithm \ref{alg:gradient-compute}.}
\end{algorithm}

\begin{algorithm}[H]
    \label{alg:gradient-compute}
    \caption{
    Computation of the approximate log-marginal likelihood, $\approxloglikik{\quadnum}{}(\params)$, and its exact gradient, $\nabla_{\params}\approxloglikik{\quadnum}{}(\params)$; see Algorithm \ref{alg:gradient-input} for setup and definitions.
    }
    \KwData{Output of Algorithm \ref{alg:gradient-input}.}
    \KwResult{$\approxloglikik{\quadnum}{}(\params)\in\Reals$, $\nabla_{\params}\approxloglikik{\quadnum}{}(\params) = \partial\approxloglikik{\quadnum}{}(\params)/\partial\params\Tr\in\Reals^{\paramdim}$}
    \For{$i=1,\ldots,\numgroups$}{
        Let $t_1 = 0$. \\
        Compute $\condmodei{i}(\params)$ via Newton's method. \\
        Compute $\jointhessi{i} = -\partial^2\log\margdens_i(\params,\condmodei{i}(\params)) / \partial\re_i\partial\re_i\Tr$; \\
        Compute $\jointcholi{i}$, lower Cholesky triangle of $\jointhessi{i}$; \\
        Compute $\log|\jointcholi{i}| = \log\jointcholi{i}_{11} + \cdots + \log\jointcholi{i}_{\redim\redim}$; \\
        Compute $\hessD(\params,\condmodei{i}(\params);\obsall_i)$; \\
        Set $\mb{w}^L = \zero$. \\
        \For{$j=1,\ldots,|\quadpointset(\quadnum,\redim)|$}{
            Compute $\widehat{\quadpoint}_j(\params) = \jointcholi{i}\inv\quadpoint_j + \condmodei{i}(\params)$; \\
            Set $\mb{v}_j = \log\margdens_i(\params,\widehat{\quadpoint}_j(\params);\obsall_i) + \log\weight_k(\quadpoint_j) - \log|\jointcholi{i}|$; \\
            Set $\mb{w}^u_{j,\cdot} = \partial\log\margdens_i(\params,\widehat{\quadpoint}_j(\params);\obsall_i)/\partial\re_i$; \\
            Compute $\jointcholi{i}\inv\quadpointvec$ and $\mb{t}_2\equiv(\partial\jointcholi{i}\inv\quadpointvec / \partial L_{11},\ldots,\partial\jointcholi{i}\inv\quadpointvec / \partial L_{\redim\redim})\Tr$ via Algorithm \ref{alg:forwardsub}; \\
            Set $\mb{w}^L_{j,\cdot} = \mb{t}_2\Tr\mb{w}^u_{j,1:\redim}$; \\
        }
        Set $t_1 = \log\sum_{j=1}^{|\quadpointset(\quadnum,\redim)|}\exp(\mb{v}_j)$; \\
        Increment $\approxloglikik{\quadnum}{}(\params)\mathrel{{+}{=}}t_1$; \\
        Set $\mb{g}_1 = \sum_{j=1}^{|\quadpointset(\quadnum,\redim)|}\mb{w}^u_{j,\cdot}\exp(\mb{v}_j)$; \\
        Set $\texttt{lower.tri}(\cholD) = \sum_{j=1}^{|\quadpointset(\quadnum,\redim)|}\mb{w}^L_{j,\cdot}\exp(\mb{v}_j)$; \\
        $\mb{g}_2 = \partial\approxloglikik{\quadnum}{i}\left(\chol(\params,\condmodei{i}(\params));\params,\condmodei{i}(\params)\right)/\partial(\params,\re_i)$, via the algorithm from Section 2.3.1 of Smith (1995) (see Section 3.1.2 of the main manuscript); \\
        Set $\mb{g}_3 = \mb{g}_1 + \mb{g}_2$; \\
        Set $\mb{g}_4 = (\mb{g}_3)_{(\redim+1):(\redim+\paramdim)} + \left[\jointcholi{i}\inv\jointhessi{i}_{1:\redim,(\redim+1):(\redim+\paramdim)}\right]\Tr$; \\
        Increment $\nabla_{\params}\approxloglikik{\quadnum}{}(\params)\mathrel{{+}{=}}\mb{g}_4\times\exp(-t_1)$.
    }

\end{algorithm}

\subsection{Forward substitution}

\begin{algorithm}[H]
    \label{alg:forwardsub}
    \caption{Forward substitution with derivative. The input vector $\quadpointvec$ is overwritten with $\chol\inv\quadpointvec$, and the vector $\gradvec$ contains the $\redim$-dimensional vector $\partial_{L_{rs}}\chol\inv\quadpointvec$ where $r,s$ index the lower triangle of $\chol$ in column-major order, $L_{11},L_{1,2},\ldots,L_{dd}$.
    }
    \KwData{$\chol\in\Reals^{\redim\times\redim}$, lower triangular, $\quadpointvec\in\Reals^{\redim}$, $r\geq s\in\{1,\ldots,\redim\}$.}
    \KwResult{$\quadpointvec\to\chol\inv\quadpointvec$, $\gradvec = \partial_{L_{kl}}\chol\inv\quadpointvec$}
    Set $\gradvec=\zero\in\Reals^{\redim}$\;
    \For{$i=1,\ldots,\redim$}{
        \For{$j=1,\dots,i-1$}{
            $z_i\leftarrow z_i-L_{ij}z_j$\; 
            $g_i\leftarrow g_i-L_{ij}g_j$\;
            \If{$i=r,j=s$}{
                $g_i\leftarrow g_i-z_j$\;
            }
        }
        $z_i = z_i/L_{ii}$\;
        $g_i = g_i/L_{ii}$\;
        \If{$i = r = s$}{$g_i = g_i - z_i/L_{ii}$}
    }
\end{algorithm}

\subsection{Multivariate Gaussian quantities}

This section contains algorithms for evaluating
\begin{itemize}
    \item The selection matrix, $\Su$, such that $\normalchol\Tr\re = \re + \Su\corparam$ (Algorithm \ref{alg:Su}),
    \item The derivative matrix $\partial\Su/\partial\reidx_j$, $j=1,\ldots,\redim$ (Algorithm \ref{alg:dSu}), and
    \item The derivative matrix $\partial\normalchol/\partial\phi_j$, $j=1,\ldots,\redim(\redim-1)/2$ (Algorithm \ref{alg:getij}).
\end{itemize}
These details are required to implement the multivariate Gaussian calculations from Section 4.2 of the main manuscript.

Algorithm \ref{alg:Su} shows how to obtain the selection matrix, $\Su$, such that
$$
\normalchol\Tr\re = \re + \Su\corparam,
$$
where
$$
\normalchol = \begin{pmatrix} 1 & & & & \\ \phi_1 & 1 & & & \\ \phi_2 & \phi_{\redim} & 1 & & \\ \vdots & \vdots & \vdots & 1 & \\ \phi_{\redim-1} & \cdots & \cdots & \phi_{\corparamdim} & 1 \\ \end{pmatrix}.
$$
An example is helpful to illustrate this construction.
Let $\redim=3$ so $\re=(\reidx_1,\reidx_2,\reidx_3)\Tr$,
$\corparamdim=3(3-1)/2=3$ and hence $\corparam=(\phi_1,\phi_2,\phi_3)\Tr$.
Then
\begin{align*}
\normalchol\Tr\re &= 
\begin{pmatrix}
1 & \phi_1 & \phi_2 \\
0 & 1      & \phi_3 \\
0 & 0      & 1      \\
\end{pmatrix}
\begin{pmatrix}
\reidx_1 \\
\reidx_2 \\
\reidx_3 \\
\end{pmatrix} \\
&=
\begin{pmatrix}
\reidx_1 + \phi_1\reidx_2 + \phi_2\reidx_3 \\
                 \reidx_2 + \phi_3\reidx_3 \\
                                  \reidx_3 \\
\end{pmatrix} \\
&=
\re + 
\begin{pmatrix}
\reidx_2 & \reidx_3 & 0 \\
0        & 0        & \reidx_3 \\
0        & 0        & 0        \\
\end{pmatrix}\corparam.
\end{align*}
Hence, with $\redim=3$, we have
$$
\Su = \begin{pmatrix}
\reidx_2 & \reidx_3 & 0 \\
0        & 0        & \reidx_3 \\
0        & 0        & 0        \\
\end{pmatrix}.
$$
Further, it is clear that $\partial\Su/\partial\reidx_j$ will be a
matrix containing $0$ and $1$ only, with a $1$ located wherever $\reidx_j$ is.
An advantage of the following algorithmic construction
of $\Su$ is that differentiation of this algorithm immediately
yields an algorithm for computing the matrix $\partial\Su/\partial\reidx_j$ for any $j=1,\ldots,\redim$; see Algorithm \ref{alg:dSu}.

\begin{algorithm}[H]
    \label{alg:Su}
    \caption{Obtain the matrix, $\Su$, such that $\normalchol\Tr\re = \re + \Su\corparam$. \emph{Note: $0$-based indexing.}}
    \KwData{$\re\in\Reals^\redim$}
    \KwResult{$\Su\in\Reals^{\redim\times\redim(\redim-1)/2}$}
    Set $\corparamdim = \redim(\redim-1)/2$;
    \For{$k=0,\ldots,\corparamdim-1$}{
            \If{$(k+1)\%(\redim-i) == 0$}{
                $i\leftarrow i+1$; \\
                $j\leftarrow i+1$; \\
            }
            $\Su_{i,k}\leftarrow \reidx_j$; \\
            $j\leftarrow j+1$; \\
        }
\end{algorithm}

\begin{algorithm}[H]
    \label{alg:dSu}
    \caption{Obtain the matrix $\partial\Su/\partial\reidx_l$ for $l=0,\ldots,\redim-1$. \emph{Note: $0$-based indexing.}}
    \KwData{$\re\in\Reals^\redim$, $l\in\left\{0,\ldots,\redim-1\right\}.$}
    \KwResult{$\partial\Su/\partial\reidx_l\in\Reals^{\redim\times\redim(\redim-1)/2}$}
    Set $s = \redim(\redim-1)/2$\;
    \For{$k=0,\ldots,s-1$}{
            \If{$(k+1)\%(\redim-i) == 0$}{
                $i\leftarrow i+1$; \\
                $j\leftarrow i+1$; \\
            }
            \If{$j==l$} {
                $\Su_{i,k}\leftarrow 1$; \\
            }
            $j\leftarrow j+1$; \\
        }
\end{algorithm}

The final details required are the derivative matrices,
$\partial\normalchol/\partial\phi_l$ and $\partial\normalchol\Tr/\partial\phi_l$.
These are also matrices of only $0$ and $1$ and we determine
them algorithmically.
Specifically, the matrix $\partial\normalchol/\partial\phi_l$ is a matrix of all $0$ except for a $1$ at index $(i,j)$ such that $d(j-1) - j(j-1)/2 + (i-j) = l$.
To determine the $(i,j)$ location of $\phi_l$, we simply invert this relation;
see Algorithm \ref{alg:getij}.

A note on implementation: we do not actually compute the matrix $\partial\normalchol/\partial\phi_l$
when computing products of the form
$\partial\normalchol/\partial\phi_l\D\normalchol\Tr$.
Instead we compute $(i,j)$ from Algorithm \ref{alg:getij} and
use that the product $\partial\normalchol/\partial\phi_l\D\normalchol\Tr$
is the $\redim\times\redim$ matrix formed by placing the $j^{th}$ row
of $\D\normalchol\Tr$ in the $i^{th}$ row of the $\redim\times\redim$-dimensional
zero matrix, and $\normalchol\D\partial\normalchol\Tr/\partial\phi_l$ is its transpose.
These are formed directly.

\begin{algorithm}[H]
    \label{alg:getij}
    \caption{Return the location, $(i,j)$, of $\phi_l$ in $\normalchol$, for $l=1,\ldots,d(d-1)/2$.}
    \KwData{$d\in\Nats,l\in\left\{1,\ldots,d\right\}$}
    \KwResult{$(i,j)$ such that $j\in\{1,\ldots,d-1\}, i\in\{j+1,\ldots,d\}$, and $l = d(j-1) - j(j-1)2 + (i-j)$}
    Set $i=1,j=0$; \\
    \While{$j<d$}{
        \If{$l == d(j-1) - j(j-1)/2 + (i-j)$}{
            \Return $(i,j)$; \\
        }
        $i\leftarrow i+1$;\\
        \If{$i>d$}{
            $j\leftarrow j+1$;\\
            $i\leftarrow j+1$; \\
        }
    }
\end{algorithm}

\section{Exponential family calculations}\label{supp:exponentialfamily}

The calculations for Bernoulli generalized linear mixed models given in Section 4.1 of the main manuscript are an important special case of the generalized linear mixed model with exponential family response. While the Bernoulli case is of specific interest to readers and the model used in all the experiments in the paper, the strategy for obtaining the calculations for general exponential family responses is identical; the former really is a special case of the latter.
We give those calculations here, as follows.

The model is:
\begin{equation}\begin{aligned}\label{eqn:mixedmodelappendix}
    \response_{ij} \setdelim \re_i &\indsim \responsedist(\mean_{ij},\obssd), \ \re_i \iidsim \redist(\resd), \\
    \pred_{ij} = \linkfunc(\mean_{ij}) &=  \cov_{ij}\Tr\regparam + \recov_{ij}\Tr\re_i.
\end{aligned}\end{equation}
The response distribution, $\responsedist(\mean_{ij},\obssd)$, belongs to the exponential family
if its density is
\begin{equation}\label{eqn:exponentialfamilydensity}
    \responsedens(\obs_{ij};\mean_{ij},\obssdi)  = \exp\left\{\frac{\obs_{ij}\pred_{ij} - b(\pred_{ij})}{\obssdi}\right\}c(\obs_{ij};\obssd),
\end{equation}
for known functions $b,c$, where $\obssd = \obssdi$ is a single dispersion parameter, and $\pred_{ij} = \linkfunc(\mean_{ij})$ as in Eq. \ref{eqn:mixedmodelappendix}. 
We treat the dispersion parameter, $\obssdi$, as fixed and known; the most common case where this must be
estimated is when the response is Gaussian, in which case the methods in this paper are not relevant, as
the marginal likelihood is tractable.

We have
\begin{align*}
    \log\responsedens(\obs_{ij};\mean_{ij},\obssdi) &= \frac{\obs_{ij}\pred_{ij} - b(\pred_{ij})}{\obssdi} + \log c(\obs_{ij};\obssd), \\
    \nabla_{\params}\log\responsedens(\obs_{ij};\mean_{ij},\obssdi) &= \frac{\obs_{ij} - b^{\prime}(\pred_{ij})}{\obssdi}\cov_{ij}, \\
    \nabla_{\re_i}\log\responsedens(\obs_{ij};\mean_{ij},\obssdi) &= \frac{\obs_{ij} - b^{\prime}(\pred_{ij})}{\obssdi}\recov_{ij}, \\
    \nabla^2_{\params\params\Tr}\log\responsedens(\obs_{ij};\mean_{ij},\obssdi) &= -\frac{b^{\prime\prime}(\pred_{ij})}{\obssdi}\cov_{ij}\cov_{ij}\Tr, \\ 
    \nabla^2_{\re_i\re_i\Tr}\log\responsedens(\obs_{ij};\mean_{ij},\obssdi) &= -\frac{b^{\prime\prime}(\pred_{ij})}{\obssdi}\recov_{ij}\recov_{ij}\Tr, \\ 
    \nabla^2_{\re_i\params\Tr}\log\responsedens(\obs_{ij};\mean_{ij},\obssdi) &= -\frac{b^{\prime\prime}(\pred_{ij})}{\obssdi}\recov_{ij}\cov_{ij}\Tr. \\ 
\end{align*}
The above expressions are standard and immediate, and
recover the corresponding expressions in Section 4.1 of the
main manuscript.
We also require third derivatives of the log-likelihood.
We seek the $(\paramdim+\redim)\times(\paramdim+\redim)\times(\paramdim+\redim)$-dimensional tensor
$$
\hessD(\params,\re)  = \partial_{(\params,\re)}\hess^{\obs}_{ij}(\params,\re_i),
$$
where 
$$
    \hess^{\obs}_{ij}(\params,\re_i) = 
         -\nabla^2_{\re_i\re_i\Tr}\log\responsedens(\obs_{ij};\mean_{ij},\obssdi) = \nabla^2_{\re_i\re_i\Tr}\frac{b^{\prime\prime}(\pred_{ij})}{\obssdi}\recov_{ij}\recov_{ij}\Tr.
$$
We have
\begin{equation*}
    \begin{aligned}
        \frac{\partial}{\partial\regparamidx_l}\hess_{ij}^{\obs}(\params,\re_i) &= -\frac{b^{\prime\prime\prime}(\pred_{ij})}{\obssdi}
        \recov_{ij}\recov_{ij}\Tr\covi_{ijl} \\
        \frac{\partial}{\partial\reidx_l}\hess_{ij}^{\obs}(\params,\re_i) &= -\frac{b^{\prime\prime\prime}(\pred_{ij})}{\obssdi}
                            \recov_{ij}\recov_{ij}\Tr\recovi_{ijl}. \\
    \end{aligned}
\end{equation*}
This completes the calculations required to implement
Algorithm \ref{alg:gradient-compute} for any exponential
family distribution.

\section{Complete Simulation Results}\label{supp:simulations}

\subsection{Simulation 1: absolute performance of the new method}

In Section 4 we performed the following simulation study. For each combination of $\numgroups=\{100,200,500,1000\}$, $\numpergroup=\{3,5,7,9\}$, and
$\quadnum = \{1,3,.\ldots,23,25\}$, we generated $1000$ datasets from the model
\begin{equation}\label{eqn:simmodel1}
  \response_{ij} \setdelim \re_i \indsim \text{Bern}(p_{ij}), \ \re_i \iidsim \text{N}\{\zero,\Varmat(\resd)\}, \ 
  \log\frac{p_{ij}}{1-p_{ij}} = \beta_0 + \beta_1\covi_i + \beta_2 t_j + \beta_3\covi_i t_j + \reidx_{i1} + \reidx_{i2}t_j,
\end{equation}
where $\covi_i$ takes value $1$ for half the groups and value $0$ for the other half,
and $t_j$ takes values on an equally-spaced grid of length $\numpergroup_i$ from $-3$ to $3$.
We used $\regparam = (-2.5,-.15,.1,.2)\Tr$ and
$$
\Varmat(\resd) = \begin{pmatrix} \sigma^2_{1} & \sigma_{12} \\ \sigma_{12} & \sigma^2_{2} \end{pmatrix} = \begin{pmatrix} 2 & 1 \\ 1 & 1 \end{pmatrix}.
$$
We fit the model using the new procedure, and
compute and report bias (for $\regparam$ and $\sigma_{12}$) and relative bias (for $\sigma^2_1$ and $\sigma^2_2$), 
as well as the lengths and empirical coverage proportions of Wald confidence intervals.
The complete simulation results are too extensive to report in the main manuscript, so we report them here.

\begin{figure}[p]
\centering
\includegraphics{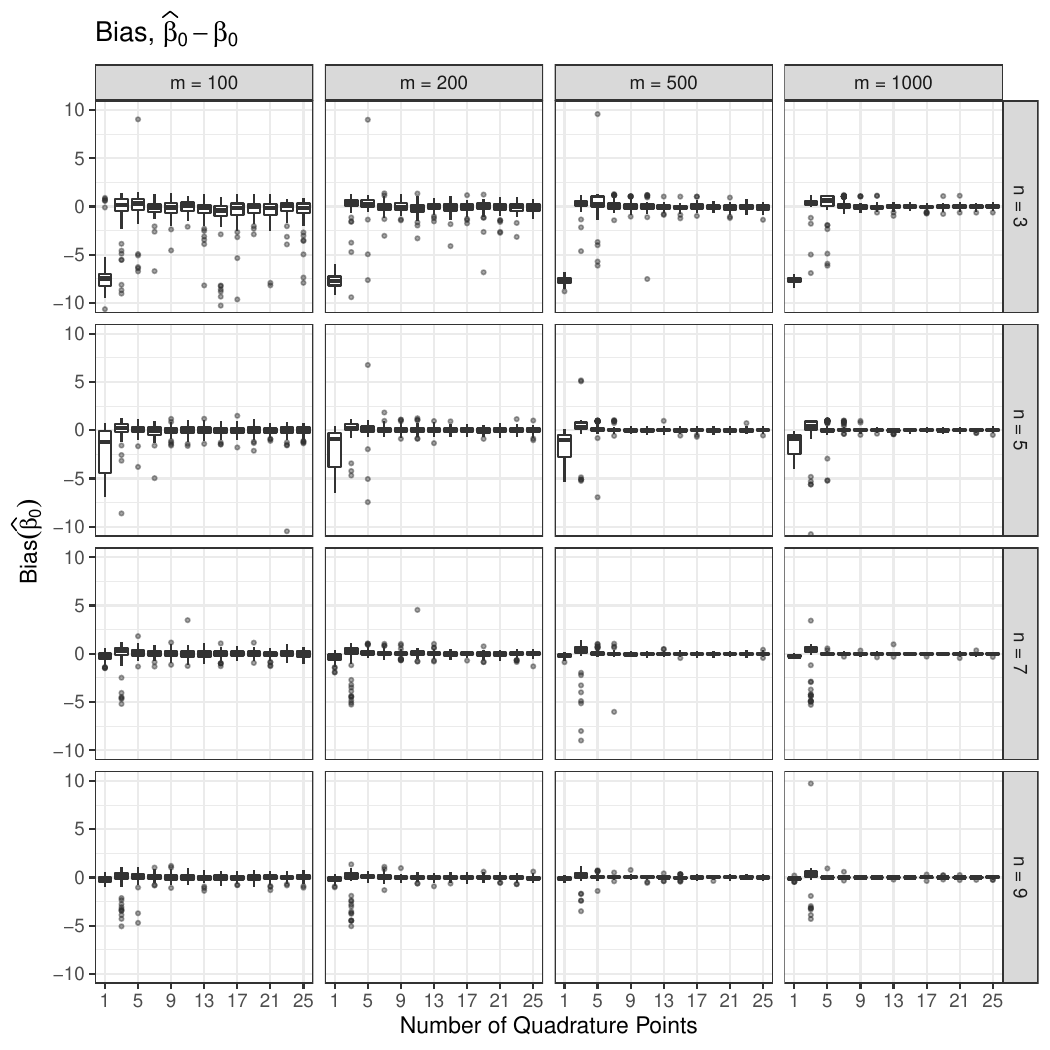}
\caption{Empirical bias, $\widehat{\beta_0} - \beta_0$, for $\beta_0$ in the simulation study of Section 4.3 in the main manuscript, where
$1000$ sets of data were generated from the random-slopes model (\ref{eqn:simmodel1}). The Laplace approximation ($\quadnum=1$)
exhibits high bias for lower $\numpergroup=3,5$. Small numbers of quadrature points, $\quadnum<5$ or so, appear to yield a higher number of
outlying bias results. Higher numbers of quadrature points appear to give accurate integral approximations: even in the challenging case of $\numgroups=1000$
and $\numpergroup=3$, the larger $\quadnum$ results appear to have low bias. For $\numpergroup=3$ and lower $\numgroups$, the results are less stable, 
although this case would be expected to be challenging even if an exact likelihood could be used.}
\label{fig:beta0bias}
\end{figure}

\begin{figure}[p]
\centering
\includegraphics{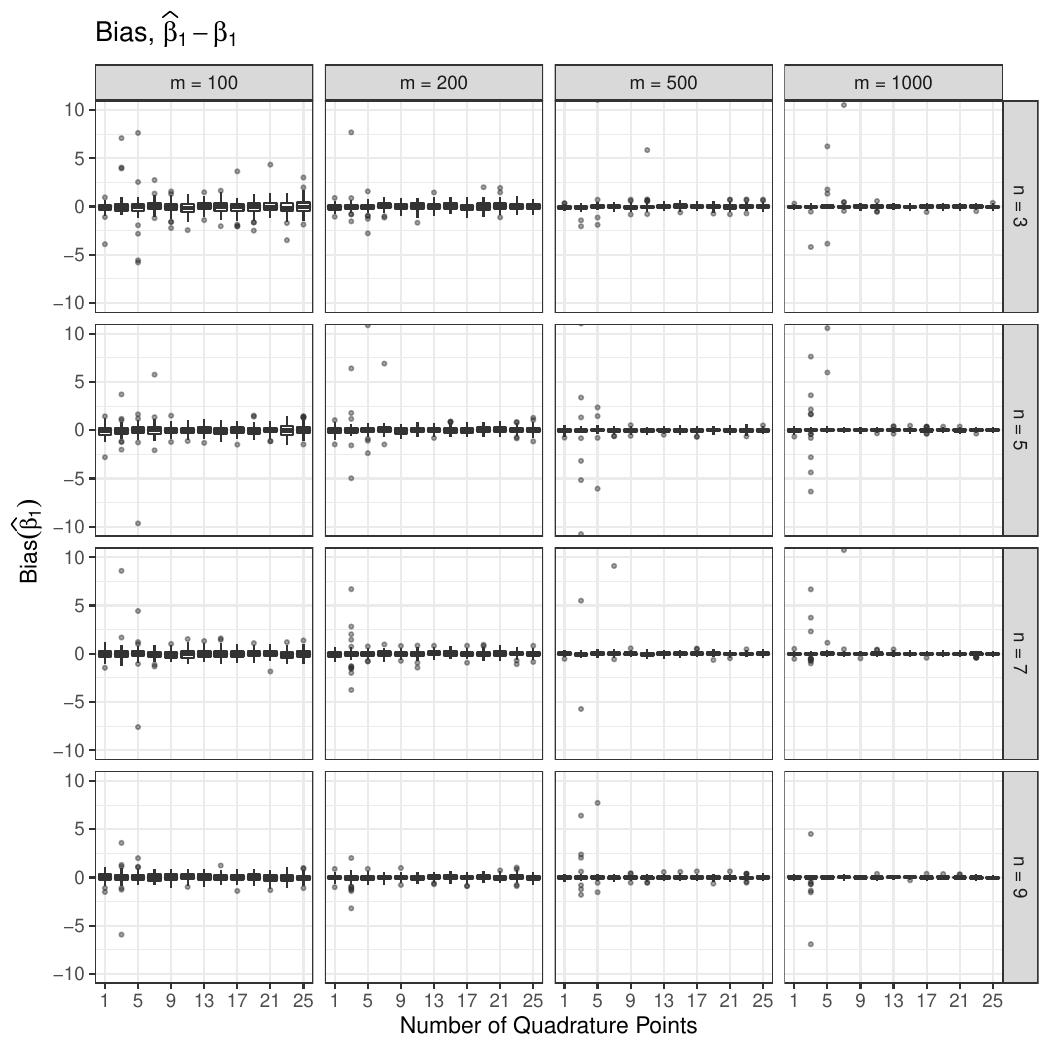}
\caption{Empirical bias, $\widehat{\beta_1} - \beta_1$, for $\beta_1$ in the simulation study of Section 4.3 in the main manuscript, where
$1000$ sets of data were generated from the random-slopes model (\ref{eqn:simmodel1}). All combinations of $\numgroups,\numpergroup,\quadnum$
appear to yield generally low bias, with a small number of outliers. Inference for $\beta_1$ is not expected to be very challenging in
this model, and these results are not surprising.}
\label{fig:beta1bias}
\end{figure}

\begin{figure}[p]
\centering
\includegraphics{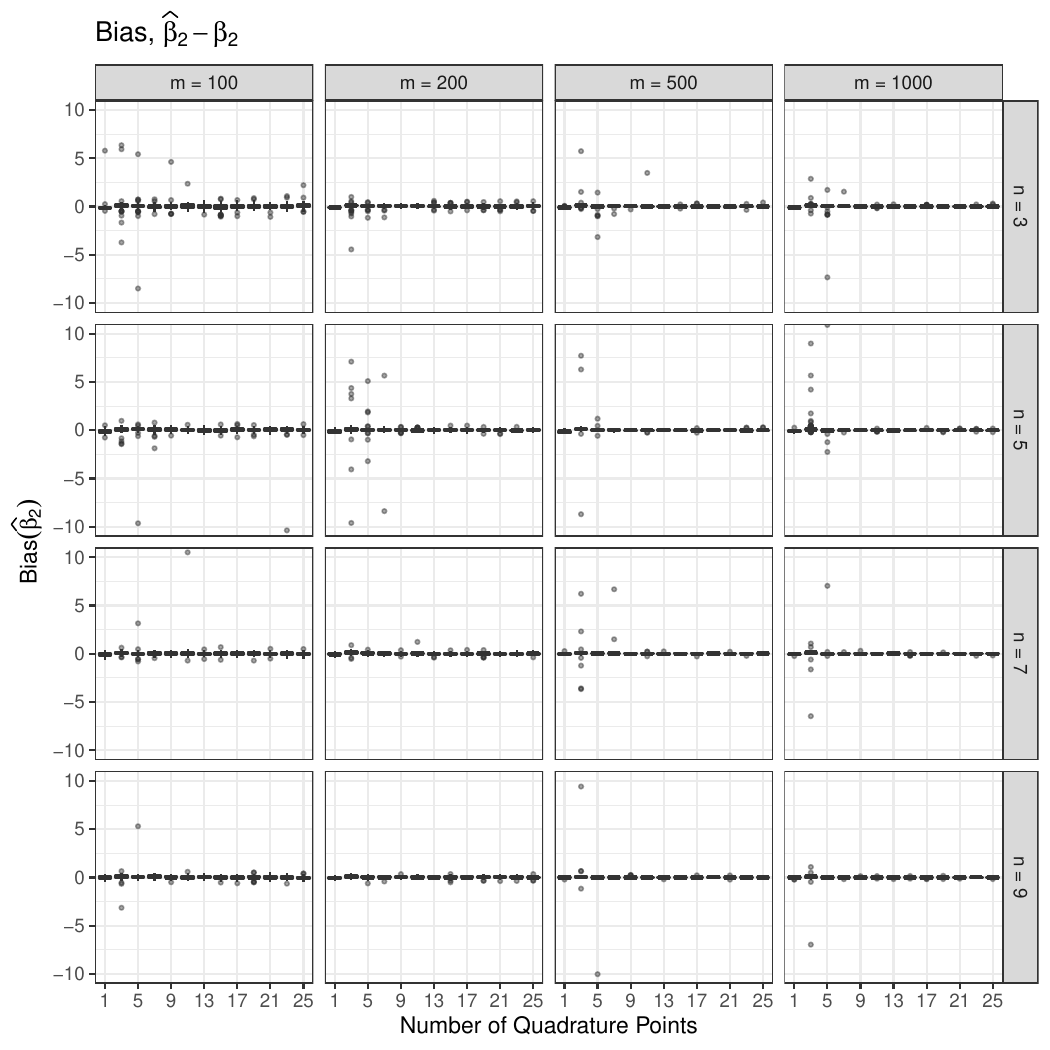}
\caption{Empirical bias, $\widehat{\beta_2} - \beta_2$, for $\beta_2$ in the simulation study of Section 4.3 in the main manuscript, where
$1000$ sets of data were generated from the random-slopes model (\ref{eqn:simmodel1}). All combinations of $\numgroups,\numpergroup,\quadnum$
appear to yield generally low bias, with a small number of outliers. Inference for $\beta_2$ is not expected to be very challenging in
this model, and these results are not surprising.}
\label{fig:beta2bias}
\end{figure}

\begin{figure}[p]
\centering
\includegraphics{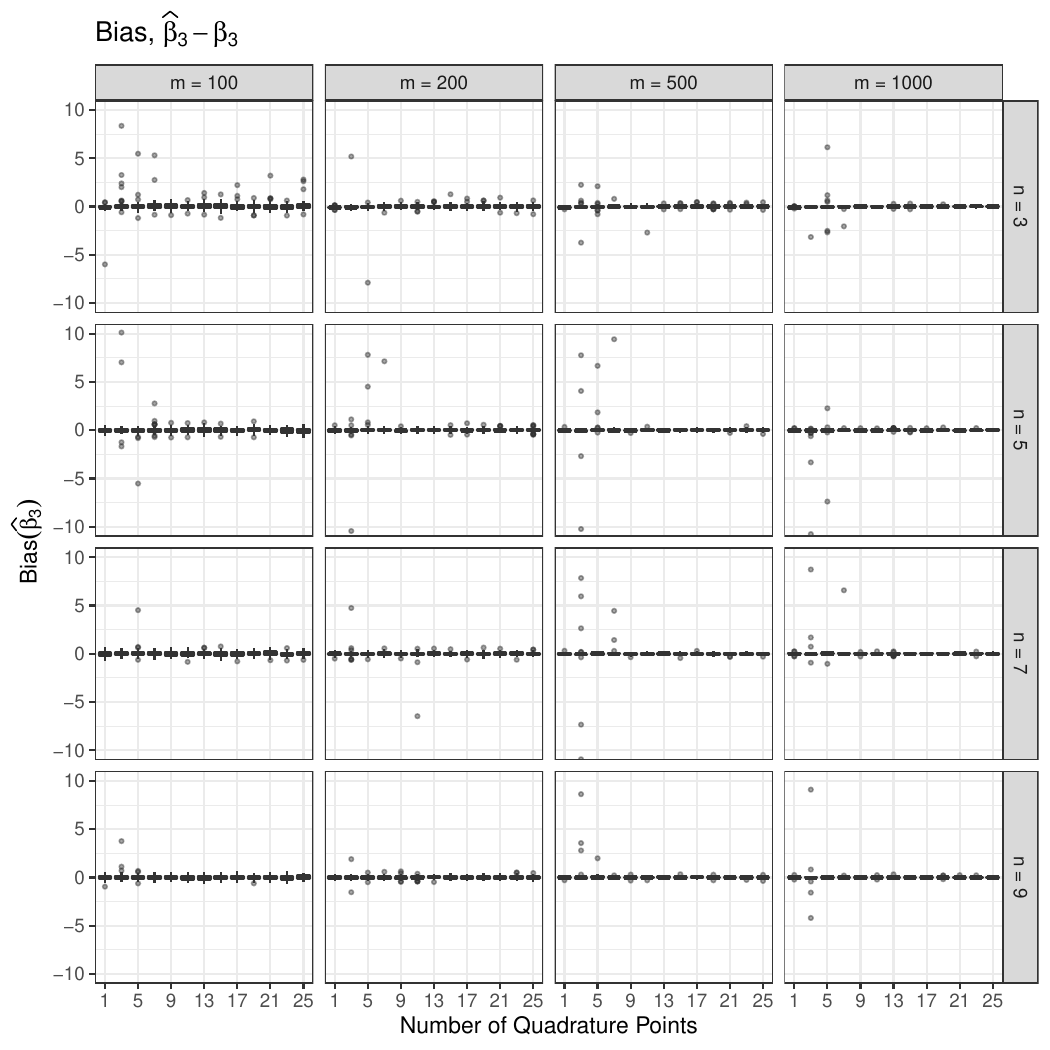}
\caption{Empirical bias, $\widehat{\beta_3} - \beta_3$, for $\beta_3$ in the simulation study of Section 4.3 in the main manuscript, where
$1000$ sets of data were generated from the random-slopes model (\ref{eqn:simmodel1}). All combinations of $\numgroups,\numpergroup,\quadnum$
appear to yield generally low bias, with a small number of outliers. Inference for $\beta_3$ is not expected to be very challenging in
this model, and these results are not surprising.}
\label{fig:beta3bias}
\end{figure}

\begin{figure}[p]
\centering
\includegraphics{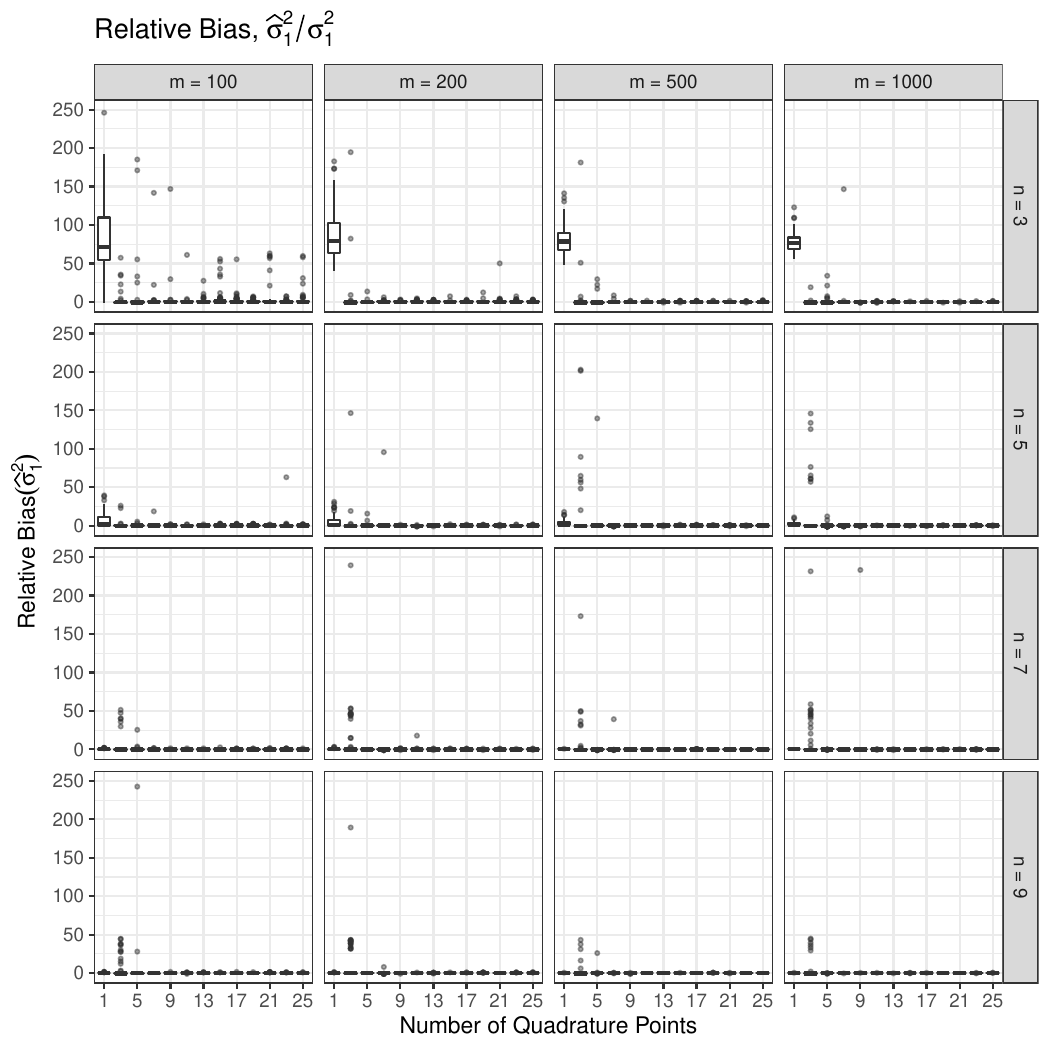}
\caption{Empirical relative bias, $\widehat{\sigma}^2_1 / \sigma^2_1$, for $\sigma^2_1$ in the simulation study of Section 4.3 in the main manuscript, where
$1000$ sets of data were generated from the random-slopes model (\ref{eqn:simmodel1}). The y-axis range is very large so that the scale of the relative bias for 
the Laplace approximation with low $\numpergroup$ is visible. A larger $\quadnum$ leads to greatly reduced bias across values of $\numgroups$ and $\numpergroup$.}
\label{fig:sigmasq1bias}
\end{figure}

\begin{figure}[p]
\centering
\includegraphics{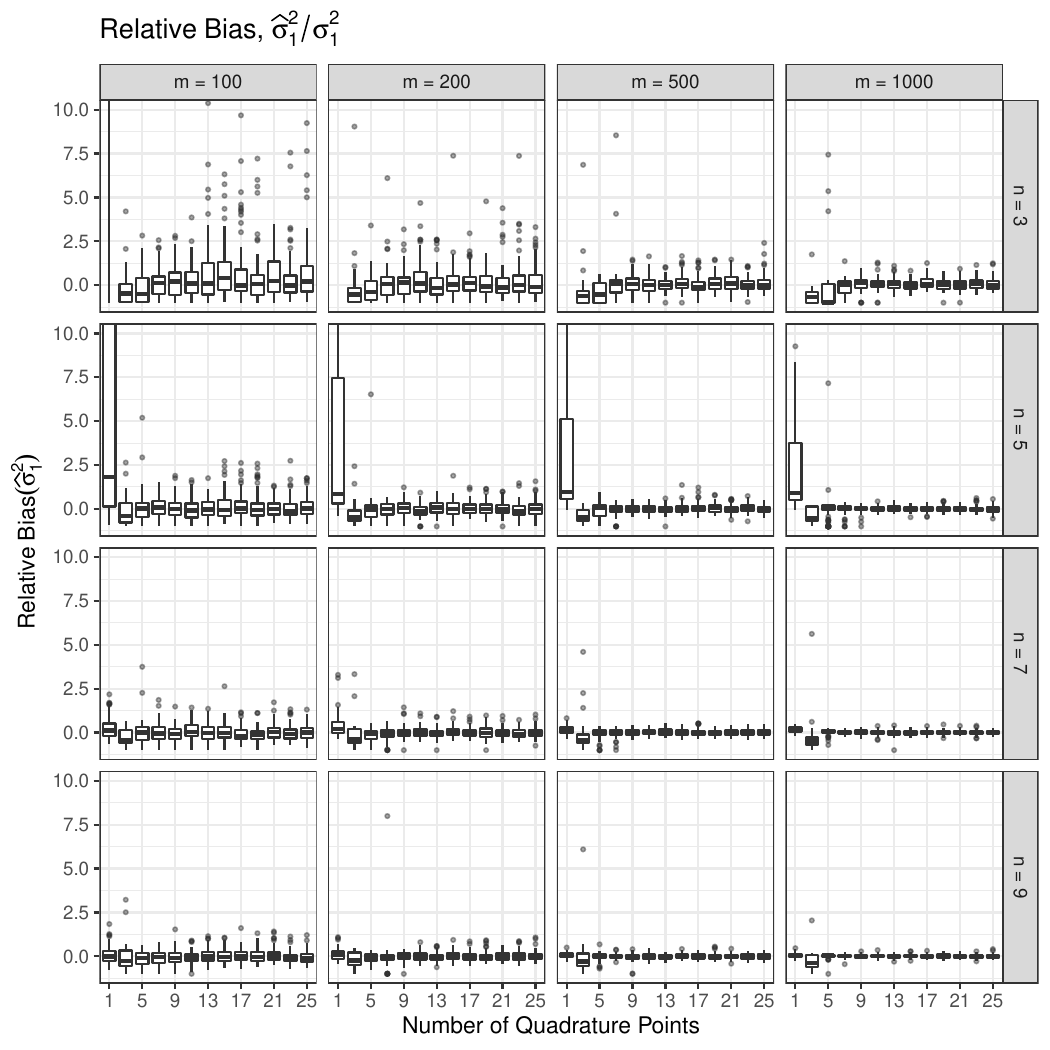}
\caption{Empirical relative bias, $\widehat{\sigma}^2_1 / \sigma^2_1$, for $\sigma^2_1$ in the simulation study of Section 4.3 in the main manuscript, where
$1000$ sets of data were generated from the random-slopes model (\ref{eqn:simmodel1}). The y-axis range is zoomed in so that the pattern in bias across
all $\numgroups,\numpergroup,\quadnum$ is visible, except for the massive biases incurred with $\numpergroup=3$ and $\quadnum=1$. For all $\numpergroup$,
 $\quadnum=1$ exhibits positive bias and $\quadnum=3$ exhibits negative bias, on average, with larger $\numpergroup$ diminishing this effect. 
However, larger $\quadnum$ yields small bias for estimating $\sigma^2_1$ for all values of $\numgroups,\numpergroup$, even large $\numgroups$ with small $\numpergroup$,
where the likelihood should be the most difficult to approximate accurately. The true value, $\sigma^2_1=2$ on the logit scale, can be regarded as very large and
difficult to estimate accurately.}
\label{fig:sigmasq1biaszoom}
\end{figure}

\begin{figure}[p]
\centering
\includegraphics{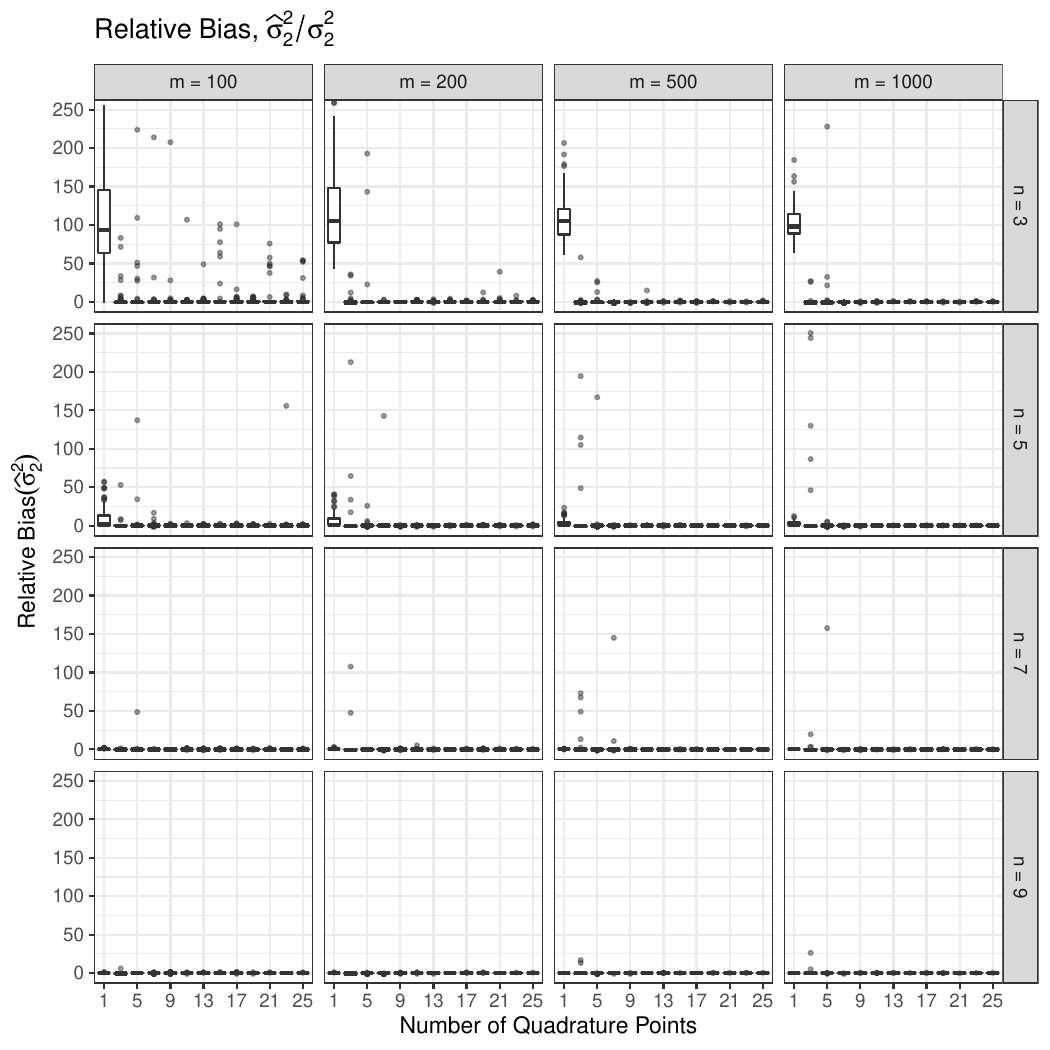}
\caption{Empirical relative bias, $\widehat{\sigma}^2_2 / \sigma^2_2$, for $\sigma^2_2$ in the simulation study of Section 4.3 in the main manuscript, where
$1000$ sets of data were generated from the random-slopes model (\ref{eqn:simmodel1}). The y-axis range is very large so that the scale of the relative bias for 
the Laplace approximation with low $\numpergroup$ is visible. A larger $\quadnum$ leads to greatly reduced bias across values of $\numgroups$ and $\numpergroup$.}
\label{fig:sigmasq2bias}
\end{figure}

\begin{figure}[p]
\centering
\includegraphics{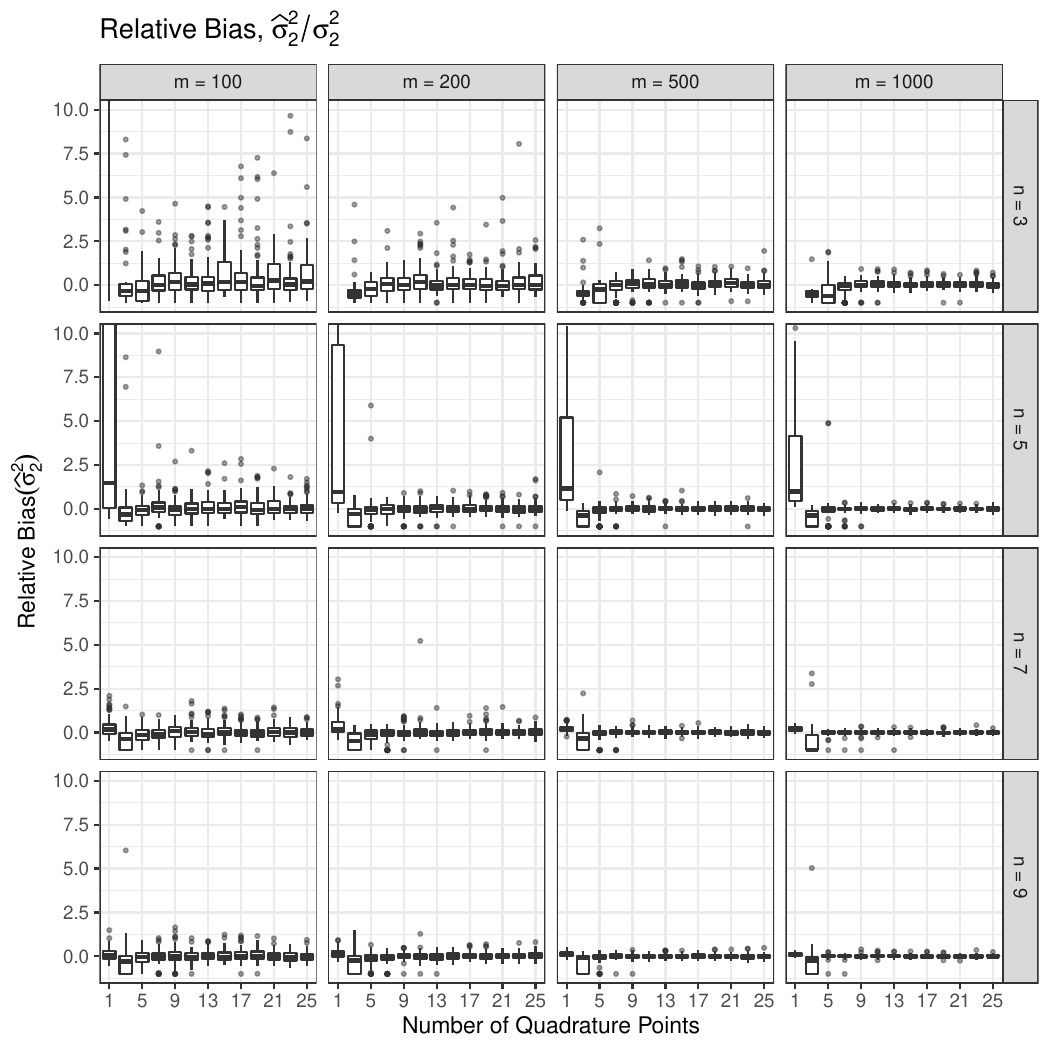}
\caption{Empirical relative bias, $\widehat{\sigma}^2_2 / \sigma^2_2$, for $\sigma^2_2$ in the simulation study of Section 4.3 in the main manuscript, where
$1000$ sets of data were generated from the random-slopes model (\ref{eqn:simmodel1}). The y-axis range is zoomed in so that the pattern in bias across
all $\numgroups,\numpergroup,\quadnum$ is visible, except for the massive biases incurred with $\numpergroup=3$ and $\quadnum=1$. For all $\numpergroup$,
 $\quadnum=1$ exhibits positive bias and $\quadnum=3$ exhibits negative bias, on average, with larger $\numpergroup$ diminishing this effect. 
However, larger $\quadnum$ yields small bias for estimating $\sigma^2_2$ for all values of $\numgroups,\numpergroup$, even large $\numgroups$ with small $\numpergroup$,
where the likelihood should be the most difficult to approximate accurately.}
\label{fig:sigmasq2biaszoom}
\end{figure}

\begin{figure}[p]
\centering
\includegraphics{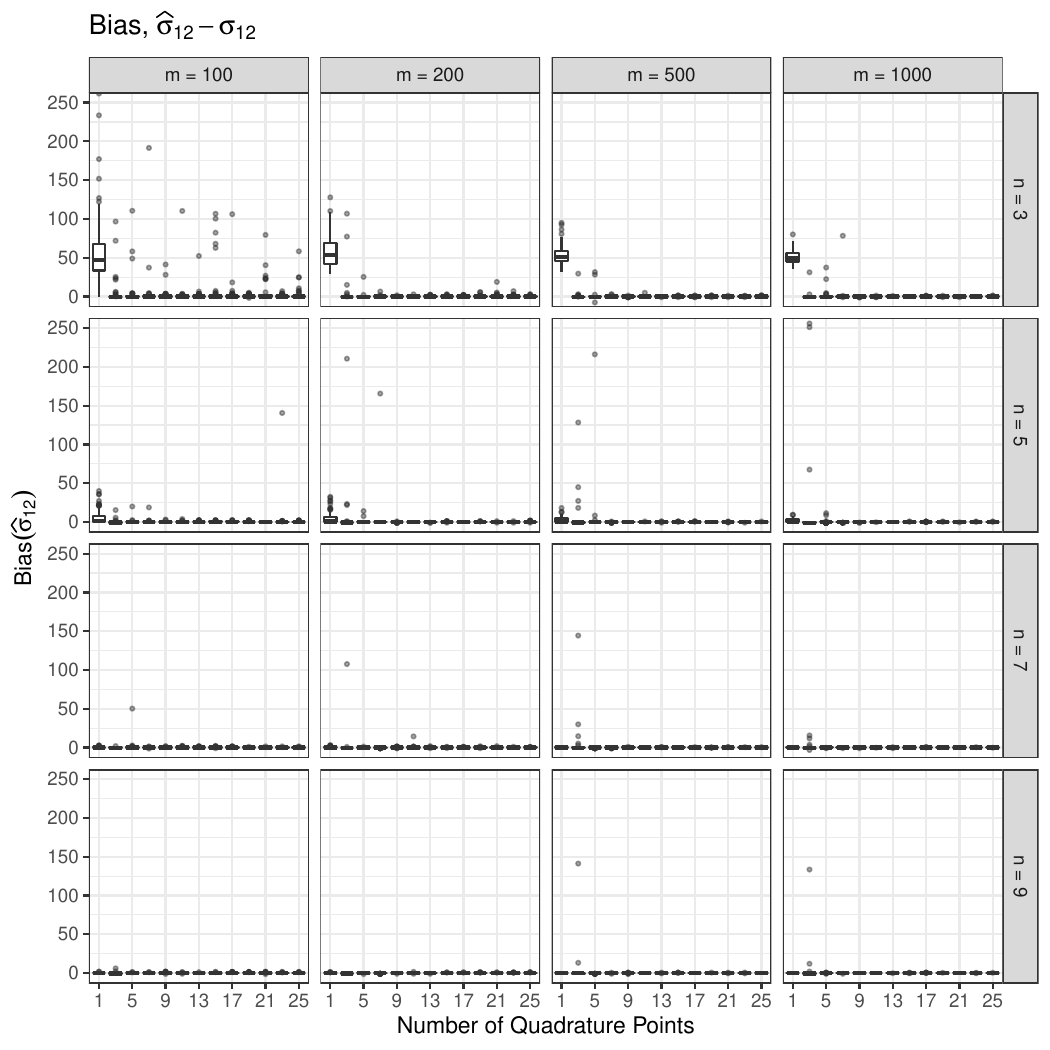}
\caption{Empirical bias, $\widehat{\sigma}_{12} - \sigma_{12}$, for $\sigma_{12}$ in the simulation study of Section 4.3 in the main manuscript, where
$1000$ sets of data were generated from the random-slopes model (\ref{eqn:simmodel1}). The y-axis range is very large so that the scale of the relative bias for 
the Laplace approximation with low $\numpergroup$ is visible. A larger $\quadnum$ leads to greatly reduced bias across values of $\numgroups$ and $\numpergroup$.}
\label{fig:sigmacov1bias}
\end{figure}

\begin{figure}[p]
\centering
\includegraphics{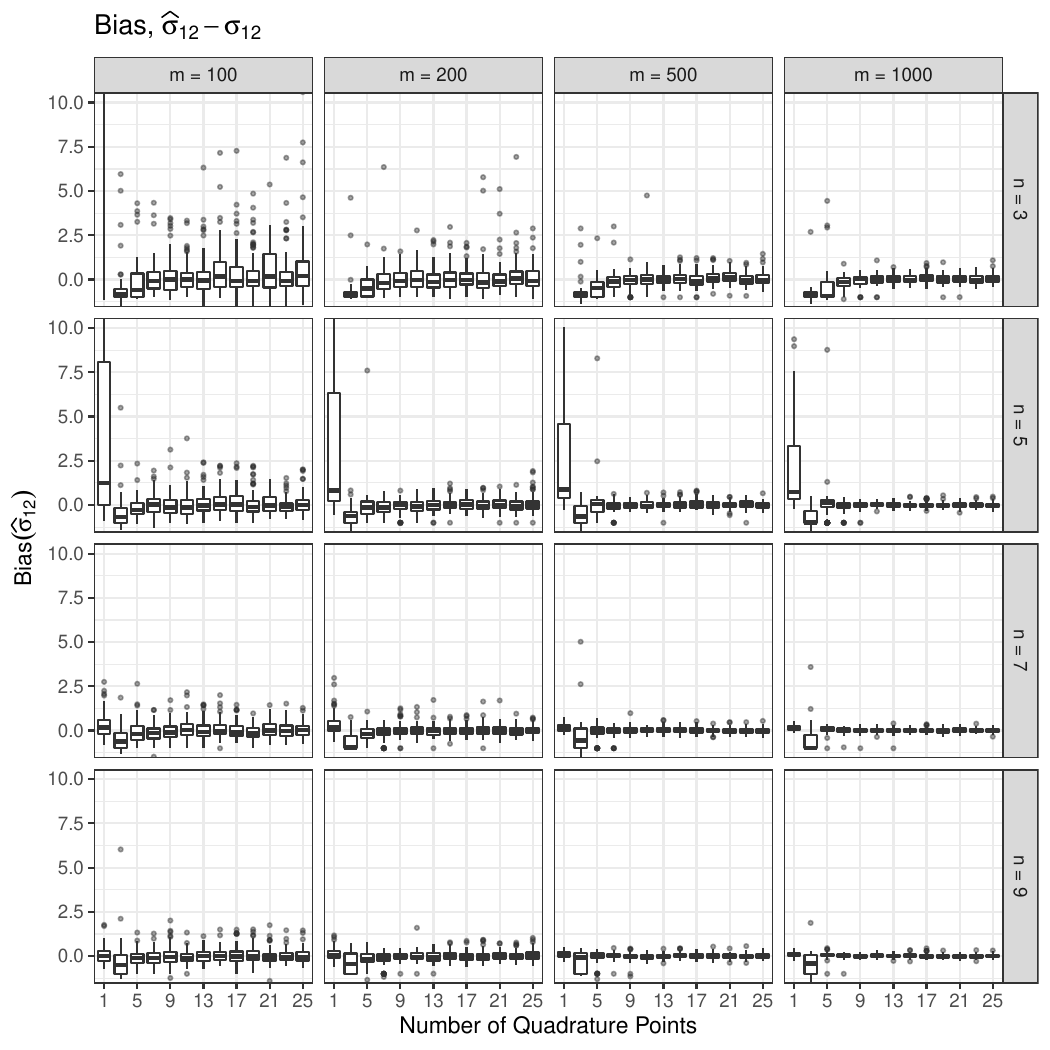}
\caption{Empirical bias, $\widehat{\sigma}_{12} - \sigma_{12}$, for $\sigma_{12}$ in the simulation study of Section 4.3 in the main manuscript, where
$1000$ sets of data were generated from the random-slopes model (\ref{eqn:simmodel1}). The y-axis range is zoomed in so that the pattern in bias across
all $\numgroups,\numpergroup,\quadnum$ is visible, except for the massive biases incurred with $\numpergroup=3$ and $\quadnum=1$. For all $\numpergroup$,
 $\quadnum=1$ exhibits positive bias and $\quadnum=3$ exhibits negative bias, on average, with larger $\numpergroup$ diminishing this effect. 
However, larger $\quadnum$ yields small bias for estimating $\sigma^2_2$ for all values of $\numgroups,\numpergroup$, even large $\numgroups$ with small $\numpergroup$,
where the likelihood should be the most difficult to approximate accurately.}
\label{fig:sigmacov1biaszoom}
\end{figure}

The bias plots, Figures \ref{fig:beta0bias} -- \ref{fig:sigmacov1biaszoom}, report boxplots of $\widehat{\theta} - \theta$ (bias)
or $\widehat{\theta}/\theta$ (relative bias) as appropriate, where each point is the realized value of this statistic from one
simulated set of data and model fit. 

\begin{figure}[p]
\centering
\includegraphics{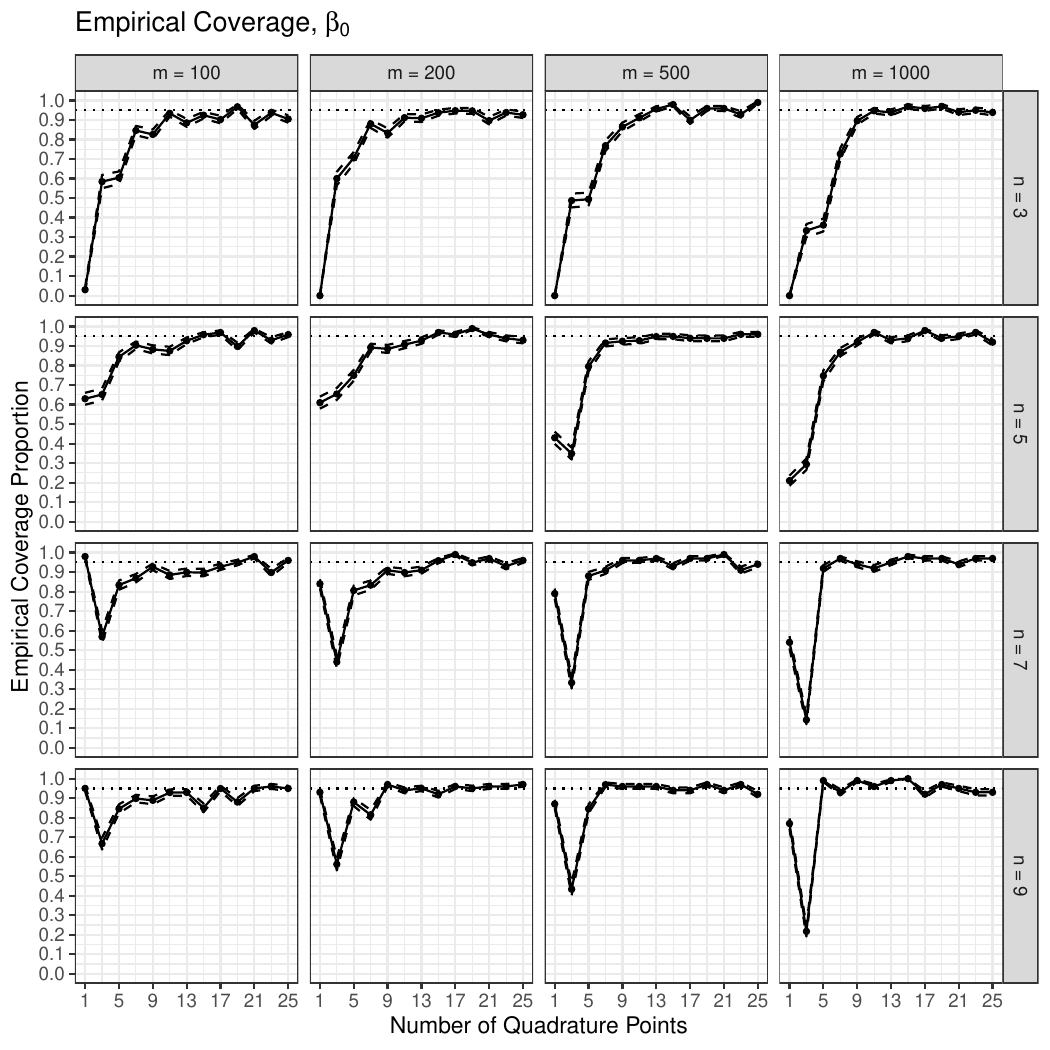}
\caption{Empirical coverage proportion for $\widehat{\beta}_0$ in the simulation study of Section 4.3 in the main manuscript, where
$1000$ sets of data were generated from the random-slopes model (\ref{eqn:simmodel1}).
The Laplace ($\quadnum=1$) coverage is low for $\numpergroup=3,5$ due to high bias and poor standard error estimation. Although the coverage appears nominal for $\numpergroup\geq7$, 
the intervals are also much wider than those for higher $\quadnum$ (Figure \ref{fig:beta0length}). 
The $\quadnum=3$ coverages are low due to strong negative bias. In all cases,
taking $\quadnum$ large enough leads to nominal empirical coverage, motivating the
use of adaptive quadrature for fitting these models.}
\label{fig:beta0covr}
\end{figure}

\begin{figure}[p]
\centering
\includegraphics{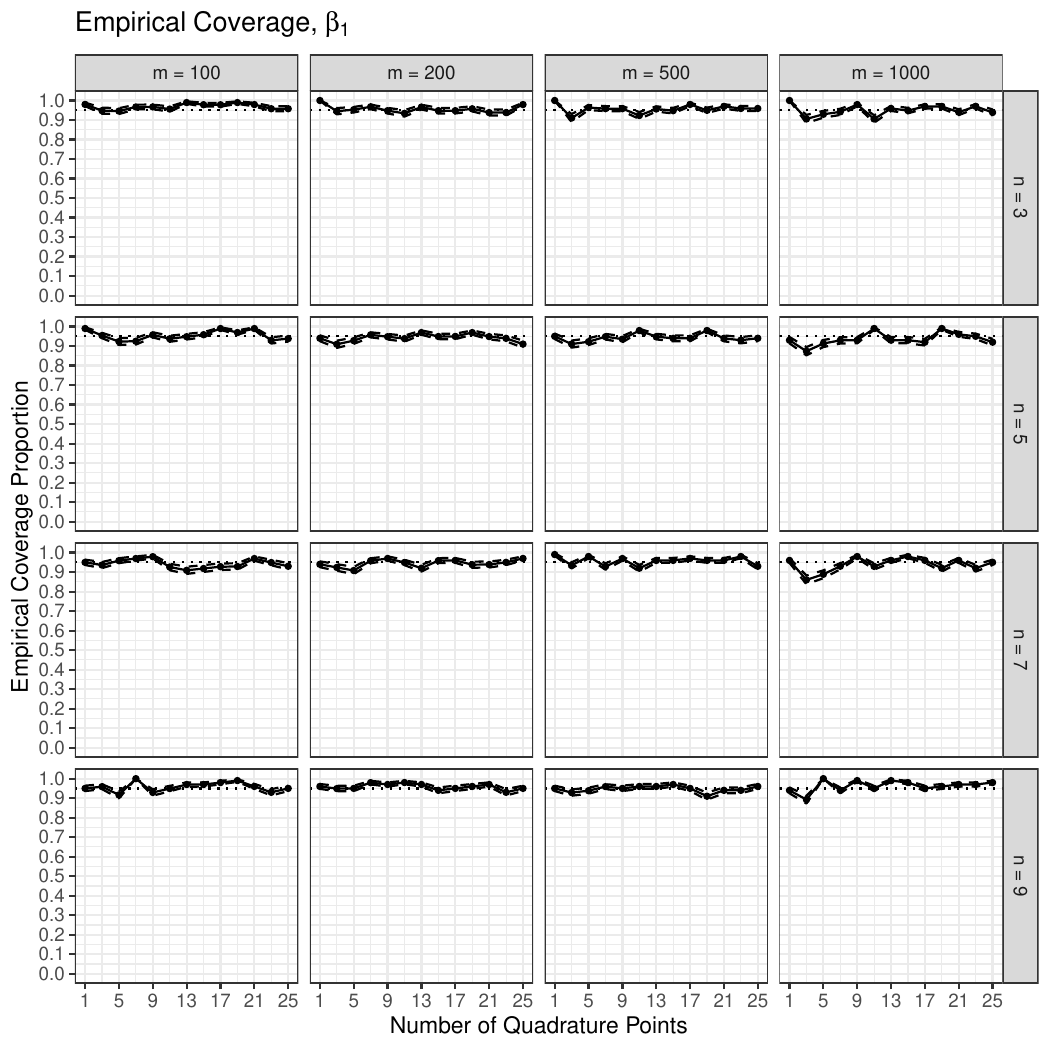}
\caption{Empirical coverage proportion for $\widehat{\beta}_1$ in the simulation study of Section 4.3 in the main manuscript, where
$1000$ sets of data were generated from the random-slopes model (\ref{eqn:simmodel1}).
The coverage is nominal across all values of $\numgroups,\numpergroup,\quadnum$,
which is expected for this parameter in this model.}
\label{fig:beta1covr}
\end{figure}

\begin{figure}[p]
\centering
\includegraphics{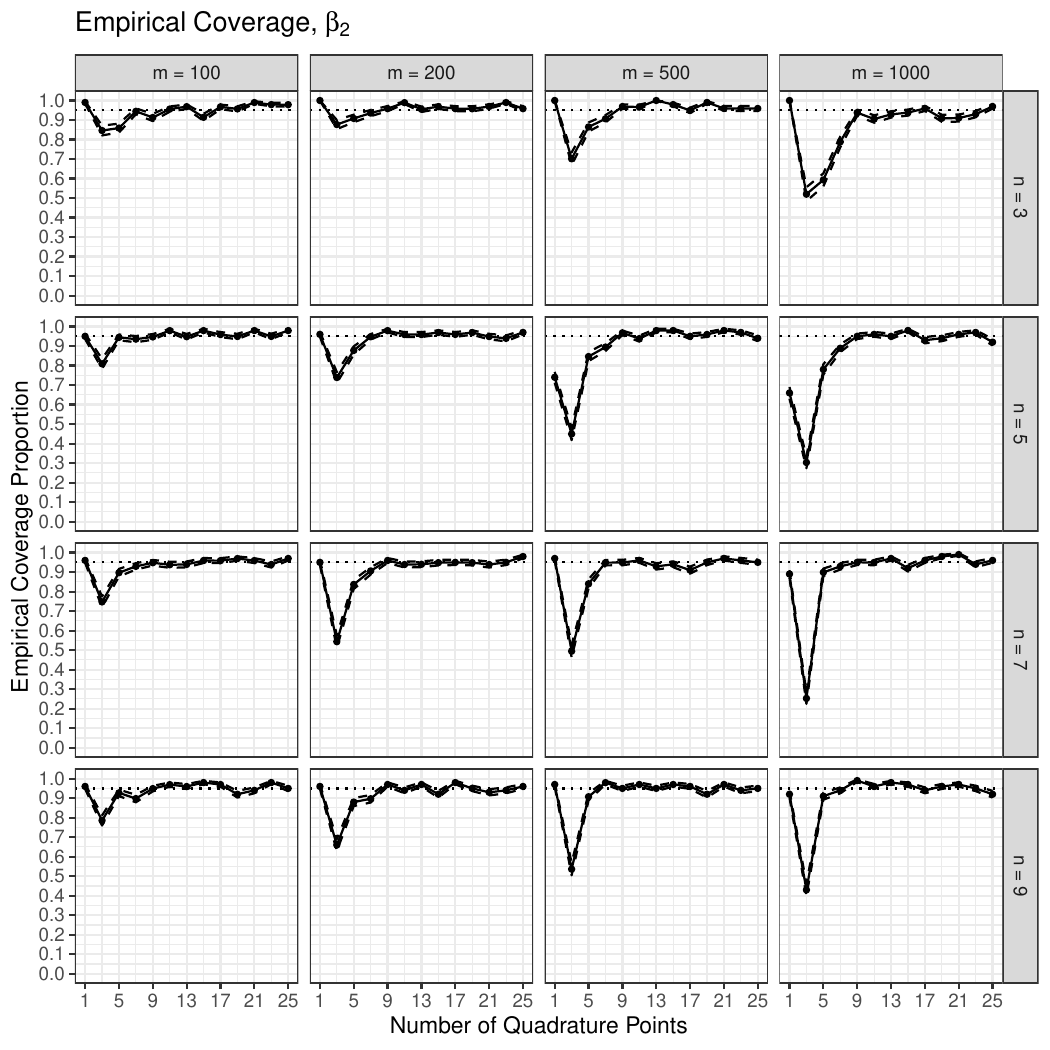}
\caption{Empirical coverage proportion for $\widehat{\beta}_2$ in the simulation study of Section 4.3 in the main manuscript, where
$1000$ sets of data were generated from the random-slopes model (\ref{eqn:simmodel1}).
The coverages are nominal except for $\quadnum=3,5$ for certain $\numgroups,\numpergroup$.
Examining the lengths of the intervals (Figure \ref{fig:beta2length}) shows that this is due to under-estimation
of standard error; increasing $\quadnum$ mitigates this and yields nominal coverage for all
$\numgroups,\numpergroup$.}
\label{fig:beta2covr}
\end{figure}

\begin{figure}[p]
\centering
\includegraphics{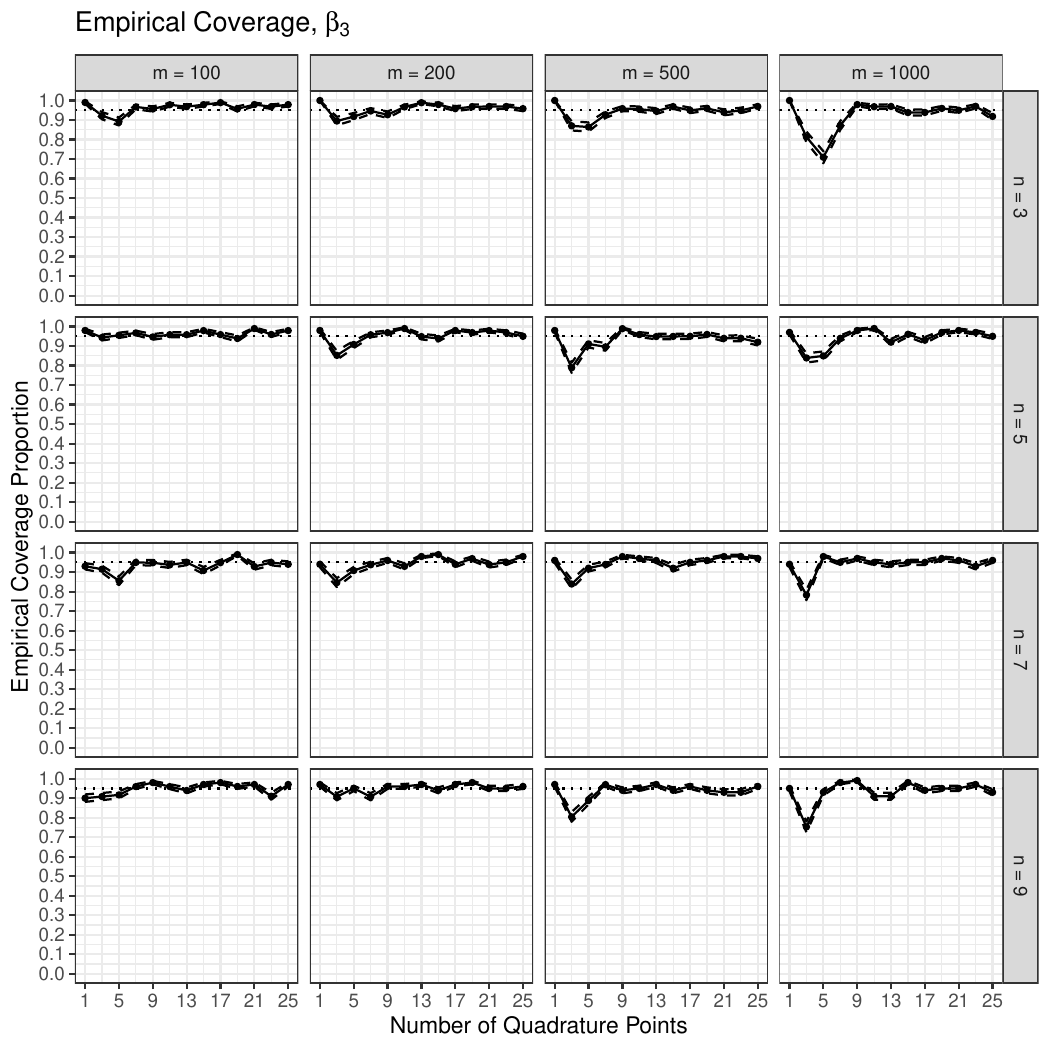}
\caption{Empirical coverage proportion for $\widehat{\beta}_3$ in the simulation study of Section 4.3 in the main manuscript, where
$1000$ sets of data were generated from the random-slopes model (\ref{eqn:simmodel1}).
The coverages are nominal except for $\quadnum=3,5$ for certain $\numgroups,\numpergroup$.
Examining the lengths of the intervals (Figure \ref{fig:beta3length}) shows that this is due to under-estimation
of standard error; increasing $\quadnum$ mitigates this and yields nominal coverage for all
$\numgroups,\numpergroup$.}
\label{fig:beta3covr}
\end{figure}

\begin{figure}[p]
\centering
\includegraphics{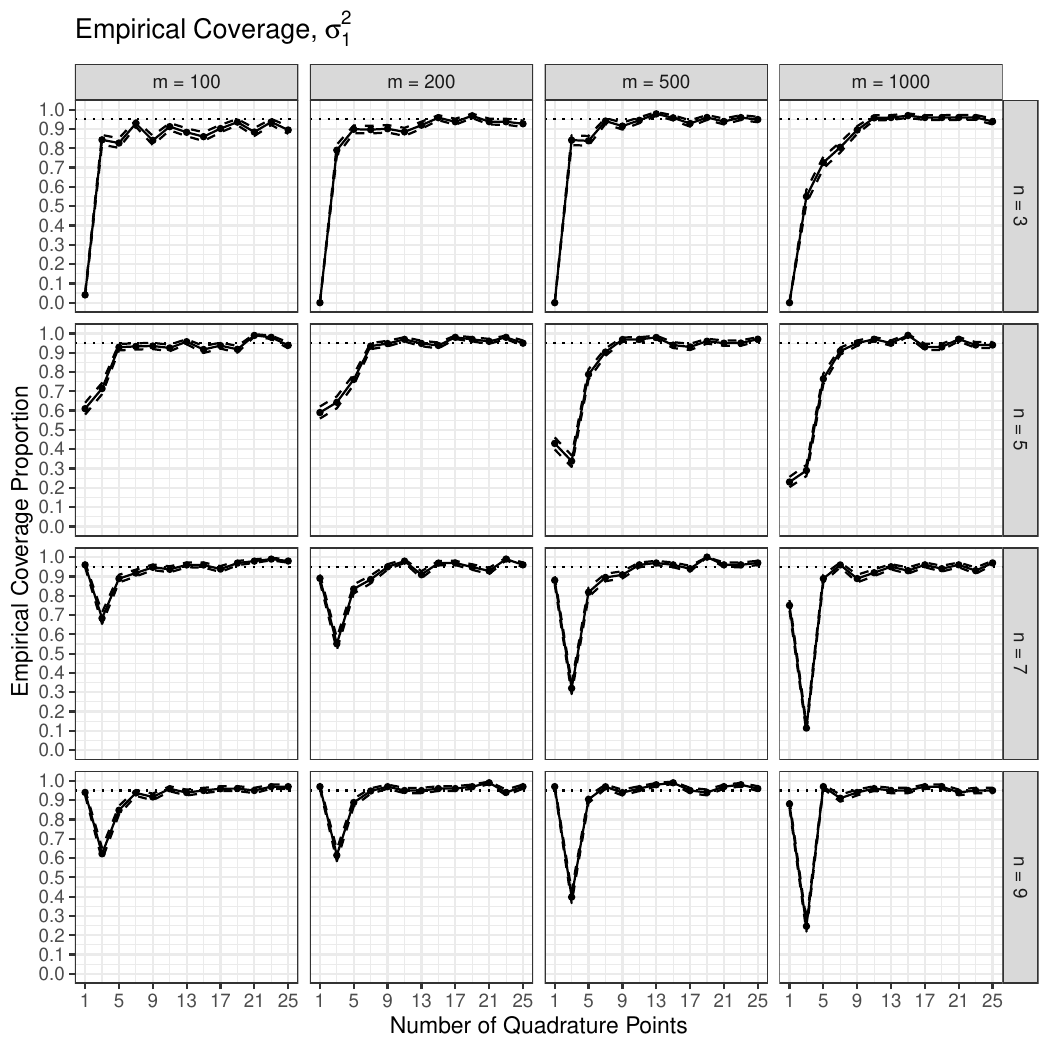}
\caption{Empirical coverage proportion for $\widehat{\sigma}^2_1$ in the simulation study of Section 4.3 in the main manuscript, where
$1000$ sets of data were generated from the random-slopes model (\ref{eqn:simmodel1}).
The Laplace ($\quadnum=1$) coverage is low for $\numpergroup=3,5$ due to high bias and poor standard error estimation. Although the coverage appears nominal for $\numpergroup\geq7$, 
the intervals are also much wider than those for higher $\quadnum$ (Figure \ref{fig:sigmasq1length}). 
The $\quadnum=3$ coverages are low due to strong negative bias. In all cases,
taking $\quadnum$ large enough leads to nominal empirical coverage, motivating the
use of adaptive quadrature for fitting these models.}
\label{fig:sigmasq1covr}
\end{figure}

\begin{figure}[p]
\centering
\includegraphics{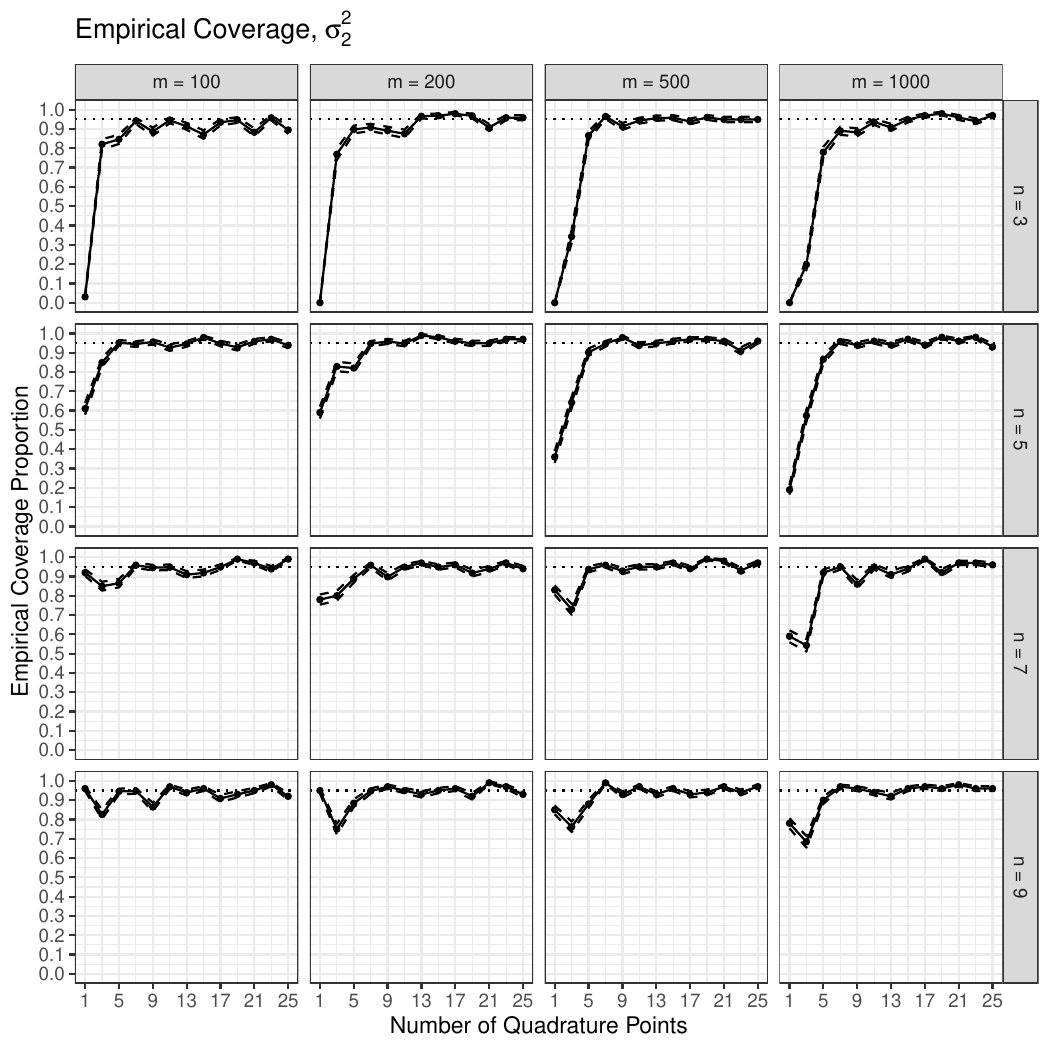}
\caption{Empirical coverage proportion for $\widehat{\sigma}^2_2$ in the simulation study of Section 4.3 in the main manuscript, where
$1000$ sets of data were generated from the random-slopes model (\ref{eqn:simmodel1}).
The Laplace ($\quadnum=1$) coverage is low for $\numpergroup=3,5$ due to high bias and poor standard error estimation. Although the coverage appears nominal for $\numpergroup\geq7$, 
the intervals are also much wider than those for higher $\quadnum$ (Figure \ref{fig:sigmasq2length}). 
The $\quadnum=3$ coverages are low due to strong negative bias. In all cases,
taking $\quadnum$ large enough leads to nominal empirical coverage, motivating the
use of adaptive quadrature for fitting these models.}
\label{fig:sigmasq2covr}
\end{figure}

\begin{figure}[p]
\centering
\includegraphics{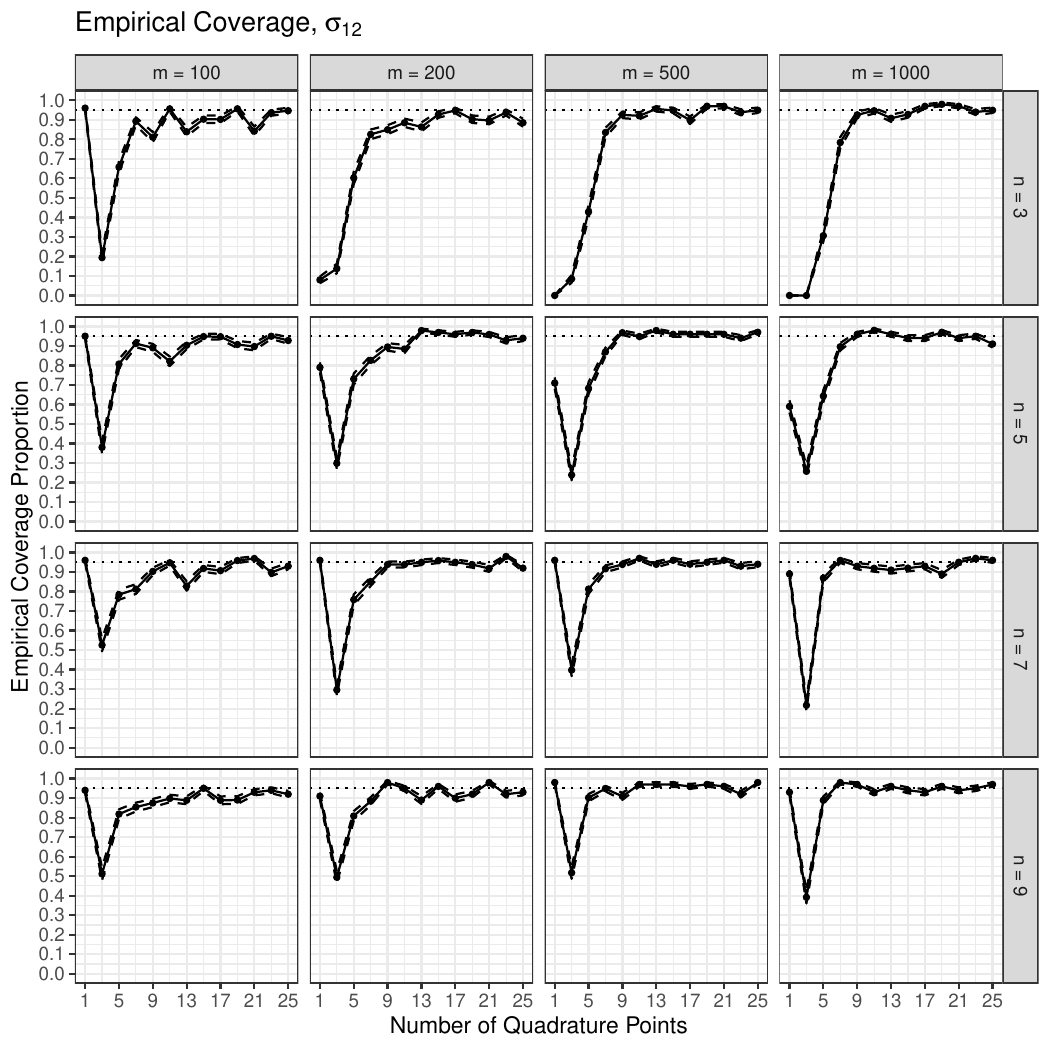}
\caption{Empirical coverage proportion for $\widehat{\sigma}_{12}$ in the simulation study of Section 4.3 in the main manuscript, where
$1000$ sets of data were generated from the random-slopes model (\ref{eqn:simmodel1}).
The Laplace ($\quadnum=1$) coverage is low for $\numpergroup=3,5$ due to high bias and poor standard error estimation. Although the coverage appears nominal for $\numpergroup\geq7$, 
the intervals are also much wider than those for higher $\quadnum$ (Figure \ref{fig:sigmacov1length}). 
The $\quadnum=3$ coverages are low due to strong negative bias. In all cases,
taking $\quadnum$ large enough leads to nominal empirical coverage, motivating the
use of adaptive quadrature for fitting these models.}
\label{fig:sigmacov1covr}
\end{figure}

\begin{figure}[p]
\centering
\includegraphics{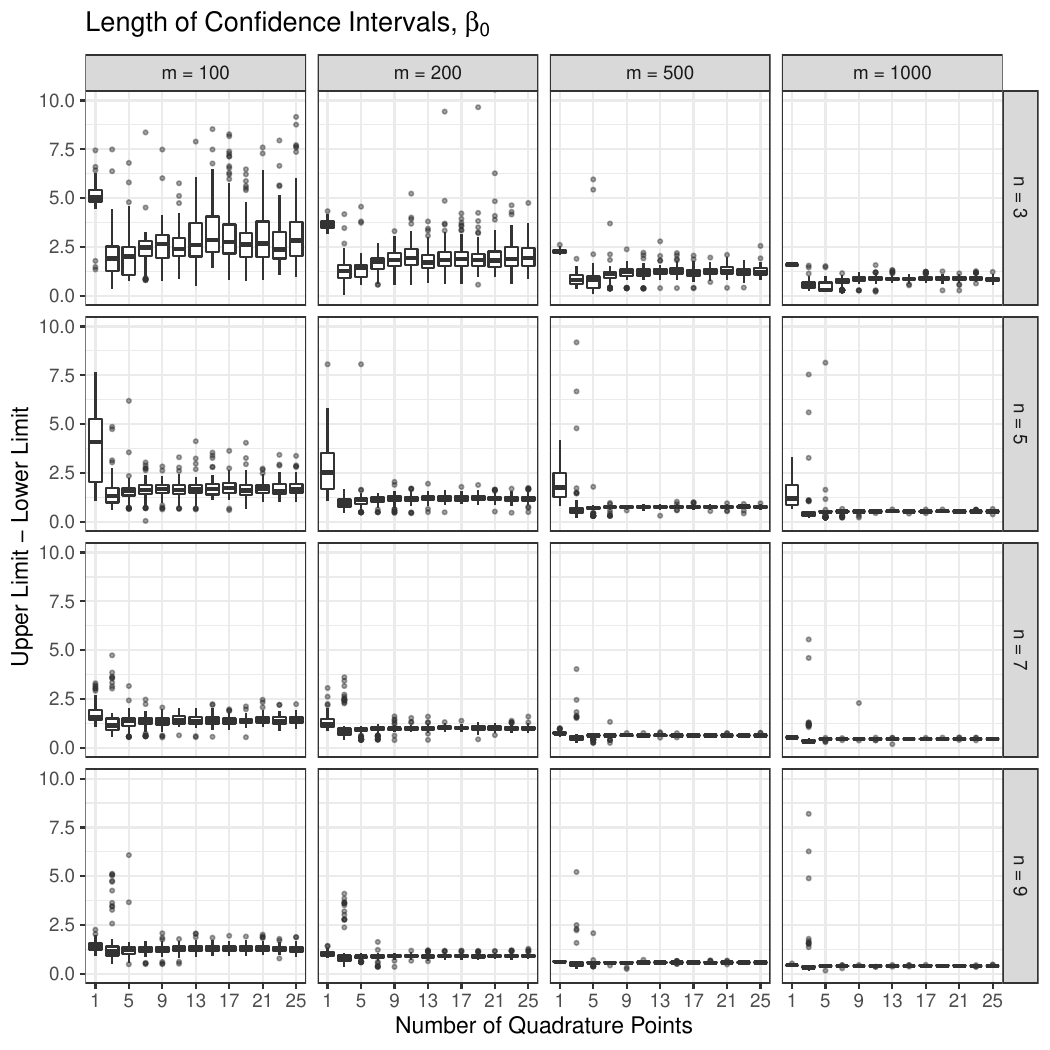}
\caption{Length of the Wald intervals for $\widehat{\beta}_{0}$ in the simulation study of Section 4.3 in the main manuscript, where
$1000$ sets of data were generated from the random-slopes model (\ref{eqn:simmodel1}).
The Laplace ($\quadnum=1$) intervals are substantially wider than those for larger $\quadnum$.
Those for $\quadnum=3,5$ are narrower, and then the lengths level off for $\quadnum\geq7$.
This pattern coincides with coverages converging to nominal (Figure \ref{fig:beta0covr}) as $\quadnum$ increases, across values
of $\numgroups,\numpergroup$.}
\label{fig:beta0length}
\end{figure}

\begin{figure}[p]
\centering
\includegraphics{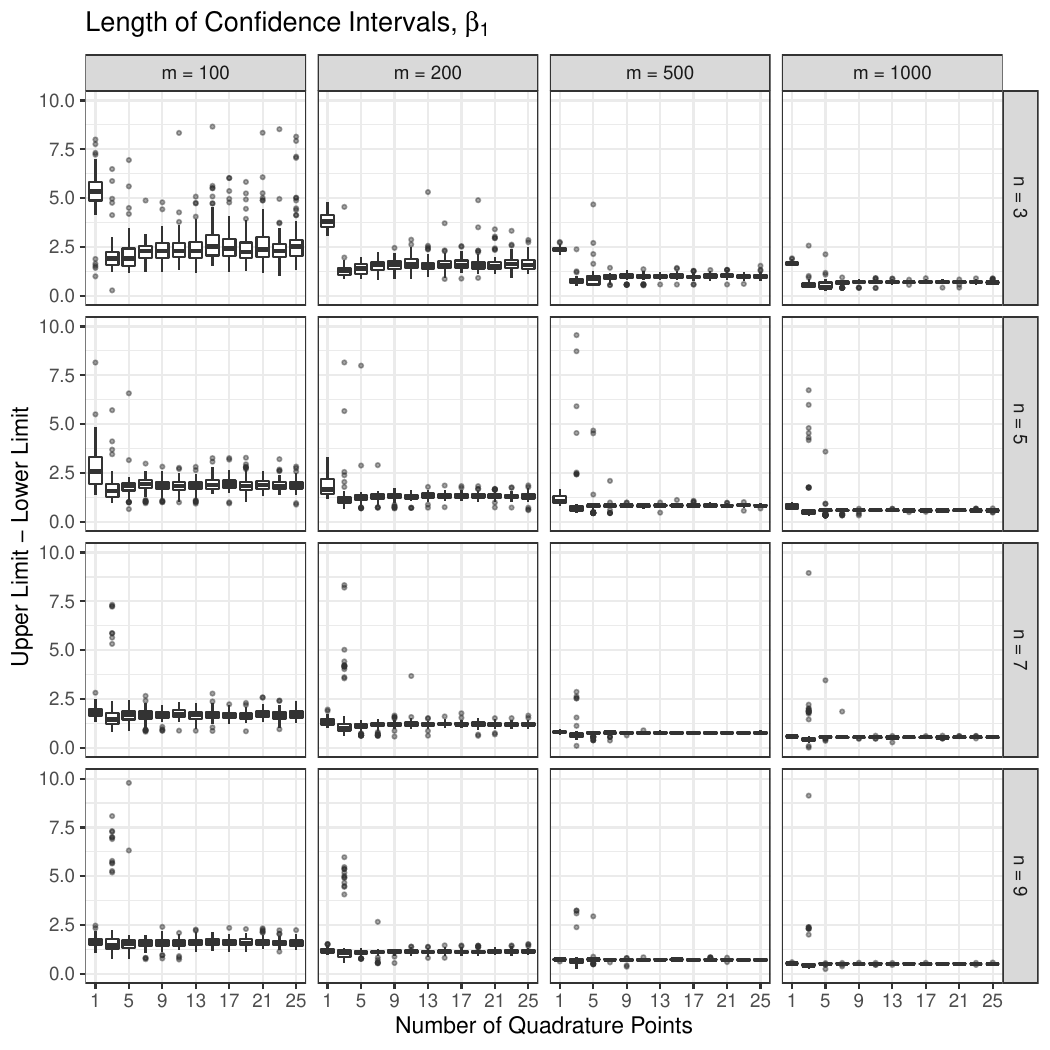}
\caption{Length of the Wald intervals for $\widehat{\beta}_{1}$ in the simulation study of Section 4.3 in the main manuscript, where
$1000$ sets of data were generated from the random-slopes model (\ref{eqn:simmodel1}).
The Laplace ($\quadnum=1$) intervals are substantially wider than those for larger $\quadnum$.
Those for $\quadnum=3,5$ are narrower, and then the lengths level off for $\quadnum\geq7$.
This pattern coincides with coverages converging to nominal (Figure \ref{fig:beta1covr}) as $\quadnum$ increases, across values
of $\numgroups,\numpergroup$.}
\label{fig:beta1length}
\end{figure}

\begin{figure}[p]
\centering
\includegraphics{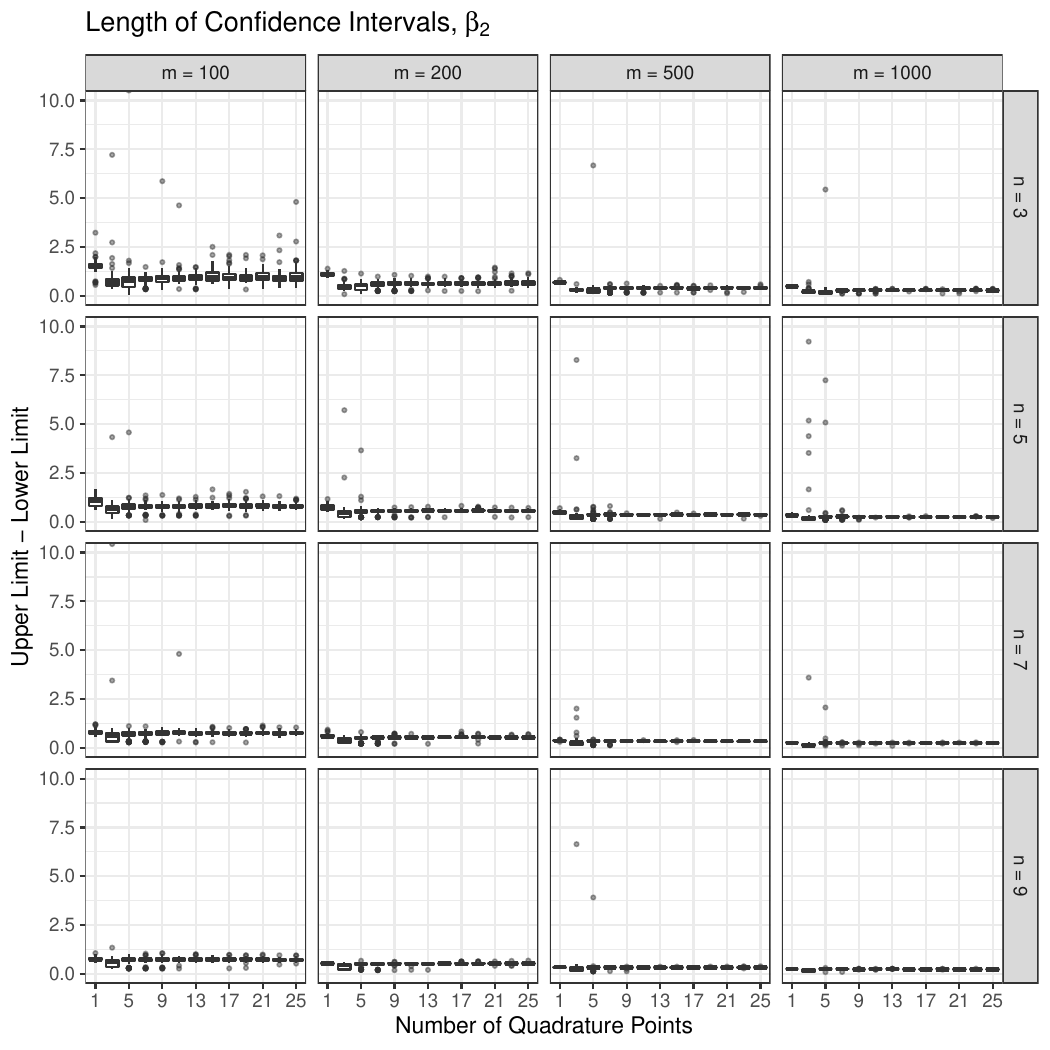}
\caption{Length of the Wald intervals for $\widehat{\beta}_{2}$ in the simulation study of Section 4.3 in the main manuscript, where
$1000$ sets of data were generated from the random-slopes model (\ref{eqn:simmodel1}).
The Laplace ($\quadnum=1$) intervals are somewhat wider than those for larger $\quadnum$.
Those for $\quadnum=3,5$ are narrower, and then the lengths level off for $\quadnum\geq7$.
This pattern coincides with coverages converging to nominal (Figure \ref{fig:beta2covr}) as $\quadnum$ increases, across values
of $\numgroups,\numpergroup$.}
\label{fig:beta2length}
\end{figure}

\begin{figure}[p]
\centering
\includegraphics{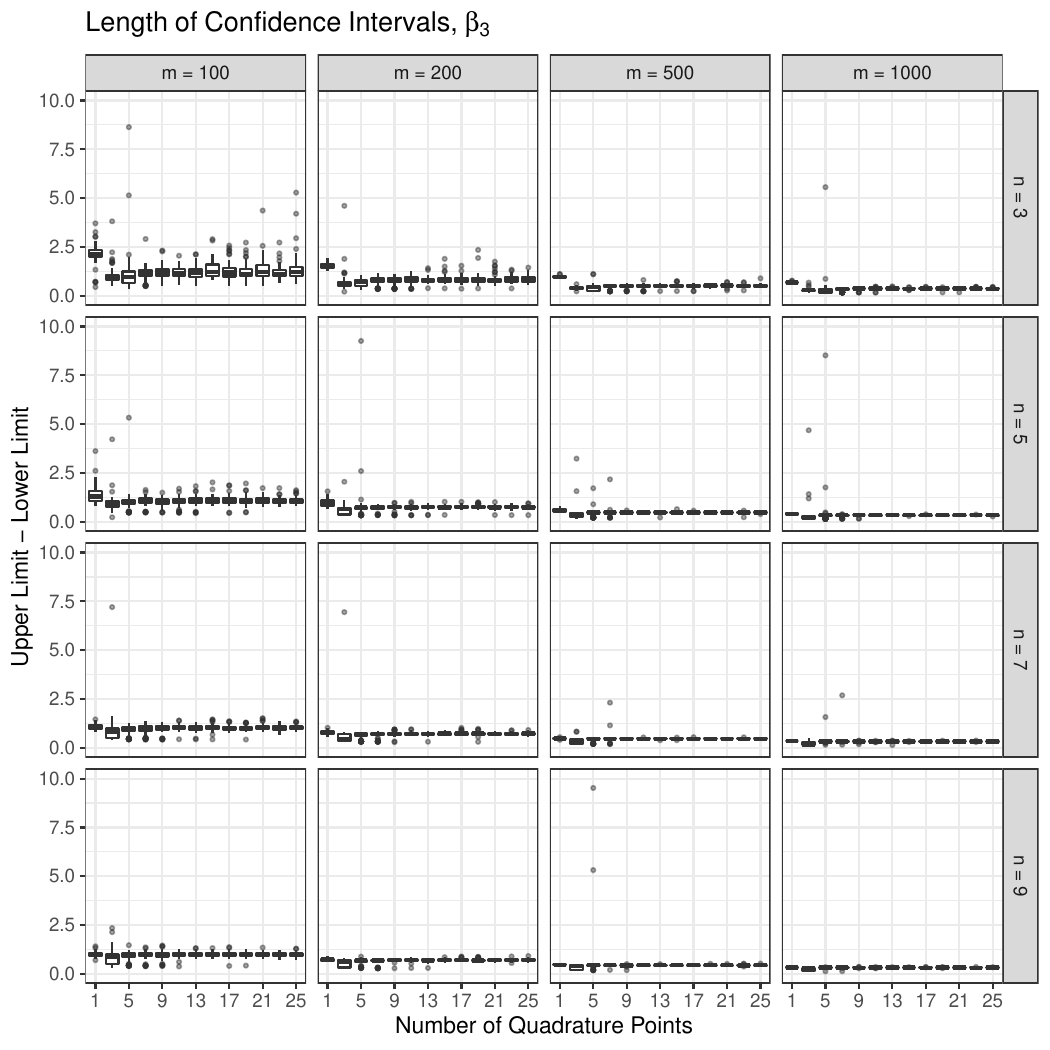}
\caption{Length of the Wald intervals for $\widehat{\beta}_{3}$ in the simulation study of Section 4.3 in the main manuscript, where
$1000$ sets of data were generated from the random-slopes model (\ref{eqn:simmodel1}).The Laplace ($\quadnum=1$) intervals are somewhat wider than those for larger $\quadnum$.
Those for $\quadnum=3,5$ are narrower, and then the lengths level off for $\quadnum\geq7$.
This pattern coincides with coverages converging to nominal (Figure \ref{fig:beta3covr}) as $\quadnum$ increases, across values
of $\numgroups,\numpergroup$.}
\label{fig:beta3length}
\end{figure}

\begin{figure}[p]
\centering
\includegraphics{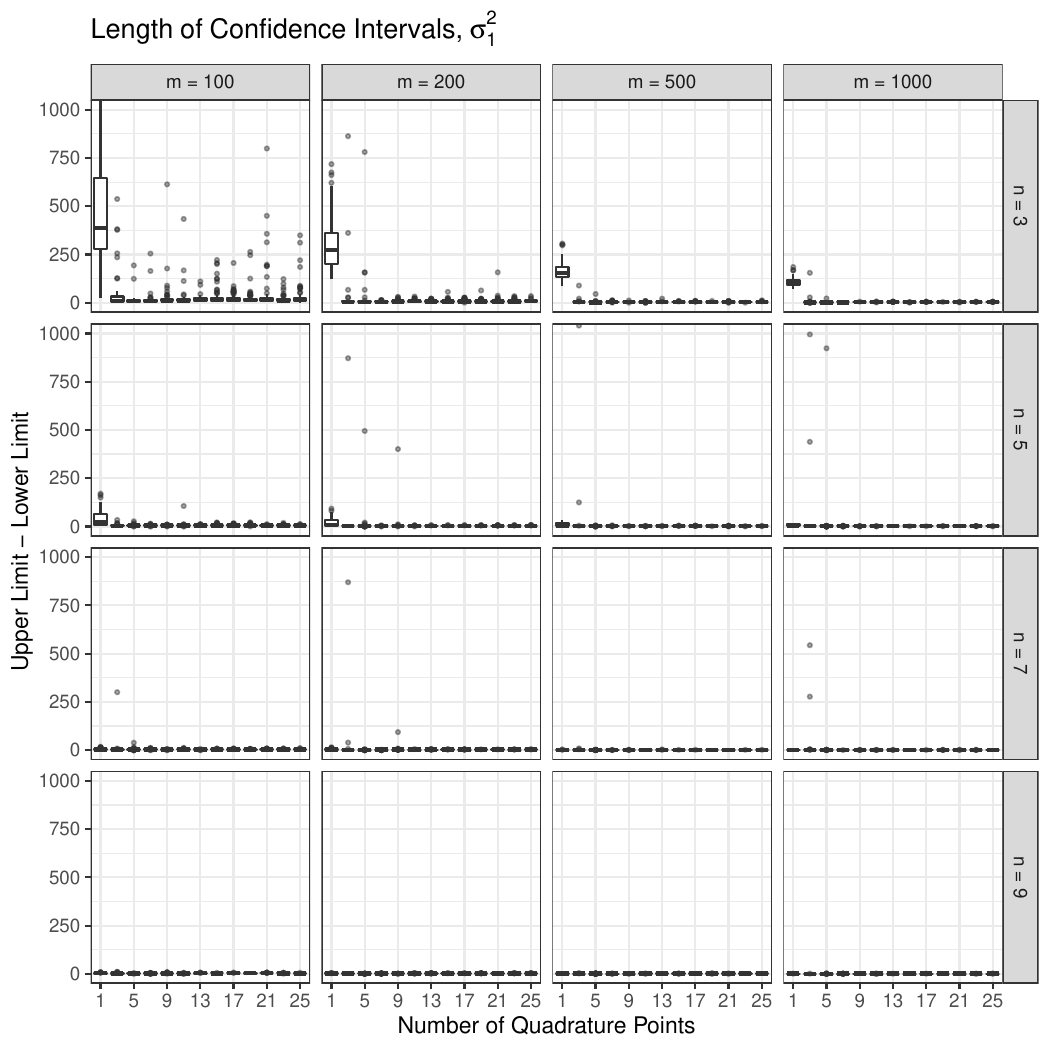}
\caption{Length of the Wald intervals for $\widehat{\sigma}^2_{1}$ in the simulation study of Section 4.3 in the main manuscript, where
$1000$ sets of data were generated from the random-slopes model (\ref{eqn:simmodel1}).
The Laplace ($\quadnum=1$) intervals are orders of magnitude wider than those for larger $\quadnum$; the y-axis is zoomed out to capture this.}
\label{fig:sigmasq1length}
\end{figure}

\begin{figure}[p]
\centering
\includegraphics{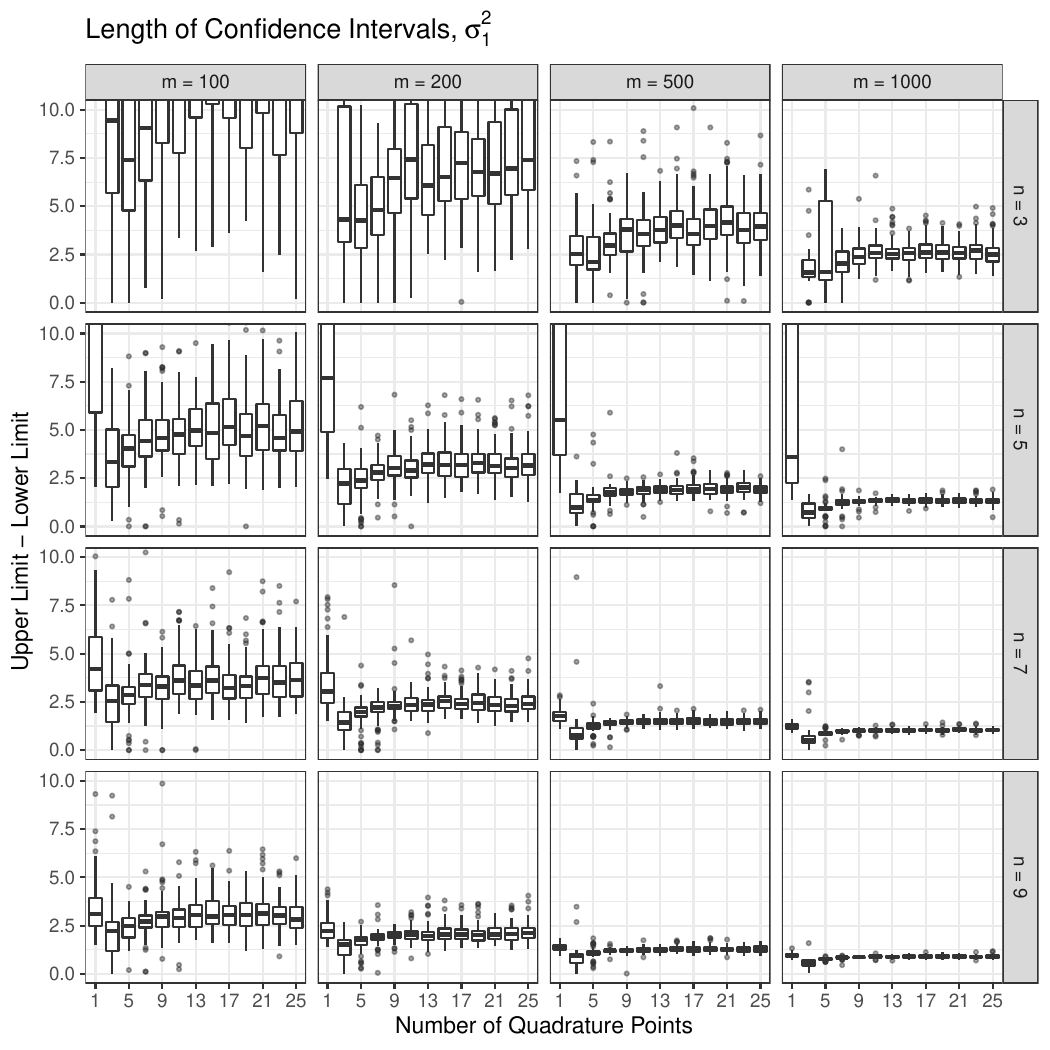}
\caption{Length of the Wald intervals for $\widehat{\sigma}^2_{1}$ in the simulation study of Section 4.3 in the main manuscript, where
$1000$ sets of data were generated from the random-slopes model (\ref{eqn:simmodel1}).
The y-axis is zoomed in, which obscures the massive length of the Laplace ($\quadnum=1$) intervals.
Those for $\quadnum=3,5$ are narrower, and then the lengths level off for $\quadnum\geq7$.
This pattern coincides with coverages converging to nominal (Figure \ref{fig:sigmasq1covr}) as $\quadnum$ increases, across values
of $\numgroups,\numpergroup$.}
\label{fig:sigmasq1lengthzoom}
\end{figure}

\begin{figure}[p]
\centering
\includegraphics{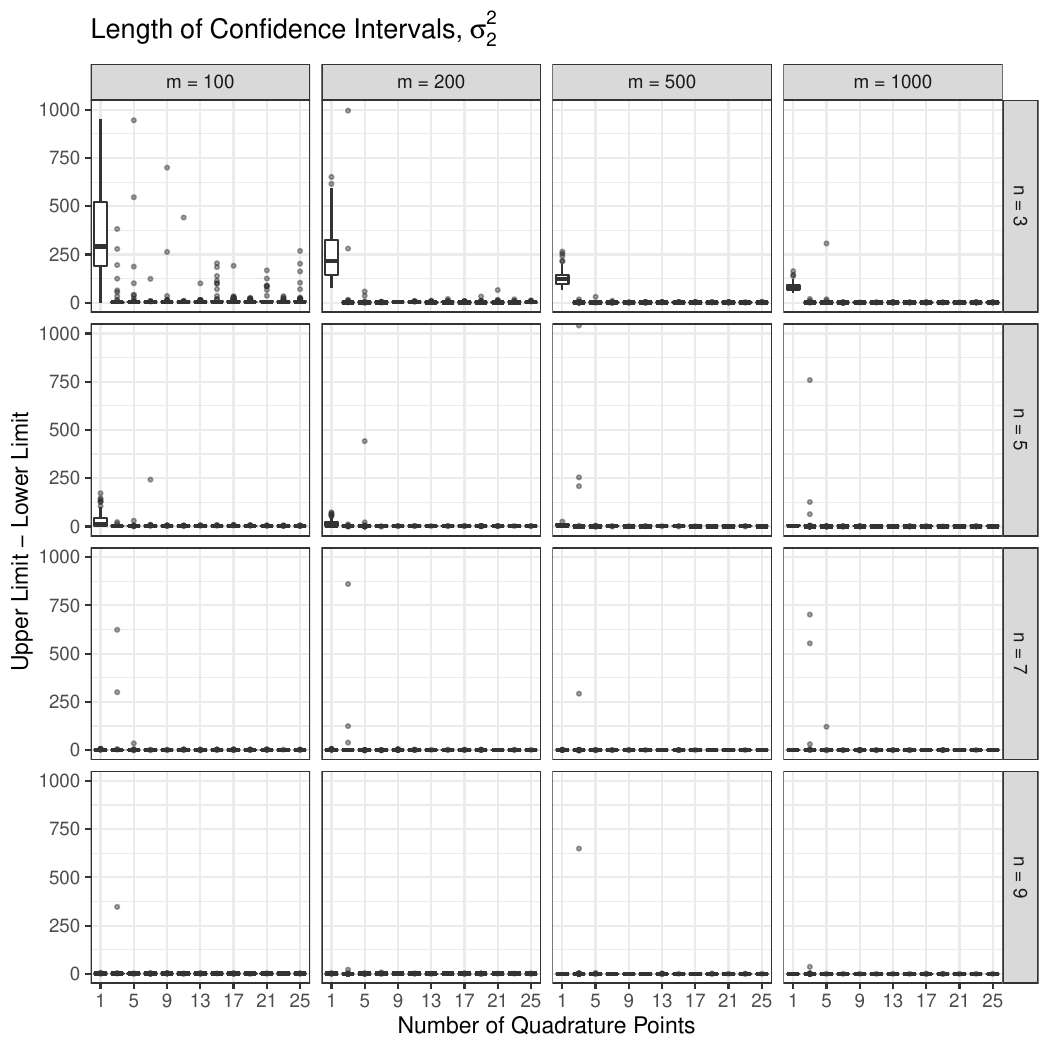}
\caption{Length of the Wald intervals for $\widehat{\sigma}^2_{2}$ in the simulation study of Section 4.3 in the main manuscript, where
$1000$ sets of data were generated from the random-slopes model (\ref{eqn:simmodel1}).
The Laplace ($\quadnum=1$) intervals are orders of magnitude wider than those for larger $\quadnum$; the y-axis is zoomed out to capture this.}
\label{fig:sigmasq2length}
\end{figure}

\begin{figure}[p]
\centering
\includegraphics{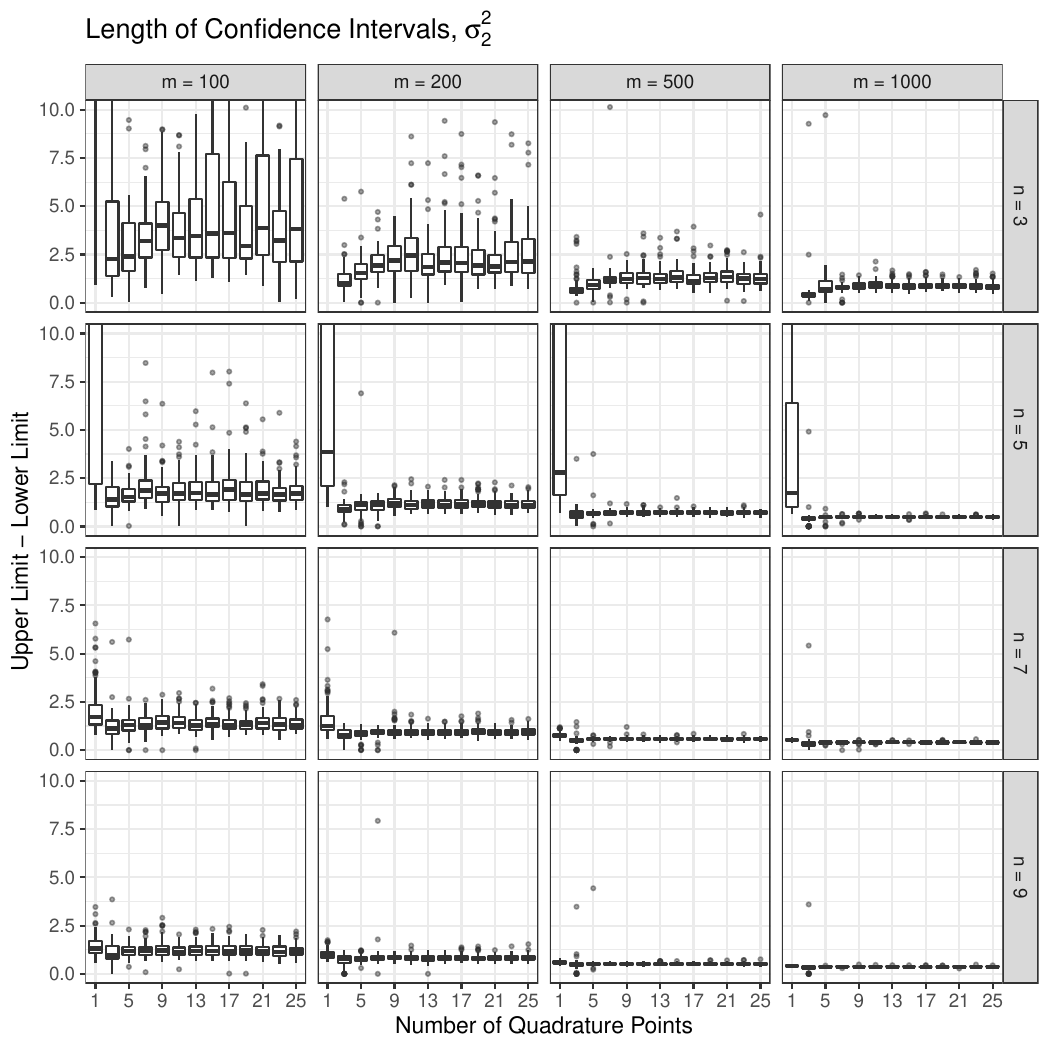}
\caption{Length of the Wald intervals for $\widehat{\sigma}^2_{2}$ in the simulation study of Section 4.3 in the main manuscript, where
$1000$ sets of data were generated from the random-slopes model (\ref{eqn:simmodel1}).
The y-axis is zoomed in, which obscures the massive length of the Laplace ($\quadnum=1$) intervals.
Those for $\quadnum=3,5$ are narrower, and then the lengths level off for $\quadnum\geq7$.
This pattern coincides with coverages converging to nominal (Figure \ref{fig:sigmasq2covr}) as $\quadnum$ increases, across values
of $\numgroups,\numpergroup$.}
\label{fig:sigmasq2lengthzoom}
\end{figure}

\begin{figure}[p]
\centering
\includegraphics{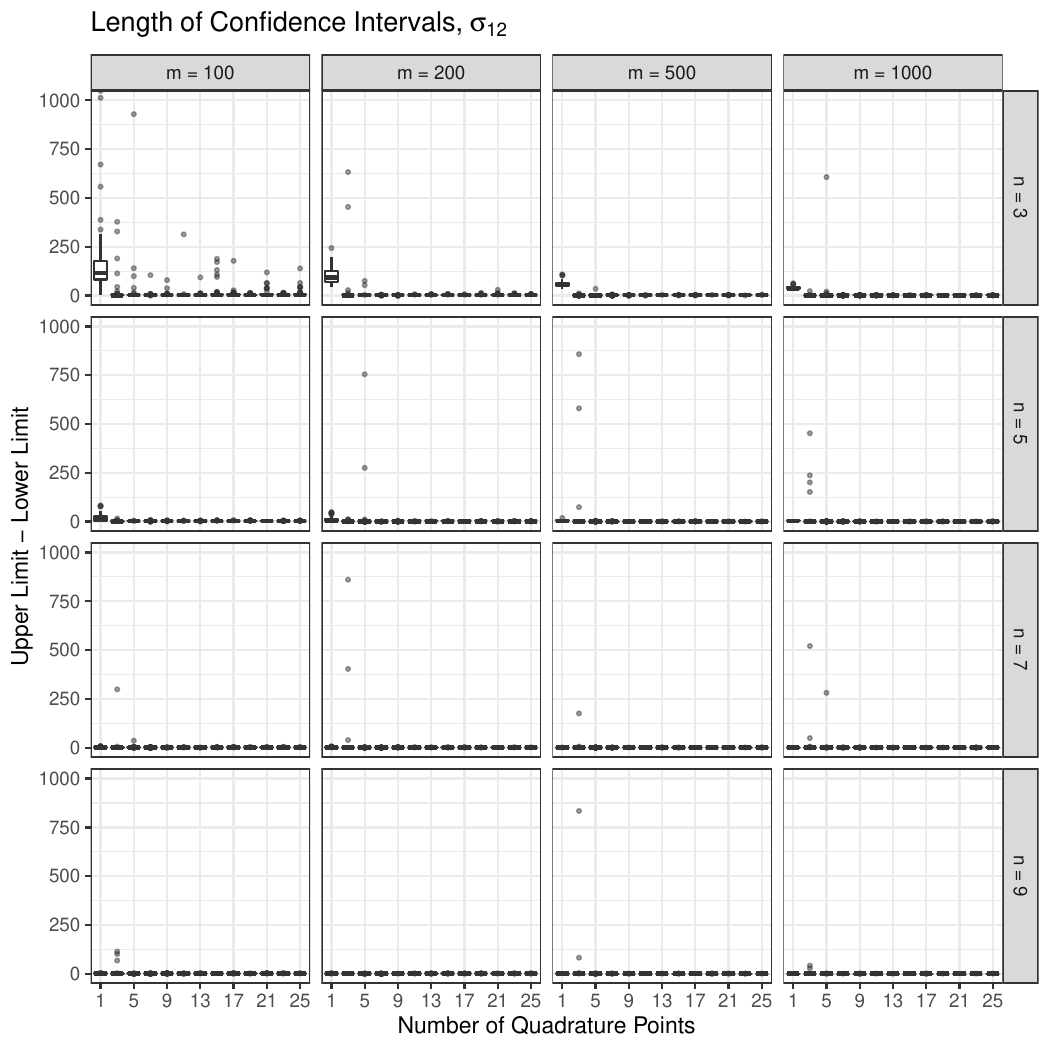}
\caption{Length of the Wald intervals for $\widehat{\sigma}_{12}$ in the simulation study of Section 4.3 in the main manuscript, where
$1000$ sets of data were generated from the random-slopes model (\ref{eqn:simmodel1}).
The Laplace ($\quadnum=1$) intervals are orders of magnitude wider than those for larger $\quadnum$; the y-axis is zoomed out to capture this.}
\label{fig:sigmacov1length}
\end{figure}

\begin{figure}[p]
\centering
\includegraphics{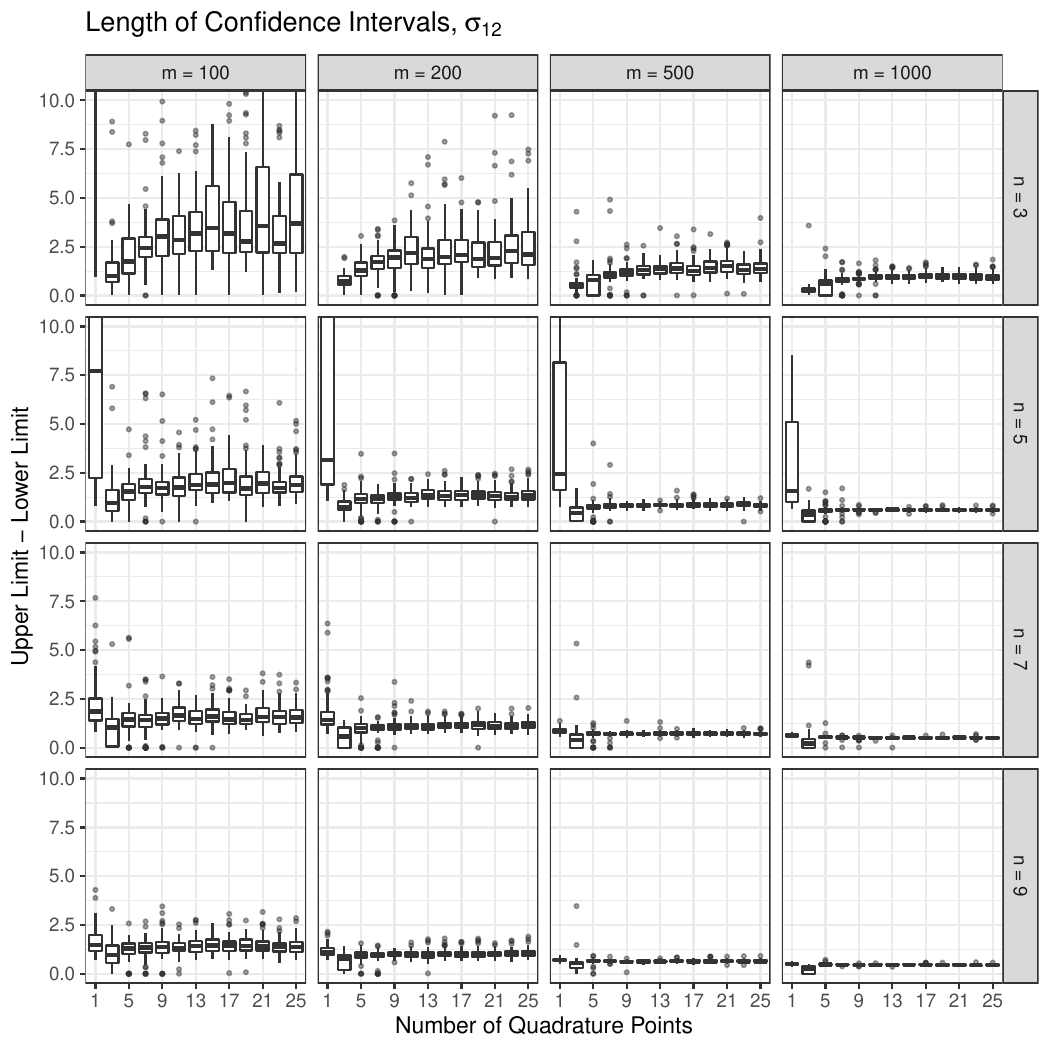}
\caption{Length of the Wald intervals for $\widehat{\sigma}_{12}$ in the simulation study of Section 4.3 in the main manuscript, where
$1000$ sets of data were generated from the random-slopes model (\ref{eqn:simmodel1}).
The y-axis is zoomed in, which obscures the massive length of the Laplace ($\quadnum=1$) intervals.
Those for $\quadnum=3,5$ are narrower, and then the lengths level off for $\quadnum\geq7$.
This pattern coincides with coverages converging to nominal (Figure \ref{fig:sigmacov1covr}) as $\quadnum$ increases, across values
of $\numgroups,\numpergroup$.}
\label{fig:sigmacov1lengthzoom}
\end{figure}

The coverage plots, Figures \ref{fig:beta0covr} -- \ref{fig:sigmacov1covr}, report $X(\theta)/B$ where $B=1000$ and 
$$
X(\theta) = \sum_{b=1}^{B}\one\left[ \theta\in\left\{\widehat{\theta} - 2\times\text{s.e.}\left(\widehat{\theta}\right),\widehat{\theta} + 2\times\text{s.e.}\left(\widehat{\theta}\right)\right\} \right]
$$
is the number of simulated sets of data for which the Wald interval contained the true value of $\theta$.
The intervals drawn around the lines are confidence intervals for this Binomial proportion and are calculated as
$$
\{X(\theta)/B\} \pm 2\times\sqrt{\frac{\{X(\theta)/B\}(1-\{X(\theta)/B\})}{B}}.
$$
The length plots, Figures \ref{fig:beta0length} -- \ref{fig:sigmacov1lengthzoom}, report 
$$
4\times\text{s.e.}\left(\widehat{\theta}\right),
$$
which is the length of the Wald confidence interval for $\theta$.

%%%%%% BIAS %%%%%%

\clearpage

\clearpage

\clearpage

\clearpage

\clearpage

\clearpage

\clearpage

\clearpage

\clearpage

\clearpage

%%%%%% COVERAGE %%%%%%

\clearpage

\clearpage

\clearpage

\clearpage

\clearpage

\clearpage

\clearpage

%%%%%% LENGTH %%%%%%

\clearpage

\clearpage

\clearpage

\clearpage

\clearpage

\clearpage

\clearpage

\clearpage

\clearpage

\clearpage

\subsection{Simulation 2: relative performance of the new method}\label{subsec:glmmadaptivecompare}

We now assess the relative performance of the new method compared to
the \texttt{R} packages
\texttt{GLMMadaptive} and \texttt{lme4}.
A method can be considered advantageous when compared to another
method if it returns similar results in a faster time, and/or superior
results in a similar amount of time. Accordingly, we compare computation times,
proportion of successful model fits, computed minimized negative log-likelihood
values, and computed gradient norms for each method.
We broadly find that the new approach tends to return results at least as favourable or superior
to other methods in terms
of computed log-likelihood and gradient norm values in faster times, 
and has a higher empirical probability of providing a 
successful model fit. For all methods, computations at higher $\quadnum$
take a longer amount of time but appear more stable.

\subsubsection{Comparison with \texttt{GLMMadaptive}}

We generated $500$ datasets from model (\ref{eqn:simmodel1})
and fit them using our procedure and 

\noindent\texttt{GLMMadaptive::mixed\_model()}
with the following non-default control options:
\begin{itemize}
  \item $\texttt{update\_GH\_every} = 1$: update $\condmodei{i}(\params)$ for every new value of $\params$,
  \item $\texttt{iter\_EM} = 0$: do not use the EM algorithm.
\end{itemize}
These options render \texttt{GLMMadaptive} as similar as possible to our approach,
except for its use of finite-differenced gradients: it uses L-BFGS to directly minimize
the AQ-approximate log-marginal likelihood, as does the new approach, which uses exact gradients.

\begin{figure}[p]
\centering
\includegraphics{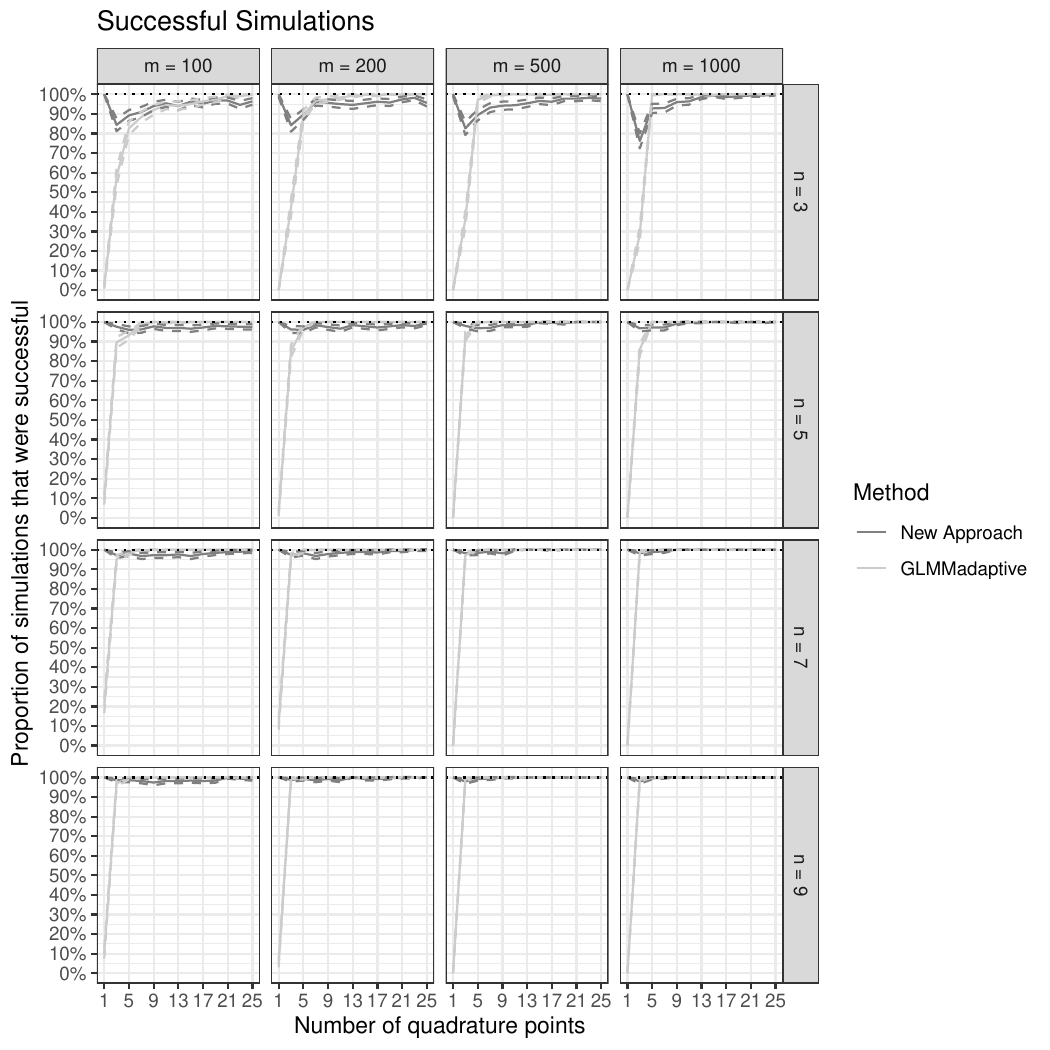}
\caption{Proportion of successful simulation runs for the new method and \texttt{GLMMadaptive} out of $500$ simulated data sets.
Both methods achieve nearly $100\%$ success for larger $\quadnum$ at all combinations of $\numgroups$ and $\numpergroup$.
For low $\numpergroup$, moderate $\quadnum$, and all values of $\numgroups$, the new method achieves slightly fewer successful runs,
however inspection of Figures \ref{fig:glmmadaptive-nlldiffplot} and \ref{fig:glmmadaptive-normdiffplot} indicate that the runs that were successful
tended to yield superior estimates.
For the Laplace approximation ($\quadnum=1$), \texttt{GLMMadaptive} achieves a low success rate for every $\numgroups,\numpergroup$.
}
\label{fig:glmmadaptive-successplot}
\end{figure}

\begin{figure}[p]
\centering
\includegraphics{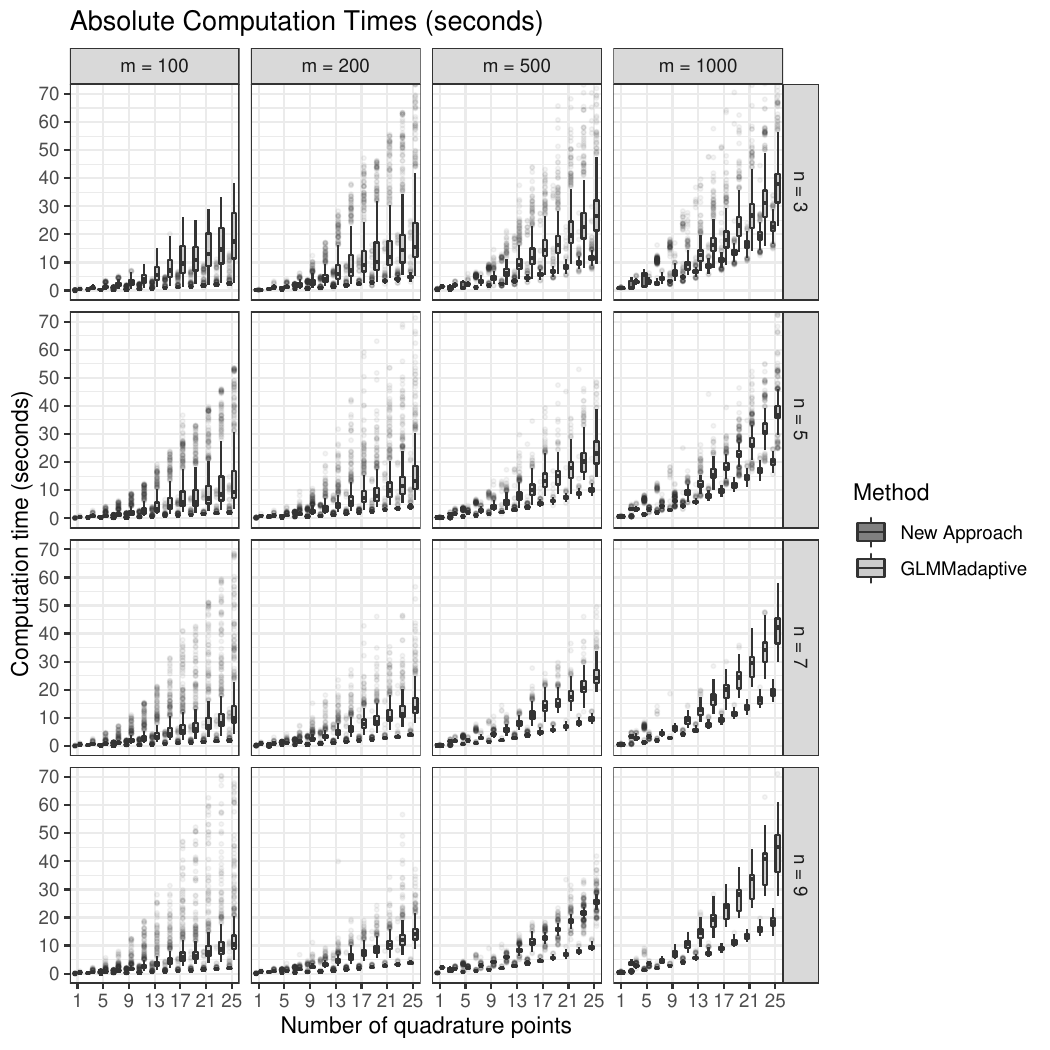}
\caption{Absolute computation times in seconds for the new method and \texttt{GLMMadaptive}.
The latter takes more computation time at all values of $\numgroups$, $\numpergroup$, and $\quadnum$.
The new method appears more stable than \texttt{GLMMadaptive} in the sense that the distributions of its
run times appear more concentrated and have fewer outlying run times.}
\label{fig:glmmadaptive-absolutetimeplot}
\end{figure}

\begin{figure}[p]
\centering
\includegraphics{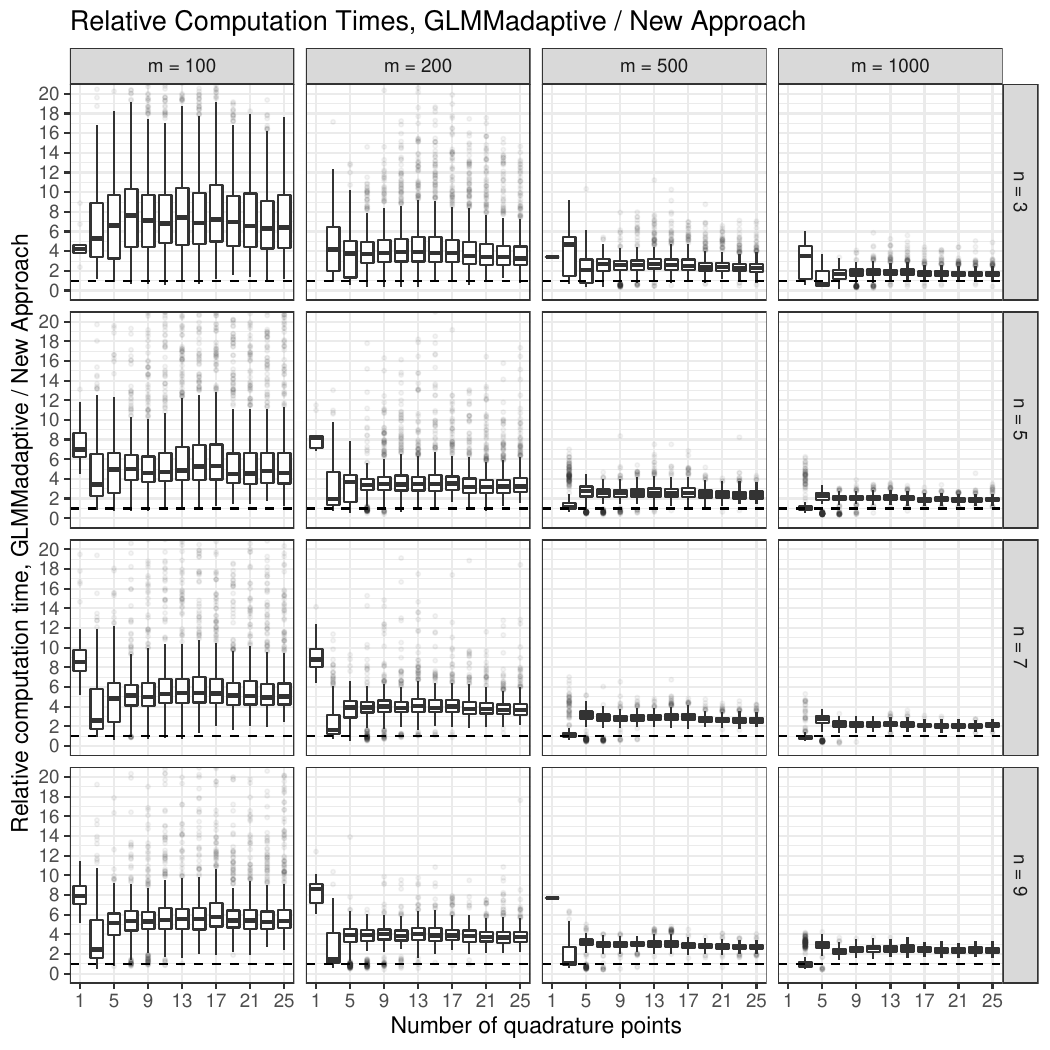}
\caption{Relative computation times for \texttt{GLMMadaptive} compared to the new method; for example, a value of $2$ indicates that \texttt{GLMMadaptive} ran in twice as much time as the new method.
The new method appears to run much faster than \texttt{GLMMadaptive} for lower $\numgroups$, and then settles at about $3$ times faster for large values of $\numgroups$ and $\quadnum$.
The total run time should be affected by data size, but also by accuracy of the approximation, with less accurate likelihood and gradient approximations leading to slower optimization,
even while smaller data size leads to faster evaluation; this may partly explain why the relative computation times appear to settle around a common value at all of $\numgroups$,
$\numpergroup$, and $\quadnum$ increase.}
\label{fig:glmmadaptive-relativetimeplot}
\end{figure}

\begin{figure}[p]
\centering
\includegraphics{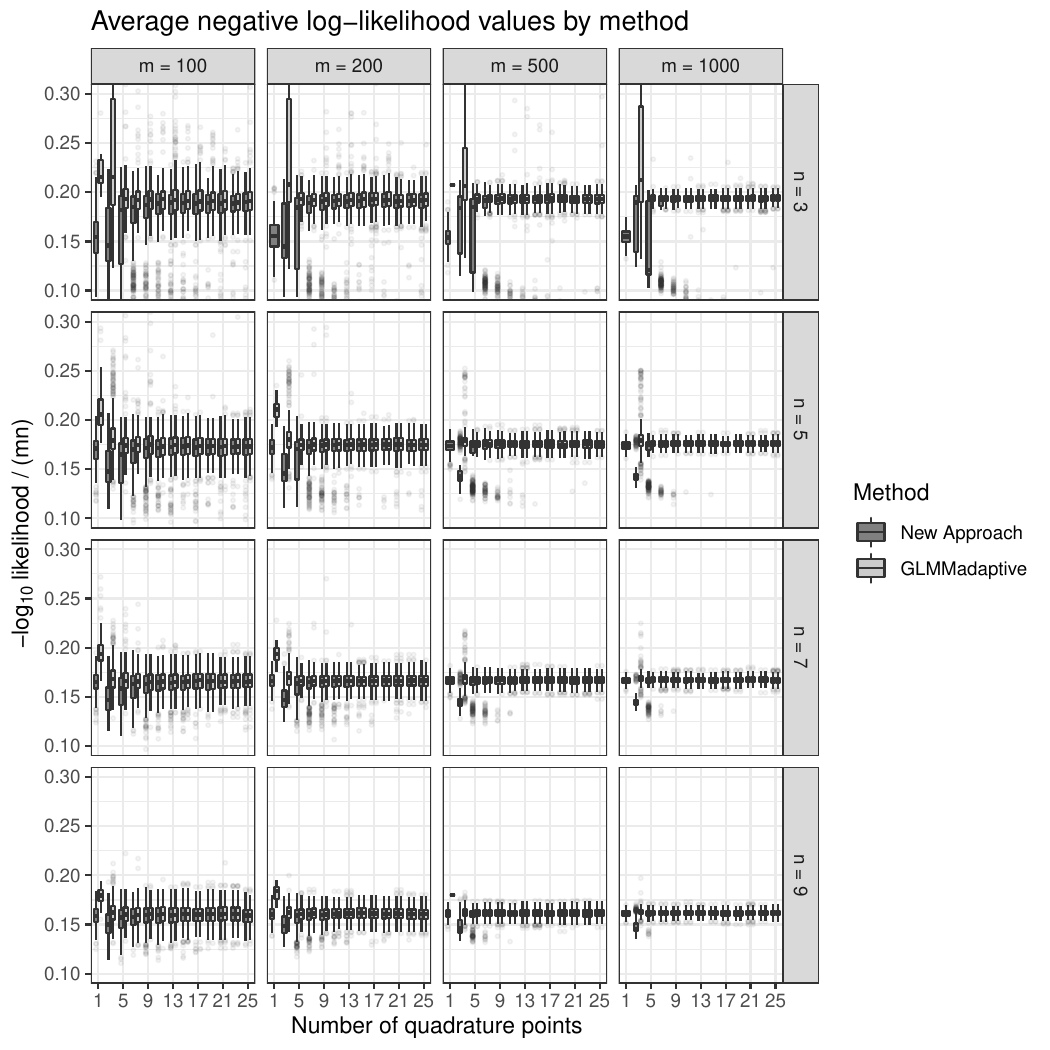}
\caption{Values of $-\approxloglikAQMLE{\quadnum} / (\numtotal\log10)$ for each method. Lower values are more desirable. The two methods 
return broadly the same values for moderate $\quadnum$ and higher. While the new method appears to return lower values for lower $\quadnum$,
this is not a substantial practical advantage since the practical recommendation is to increase $\quadnum$ until these values stop changing,
which will yield similar results from both methods. However, we point out that as substantiated in Figure \ref{fig:glmmadaptive-relativetimeplot},
the new method returns these results several times faster.}
\label{fig:glmmadaptive-rawnllplot}
\end{figure}

\begin{figure}[p]
\centering
\includegraphics{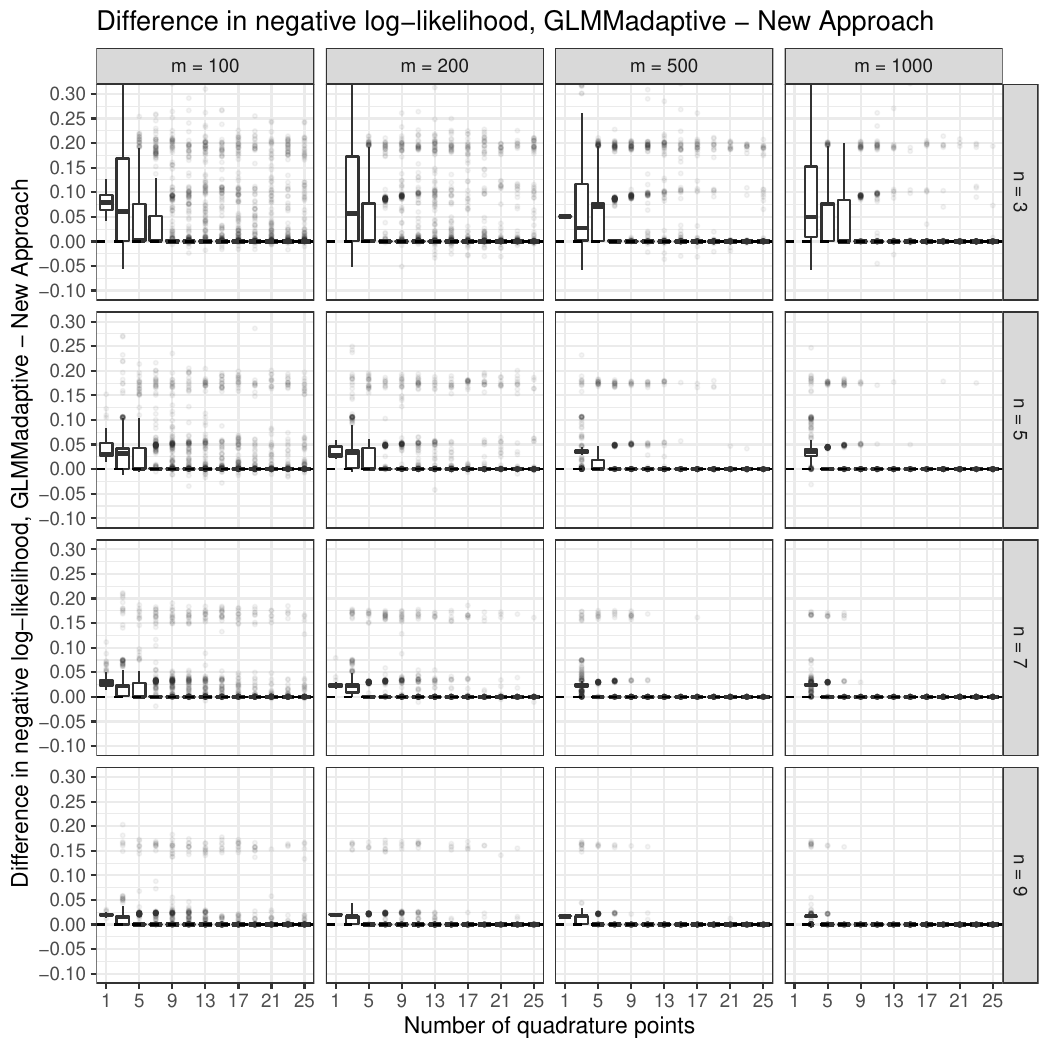}
\caption{Difference of $-\approxloglikAQMLE{\quadnum} / (\numtotal\log10)$ for \texttt{GLMMadaptive} minus the new method. 
A positive value indicates that the new method achieved a lower value of $-\approxloglikAQMLE{\quadnum}$ than \texttt{GLMMadaptive}, and hence superior
performance. The two methods 
return broadly the same values for moderate $\quadnum$ and higher, with \texttt{GLMMadaptive} returning higher values in a small number of simulations. 
While the new method appears to return lower values for lower $\quadnum$,
this is not a substantial practical advantage since the practical recommendation is to increase $\quadnum$ until these values stop changing,
which will yield similar results from both methods. However, we point out that as substantiated in Figure \ref{fig:glmmadaptive-relativetimeplot},
the new method returns these results several times faster.}
\label{fig:glmmadaptive-nlldiffplot}
\end{figure}

\begin{figure}[p]
\centering
\includegraphics{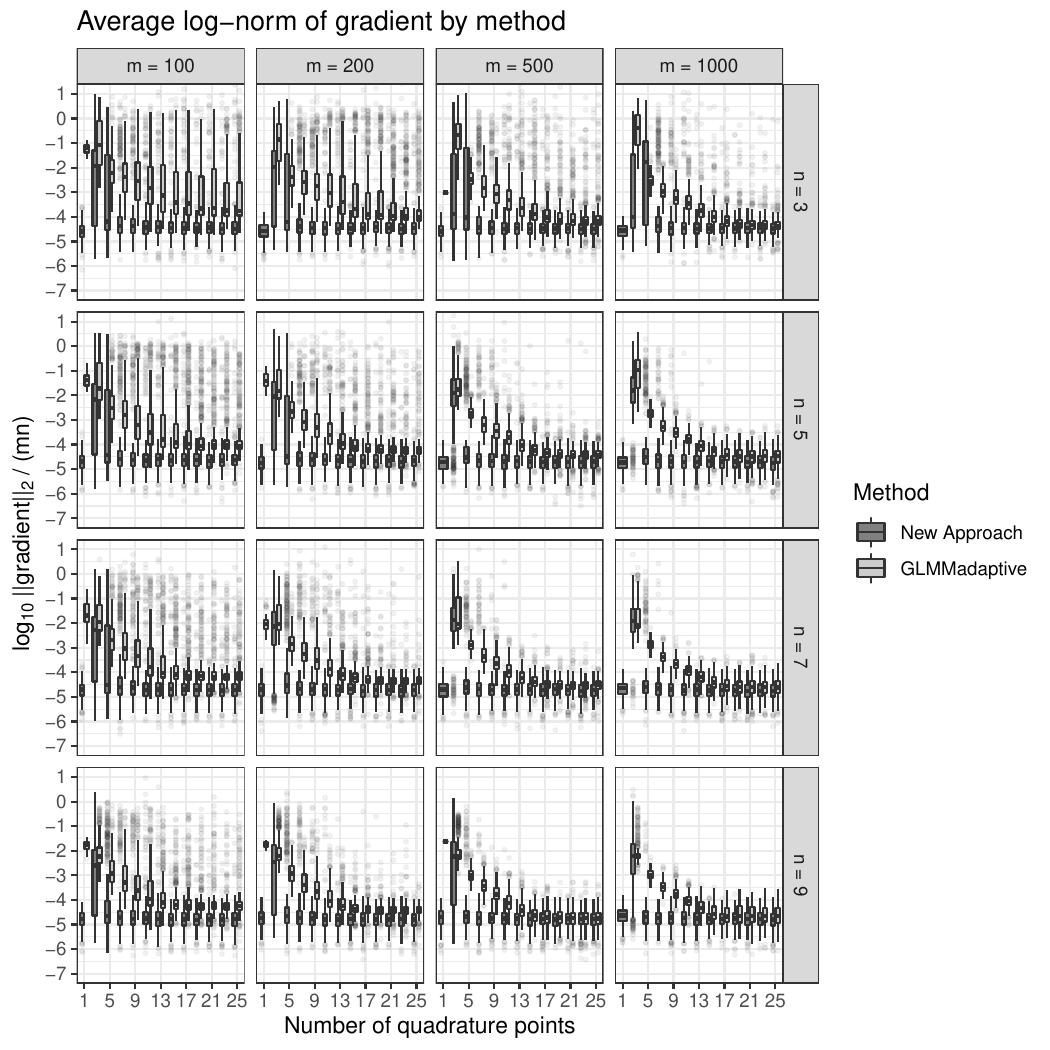}
\caption{Values of $\log_{10}\|\nabla\approxloglikAQMLE{\quadnum}\| / \numtotal$ for each method. Lower values are more desirable.
The new method appears very stable, achieving approximately the same values for all $\numgroups$, $\numpergroup$, and $\quadnum$
with concentrated distributions across simulations; the exception to the last point is for either small $\numgroups$ and/or small $\numpergroup$
combined with lower $\quadnum$, where the distribution across simulations is right-skewed.
In contrast, \texttt{GLMMadaptive} achieves results that clearly improve with increasing $\quadnum$, and stabilize at just slightly higher
than the new method. Because \texttt{GLMMadaptive} minimizes the norm of a finite-differenced gradient, it is not surprising that it would not
achieve as favourable a result on this metric as the new approach, which uses exact gradients.
In practice, because the recommendation is to increase $\quadnum$ until results stop changing, both approaches would end up giving similar
results on this metric; however, Figure \ref{fig:glmmadaptive-relativetimeplot} shows that the new method returns these results several times faster.
}
\label{fig:glmmadaptive-rawnormplot}
\end{figure}

\begin{figure}[p]
\centering
\includegraphics{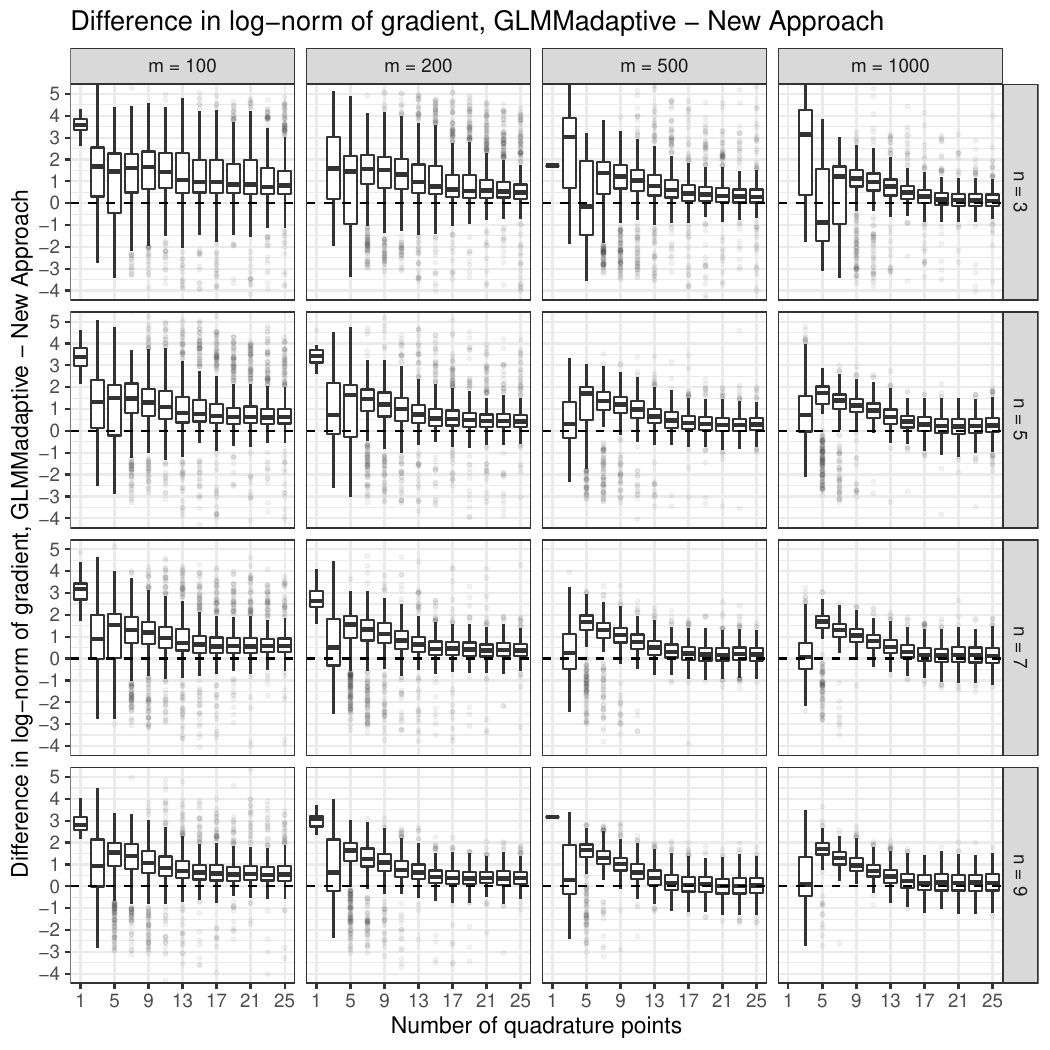}
\caption{Difference of $\log_{10}\|\nabla\approxloglikAQMLE{\quadnum}\|$ for \texttt{GLMMadaptive} minus the new method. 
A positive value indicates that the new method achieved a lower value of $-\approxloglikAQMLE{\quadnum}$ than \texttt{GLMMadaptive} and hence performed favourably.
The distribution of this difference is concentrated on positive values for nearly all combinations of $\numgroups$, $\numpergroup$, and $\quadnum$, indicating
broadly superior performance. For all $\numgroups$ and $\numpergroup$, the difference appears to settle around $0-1$ as $\quadnum$ is increased; since the
practical recommendation is to increase $\quadnum$ until results stop changing, this indicates that both approaches will end up yielding similar results,
with the new approach often slightly favourable. Figure \ref{fig:glmmadaptive-relativetimeplot} shows that the new method returns these favourable results several times faster.}
\label{fig:glmmadaptive-normdiffplot}
\end{figure}

We compute four summaries across simulations, shown in Figures \ref{fig:glmmadaptive-successplot} -- \ref{fig:glmmadaptive-normdiffplot}:
\begin{enumerate}
  \item \textbf{Computation times}: we report the absolute computation time of each method in seconds and the relative computation time of \texttt{GLMMadaptive}, which is defined as the absolute computation time of \texttt{GLMMadaptive} divided by the absolute computation time of the new method. Methods with lower absolute computation times for producing the same estimates are considered favourable. Values of relative computation time that are greater than $1$ indicate that the new method produced a result in less time than \texttt{GLMMadaptive}.
  \item \textbf{Successful runs}: a successful run is defined as a method returning finite numeric values for the point and interval estimates for all parameters. We report the proportion of successful runs for each method across simulations. It is favourable for a method to achieve a higher proportion of successful runs.
  \item \textbf{Negative log-likelihood values}: the methods each terminate at a point estimate, $\AQmle{\quadnum}$, which corresponds to the smallest computed value for the negative log-likelihood, $-\approxloglikAQMLE{\quadnum}$. We report the average base-$10$ log-likelihood, $-\approxloglikAQMLE{\quadnum} / (\numtotal\log10)$ where $\numtotal = \numgroups\numpergroup$, for each simulation and each method, as well as the difference in these values for \texttt{GLMMadaptive} minus that from the new method. Methods returning smaller such values are favourable. A positive value of the difference indicates that the new method performed favourably on this metric.
    We use our own implementation to compute $\approxloglikAQMLE{\quadnum}$ for all methods.
  \item \textbf{Gradient norm values}: the optimization procedures all attempt to find $\AQmle{\quadnum}$ such that $\|\nabla\approxloglikAQMLE{\quadnum}\|_2=0$. We report the average base-$10$ log of the norm of the gradient at the computed maximum, $\log_{10}\|\nabla\approxloglikAQMLE{\quadnum}\|_2/\numtotal$, for each simulation and each method. 
    It is favourable for a method to return a smaller value of this metric. 
    We also compute the difference in these values for \texttt{GLMMadaptive} minus that from the new method.
    A positive value of this difference indicates that the new method performed favourably on this metric. 
    We use our own implementation to compute $\nabla\approxloglikAQMLE{\quadnum}$ for all methods.
\end{enumerate}

Relative computation times are more relevant to comparison of methods because they are
less sensitive to hardware, and their interpretation generalizes more readily to data and models
other than the specific one(s) considered in a particular simulation study.
Absolute computation times are of passing interest only, and should not be considered
representative of how each method would perform using different hardware and data.
We report only relative computation times in the manuscript, but report both relative and absolute computation times in this supplement. 
All computations were performed
on a 2021 M1 Mac Book Pro with $64$ Gb of RAM, with individual simulations run in parallel across $10$ CPU cores.

%%%%%%%%%% glmmadaptive %%%%%%%%%%

%%% Successful simulation runs %%%
\clearpage

%%% Computation times %%%
\clearpage

\clearpage

%%% Negative log-likelihood values %%%
\clearpage

\clearpage

%%% Gradient norm values %%%
\clearpage

\clearpage

%%%%%%%%%% lme4 %%%%%%%%%%

\subsubsection{Comparison with \texttt{lme4}}\label{subsec:lme4compare}

Because \texttt{lme4} cannot fit model (\ref{eqn:simmodel1}) with $\quadnum>1$, we instead simulate from the following model:
\begin{equation}\label{eqn:simmodel2}
    \response_{ij} \setdelim \reidx_i \indsim \text{Bern}(p_{ij}), \ \reidx_i \iidsim \text{N}(0,\sigma^2), \ 
    \log\frac{p_{ij}}{1-p_{ij}} = \beta_0 + \beta_1\covi_i + \reidx_{i}.
\end{equation}
We simulate $1000$ sets of data from model (\ref{eqn:simmodel2}) with $\regparam = (-2.5,-.15)$ and $\sigma^2 = 2$.
We compute and report the same statistics with the same interpretations as for the comparison to \texttt{GLMMadaptive} in Section \ref{subsec:glmmadaptivecompare}.

%%% Successful simulation runs %%%
\begin{figure}[p]
\centering
\includegraphics{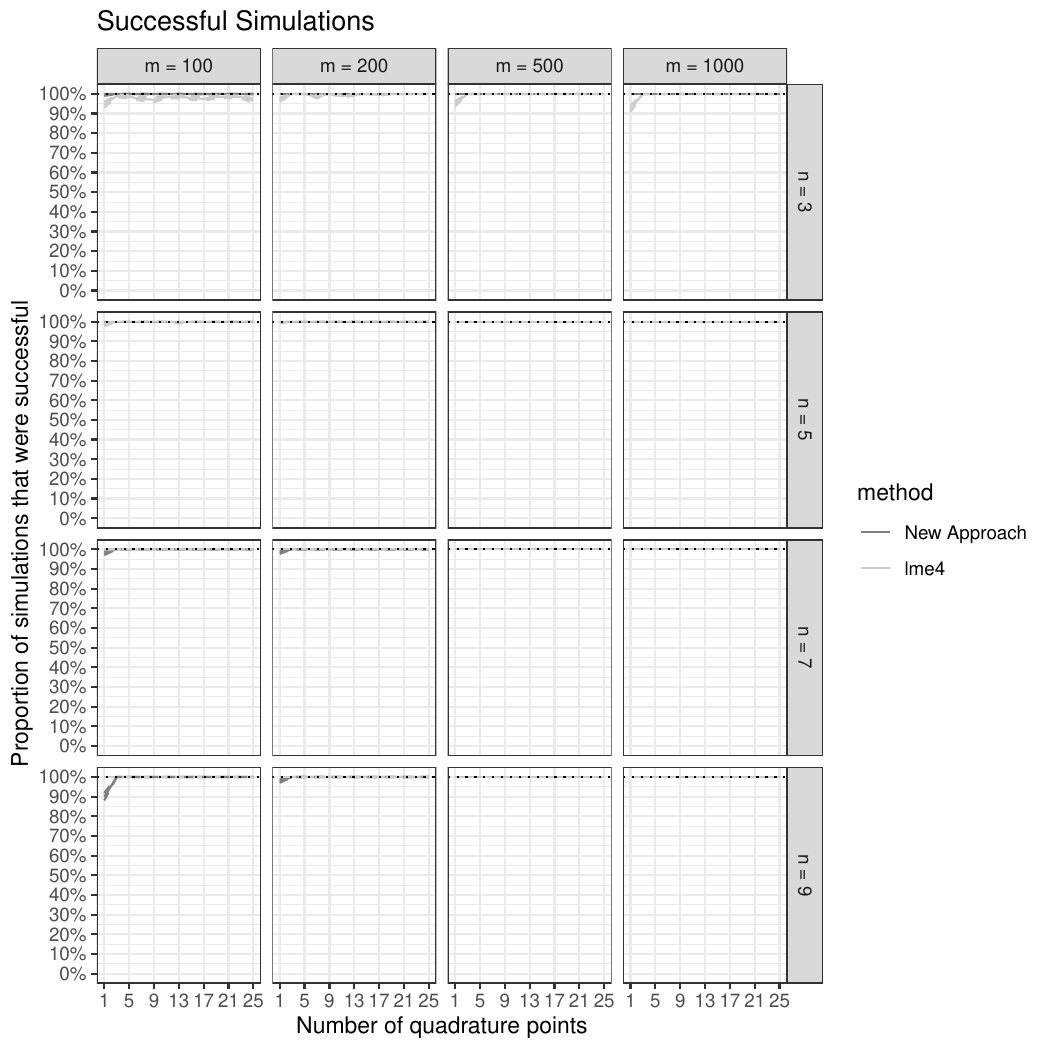}
\caption{Proportion of successful simulation runs for the new method and \texttt{lme4}.
Both methods appear highly stable at all values of $\numgroups$, $\numpergroup$, and $\quadnum$,
with \texttt{lme4} achieving a very slightly lower proportion of successful runs in the very difficult $\numgroups=100$, $\numpergroup=3$ case.}
\label{fig:lme4-successplot}
\end{figure}
\clearpage

%%% Computation times %%%
\begin{figure}[p]
\centering
\includegraphics{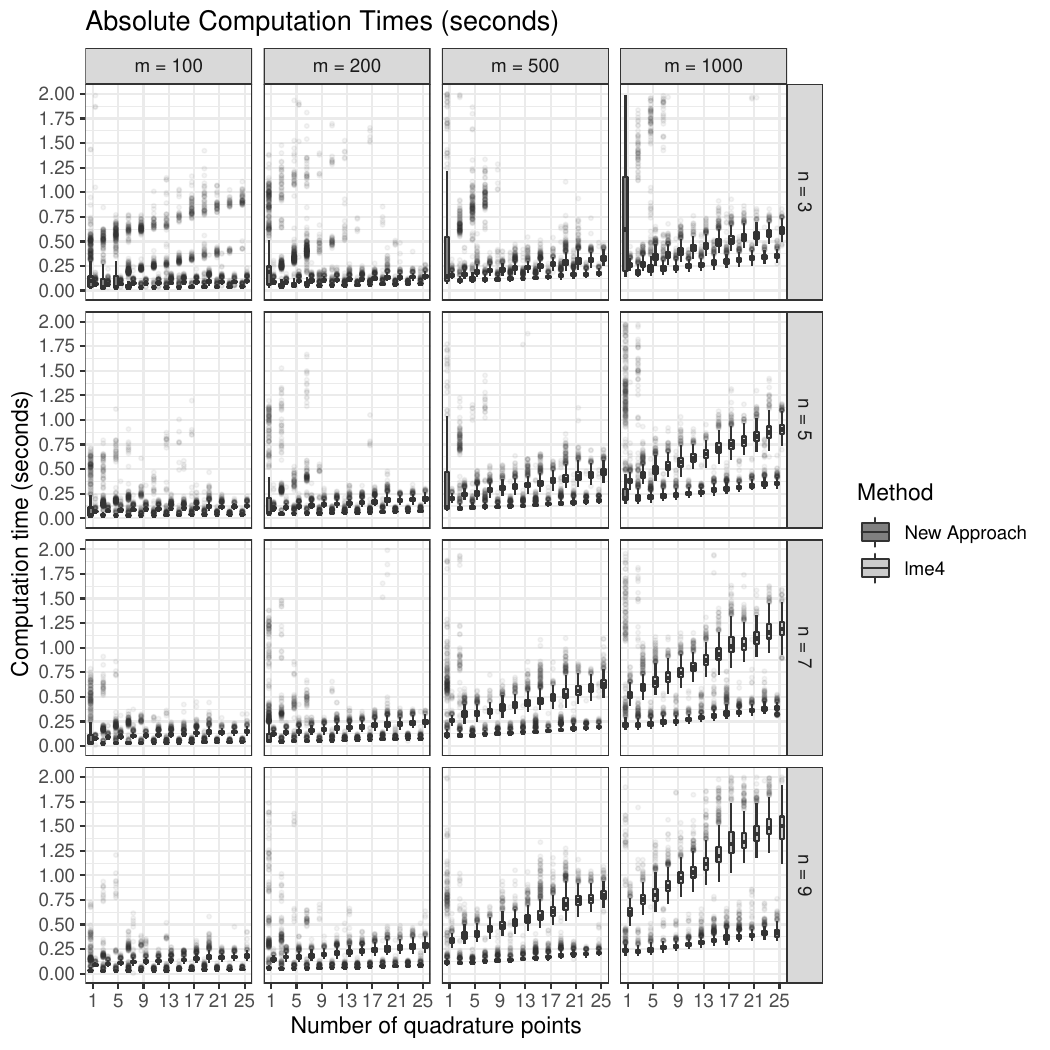}
\caption{Absolute computation times in seconds for the new method and \texttt{lme4}.
The new approach appears broadly faster and more stable than \texttt{lme4}, defined as having a distribution of run times across simulations centred at a lower value
and exhibiting less variability. The $\quadnum=1$ case appears challenging for the new approach, while \texttt{lme4} performs well in this case;
however, the new approach appears much faster and less variable at higher $\numgroups$. We point out that the practitioner can choose $\quadnum$, but often cannot choose $\numgroups$.}
\label{fig:lme4-absolutetimeplot}
\end{figure}
\clearpage

\begin{figure}[p]
\centering
\includegraphics{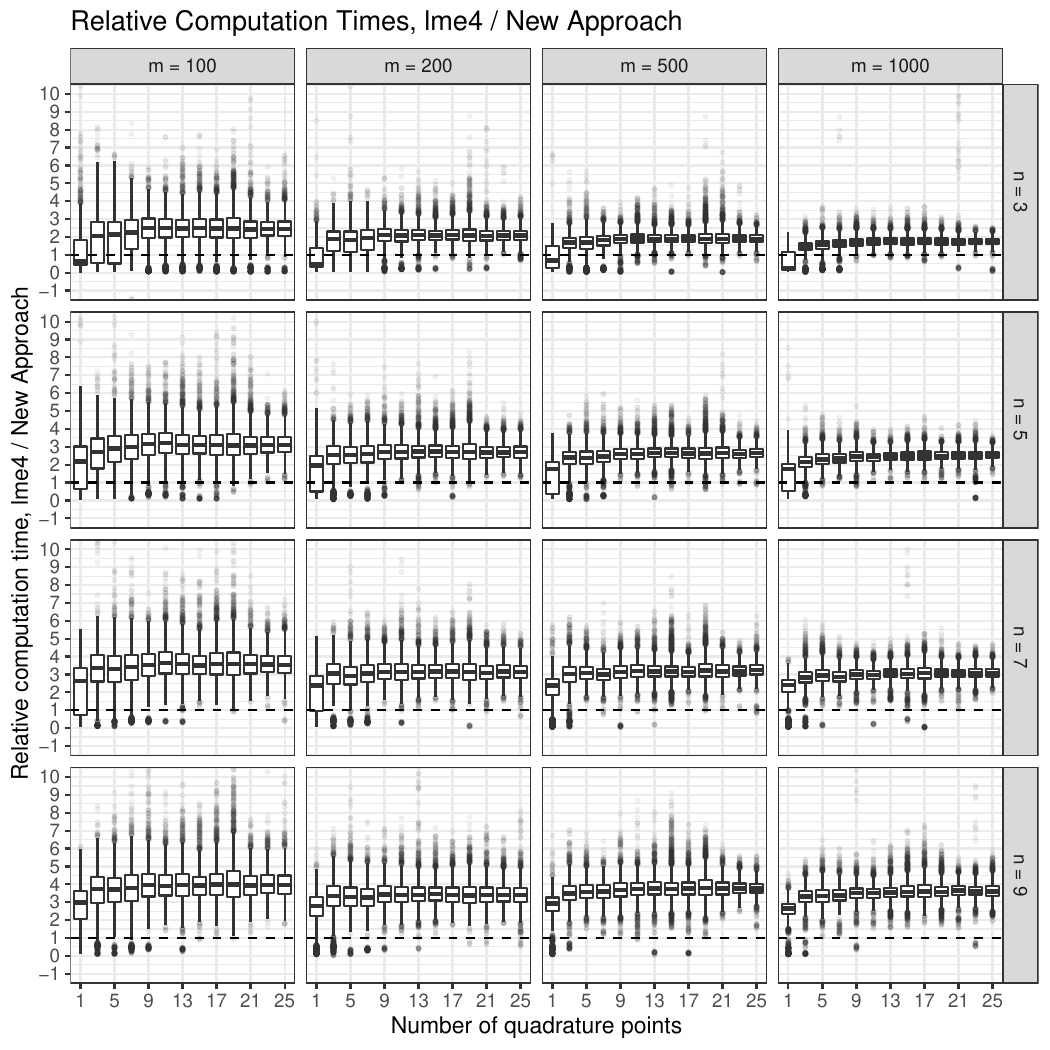}
\caption{Relative computation times for \texttt{lme4} compared to the new method; for example, a value of $2$ indicates that \texttt{lme4} ran in twice as much time as the new method.
The new method appears to be about $2-4$ times faster than \texttt{lme4}, with concentrated, symmetric distributions of run times, for almost all $\numgroups$, $\numpergroup$, and $\quadnum$.
For $\quadnum=1$, \texttt{lme4} is sometimes faster for lower $\numpergroup$, and the $\quadnum=1,\numpergroup=3$ case is the only case for which \texttt{lme4} appears superior overall in terms
of relative computation time. The practical implications of this are limited, as the method is not expected to be accurate for low $\quadnum$ combined with low $\numpergroup$. 
Further, Figures \ref{fig:lme4-rawnllplot} -- \ref{fig:lme4-normdiffplot} show that the computed results from both methods change as $\quadnum$ is increased beyond $1$, and hence the practical
advice to increase $\quadnum$ until results stop changing would yield faster run times from the new method.
}
\label{fig:lme4-relativetimeplot}
\end{figure}
\clearpage

%%% Negative log-likelihood values %%%
\begin{figure}[p]
\centering
\includegraphics{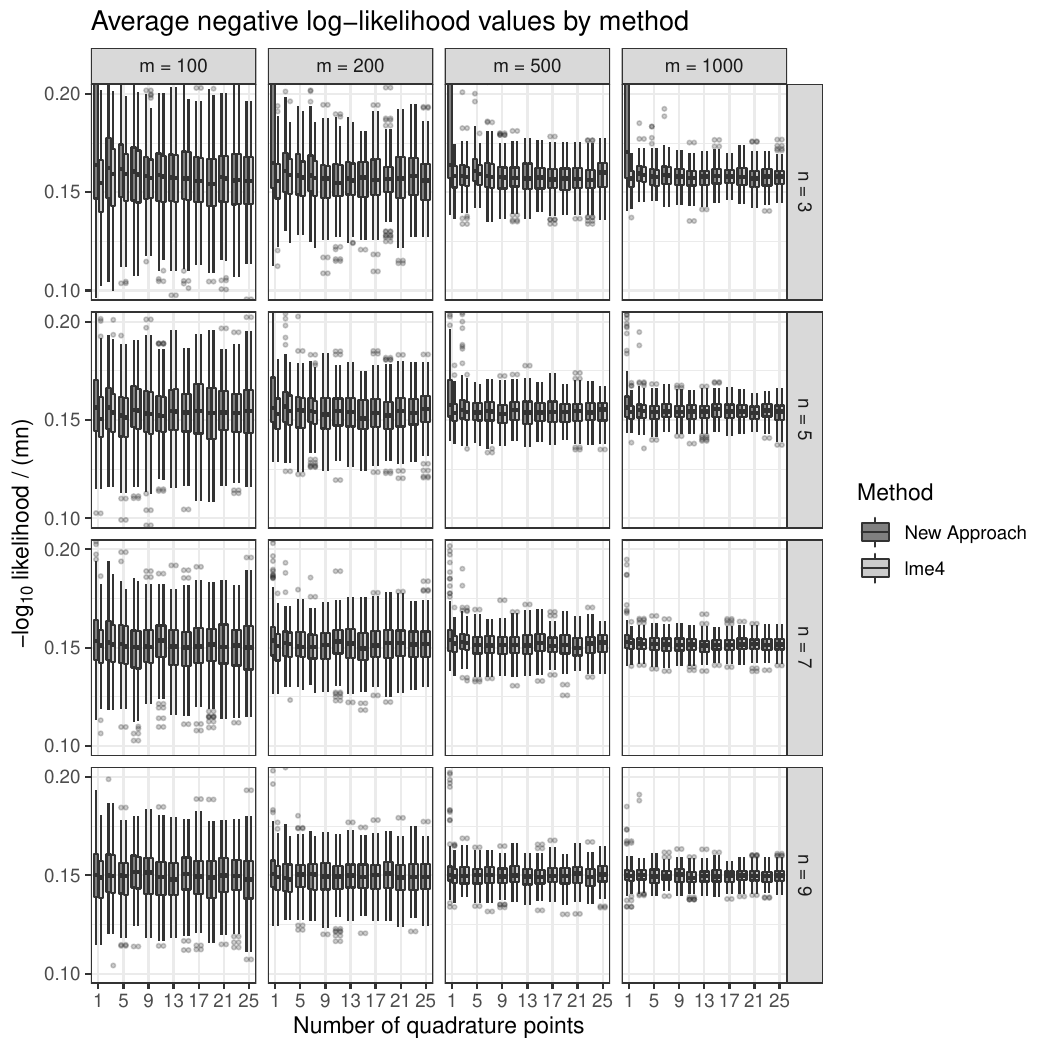}
\caption{Values of $-\approxloglikAQMLE{\quadnum} / (\numtotal\log10)$ for each method.
The methods broadly return comparable results. The $\quadnum=1$ case is better handled by \texttt{lme4} when $\numpergroup$ is low.
The practical implications of this are limited, as the method is not expected to be accurate for low $\quadnum$ combined with low $\numpergroup$.
The practical advice is to $\quadnum$ until results stop changing, and it is clear that this would yield similar results from both methods;
Figure \ref{fig:lme4-relativetimeplot} shows that these results are obtained several times faster using the new method.
}
\label{fig:lme4-rawnllplot}
\end{figure}
\clearpage

\begin{figure}[p]
\centering
\includegraphics{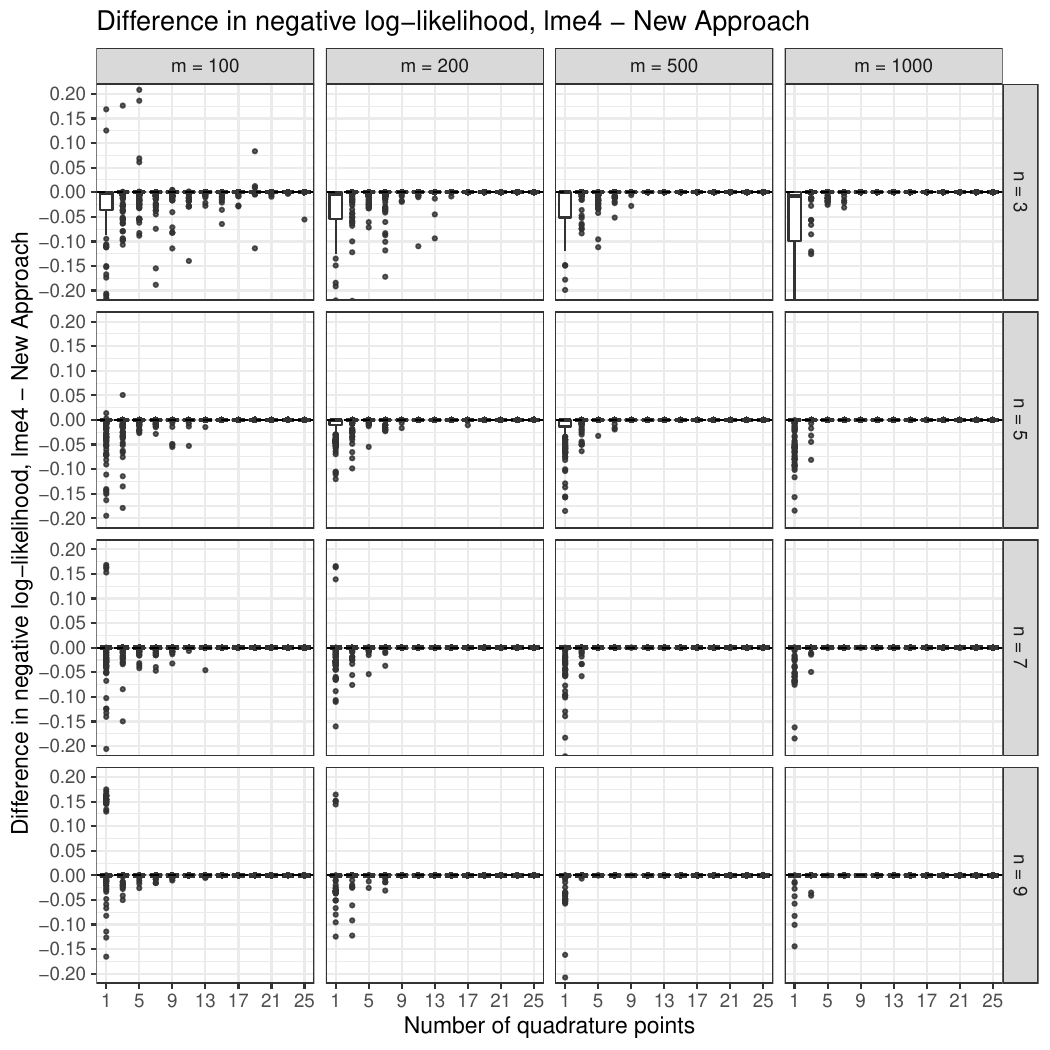}
\caption{Difference of $-\approxloglikAQMLE{\quadnum} / \numtotal$ for \texttt{lme4} minus the new method. 
A positive value indicates that the new method achieved a lower value of $-\approxloglikAQMLE{\quadnum}$ than \texttt{lme4}.
The methods broadly return similar results, with \texttt{lme4} achieving superior results in a small number of the $1000$ simulated data sets,
for smaller values of $\quadnum$.
The practical implications of this are limited, as the advice is to $\quadnum$ until results stop changing, and it is clear that this would yield similar results from both methods;
Figure \ref{fig:lme4-relativetimeplot} shows that these results are obtained several times faster using the new method.
}
\label{fig:lme4-nlldiffplot}
\end{figure}
\clearpage

%%% Gradient norm values %%%
\begin{figure}[p]
\centering
\includegraphics{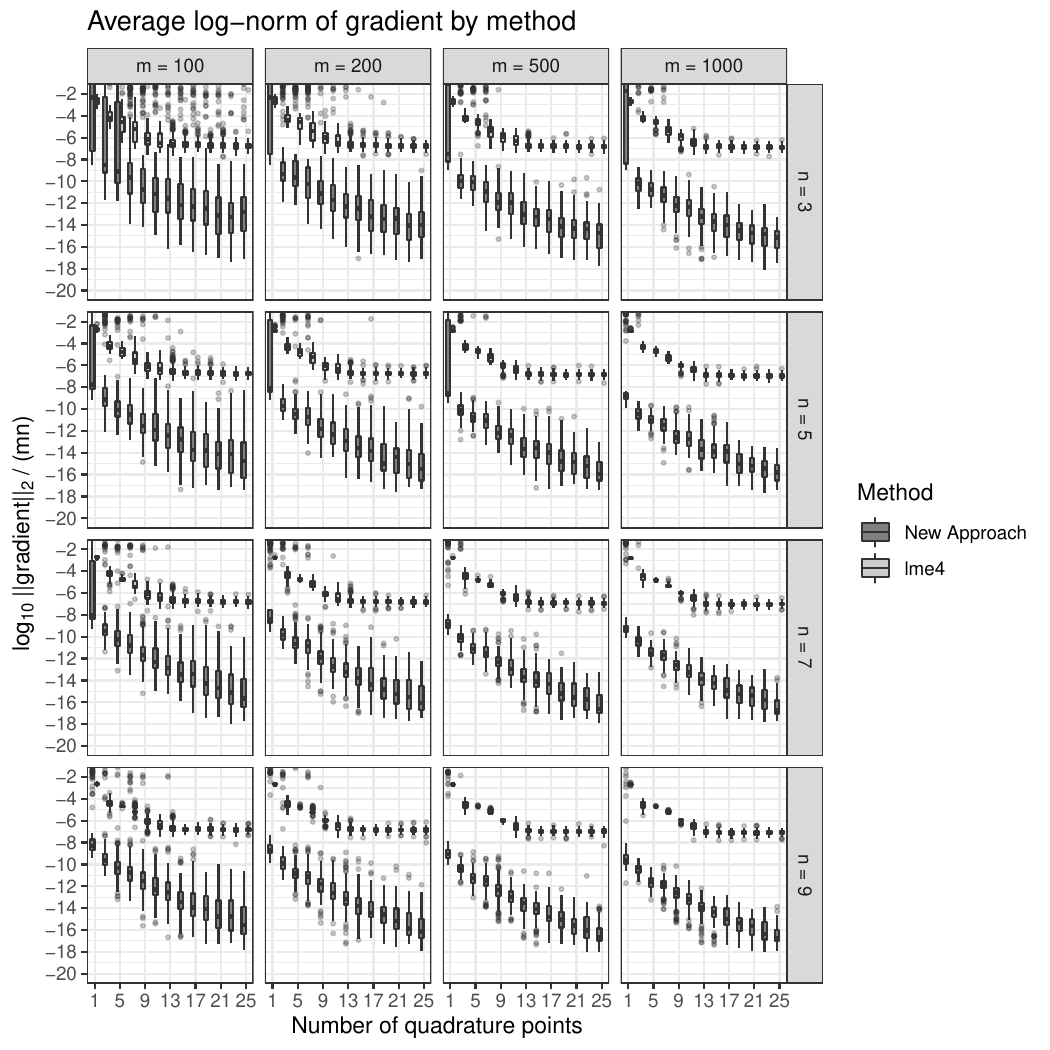}
\caption{Values of $\log_{10}\|\nabla\approxloglikAQMLE{\quadnum}\| / \numtotal$ for each method.
The new method appears to return smaller gradient norms than \texttt{lme4} overall, and hence achieves superior performance on this metric. 
% Because machine epsilon (the smallest positive floating point number, $\epsilon>0$, for which $1+\epsilon>1$ on the hardware used) satisfies $\log_{10}\epsilon\approx-15.654$
% on the hardware used, we see that the new method performed about as well as could have been achieved on this metric.
We emphasize that the performance of \texttt{lme4} is also quite satisfactory, and when combined with Figure \ref{fig:lme4-relativetimeplot}, the conclusion is
that the two methods performed similarly with the new approach returning results several times faster; this figure serves to confirm that the new approach
does not achieve these higher run times at the expense of satisfactory performance.}
\label{fig:lme4-rawnormplot}
\end{figure}
\clearpage

\begin{figure}[p]
\centering
\includegraphics{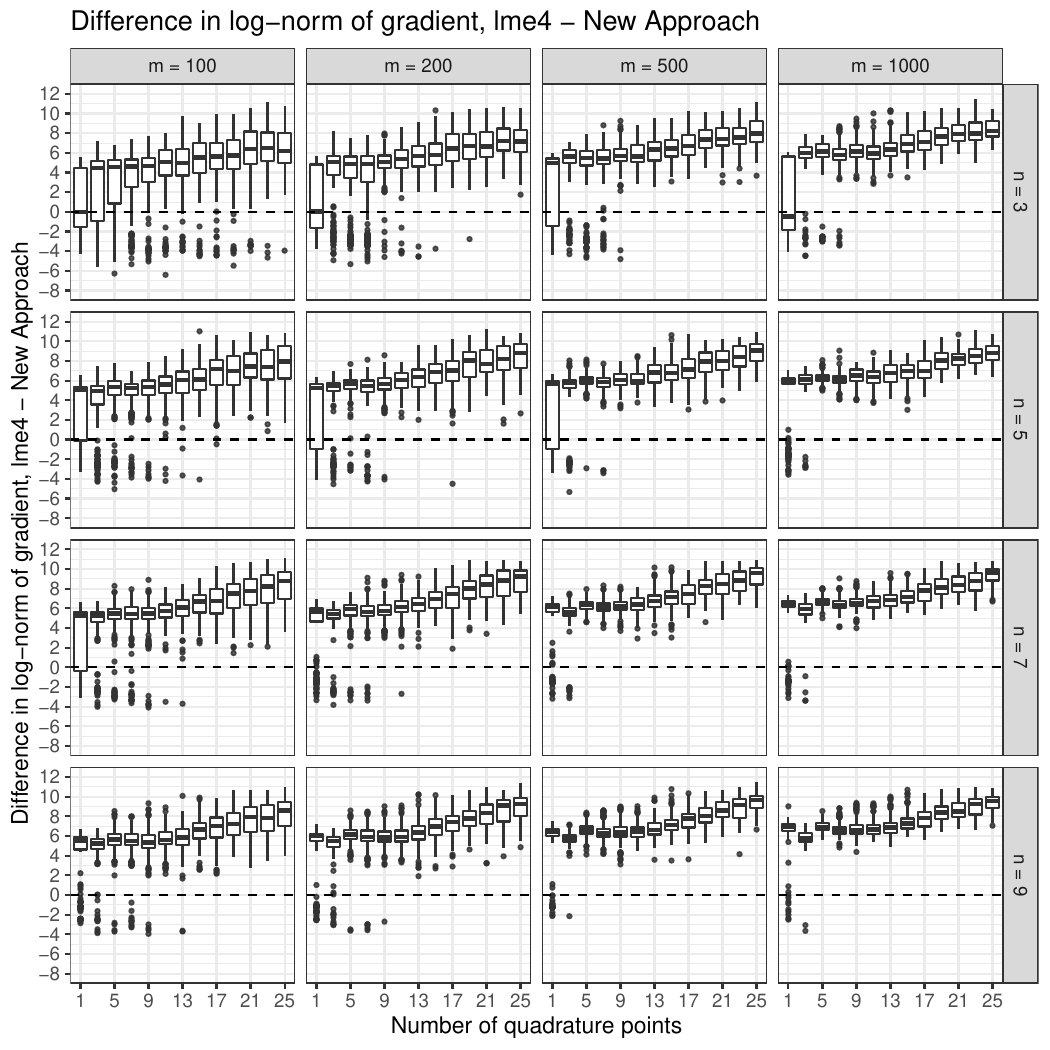}
\caption{Difference of $\log_{10}\|\nabla\approxloglikAQMLE{\quadnum}\|$ for \texttt{lme4} minus the new method. A positive value indicates that the new method achieved a lower value of $-\approxloglikAQMLE{\quadnum}$ than \texttt{lme4}.
The new method appears to return smaller gradient norms than \texttt{lme4} overall, and hence achieves superior performance on this metric. 
Combined with Figure \ref{fig:lme4-relativetimeplot}, we conclude that the new approach achieves faster computation times than \texttt{lme4} without sacrificing performance.
}
\label{fig:lme4-normdiffplot}
\end{figure}
\clearpage

\subsubsection{Comparison with \texttt{glmmTMB}}

The \texttt{glmmTMB} package implements the Laplace approximation via automatic differentiation using the \texttt{TMB} package.
While we do not advocate the use of the Laplace approximation to fit the mixed models
discussed in this paper, it is of interest to compare how a partial implementation of the gradient-based methods we investigate here via automatic differentiation compares to our present implementation.
We find that our implementation is slightly faster except for the case of a large number of small groups ($\numgroups$ large and $\numpergroup$ small).
Care should be taken to not over-interpret these results, as the efficiency
of both the new method and any method based on automatic differentiation
will be sensitive to implementation details.

\begin{figure}[p]
\centering
\includegraphics{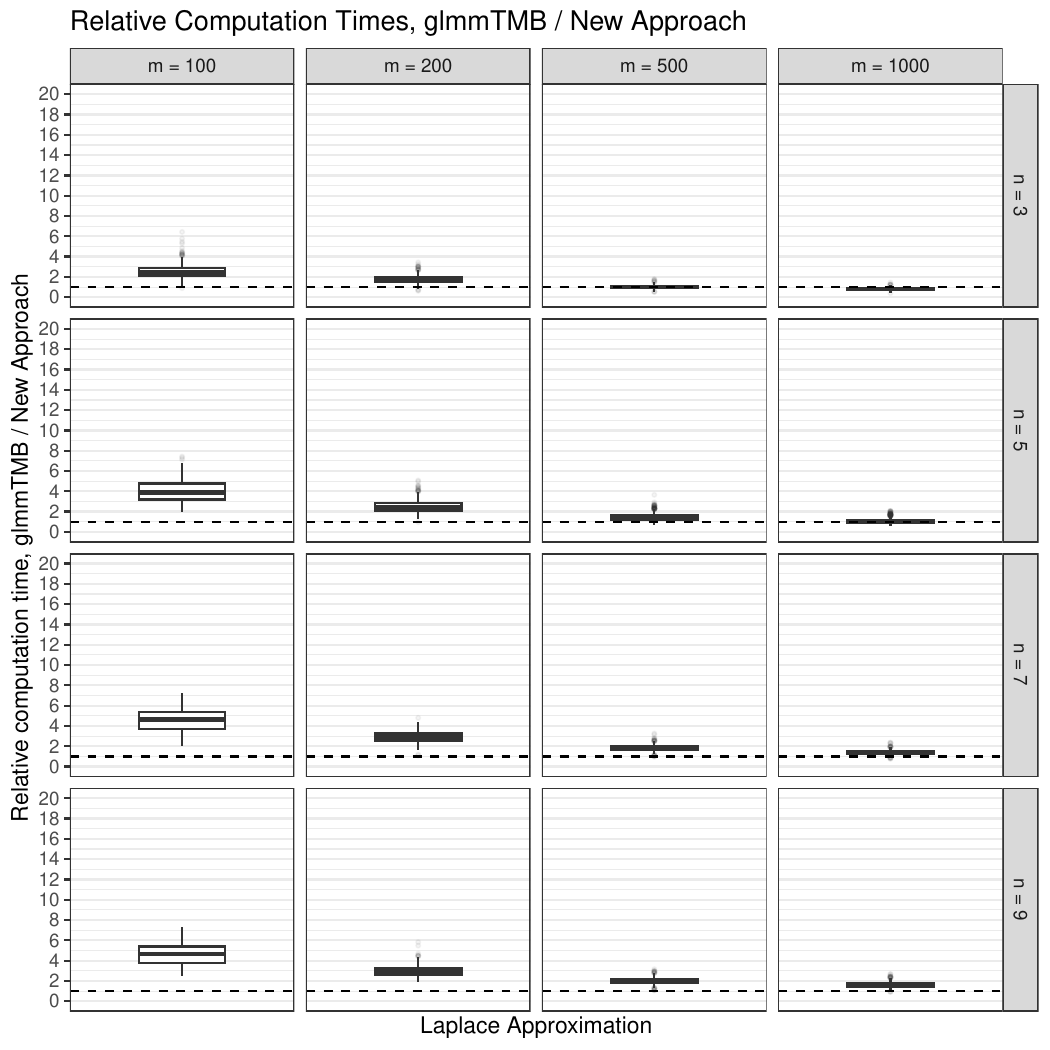}
\caption{Relative computation times for \texttt{glmmTMB} compared to the new method; for example, a value of $2$ indicates that \texttt{glmmTMB} ran in twice as much time as the new method.
Only the Laplace approximation ($\quadnum=1$) is shown because this is what \texttt{glmmTMB} computes.
}
\label{fig:glmmTMB-reltimeplot}
\end{figure}

\subsection{Additional simulation: scalar random effects}

\begin{figure}[p]
\centering
\includegraphics{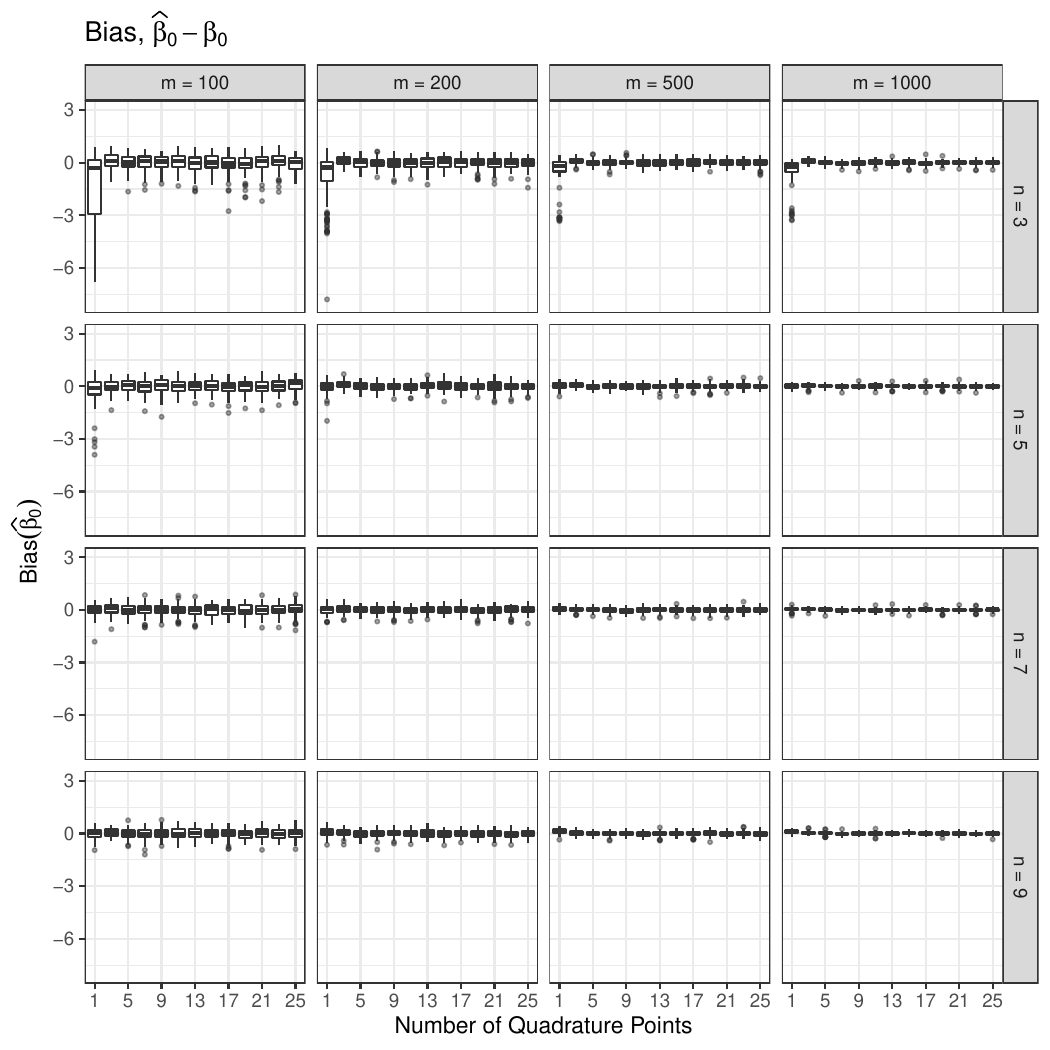}
\caption{Empirical bias, $\widehat{\beta_0} - \beta_0$, for $\beta_0$ from the random intercepts model (\ref{eqn:simmodel2}),
based on $1000$ sets of simulated data. The Laplace approximation exhibits high bias for low to moderate $\numpergroup$.
This is mitigated by increasing $\quadnum$.
}
\label{fig:beta0biasscalar}
\end{figure}

\begin{figure}[p]
\centering
\includegraphics{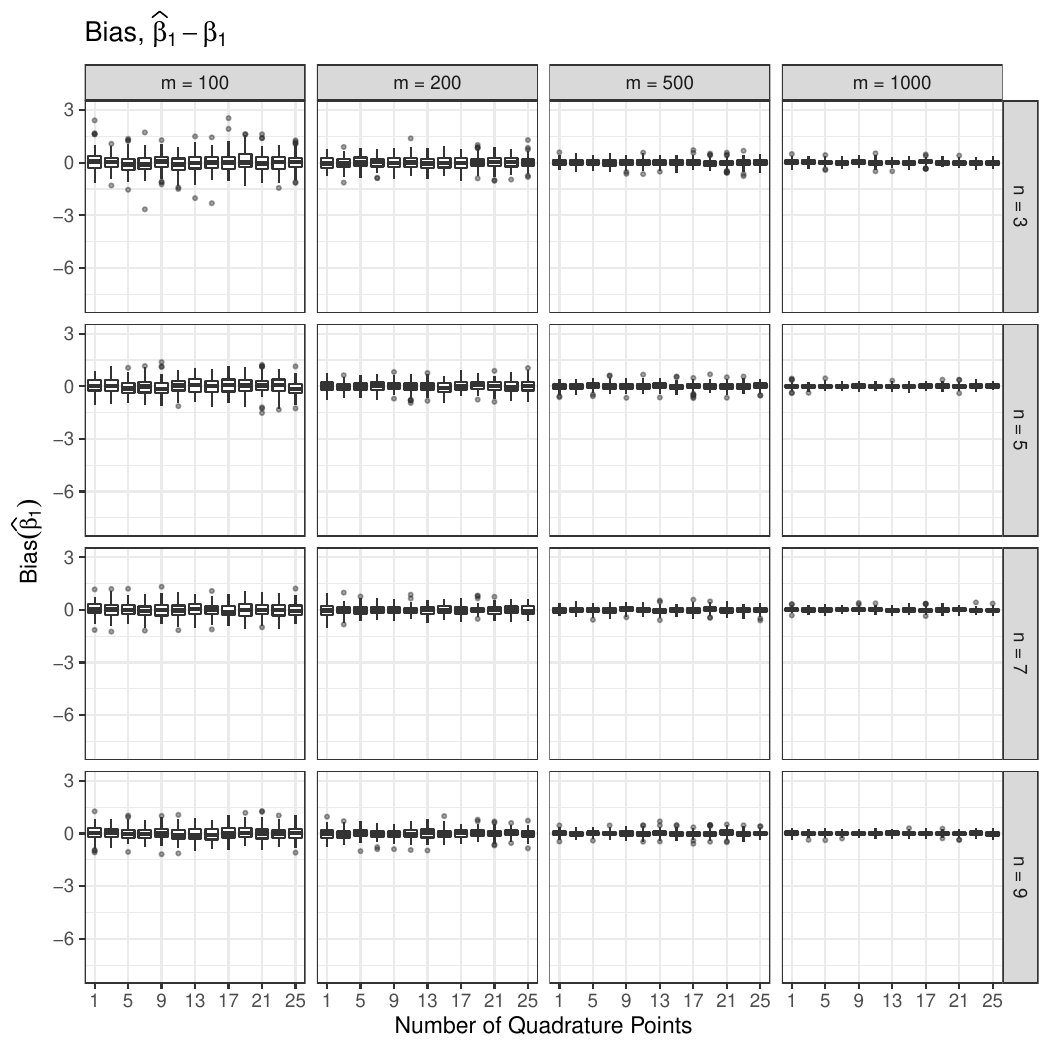}
\caption{Empirical bias, $\widehat{\beta_1} - \beta_1$, for $\beta_1$ from the random intercepts model (\ref{eqn:simmodel2}),
based on $1000$ sets of simulated data. All values of $\quadnum$ appear to yield low bias estimates.
}
\label{fig:beta1biasscalar}
\end{figure}

\begin{figure}[p]
\centering
\includegraphics{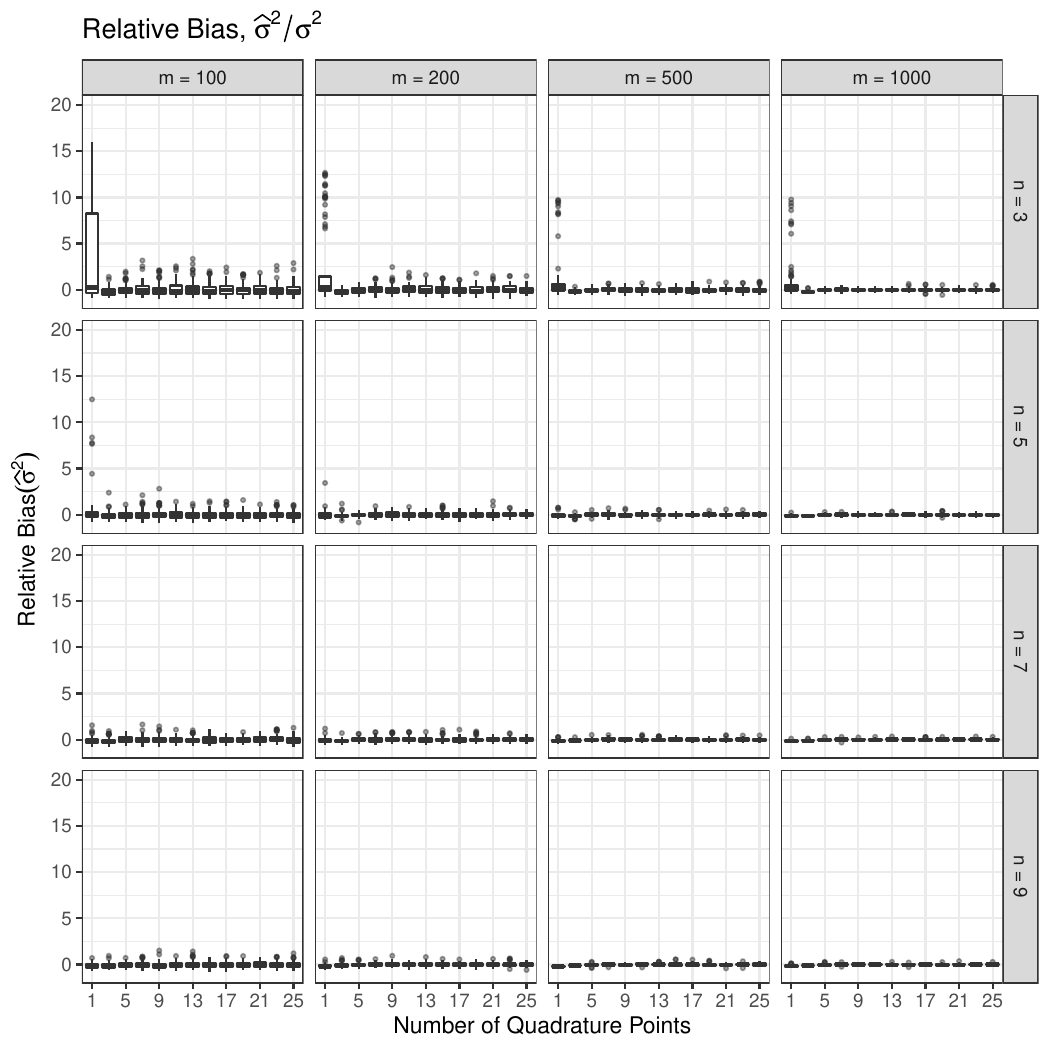}
\caption{Empirical relative bias, $\widehat{\sigma}^2 / \sigma^2$, for $\sigma^2$ from the random intercepts model (\ref{eqn:simmodel2}),
based on $1000$ sets of simulated data. 
The y-axis range is very large so that the scale of the relative bias for 
the Laplace approximation with low $\numpergroup$ is visible. 
For $\quadnum>1$, taking $\quadnum$ larger leads to somewhat reduced bias across values of $\numgroups$ and $\numpergroup$.}
\label{fig:sigmasqbiasscalar}
\end{figure}

\begin{figure}[p]
\centering
\includegraphics{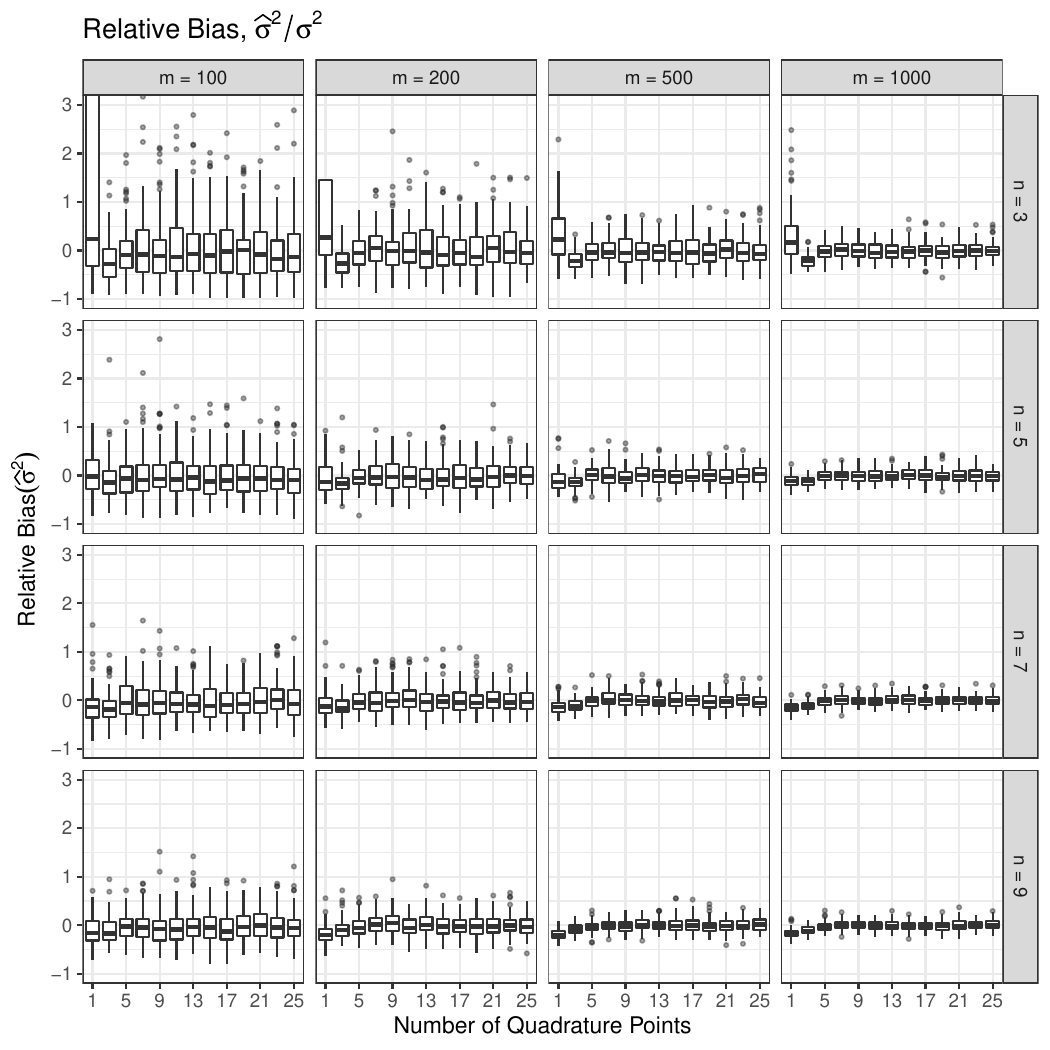}
\caption{Empirical relative bias, $\widehat{\sigma}^2 / \sigma^2$, for $\sigma^2$ from the random intercepts model (\ref{eqn:simmodel2}),
based on $1000$ sets of simulated data. 
The y-axis range is zoomed in so that results are visible for all except the Laplace approximation.
For $\quadnum>1$, taking $\quadnum$ larger leads to somewhat reduced bias across values of $\numgroups$ and $\numpergroup$.}
\label{fig:sigmasqbiasscalarzoom}
\end{figure}

\begin{figure}[p]
\centering
\includegraphics{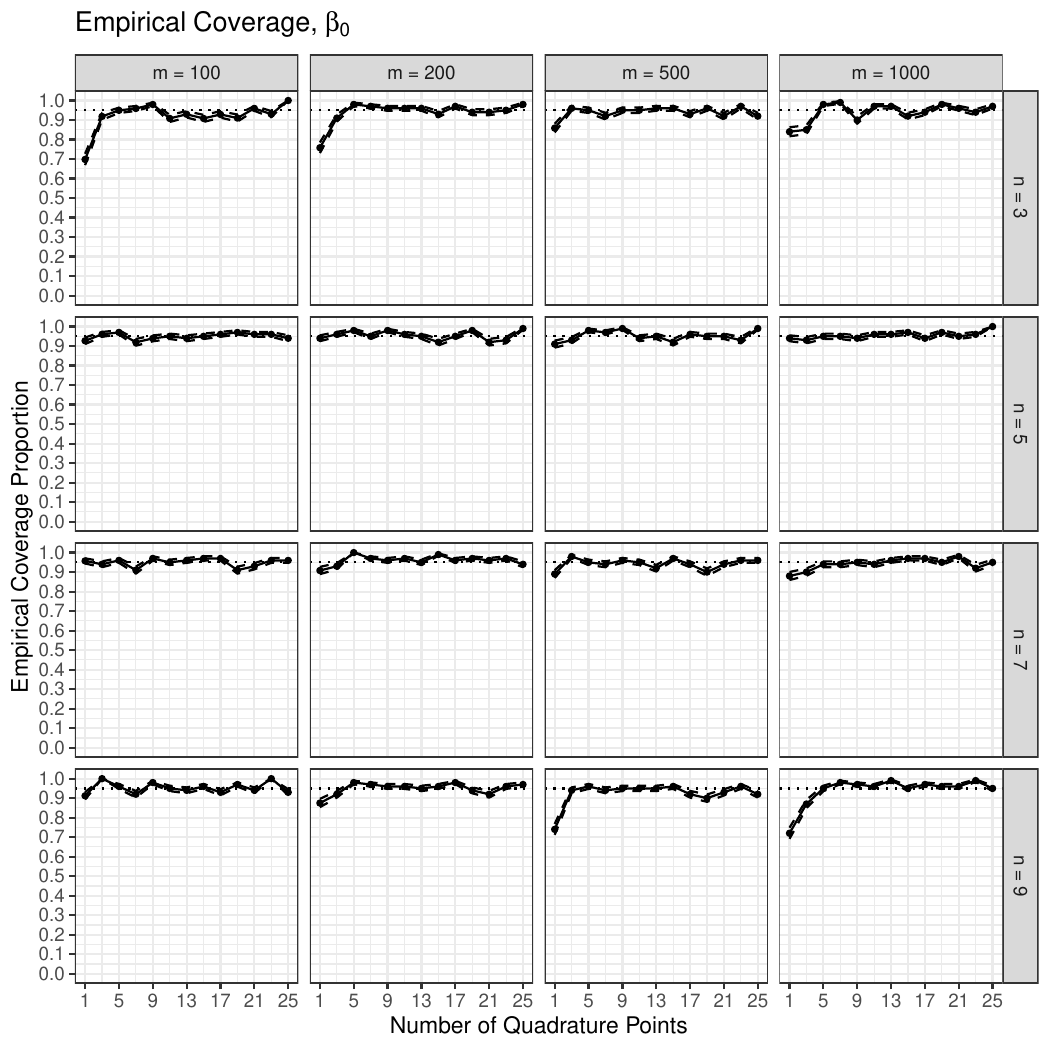}
\caption{Empirical coverage proportion for $\widehat{\beta}_0$ from the random intercepts model (\ref{eqn:simmodel2}),
based on $1000$ sets of simulated data. 
The Laplace ($\quadnum=1$) coverage is low for $\numpergroup=3$ due to high bias and poor standard error estimation. 
In all cases, taking $\quadnum$ large enough leads to nominal empirical coverage, motivating the
use of adaptive quadrature for fitting these models.}
\label{fig:beta0covrscalar}
\end{figure}

\begin{figure}[p]
\centering
\includegraphics{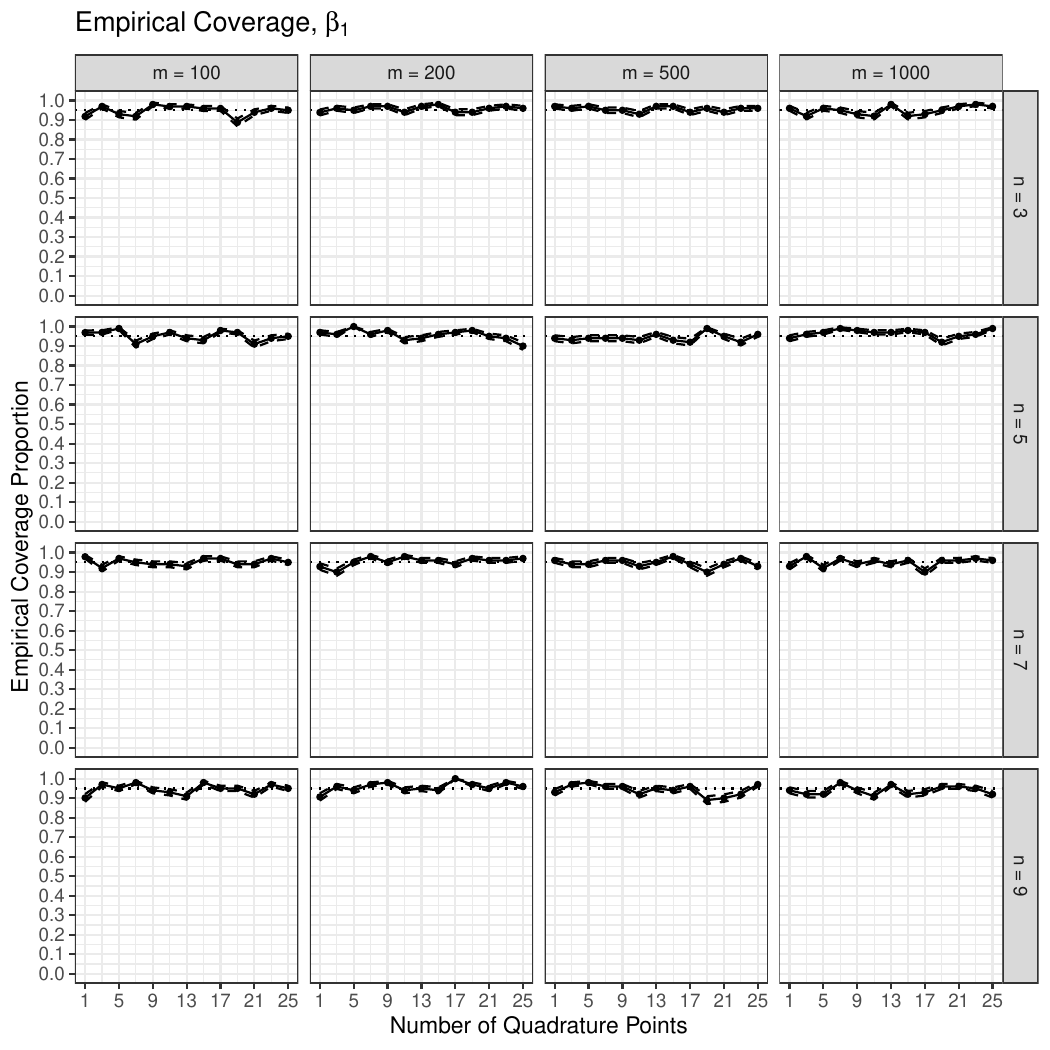}
\caption{Empirical coverage proportion for $\widehat{\beta}_1$ from the random intercepts model (\ref{eqn:simmodel2}),
based on $1000$ sets of simulated data. 
The coverage is nominal across all values of $\numgroups,\numpergroup,\quadnum$,
which is expected for this parameter in this model.}
\label{fig:beta1covrscalar}
\end{figure}

\begin{figure}[p]
\centering
\includegraphics{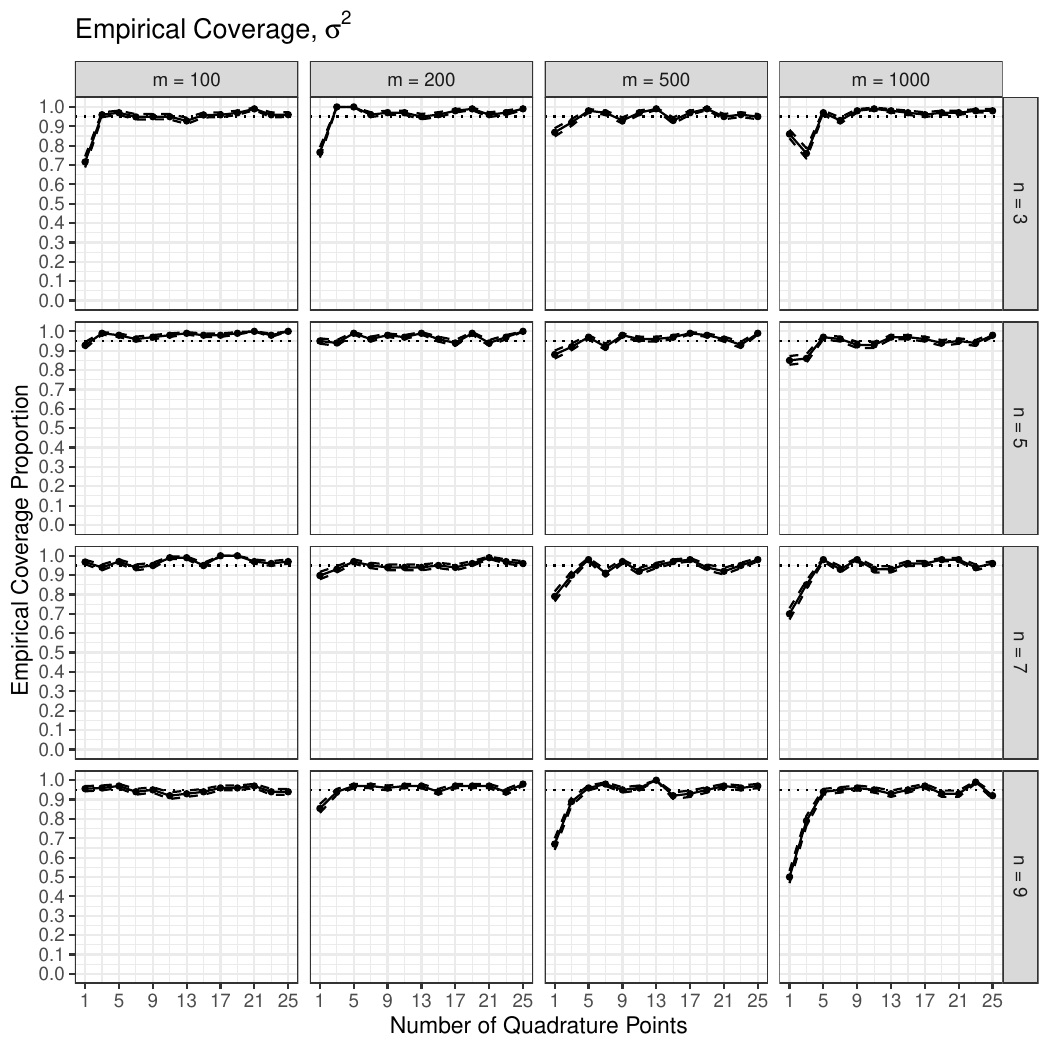}
\caption{Empirical coverage proportion for $\widehat{\sigma}^2$ from the random intercepts model (\ref{eqn:simmodel2}),
based on $1000$ sets of simulated data. 
The Laplace ($\quadnum=1$) coverage is low for $\numpergroup=3,5$ due to high bias and poor standard error estimation.
As $\numpergroup$ increases, the bias decreases but the standard error also decreases, leading to worsening coverage.
Again, the solution is to simply use a larger $\quadnum$.}
\label{fig:sigmasqcovrscalar}
\end{figure}

\begin{figure}[p]
\centering
\includegraphics{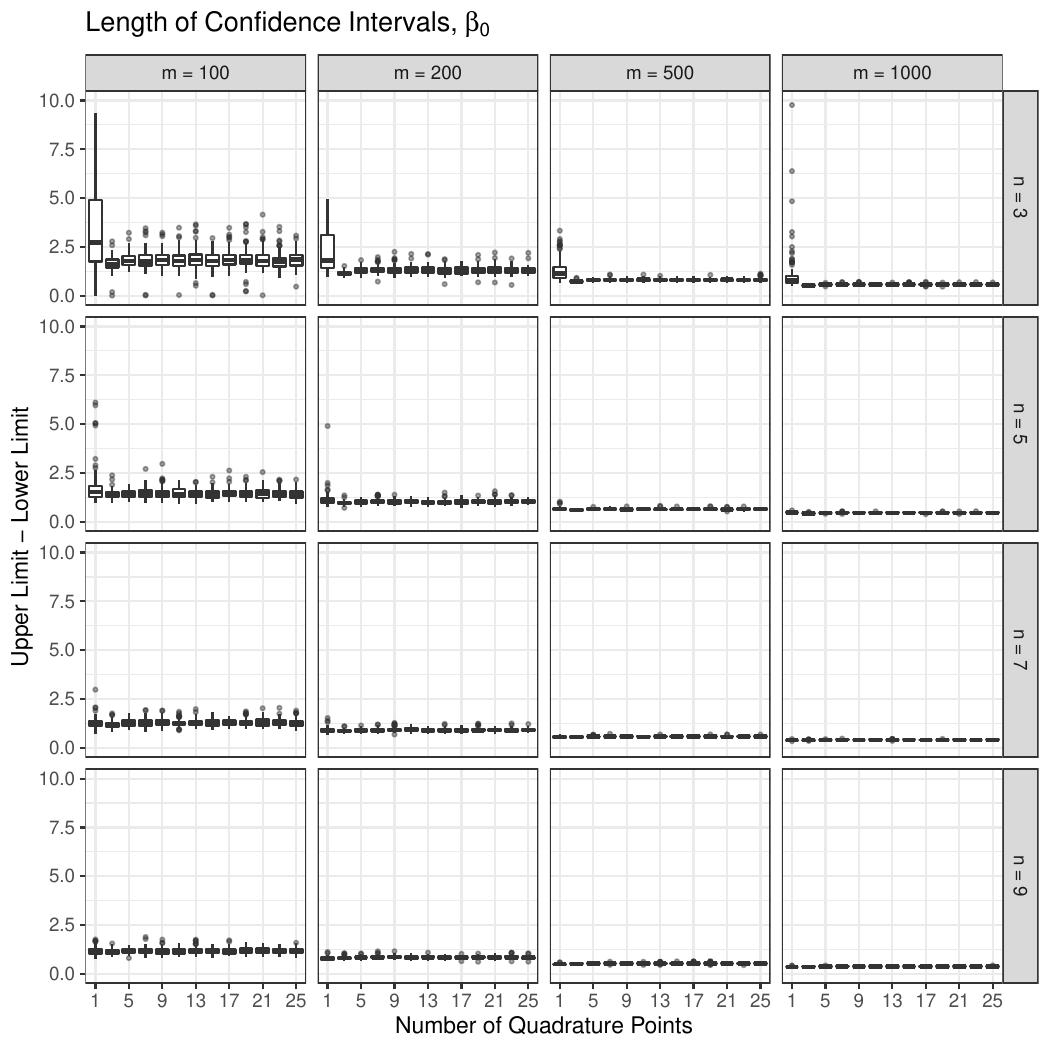}
\caption{Length of the Wald intervals for $\widehat{\beta}_{0}$ from the random intercepts model (\ref{eqn:simmodel2}),
based on $1000$ sets of simulated data. 
The Laplace ($\quadnum=1$) intervals are substantially wider than those for larger $\quadnum$.
}
\label{fig:beta0lengthscalar}
\end{figure}

\begin{figure}[p]
\centering
\includegraphics{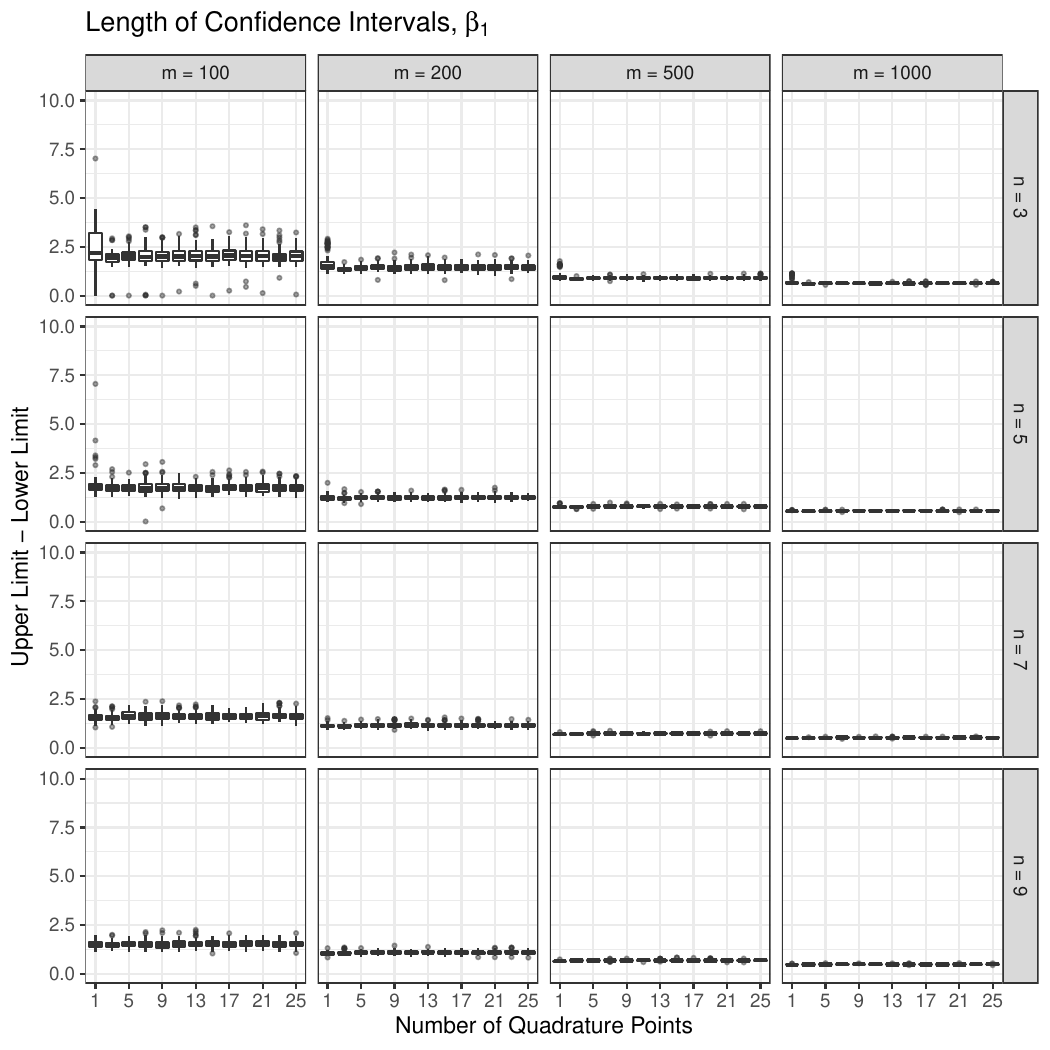}
\caption{Length of the Wald intervals for $\widehat{\beta}_{1}$ from the random intercepts model (\ref{eqn:simmodel2}),
based on $1000$ sets of simulated data. 
The Laplace ($\quadnum=1$) intervals are somewhat wider than those for larger $\quadnum$.
}
\label{fig:beta1lengthscalar}
\end{figure}

\begin{figure}[p]
\centering
\includegraphics{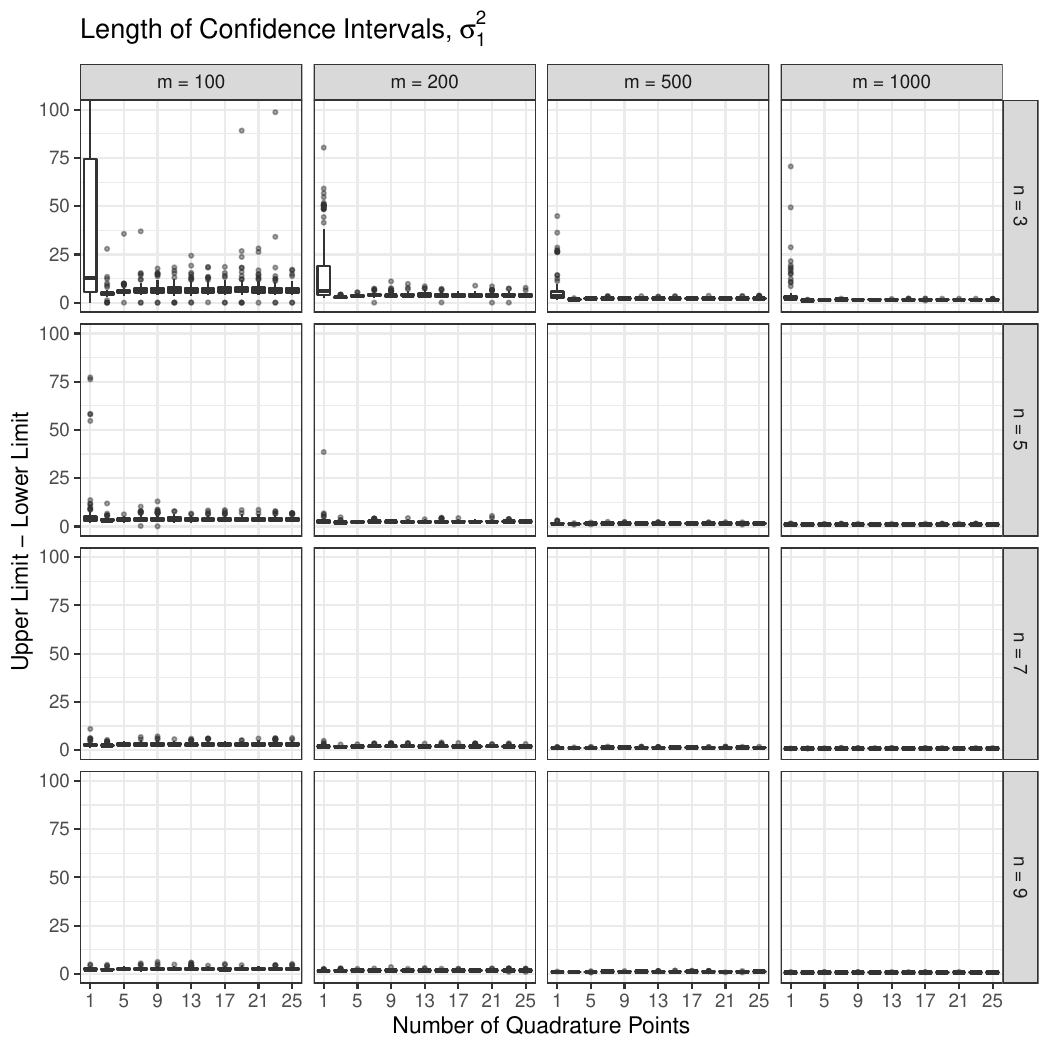}
\caption{Length of the Wald intervals for $\widehat{\sigma}^2_{1}$ from the random intercepts model (\ref{eqn:simmodel2}),
based on $1000$ sets of simulated data. 
The Laplace ($\quadnum=1$) intervals are orders of magnitude wider than those for larger $\quadnum$; the y-axis is zoomed out to capture this.}
\label{fig:sigmasqlengthscalar}
\end{figure}

\begin{figure}[p]
\centering
\includegraphics{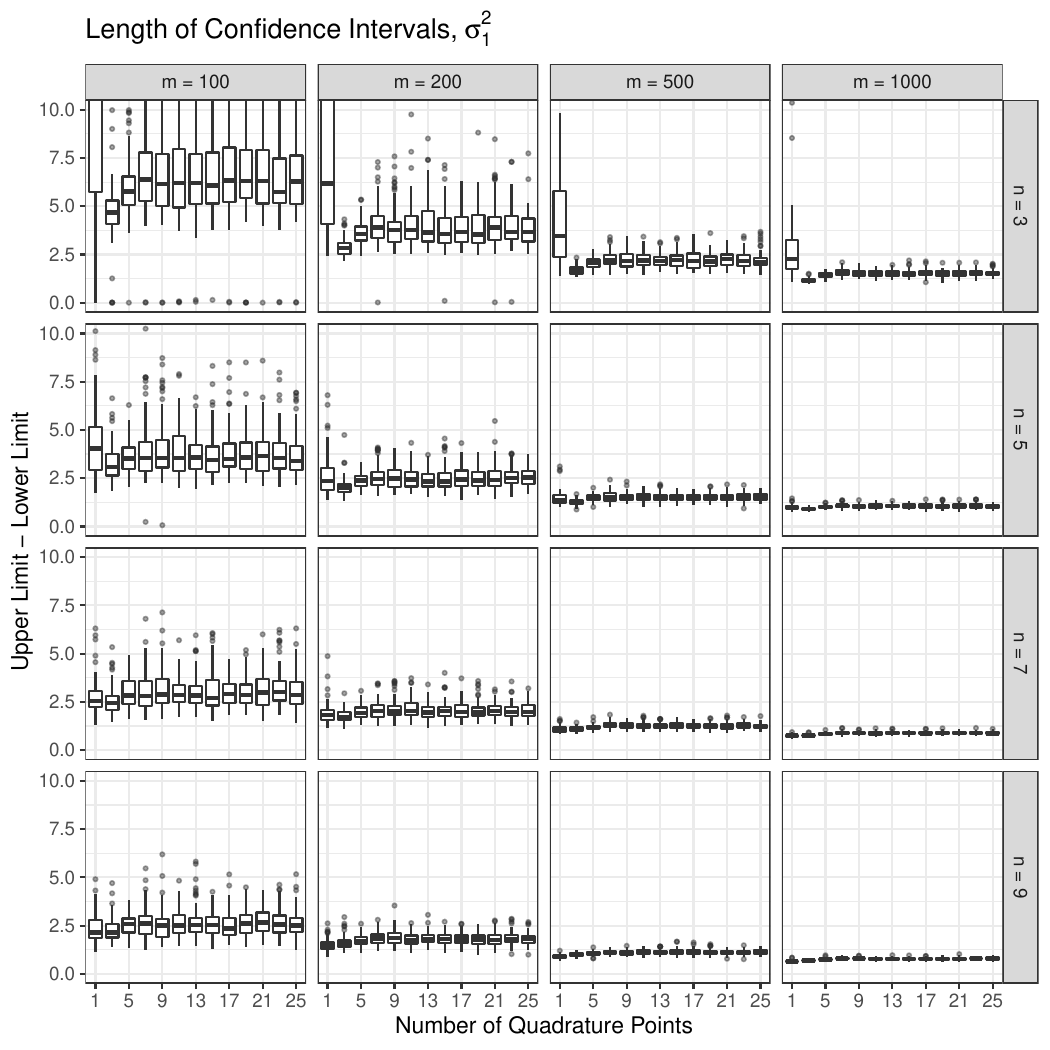}
\caption{Length of the Wald intervals for $\widehat{\sigma}^2_{1}$ from the random intercepts model (\ref{eqn:simmodel2}),
based on $1000$ sets of simulated data. 
The y-axis is zoomed in, which obscures some of the lengths of the Laplace ($\quadnum=1$) intervals.
}
\label{fig:sigmasqlengthzoomscalar}
\end{figure}

In this supplement only, we repeat Simulation 1 for the less interesting case of the scalar random effects model (\ref{eqn:simmodel2}).
Figures \ref{fig:beta0biasscalar} -- \ref{fig:sigmasqlengthzoomscalar} show the results. 
The only interesting point is that the coverage for the Laplace approximation appears to
get worse as $\numpergroup$ increases for $\beta_0$ and $\sigma$, despite the bias decreasing. 
This can be explained by looking at the plots of confidence interval lengths; for $\quadnum=1$, the lengths of the intervals
for $\beta_0$ and $\sigma$ get much smaller for larger $\numpergroup$, leading to worse coverage even with slightly less bias.
As with all such scenarios in all of the simulations here, the solution is to increase $\quadnum$.

\clearpage

\clearpage

\clearpage

\clearpage

%%% COVERAGE %%%

\clearpage

\clearpage

\clearpage

\clearpage

\clearpage

\clearpage

\clearpage

\section{Additional details for data analyses}\label{supp:dataanalysis}

\subsection{Smoking cessation}

We report parameter estimates and both absolute and relative run times for the smoking cessation data from Section 5.1.
The following model was fit with both the new procedure as well as \texttt{GLMMadaptive}:
\begin{equation}\label{eqn:smokingmodel}
    \response_{ij} \setdelim \re_i \indsim \text{Bern}(p_{ij}), \ \re_i \iidsim \text{N}\{\zero,\Varmat(\resd)\}, \ 
    \log\frac{p_{ij}}{1-p_{ij}} = \beta_0 + \beta_1\covi_i + \beta_2 t_j + \beta_3\covi_i t_j + \reidx_{i1} + \reidx_{i2}t_j.
\end{equation}
Here $\response_{ij}$ is a binary indicator where value $1$ indicates smoking cessation for subject $i$ at time $j$.
The covariate $x_i$ is a binary group indicator with value $0$ corresponding to no social support and value $1$ to
the subject receiving a type of social support.
The covariate $t_j$ is a normalized time value representing follow-up at $0$, $6$, $12$, and $24$ months post-intervention.
Some subjects had missing follow-ups.
There are hence $\numpergroup_i\in\{1,\ldots,4\}$ measurements on each of the $\numgroups=489$ subjects.
Note that this overall setup is nearly identical to model (\ref{eqn:simmodel1}) used in the simulations,
save for the varying number of measurements per subject.

\begin{figure}[p]
\centering
\includegraphics{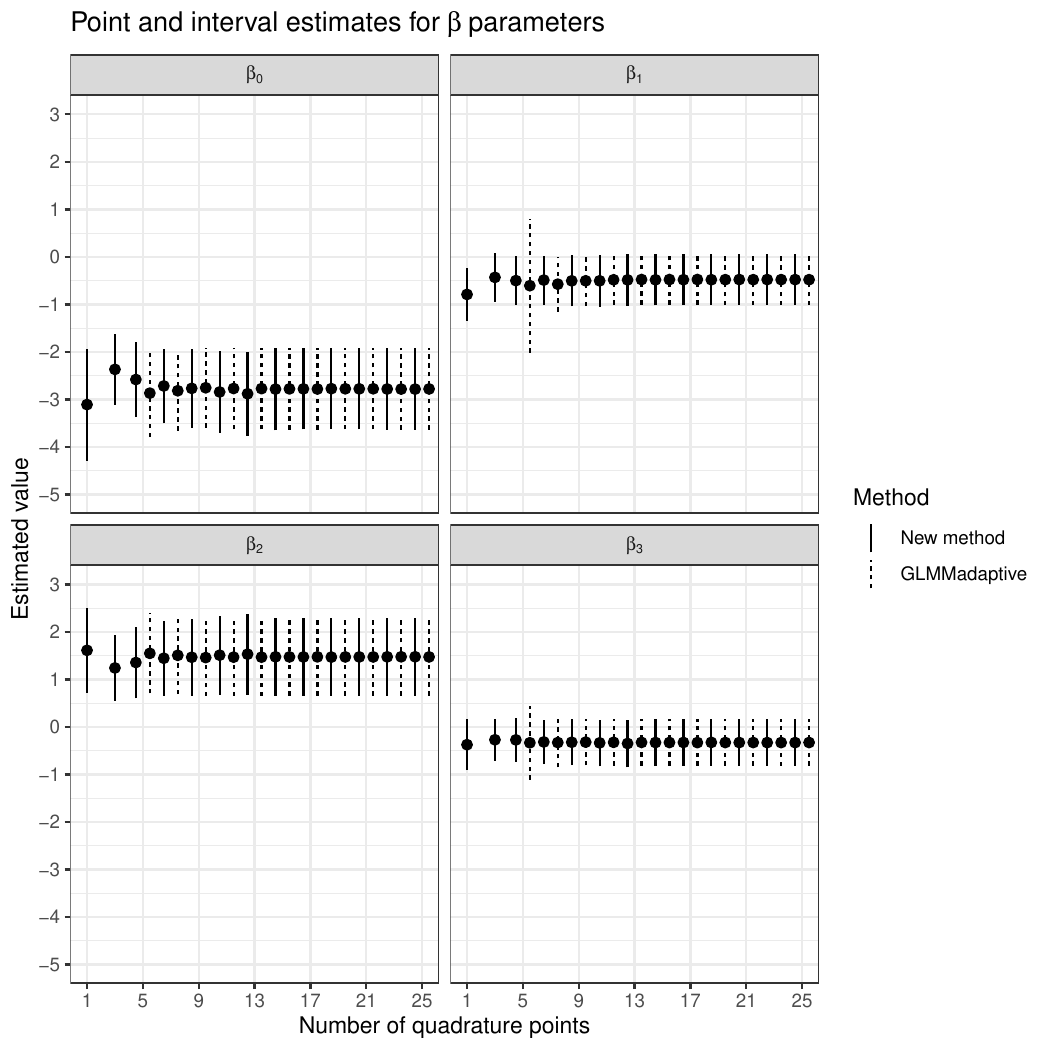}
\caption{
Point and interval estimates for regression coefficients in the smoking cessation data, for the new method and \texttt{GLMMadaptive}.
The new method successfully returns a positive-definite Hessian
matrix for all values of $\quadnum$; \texttt{GLMMadaptive} does so only for $\quadnum\geq7$ or so.
The two methods return comparable inferences at high enough values of $\quadnum$.
For all parameters, there is a clear point at which increasing $\quadnum$ stops changing the point and interval estimates.
}
\label{fig:smoking-beta}
\end{figure}

\begin{figure}[p]
\centering
\includegraphics{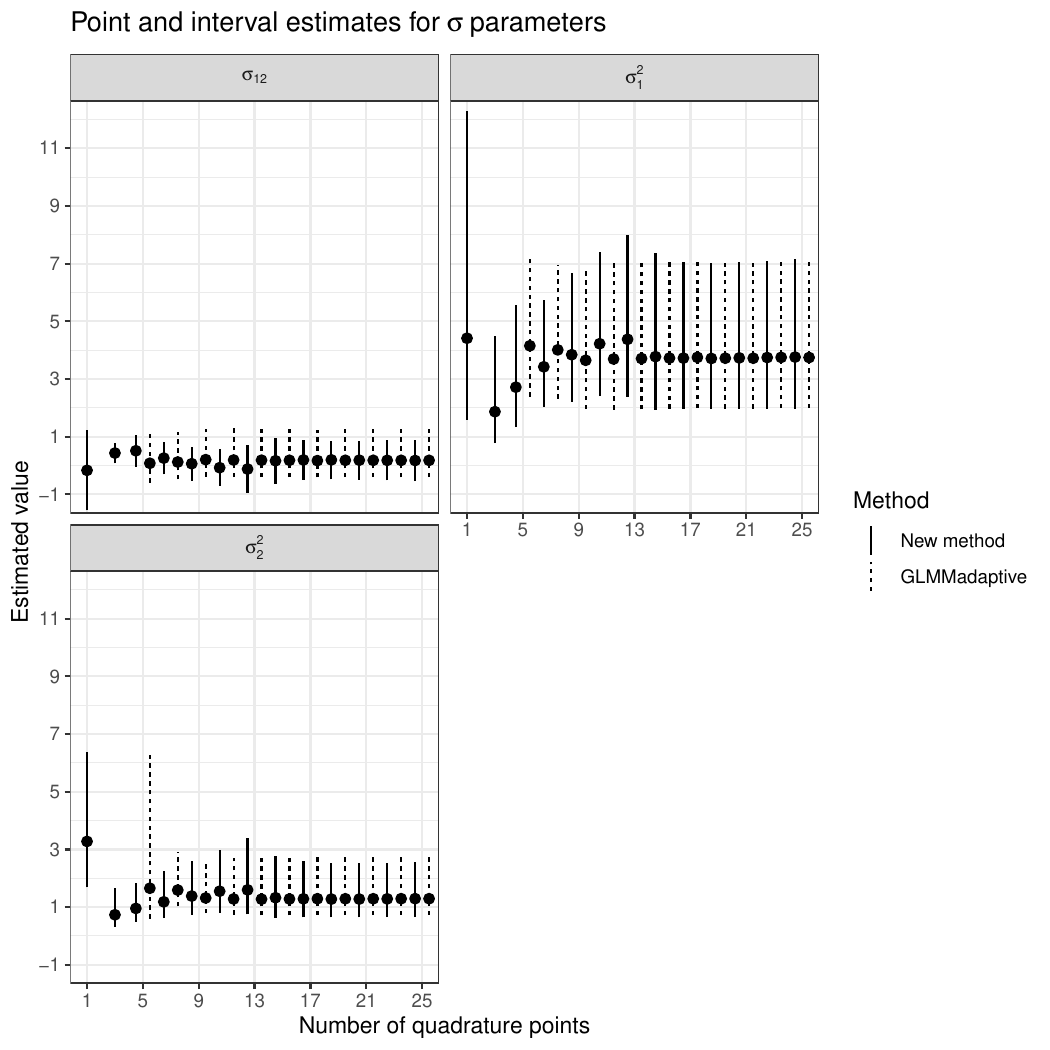}
\caption{
Point and interval estimates for variance components in the smoking cessation data, for the new method and \texttt{GLMMadaptive}.
The new method successfully returns a positive-definite Hessian
matrix for all values of $\quadnum$; \texttt{GLMMadaptive} does so only for $\quadnum\geq5$ or so.
The two methods return comparable inferences at high enough values of $\quadnum$.
For all parameters, there is a clear point at which increasing $\quadnum$ stops changing the point and interval estimates.
For $\sigma^2_2$ and $\sigma_{12}$, the final, stable intervals from the new method are narrower than those from \texttt{GLMMadaptive}.
}
\label{fig:smoking-sigma}
\end{figure}

\begin{figure}[p]
\centering
\includegraphics{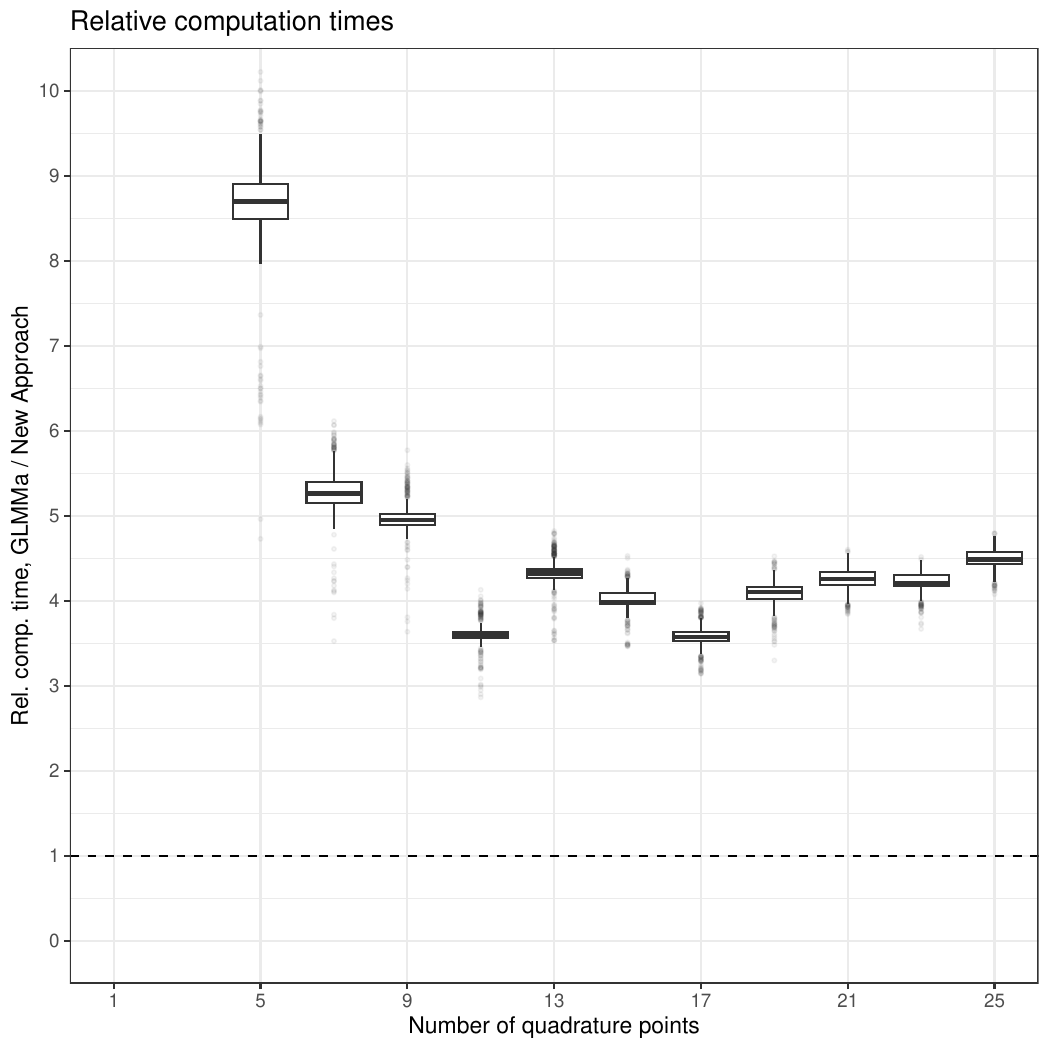}
\caption{
Relative computation times for the smoking cessation data, for the new method and \texttt{GLMMadaptive},
over $500$ repeated runs.
The new method achieves a speedup of between $3$ and $10$ times, which varies with $\quadnum$,
and is never slower than \texttt{GLMMadaptive}.
The times for $\quadnum = 1,3,5$ are not shown because \texttt{GLMMadaptive} did not return a usable
inferential result in these cases.
}
\label{fig:smoking-relcomptimes}
\end{figure}

\begin{figure}[p]
\centering
\includegraphics{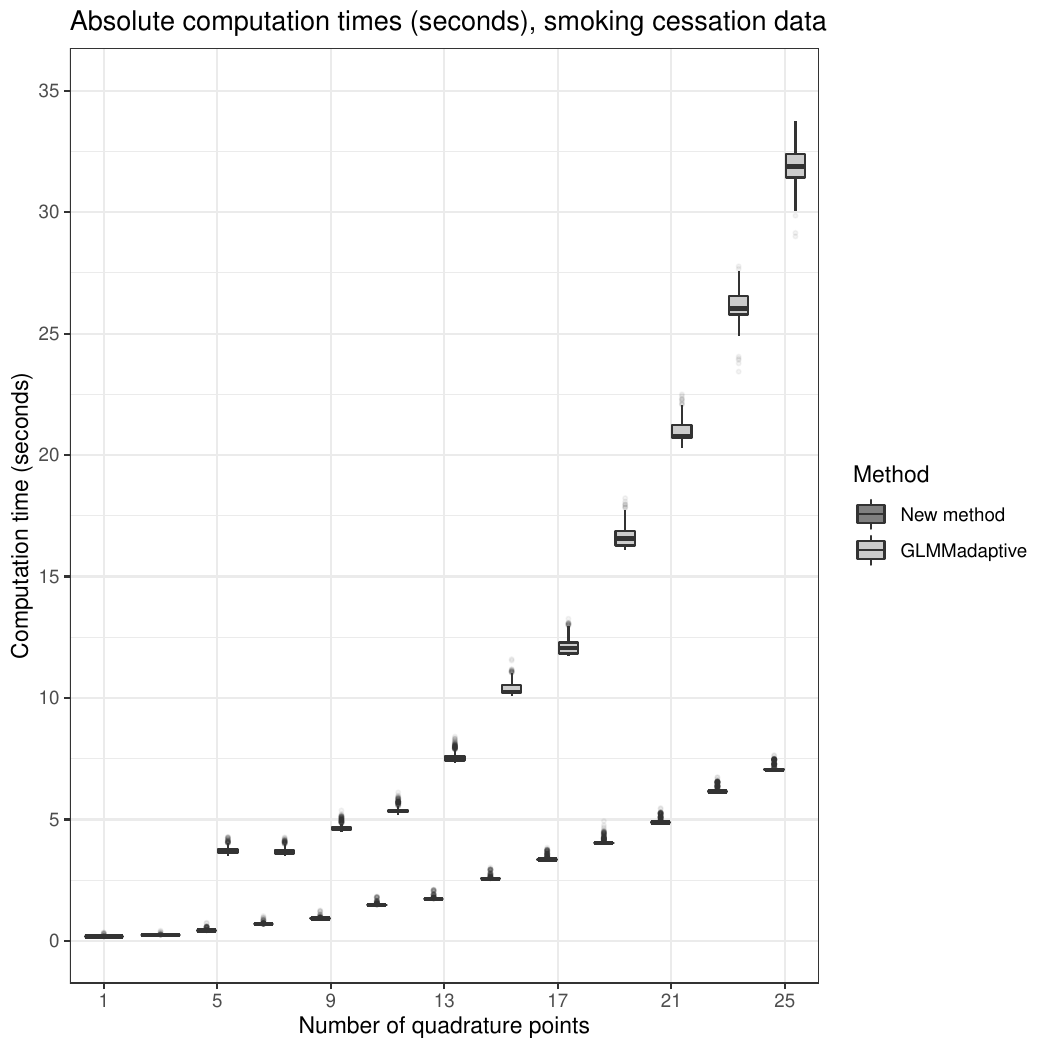}
\caption{
Absolute computation times in seconds for the smoking cessation data, for the new method and \texttt{GLMMadaptive},
over $500$ repeated runs.
Both methods generally take more time for larger $\quadnum$, which is expected.
All computations were performed on an M1 Mac book Pro laptop with 10 cores and $64$Gb of RAM.
}
\label{fig:smoking-abscomptimes}
\end{figure}

Figures \ref{fig:smoking-beta} -- \ref{fig:smoking-abscomptimes} show the parameter estimates and computation times
for the new approach and \texttt{GLMMadaptive}.
Selected such results are shown in the manuscript.
In all cases, inferences stop changing after $\quadnum$ is increased large enough.
The new approach yields comparable inferences to \texttt{GLMMadaptive} with the latter taking $3-10$ times
as long as the new method.
For $\sigma_{12}$, the Wald intervals (Figure \ref{fig:smoking-sigma}) produced by the new method are narrower than those from \texttt{GLMMadaptive};
combined with the nominal coverage for these intervals for this parameter shown in simulations (Figure \ref{fig:sigmacov1covr}),
this may be regarded as an advantage of the new approach.

\clearpage

\clearpage

\clearpage

\clearpage

\subsection{Toenail fungus treatment}

We report parameter estimates and both absolute and relative run times for the toenail fungus treatment data from Section 5.2.
The following model was fit with the new procedure as well as \texttt{GLMMadaptive} and \texttt{lme4}:
\begin{equation}\label{eqn:smokingmodel}
    \response_{ij} \setdelim \reidx_i \indsim \text{Bern}(p_{ij}), \ \reidx_i \iidsim \text{N}(0,\sigma^2), \ 
    \log\frac{p_{ij}}{1-p_{ij}} = \beta_0 + \beta_1\covi_i + \beta_2 t_j + \beta_3\covi_i t_j + \reidx_{i1}.
\end{equation}
Here $\response_{ij}$ is a binary indicator with $\response_{ij}=1$ indicating the absence of toenail
infection. 
There are $i=1,\ldots,\numgroups=294$ subjects who were given an oral treatment for toenail infection, $\covi_{i}=1$, or not, $\covi_i=0$.
The subjects were followed up at $j=1,\ldots,\numpergroup_j\in\{1,\ldots,7\}$ times, $t_j = -3,\ldots,3$.

\begin{figure}[p]
\centering
\includegraphics{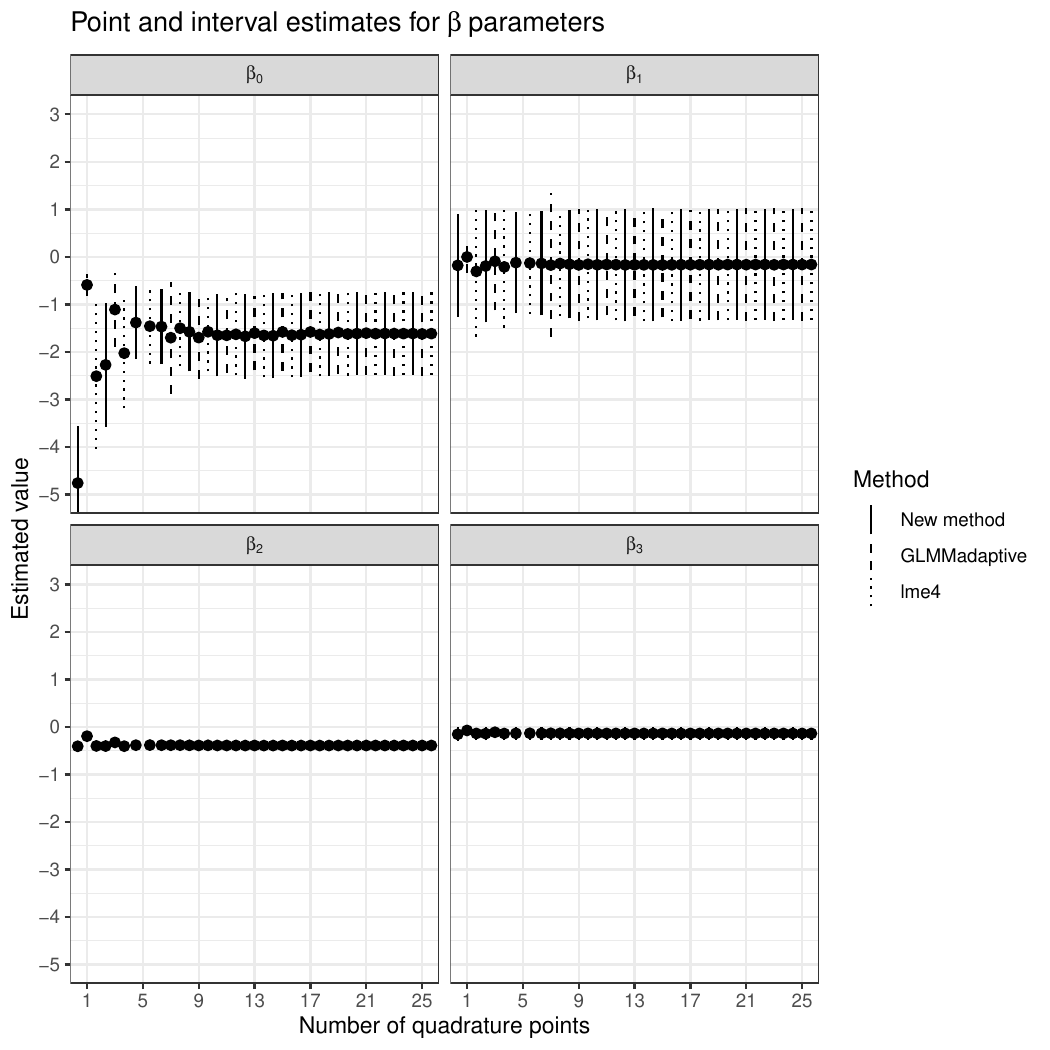}
\caption{
Point and interval estimates for regression coefficients in the toenail fungus treatment data, for the new method, \texttt{lme4}, and \texttt{GLMMadaptive}.
The methods return comparable inferences at high enough values of $\quadnum$.
For all parameters, there is a clear point at which increasing $\quadnum$ stops changing the point and interval estimates.
}
\label{fig:toenail-beta}
\end{figure}

\begin{figure}[p]
\centering
\includegraphics{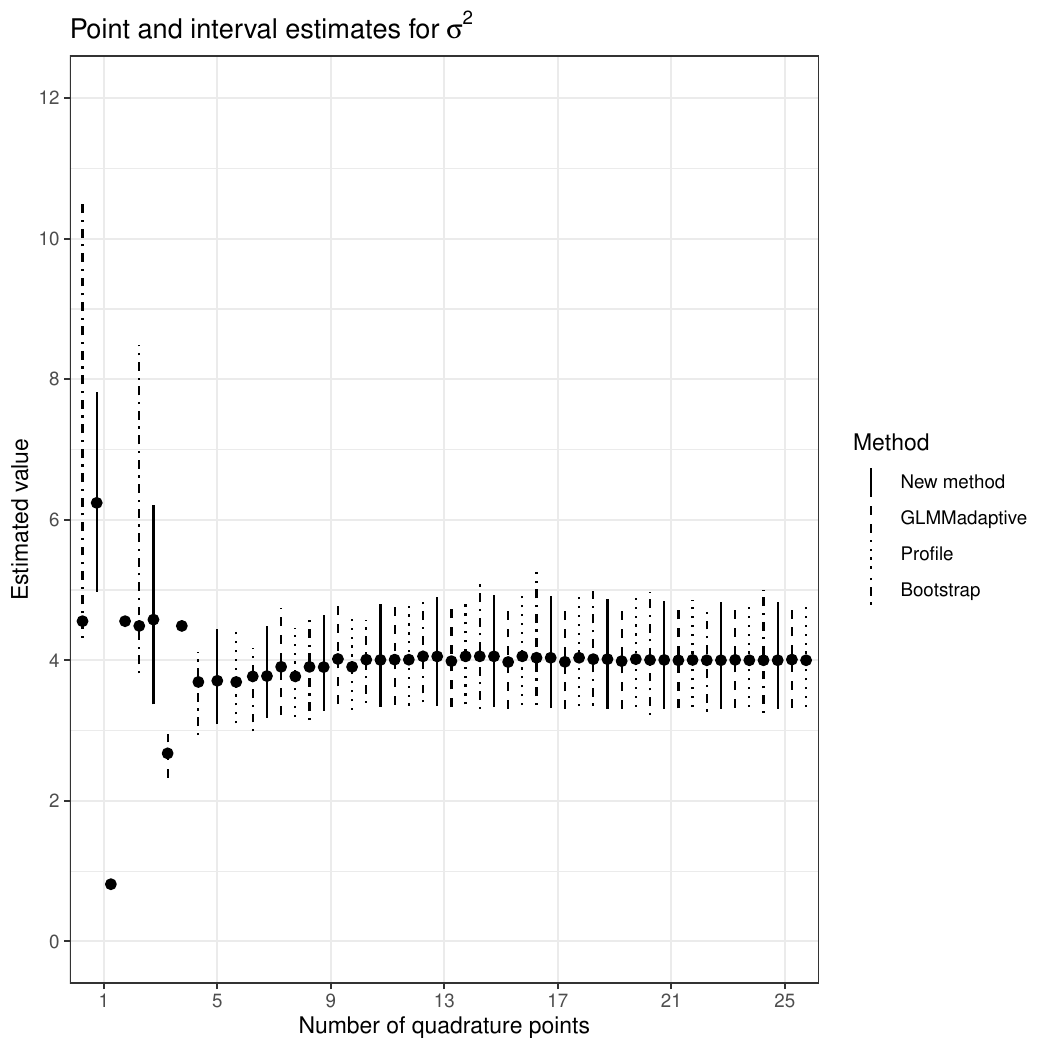}
\caption{
Point and interval estimates for $\sigma$ in the toenail fungus treatment data, for the new method, \texttt{GLMMadaptive}, and \texttt{lme4}
with both profile likelihood and bootstrap confidence intervals.
The methods return comparable inferences at high enough values of $\quadnum$.
For all parameters, there is a clear point at which increasing $\quadnum$ stops changing the point and interval estimates.
}
\label{fig:toenail-sigma}
\end{figure}

\begin{figure}[p]
\centering
\includegraphics{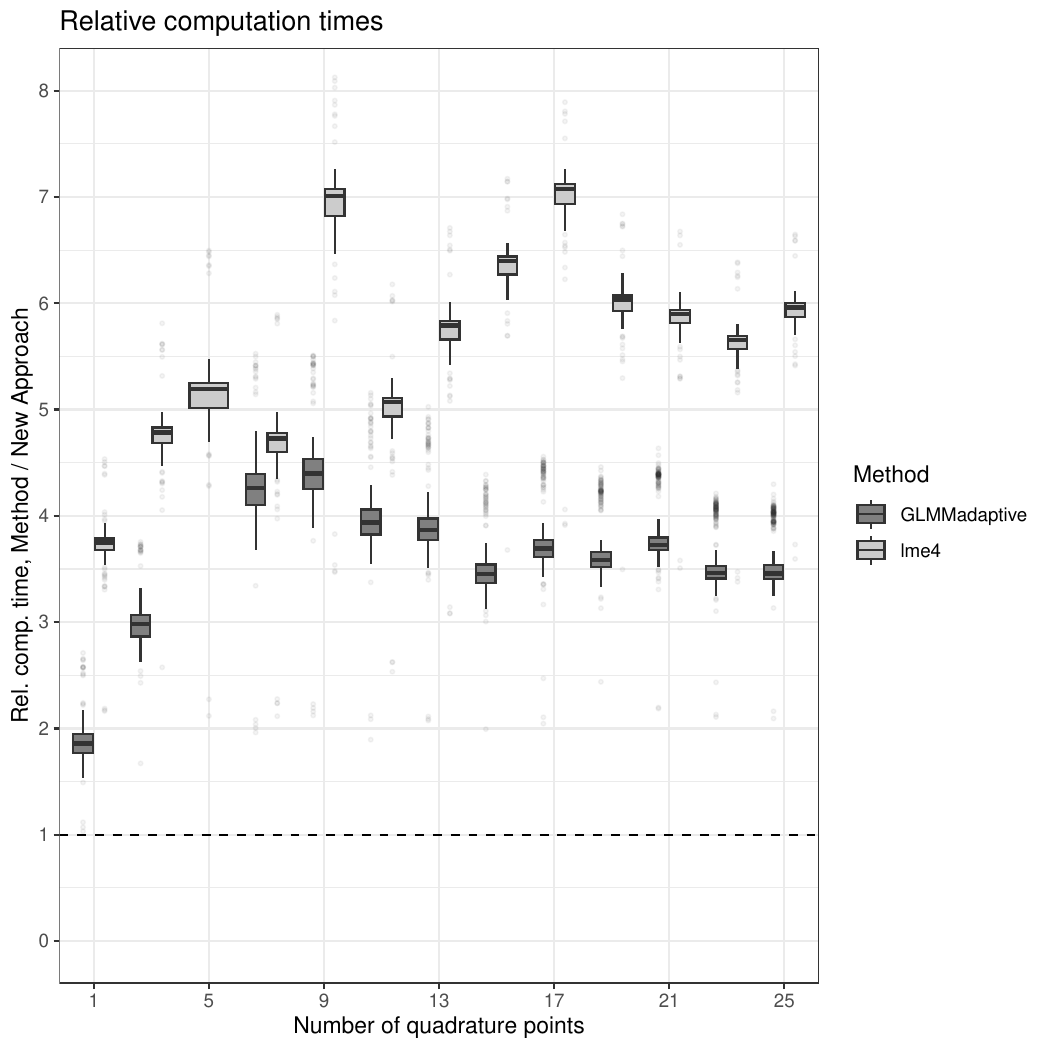}
\caption{
Relative computation times for the toenail fungus treatment data, for \texttt{GLMMadaptive} and \texttt{lme4} compared to the new method,
over $500$ repeated runs.
The new method achieves a speedup of roughyl between $2-7$ times, which varies with $\quadnum$.
}
\label{fig:toenail-relcomptimes}
\end{figure}

\begin{figure}[p]
\centering
\includegraphics{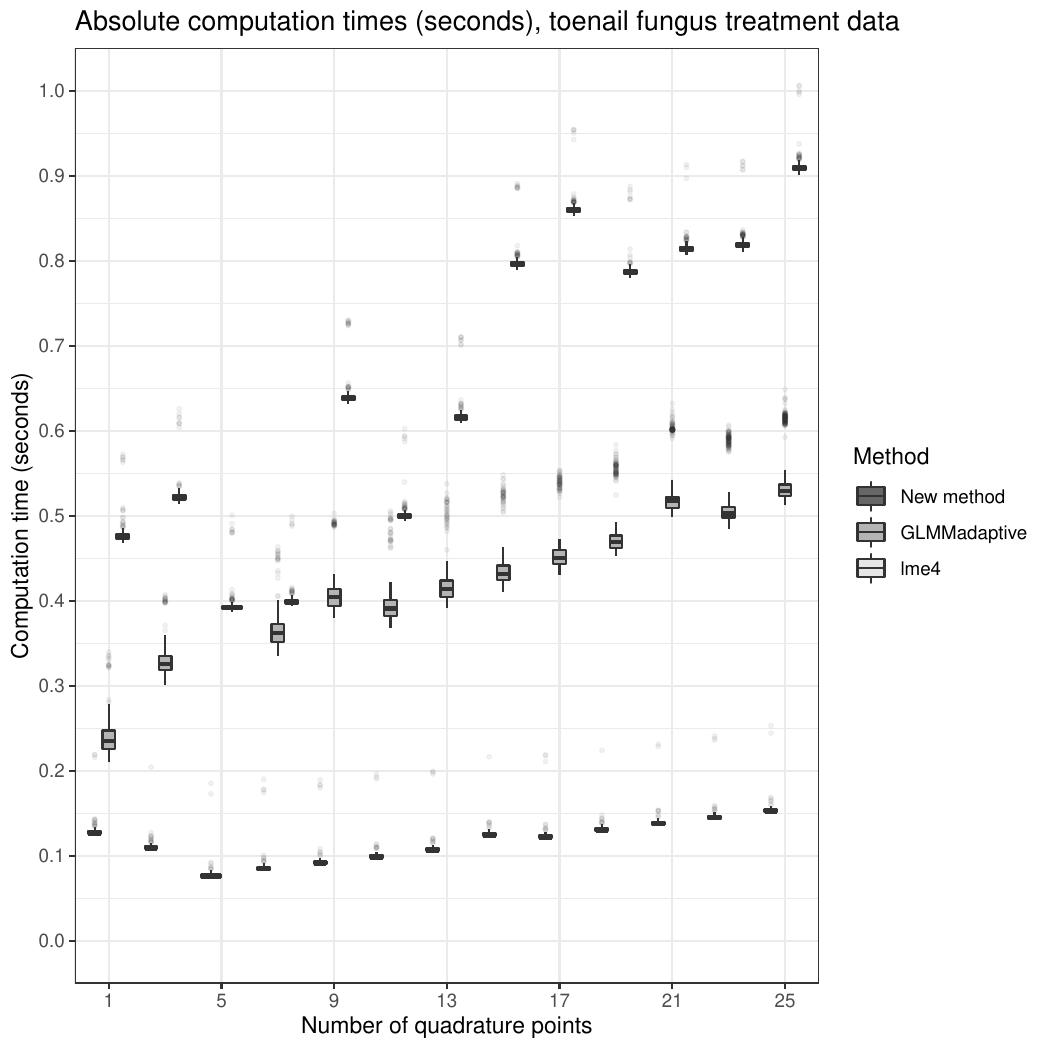}
\caption{
Absolute computation times in seconds for the toenail fungus treatment data, for the new method, \texttt{GLMMadaptive}, and \texttt{lme4}
over $500$ repeated runs.
Both methods generally take more time for larger $\quadnum$, which is expected.
All computations were performed on an M1 Mac book Pro laptop with 10 cores and $64$Gb of RAM.
}
\label{fig:toenail-abscomptimes}
\end{figure}

Figures \ref{fig:toenail-beta} -- \ref{fig:toenail-abscomptimes} show the parameter estimates and computation times
for the new approach, \texttt{GLMMadaptive}, and \texttt{lme4}.
Selected such results are shown in the manuscript.
In all cases, inferences stop changing after $\quadnum$ is increased large enough.
The new approach yields comparable inferences to \texttt{GLMMadaptive} and \texttt{lme4}; 
the former takes $1-5$ times as long as the new method and the latter takes $3-8$ times as long.
Further, the computation times for \texttt{lme4} do \emph{not} include the time taken to produce
confidence intervals for $\sigma$, because \texttt{lme4} does not provide Wald confidence intervals for $\sigma$.
Instead, it offers computationally intensive profile likelihood and bootstrap confidence intervals for $\sigma$.
These are shown in Figure \ref{fig:toenail-sigma}, and for larger $\quadnum$ are very close to the Wald intervals
provided by the new approach and \texttt{GLMMadaptive}.
In a single run, the profile intervals took roughly $60$ times longer to compute than the Wald intervals
from the new method, with slight variability across values of $\quadnum$.
The bootstrap approach requires the user to choose a number of simulated data sets to use, with $500$ the
recommended default. Using $200$ repetitions only, a single run of the procedure took between $1300-1600$
times as long as the new approach, depending on $\quadnum$, and produced very similar intervals.
We remind the reader that the coverages of the Wald intervals for this parameter are found to be
nominal in the simulations reported in Figure \ref{fig:sigmasqcovrscalar}.

\clearpage

\clearpage

\clearpage

\end{document}